\pdfoutput=1 
\documentclass[cernpreprint,english]{na61doc}

\usepackage{lineno}
\usepackage{mathptmx} 
\usepackage{enumerate}
\usepackage{amsmath}
\usepackage{amsmath}
\usepackage{placeins}
\usepackage{appendix}
\usepackage[T1]{fontenc}
\usepackage[utf8]{inputenc}
\usepackage{chngcntr}
\usepackage{microtype}
\usepackage{lineno}
\usepackage{color}
\usepackage{epstopdf}
\usepackage{colortbl}
\usepackage{tabularx}
\usepackage{multirow}

\usepackage{subcaption}
\usepackage{booktabs}
\usepackage{url}
\usepackage{cite}
\usepackage{caption}

\newcommand{\eV}{\ensuremath{\mbox{e\kern-0.1em V}}\xspace}
\newcommand{\GeV}{\ensuremath{\mbox{Ge\kern-0.1em V}}\xspace}
\newcommand{\MeV}{\ensuremath{\mbox{Me\kern-0.1em V}}\xspace}
\newcommand{\GeVc}{\ensuremath{\mbox{Ge\kern-0.1em V}\!/\!c}\xspace}
\newcommand{\GeVcc}{\ensuremath{\mbox{Ge\kern-0.1em V}\!/\!c^2}\xspace}
\newcommand{\AGeV}{\ensuremath{A\,\mbox{Ge\kern-0.1em V}}\xspace}
\newcommand{\AGeVc}{\ensuremath{A\,\mbox{Ge\kern-0.1em V}\!/\!c}\xspace}
\newcommand{\MeVc}{\ensuremath{\mbox{Me\kern-0.1em V}/c}\xspace}

\newcommand{\cm}{\ensuremath{\mbox{cm}}\xspace}

\newcommand{\y}{\ensuremath{y}\xspace}
\newcommand{\p}{\ensuremath{p}\xspace}
\newcommand{\pt}{\ensuremath{p_{T}}\xspace}
\newcommand{\plab}{\ensuremath{p_\text{lab}}\xspace}
\newcommand{\snn}{$\sqrt{s_{NN}}$}

\newcommand{\Urqmd}{{\scshape U}r{\scshape qmd}\xspace}

\newcommand{\Geant}{{\scshape Geant}\xspace}
\newcommand{\GeantThree}{{\scshape Geant3}\xspace}

\newcommand{\Epos}{{\scshape Epos}\xspace}
\newcommand{\EposLong}{{\scshape Epos1.99}\xspace}

\newcommand{\Ampt}{{\scshape Ampt}\xspace}

\newcommand{\Hijing}{{\scshape Hijing}\xspace}
\newcommand{\Phsd}{{\scshape Phsd}\xspace}
\newcommand{\Smash}{{\scshape Smash}\xspace}

\newcommand{\CernVM}{\textsc{Cern\-\kern-0.05emVM}\xspace}
\newcommand{\pp}{\mbox{\textit{p+p}}\xspace}

\newcommand{\dEdx}{$\text{d}E/\text{d}x$ }

\newcommand{\coordinate}[1]{{\fontfamily{lmss}\selectfont#1}}

\definecolor{darkred}{rgb}{0.5,0,0}
\definecolor{darkblue}{rgb}{0,0,0.5}
\definecolor{firebrick}{rgb}{0.75,0.125,0.125}
\definecolor{darkgreen}{rgb}{0,0.5,0}
\definecolor{kPink+2}{RGB}{204,102,153}
\definecolor{kOrange+8}{RGB}{255,102,51}
\definecolor{kGreen+2}{RGB}{0,153,0}
\definecolor{kCyan+2}{RGB}{0,153,153}
\definecolor{kBlue+2}{RGB}{0,0,153}
\definecolor{kRed+1}{RGB}{204,0,0}
\definecolor{kBlue}{RGB}{0,0,204}
\definecolor{kBlue-9}{RGB}{153,153,255}
\definecolor{kGreen}{RGB}{0,153,0}
\definecolor{kRed}{RGB}{204,0,0}
\definecolor{kCyan}{RGB}{51,204,204}
\definecolor{kMagenta}{RGB}{153,0,153}
\definecolor{kPink}{RGB}{204,0,102}
\definecolor{kGray}{RGB}{204,204,204}
\definecolor{kBlack}{RGB}{0,0,0}

\definecolor{kRed+3}{RGB}{102,0,0}
\definecolor{kRed+2}{RGB}{153,0,0}
\definecolor{kRed-4}{RGB}{255,51,51}
\definecolor{kRed-7}{RGB}{255,102,102}
\definecolor{kRed-9}{RGB}{255,153,153}

\ShineTitle{
Measurements of $\pi^\pm$, $K^\pm$, $p$ and $\bar{p}$ spectra in 
$^7$Be+$^9$Be collisions at beam momenta from 19$A$ to 150$A$~\GeVc 
with the \NASixtyOne spectrometer at the CERN SPS
}

\PreprintIdNumber{CERN-EP-2020-187}

\ShineJournal{Eur. Phys. J. C}
\ShineAbstract{ 
The \NASixtyOne experiment at the CERN Super Proton Synchrotron (SPS) studies the onset
of deconfinement in hadron matter by a scan of particle production in collisions of nuclei
with various sizes at a set of energies covering the SPS energy range. This paper presents results
on inclusive double-differential spectra, transverse momentum and rapidity distributions and mean multiplicities of 
$\pi^\pm$, $K^\pm$, \textit{p} and $\bar{p}$ produced in the 20\% most \textit{central} 
$^7$Be+$^9$Be collisions at beam momenta of 19$A$, 30$A$,
40$A$, 75$A$ and 150$A$~\GeVc.
The energy dependence of the $K^\pm$/$\pi^\pm$ ratios as well as of inverse slope parameters of the $K^\pm$ 
transverse mass distributions are close to those found in inelastic \textit{p+p} reactions. The new results 
are compared to the world data on \textit{p+p} and Pb+Pb collisions as well as to predictions of the \Epos, \Urqmd, \Ampt, \Phsd and \Smash models.
}
\begin{document}

\maketitle


\newpage 
{\Large The \NASixtyOne Collaboration}
\bigskip

\noindent
A.~Acharya$^{\,9}$,
H.~Adhikary$^{\,9}$,
A.~Aduszkiewicz$^{\,15}$,
K.K.~Allison$^{\,25}$,
E.V.~Andronov$^{\,21}$,
T.~Anti\'ci\'c$^{\,3}$,
V.~Babkin$^{\,19}$,
M.~Baszczyk$^{\,13}$,
S.~Bhosale$^{\,10}$,
A.~Blondel$^{\,4}$
% ...


\noindent
% ...
$^{3}$~Ru{\dj}er Bo\v{s}kovi\'c Institute, Zagreb, Croatia\\
$^{4}$~LPNHE, University of Paris VI and VII, Paris, France\\
% ...
$^{9}$~Jan Kochanowski University in Kielce, Poland\\
$^{10}$~Institute of Nuclear Physics, Polish Academy of Sciences, Cracow, Poland\\
% ...
$^{13}$~AGH - University of Science and Technology, Cracow, Poland\\
% ...
$^{15}$~University of Warsaw, Warsaw, Poland\\
% ...
$^{19}$~Joint Institute for Nuclear Research, Dubna, Russia\\
% ...
$^{21}$~St. Petersburg State University, St. Petersburg, Russia\\
% ...
$^{25}$~University of Colorado, Boulder, USA\\
% ...


\section{Introduction}

This paper presents experimental results on inclusive spectra and mean multiplicities of
 $\pi^\pm, K^\pm, p$ and $\bar{p}$  produced in the 20\% most \textit{central} $^7$Be+$^9$Be collisions
at beam momenta of 19$A$, 30$A$, 40$A$, 75$A$ and 150$A$~\GeVc (\snn = 6.1, 7.6, 8.8, 11.9 and 16.8 GeV).
These studies form part of the strong interactions programme of \NASixtyOne~\cite{Antoniou:2006mh}
investigating the properties of the onset of deconfinement and searching for the possible existence
of a critical point. This requires a two dimensional scan 
in collision energy and nuclear mass number of the colliding nuclei. Such a scan allows to explore systematically the phase diagram of strongly interacting matter~\cite{Antoniou:2006mh}. An increase of collision energy causes an increase of temperature and a decrease of baryon chemical potential of strongly
interacting matter at freeze-out
, whereas increasing the nuclear mass number of the colliding nuclei  decreases the temperature~\cite{Gazdzicki:2014sva}.

Pursuing this programme \NASixtyOne recorded data on \textit{p+p}, Be+Be, Ar+Sc, Xe+La and Pb+Pb collisions.
Moreover, further measurements of Pb+Pb interactions are planned with an upgraded detector ~\cite{PbAddendum} 
starting in 2021. 

The $^7$Be+$^9$Be collisions (see Ref.~\cite{Acharya:2020cyb} for results on $\pi^-$
production) play a special role in the \NASixtyOne scan programme. 
First, it was predicted within the statistical models~\cite{Poberezhnyuk:2015wea,Motornenko:2018gdc} 
that the yield ratio of strange hadrons to pions in these collisions should be close to those in central
Pb+Pb collisions and significantly higher than in \textit{p+p} interactions.
Second, the collision system composed of a $^7$Be and a $^9$Be nucleus has eight protons and eight neutrons, 
and thus is isospin symmetric. Within the \NASixtyOne scan programme the $^7$Be+$^9$Be collisions serve as 
the lowest mass isospin symmetric reference needed to study collisions of medium and large mass nuclei. 
This is of particular importance when data on proton-proton, neutron-proton and neutron-neutron 
are not available to construct the nucleon-nucleon reference~\cite{Gazdzicki:1991ih}.

The paper is organized as follows: after this introduction the experiment is briefly presented
in Sec.~2. The analysis procedure, as well as statistical and systematic uncertainties are
discussed in Sec.~3. Section~4 presents experimental results and compares them with measurements
of \NASixtyOne in inelastic \pp interactions~\cite{Abgrall:2013pp_pim,Aduszkiewicz:2017sei,Aduszkiewicz:2019zsv} and 
NA49 in Pb+Pb collisions~\cite{Afanasiev:2002mx,Alt:2007aa}. Section~5 discusses model
predictions. A summary in Sec.~6 closes the paper. 

The following variables and definitions are used in this paper. The particle rapidity $y$ is calculated
in the collision center of mass system (cms), $y=0.5 \cdot ln{[(E+p_{L})/(E-p_{L})]}$, where $E$
and $p_{L}$ are the particle energy and longitudinal momentum, respectively. The transverse component
of the momentum is denoted as $p_{T}$ and the transverse mass $m_{T}$ is defined as
$m_{T} = \sqrt{m^2 + (cp_{T})^2}$
where $m$ is the particle mass in GeV. The momentum in the laboratory frame is denoted $p_\text{lab}$ and the
collision energy per nucleon pair in the center of mass by \snn.

Results of the measurements correspond to collisions with low energy emitted into the forward beam spectator 
region. For $^7$Be+$^9$Be collisions this energy is not tightly correlated with geometric parameters of the interaction 
such as the collision impact parameter of the collision (see Sec.~\ref{sec:centrality}).
This is caused by the small number of nucleons and the cluster structure of the Be nucleus.
Nevertheless, following the convention widely used in the analysis of nucleus-nucleus collisions,
the term \textit{central} is used for events selected by imposing an upper limit on this energy.

\FloatBarrier
\section{Experimental setup of \NASixtyOne}

\subsection{Detector}

\begin{figure}[h] 
        \centering
        \includegraphics[width=1\linewidth]{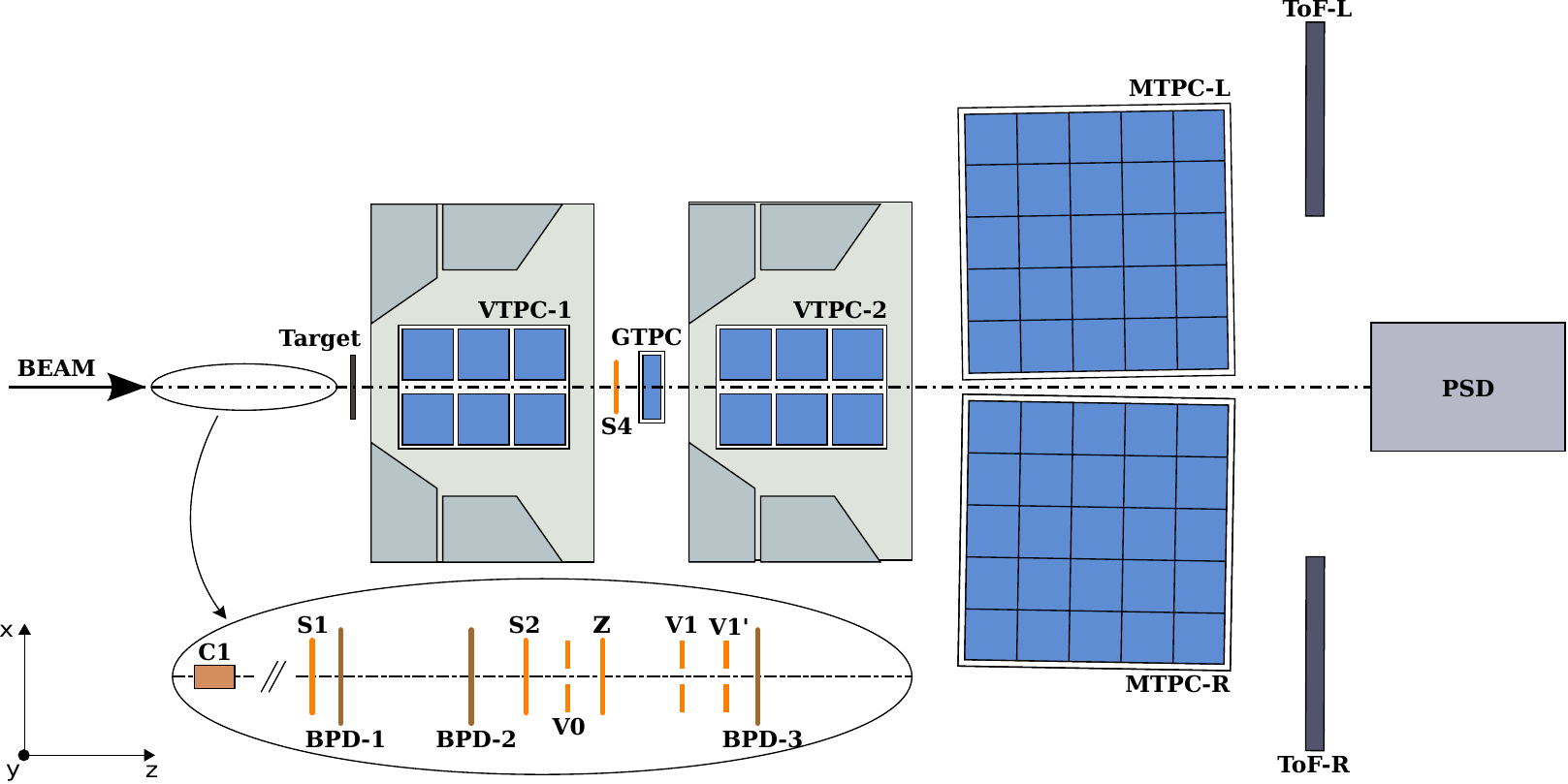}
        \caption{The schematic layout of the \NASixtyOne experiment at the CERN SPS ~\cite{Abgrall:2014fa}
                 showing the components used for the Be+Be energy scan (horizontal cut, not to scale).
                 The beam instrumentation is sketched in the inset (see also Fig.~\ref{fig:beamAndTriggerDetectors}).
                 Alignment of the chosen coordinate system as shown in the figure:
                 its origin lies in the middle of VTPC-2, on the beam axis.
                 The nominal beam direction is along the \coordinate{z}-axis. The magnetic field
                 bends charged particle trajectories in the \coordinate{x}--\coordinate{z} (horizontal) plane.
                 The drift direction in the TPCs is along the \coordinate{y} (vertical) axis.}
        \label{fig:detectorSetup}
\end{figure}

The \NASixtyOne experiment is a multi-purpose facility designed to measure particle production in
nucleus+nucleus, hadron+nucleus and \textit{p+p} interactions~\cite{Abgrall:2014fa}. The detector
is situated at the CERN Super Proton Synchrotron (SPS) in the H2 beamline of the North experimental area.
A schematic diagram of the setup is shown in Fig.~\ref{fig:detectorSetup}.
The main components of the produced particle detection system are four large volume
Time Projection Chambers (TPC). Two of them, called Vertex TPCs (VTPC), are located
downstream of the target inside superconducting magnets with maximum combined bending power of 9~Tm.
The magnetic field was scaled down in proportion to the beam momentum in order to obtain
similar phase space acceptance at all energies. The main TPCs (MTPC) and two walls of
pixel Time-of-Flight (ToF-L/R) detectors are placed symmetrically to the beamline downstream of the magnets.
The fifth small TPC (GAP-TPC) is placed between VTPC1 and VTPC2 directly on the beam line.
The TPCs are filled with Ar:CO$_{2}$ gas mixtures in proportions 90:10 for the VTPCs and the GAP-TPC, 
and 95:5 for the MTPCs. 

The Projectile Spectator Detector (PSD), which measures mainly the energy in the forward region 
of projectile spectators, is positioned 20.5 m (16.7 m) downstream of the target during measurements at 75$A$ and 150\AGeVc
(19$A$, 30$A$, 40\AGeVc) centered in the transverse plane on the position of the deflected beam. 
The PSD is used as a part of the trigger system (see Sec.\ref{subsec:trigger})
to accept collisions by imposing an upper limit on the energy measured in the
16 central modules and also to select \textit{central} events in analysis procedure (see Sec.\ref{sec:centrality}).

\begin{figure}[!ht]
\centering
\includegraphics[width=\textwidth]{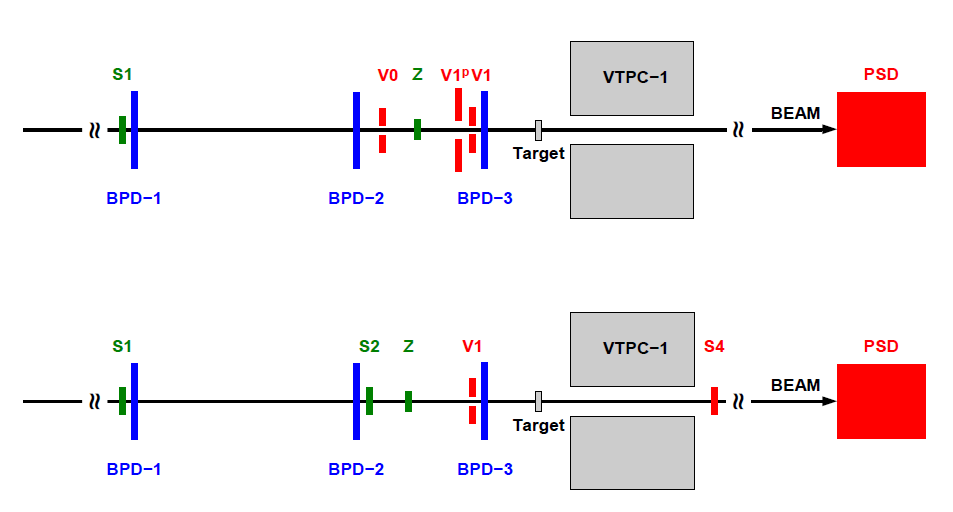}
\caption{The schematic of the placement of the beam and trigger detectors in high-momentum \emph{(top)} and 
         low-momentum \emph{(bottom)} data taking configurations showing beam counters S,
         veto counters V and beam position and charge detectors BPD, as well as a Cerenkov detector Z. Note, that the PSD calorimeter was almost 4 m closer for low momentum data taking.}
\label{fig:beamAndTriggerDetectors}
\end{figure}

The beamline instrumentation is schematically depicted in Fig.~\ref{fig:beamAndTriggerDetectors}.
It is designed for obtaining high beam purity with secondary ion beams~\cite{Abgrall:7Bebeam}
produced by fragmenting the primary Pb$^{82+}$ ions extracted from the SPS.
A detailed discussion of the properties of selected $^7$Be ions can be found in Ref.~\cite{Acharya:2020cyb}.

A set of scintillation counters as well as beam position detectors (BPDs) upstream of the spectrometer
provide timing reference, selection, identification and precise measurement of the position
and direction of individual beam particles.

The target was a plate of $^9$Be of 12 mm thickness placed $\approx$ 80 cm upstream of VTPC1. Mass concentrations
of impurities in the target were measured at 0.3\%, resulting in an estimated increase of the produced pion multiplicity by less than 0.5\% due to the small admixture of heavier elements~\cite{Banas:2018sak}. No correction was applied for this negligible contamination. Data were taken with target inserted (denoted I, 90\%) and target removed (denoted R, 10\%). 


\subsection{Trigger}
\label{subsec:trigger}
The schematic of the placement of the beam and trigger detectors can be seen in Fig.~\ref{fig:beamAndTriggerDetectors}. 
The trigger detectors consist of a set of scintillation counters recording the presence of the beam particle (S1, S2), 
a set of veto scintillation counters with a hole used to reject beam particles passing far from the centre of 
the beamline (V0, V1), and a Cherenkov charge detector (Z). Beam particles were defined by the coincidence
T1 = $\text{S1} \cdot \text{S2} \cdot \overline{\text{V1}} \cdot \text{Z(Be)}$ and
T1 = $\text{S1} \cdot \overline{\text{V0}} \cdot \overline{\text{V1}} \cdot \overline{\text{V1'}} \cdot \text{Z(Be)}$ 
for low and high momentum data taking respectively. 
An interaction trigger detector (S4) was used to check whether the beam particle changed charge after 
passing through the target. In addition, collisions were selected by requiring an energy signal 
below a set threshold from the 16 central modules of the PSD. The event trigger condition thus was 
T2 = T1$\cdot \overline{\text{S4}} \cdot \overline{\text{PSD}}$ or T2 = T1$\cdot \overline{\text{PSD}}$
for low and high beam momenta, respectively. The PSD threshold was set to
retain from $\approx$~70\% to $\approx$~40\% of inelastic collisions at low and high beam momenta,
respectively. The statistics of recorded events is summarised in Table~\ref{tab:eventStat}.

\begin{table}
   \caption{
Basic beam properties and number of events recorded and used in the analysis
for Be+Be interactions of the 20\% most \textit{central} collisions.
}
\vspace{0.5cm}
\centering
\begin{tabular}{| c | c | c | c |}
  \hline\hline
    $p_\text{beam}$~[$A$\GeVc] & \snn~$[\GeV]$ & \parbox[][1.5cm][c]{2cm}{\centering Number of event triggers} & \parbox[][1.5cm][c]{2.5cm}{\centering Number of events after selection cuts} \\
    \hline
    19  & 6.1 & $3.46\cdot10^{6}$ & $1.32\cdot10^{5}$  \\
    30  & 7.6 & $5.41\cdot10^{6}$ & $1.48\cdot10^{5}$  \\
    40  & 8.8 & $3.42\cdot10^{6}$ & $3.97\cdot10^{5}$  \\
    75  & 11.9 & $5.24\cdot10^{6}$ & $3.99\cdot10^{5}$ \\
    150 & 16.8 & $2.93\cdot10^{6}$ & $3.25\cdot10^{5}$ \\\hline\hline
  \end{tabular}
  \label{tab:eventStat}
\end{table}

\FloatBarrier
\section{Analysis procedure}

This section starts with a brief overview of the data analysis procedure and the applied corrections.
It also defines to which class of particles the final results correspond.
A description of the calibration and the track and vertex reconstruction procedures can be found in
Ref.~\cite{Abgrall:2013pp_pim}.

The analysis procedure consists of the following steps:
\begin{enumerate}[(i)]

  \item application of event and track selection criteria,
  \item determination of raw spectra of identified charged hadrons
        using the selected events and tracks,
  \item evaluation of corrections to the raw spectra based on
        experimental data and simulations,
  \item calculation of the corrected spectra and mean multiplicities,
  \item calculation of statistical and systematic uncertainties.

\end{enumerate}

Corrections for the following biases were evaluated:
\begin{enumerate}[(a)]
 \item contribution from off-target interactions,
 \item losses of in-target interactions due to the event selection criteria,
 \item geometrical acceptance,
 \item reconstruction and detector inefficiency,
 \item losses of tracks due to track selection criteria,
 \item contribution of particles other than \emph{primary} (see below)
       charged particles produced in Be+Be interactions,
 \item losses of primary charged particles
       due to their decays and secondary interactions.
\end{enumerate}

Correction (a) was not applied due to insufficient statistics of the target removed data. The
contamination of the target inserted data was estimated from the \coordinate{z} distribution of fitted
vertices 
to amount to $\approx$~0.35\%.

Corrections (b)-(g) were estimated by data and simulations. MC events were generated with the \EposLong model
(version CRMC 1.5.3)~\cite{Werner:2005jf}, passed through detector simulation employing
the \Geant 3.21 package~\cite{Geant3} and then reconstructed by the standard program chain.

The final results refer to particles produced in \textit{central} Be+Be collisions
by strong interaction processes and in electromagnetic
decays of produced hadrons. Such hadrons are referred to as \emph{primary} hadrons.
\textit{Central} collisions refer to events selected by a cut on the total
energy emitted into the forward direction as defined by the acceptance maps for the PSD
given in Ref.~\cite{PSD_acceptance}.

The analysis was performed in (\y, $p_T$) bins.
The bin size was chosen taking into account the statistical uncertainties
and the resolution of the momentum reconstruction~\cite{Abgrall:2013pp_pim}.
Corrections as well as statistical and systematic uncertainties
were calculated for each bin.

\subsection{\textit{Central} collisions}
\label{sec:centrality}

A short description of the procedure for defining \textit{central} collisions is given below.
For more details see Refs.~\cite{Acharya:2020cyb,Kaptur:2017}.

Final results presented in this paper refer to Be+Be collisions with the 20\% lowest values of 
the forward energy $E_F$ (\textit{central} collisions). The quantity $E_F$ is defined as the total energy in the laboratory system of all particles produced in a Be+Be collision via strong and electromagnetic processes in the forward momentum region defined by the acceptance map in Ref.~\cite{PSD_acceptance}. Final results on \textit{central} collisions, derived using this procedure, allow a precise comparison with predictions of models without any additional information about the \NASixtyOne setup and used magnetic field.

\begin{figure}[ht]
        \centering 
        \includegraphics[width=0.35\textwidth]{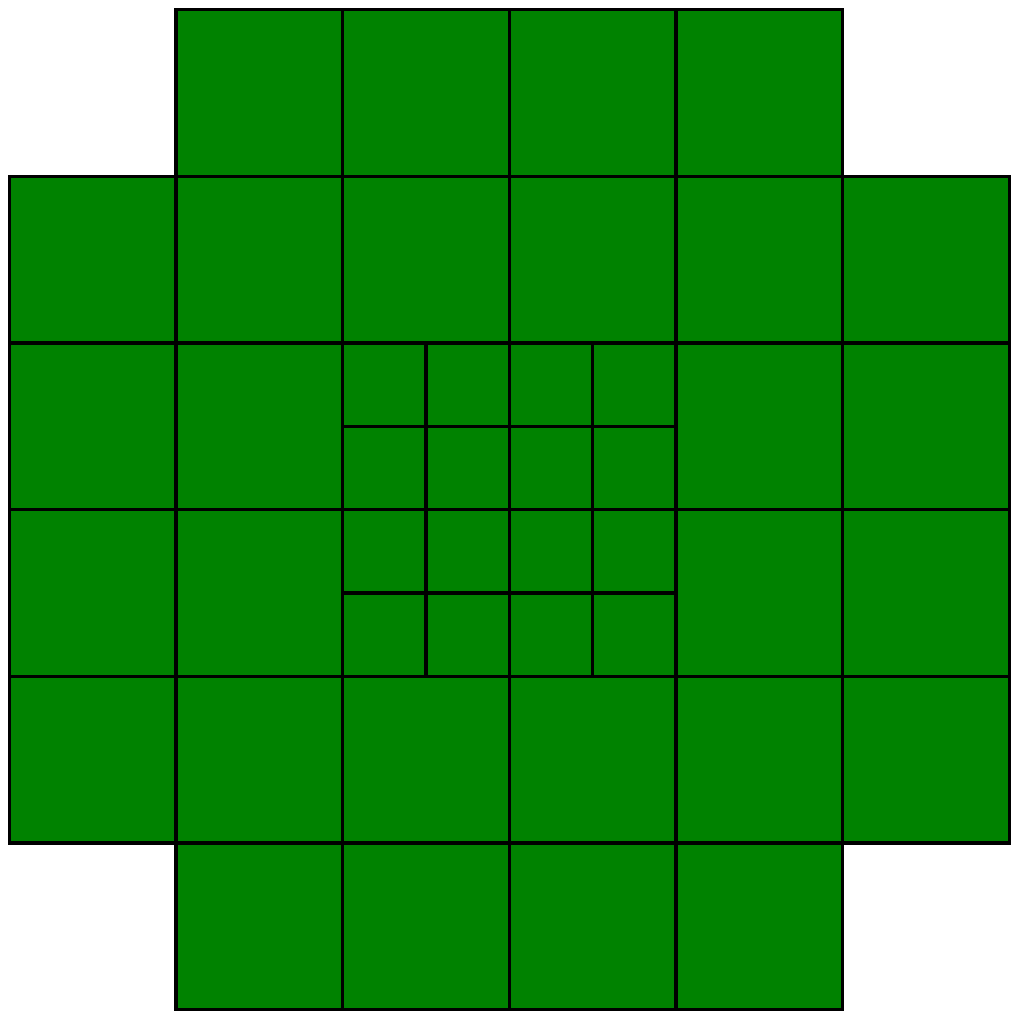}
        \includegraphics[width=0.35\textwidth]{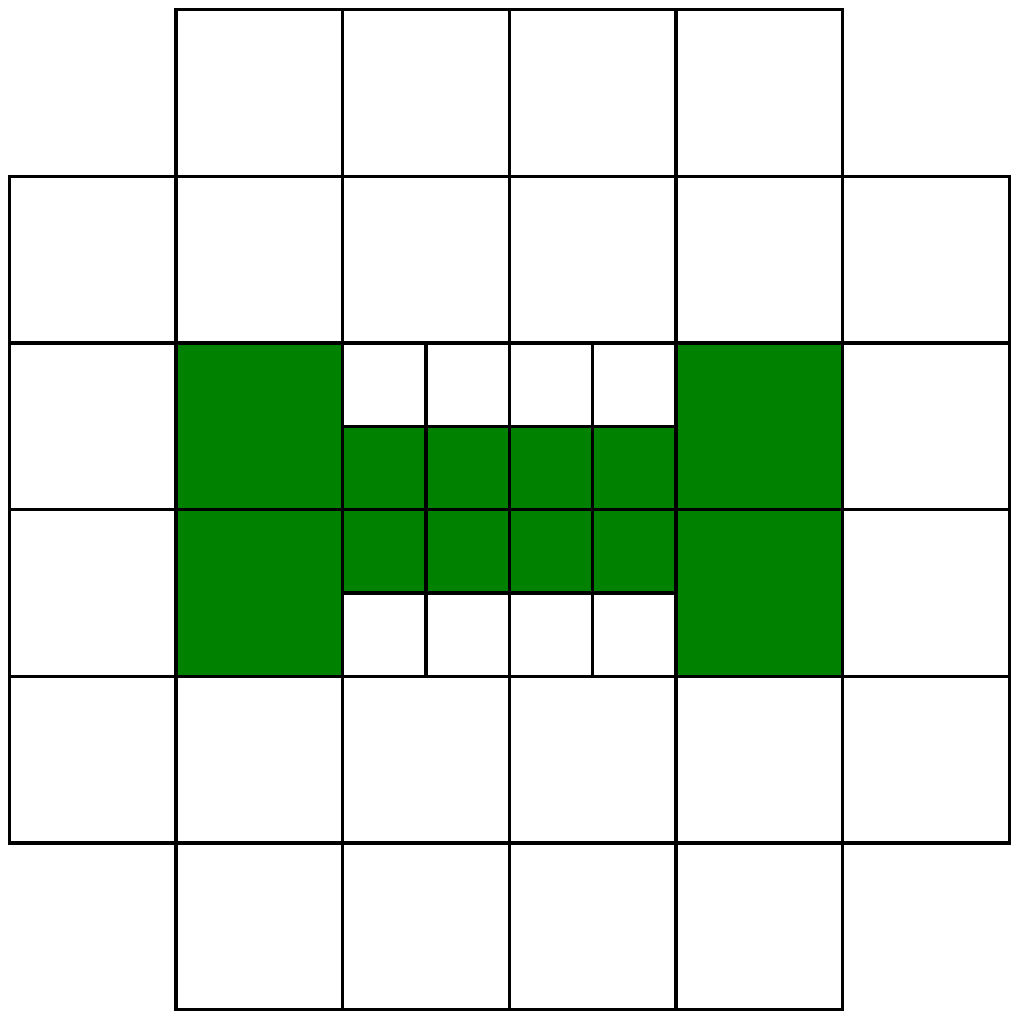}
        \caption{PSD modules included in the calculation of the projectile spectator energy $E_{PSD}$
        used for event selection  for beam momenta of 19$A$, and 30\AGeVc (\textit{left}) and for 40$A$, 75$A$ and 150\AGeVc (\textit{right})}
        \label{fig:PSDAllModuleSelections}
\end{figure}

For analysis of the data the event selection was based on the $\approx$~20\% of collisions with the lowest value of the energy $E_{PSD}$ measured by a subset of
PSD modules (see Fig.~\ref{fig:PSDAllModuleSelections}) in order to optimize the sensitivity to projectile spectators. The forward momentum acceptance in the definition of $E_F$ corresponds closely to the acceptance of this subset of PSD modules. 

Online event selection by the hardware trigger (T2) used a threshold on the electronic sum of energies over the 16 central modules of the PSD set to accept $\approx$~40\% of the inelastic interactions.
The minimum-bias distribution was obtained using the data from the beam trigger T1 with offline selection of events by requiring an event vertex in the target region and a cut on the ionisation energy detected in the GTPC to exclude Be beams. The spectrum of $E_{PSD}$ was calculated for the subset of modules and a properly normalized spectrum for target removed events was subtracted. Events for further analysis were selected by applying a cut in this distribution for the 20\% of events with the smallest values of $E_{PSD}$.
More details on the PSD modules selection optimisation and $E_{PSD}$ determination
can be found in \cite{Kaptur:2017}.


The forward energy $E_F$ cannot be measured directly. However, both $E_F$ and $E_{PSD}$ can be obtained from simulations using the 
\EposLong (version CRMC 1.5.3)~\cite{Werner:2005jf} model. A global factor $c_{cent}$ (listed in Table~\ref{tab:w}) 
was then calculated as the ratio of mean negatively charged pion multiplicities obtained with the two selection procedures 
for 20\% of all inelastic collisions. A possible dependence of the scaling factor on rapidity and transverse momentum was neglected.
The resulting factors $c_{cent}$ range from 1.00 and 1.04 which is only a small correction compared to 
the systematic uncertainties of the measured particle multiplicities. The correction was therefore not applied, but instead included
in the systematic uncertainty.

\begin{figure}[ht]
  \centering
     \includegraphics[width=0.49\textwidth]{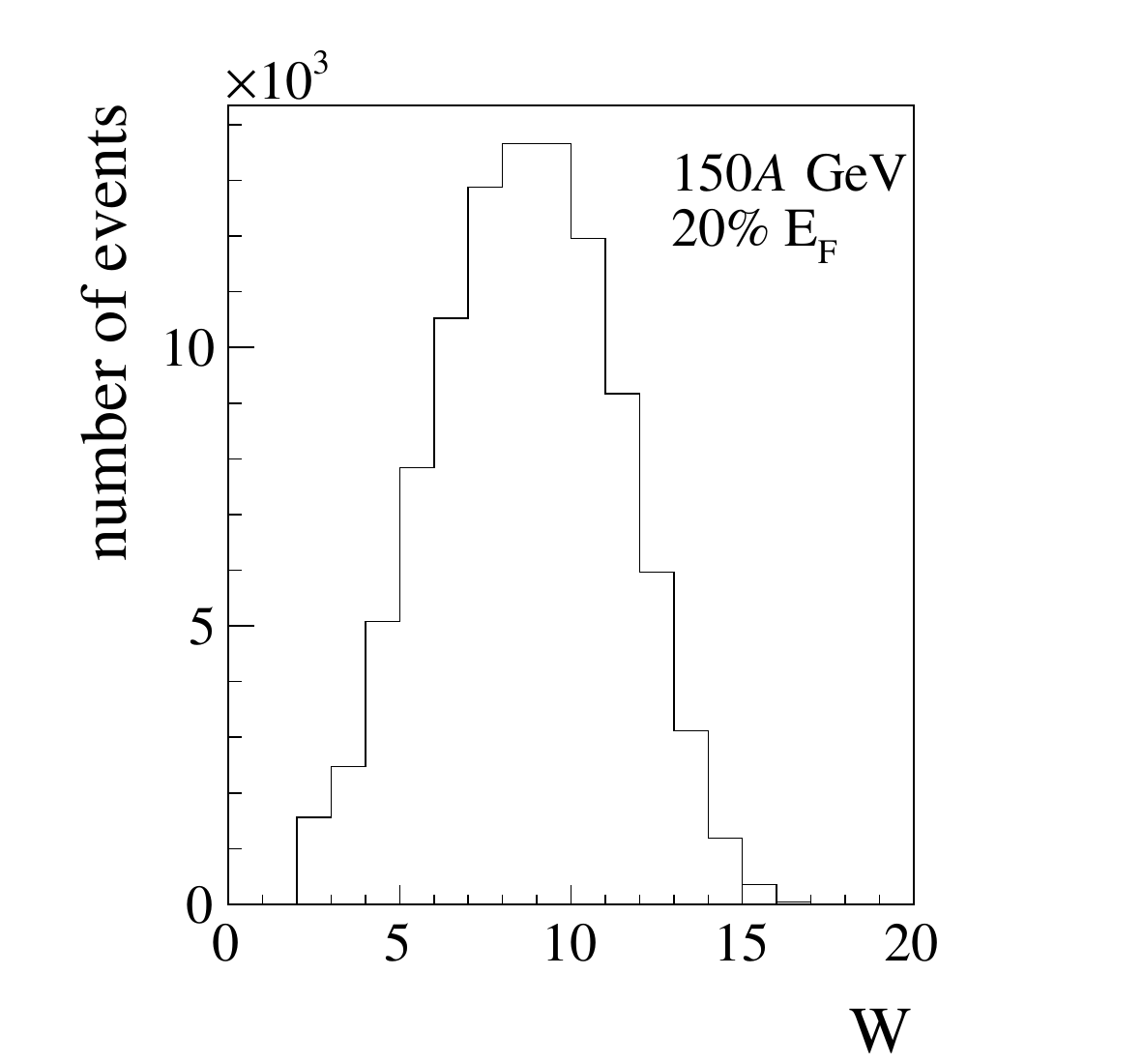}
     \includegraphics[width=0.49\textwidth]{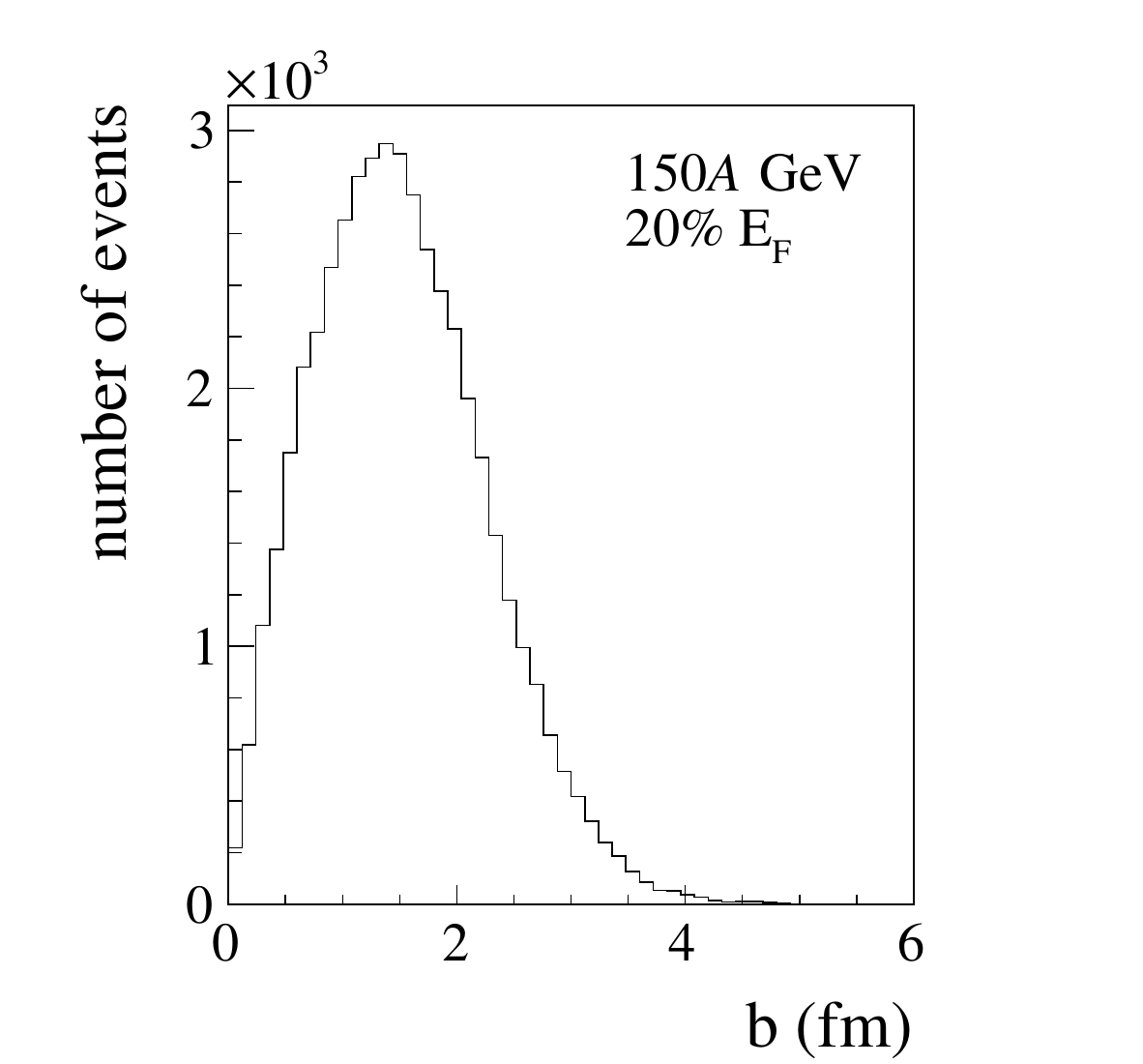}
     \includegraphics[width=0.49\textwidth]{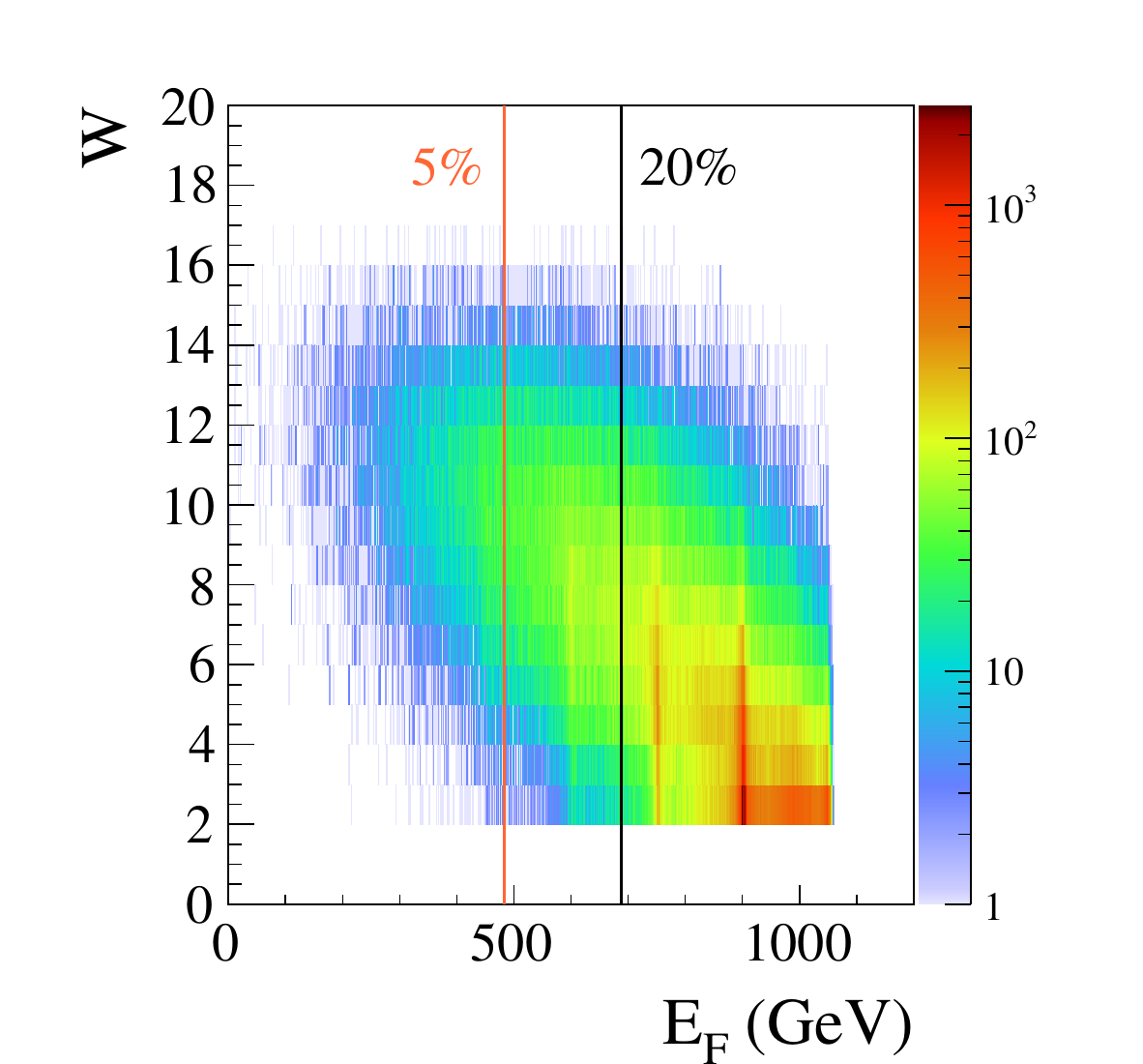}
     \includegraphics[width=0.49\textwidth]{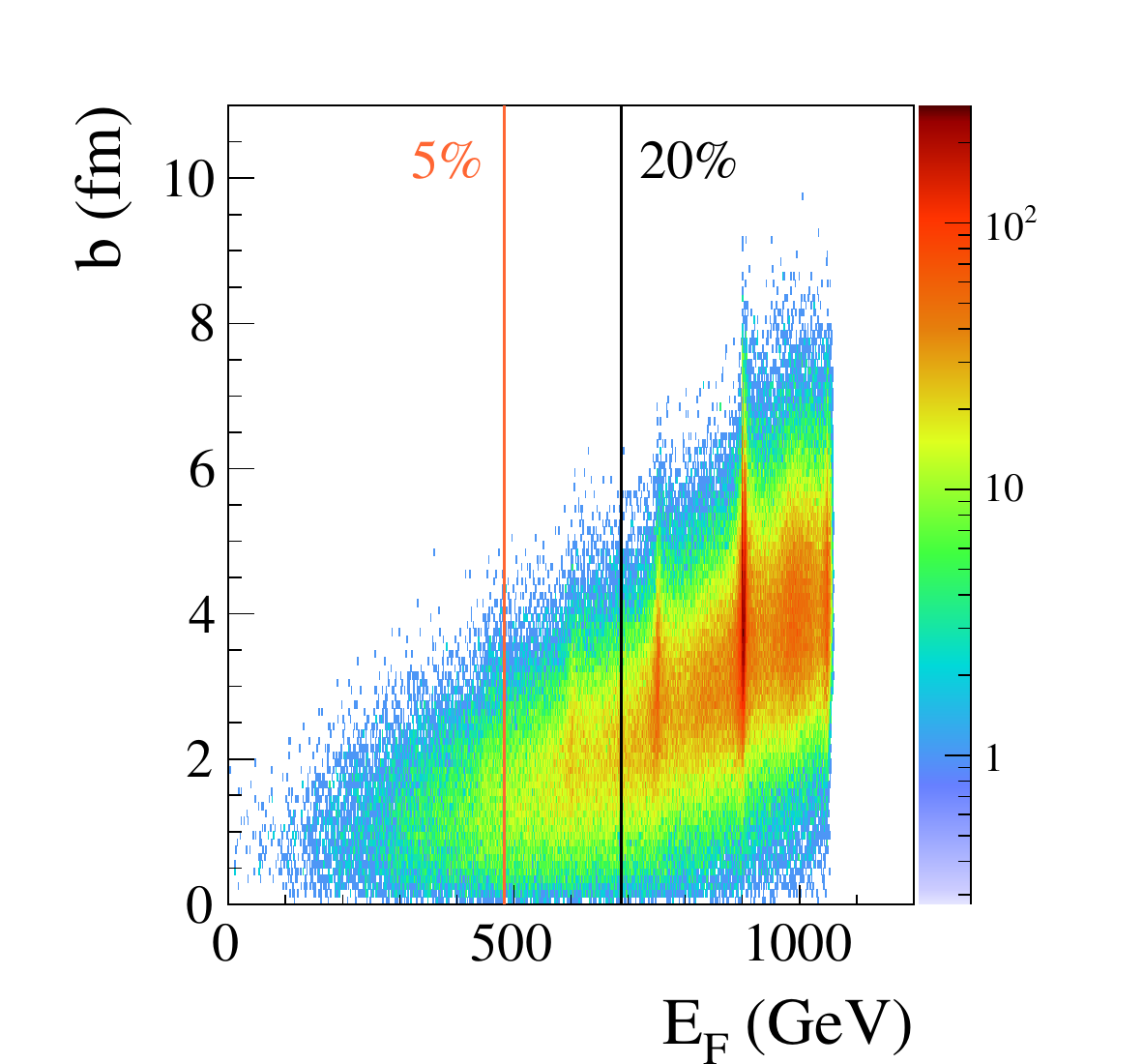}
  \caption{Examples of the distribution of the number of wounded nucleons $W$ (\textit{left}) 
           and collision impact parameter $b$ (\textit{right}) for events with the 20\%
           smallest forward energies $E_F$ \textit{(central)} at beam momentum of 150\AGeVc simulated 
           with the \Epos model using the acceptance map provided in~Ref.~\cite{PSD_acceptance}.}
  \label{fig:woundedDistribution}
\end{figure}

Finally, the average number of wounded nucleons $\langle W \rangle$ and the average collision impact parameter $\langle b \rangle$ were calculated within the Wounded Nucleon Model~\cite{Bialas:1976ed} implemented in \Epos
for events with the 20\% smallest values of $E_F$. Results are listed in Table~\ref{tab:w}. 
Example distributions for the top beam momentum are shown in Fig.~\ref{fig:woundedDistribution}.
As the Be nucleus consists of few nucleons these distributions are quite broad. For comparison 
$\langle W \rangle$ and $\langle b \rangle$ were also calculated from the GLISSANDO model
\cite{Broniowski:2007nz} which uses a different Glauber model calculation. The results, also listed
in Table~\ref{tab:w}, differ by about 10\% for $\langle W \rangle$. This discrepancy was included in
the systematic error estimate (see Sec.~\ref{sec:systematics})                                       .

\begin{table*}[ht]
 \centering
 \caption{Average number of wounded nucleons $\langle W \rangle$  and average collision impact parameter
          $\langle b\rangle$ in the 20\% most \textit{central} Be+Be collisions estimated from 
          simulations using the \Epos~\cite{Werner:2005jf} and GLISSANDO~\cite{Broniowski:2007nz} models.
          The values of $\sigma$ denote the widths of the distributions of $W$ and $b$.
          Results \Epos~WNM and Glissando are for \textit{centrality} selection using the smallest number of spectators
          in the Glauber model. \Epos~$E_F$ results correspond to selecting events by the forward energy $E_F$ 
          within the acceptance map in Ref.~\cite{PSD_acceptance}.
          Also shown are correction factors $c_{cent}$ needed to account for the different
          event selection procedures used for the data (measured $E_{PSD}$) and 
          the simulation (simulated values of $E_F$).}
 \vspace{0.2cm}
 \footnotesize
 \begin{tabular}{l|cccccc}
  Momentum (\AGeVc) & & 19 & 30 & 40 & 75 & 150\\
  \hline
  \hline
  \Epos $E_F$ \textit{(central)}& $\langle W \rangle$ & $8.04$ & $7.99$ & $8.13$ & $8.12$ & $8.15$\\
  & $\sigma$ & $2.6$ & $2.6$ & $2.6$ & $2.6$ & $2.6$ \\
  & $\langle b \rangle$ & $1.89$ & $1.93$ & $1.49$ & $1.49$ & $1.51$\\
  & $\sigma$ & $0.9$ & $0.9$ & $0.7$ & $0.7$ & $0.7$ \\
  \Epos Glauber & $\langle W \rangle$ & $8.87$ & $8.88$ & $8.89$ & $8.90$ & $8.93$\\
  & $\sigma$ & $1.8$ & $1.8$ & $1.8$ & $1.8$ & $1.8$ \\
  GLISSANDO & $\langle W \rangle$ & $8.74$ & $8.77$ & $8.76$ & $8.83$ & $8.91$\\
  & $\sigma$ & $1.78$ & $1.79$ & $1.80$ & $1.80$ & $1.81$\\
  \hline
  & $c_{cent}$ & $1.003$ & $1.043$ & $1.008$ & $1.016$ & $1.027$ \\
 \end{tabular}
 
 \label{tab:w}
\end{table*}

\subsection{Event and track selection}\label{sec:cuts}

\subsubsection{Event selection}
\label{sec:event_selection}

For further analysis Be+Be events were selected using the following criteria:
\begin{enumerate}[(i)]
    \item four units of charge measured in S1, S2, and Z counters as well as BPD3
          (this requirement also rejects most interactions upstream of the Be target),
    \item no off-time beam particle detected within a time window of $\pm$4.5$~\mu$s
          around the trigger particle,
    \item no other event trigger detected within  a time window of $\pm$25$~\mu$s
          around the trigger particle,
    \item beam particle detected in at least two planes out of four
          of BPD-1 and BPD-2 and in both planes of BPD-3,
    \item a well reconstructed interaction vertex with \coordinate{z} position (fitted using  the beam trajectory and TPC tracks) not farther away than 15~cm
          from the center of the Be target 
          (the cut removes less than 0.4\% of T2 trigger ($E_{PSD}$) selected interactions),
    \item an upper cut on the measured energy $E_{PSD}$ which selects 
          20\% of all inelastic collisions.
\end{enumerate}
The event statistics after applying the selection criteria
is summarized in Table~\ref{tab:eventStat}.


\subsubsection{Track selection}
\label{sec:track_selection}

In order to select tracks of primary charged hadrons and to reduce the contamination
by particles from secondary interactions, weak decays and off-time interactions,
the following track selection criteria were applied:

\begin{enumerate}[(i)]
    \item track momentum fit including the interaction vertex should have converged,
    \item fitted \coordinate{x} component of particle rigidity $q \cdot \plab$ is positive.
          This selection minimizes the angle between the track trajectory and the TPC
          pad direction for the chosen magnetic field direction, reducing uncertainties
          of the reconstructed cluster position, energy deposition and track parameters,
    \item total number of reconstructed points on the track should be greater than 30,
    \item sum of the number of reconstructed points in VTPC-1 and VTPC-2 should be greater than 15
          or greater than 4 in the GTPC,
    \item the distance between the track extrapolated to the interaction plane and the
          (track impact parameter) should be smaller than 4~cm in the horizontal
          (bending) plane and 2~cm in the vertical (drift) plane.
\end{enumerate}

\FloatBarrier
\subsection{Identification techniques}
\label{sec:identification}

\begin{figure}[!ht]
        \begin{center}
        \includegraphics[width=0.3\textwidth]{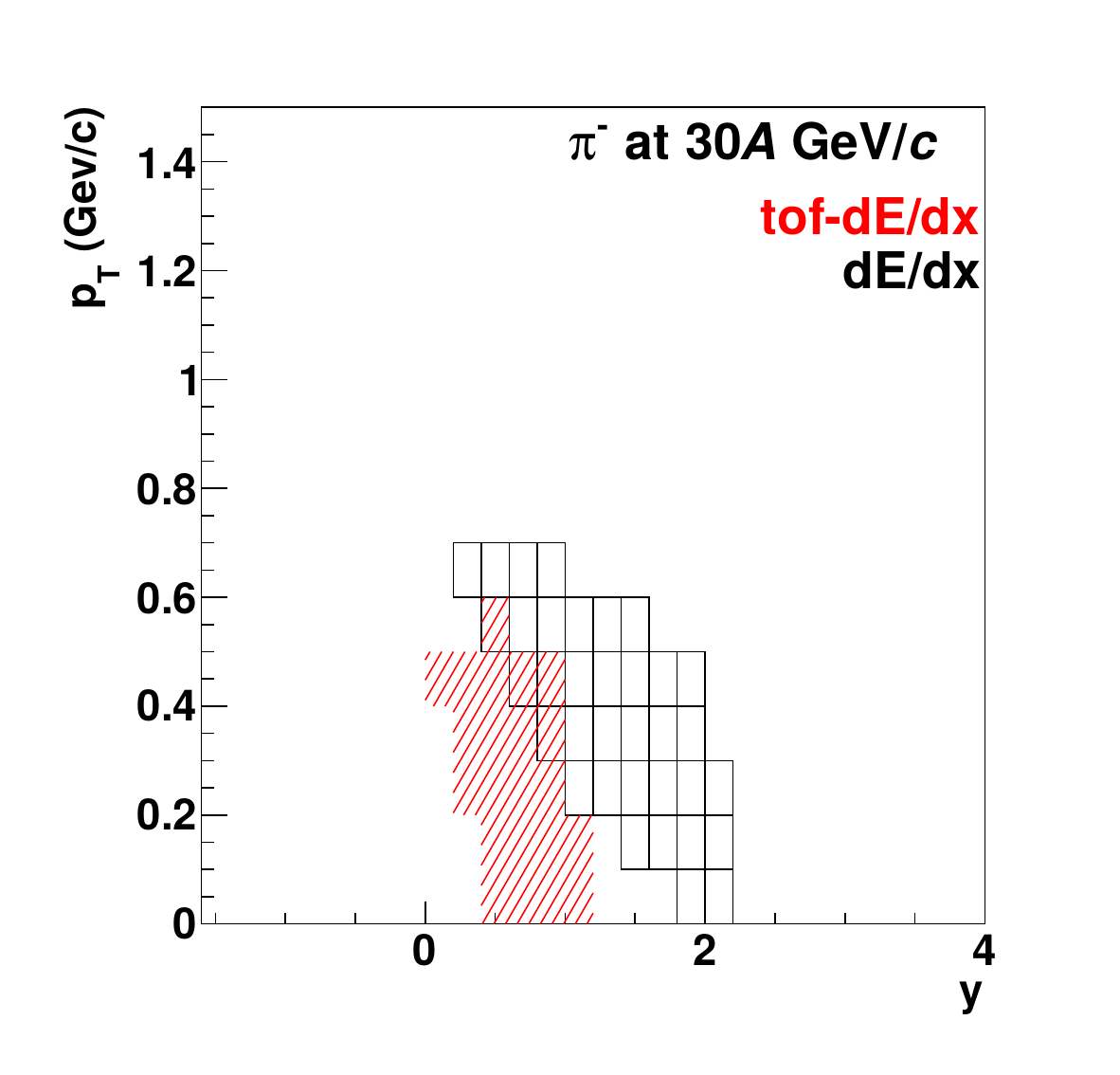}
        \includegraphics[width=0.3\textwidth]{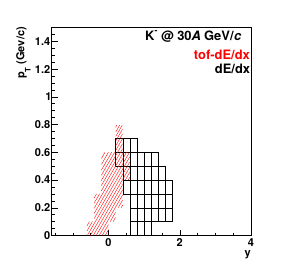}
        \includegraphics[width=0.3\textwidth]{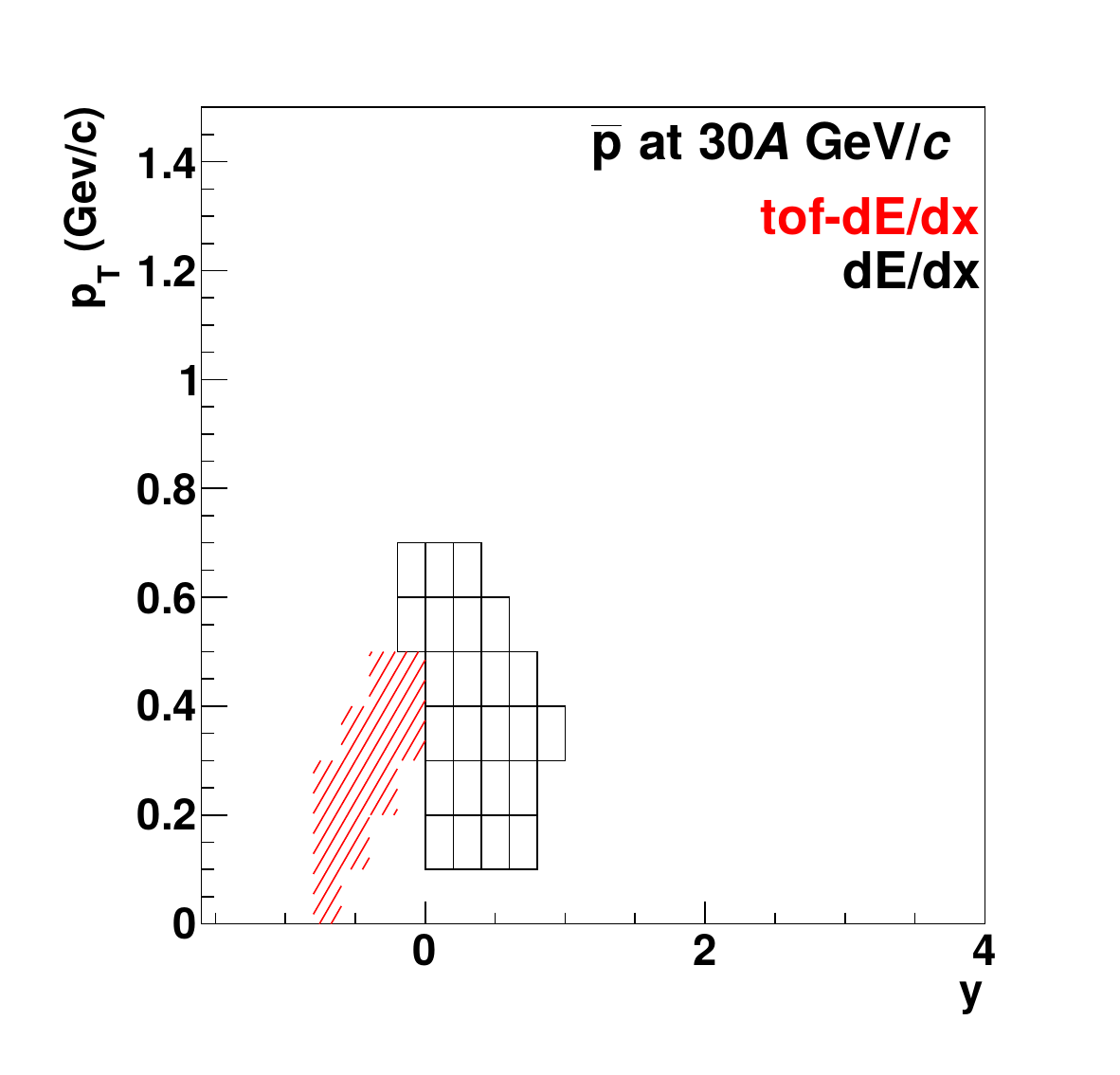}\\
        \includegraphics[width=0.3\textwidth]{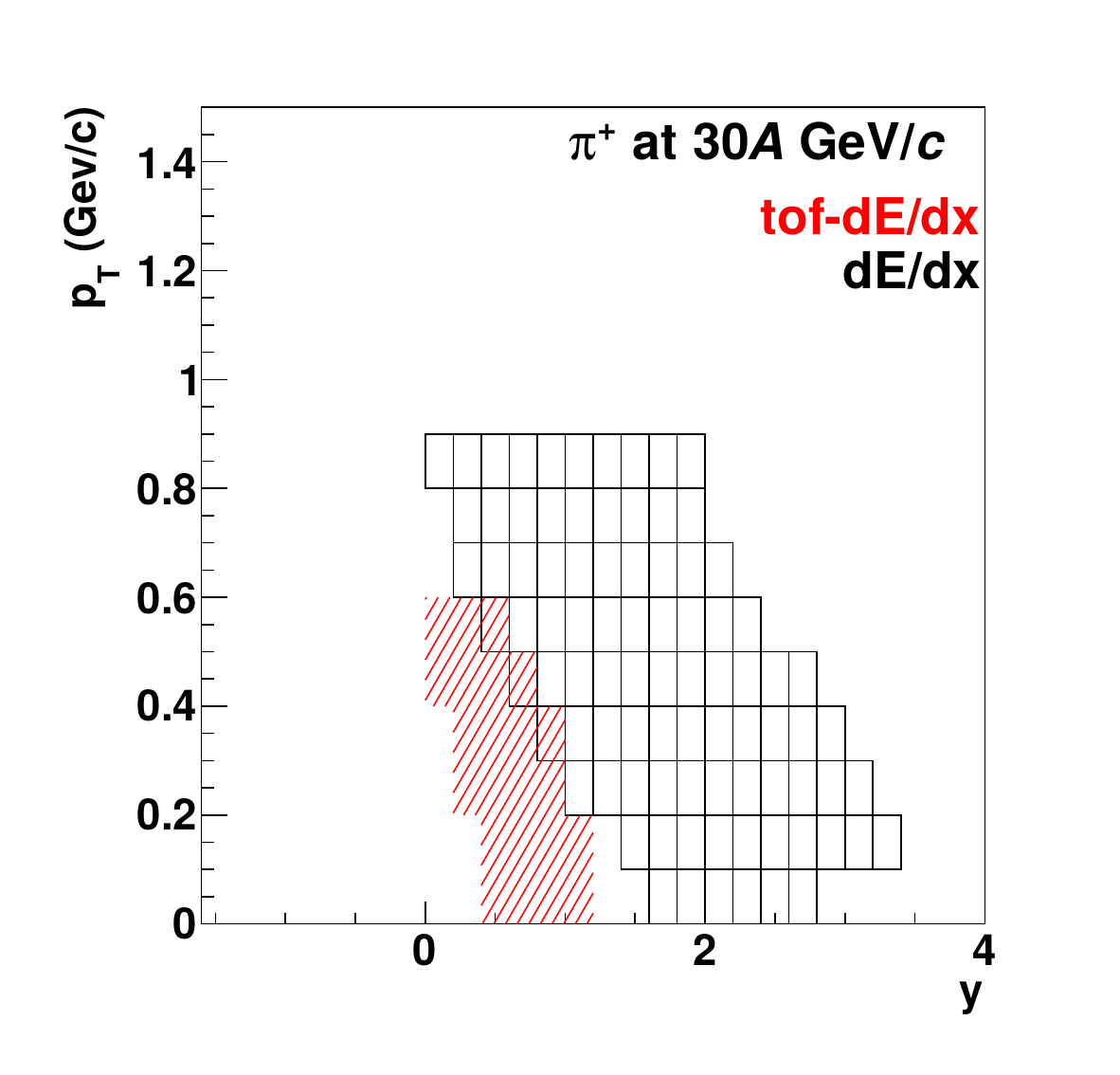}
        \includegraphics[width=0.3\textwidth]{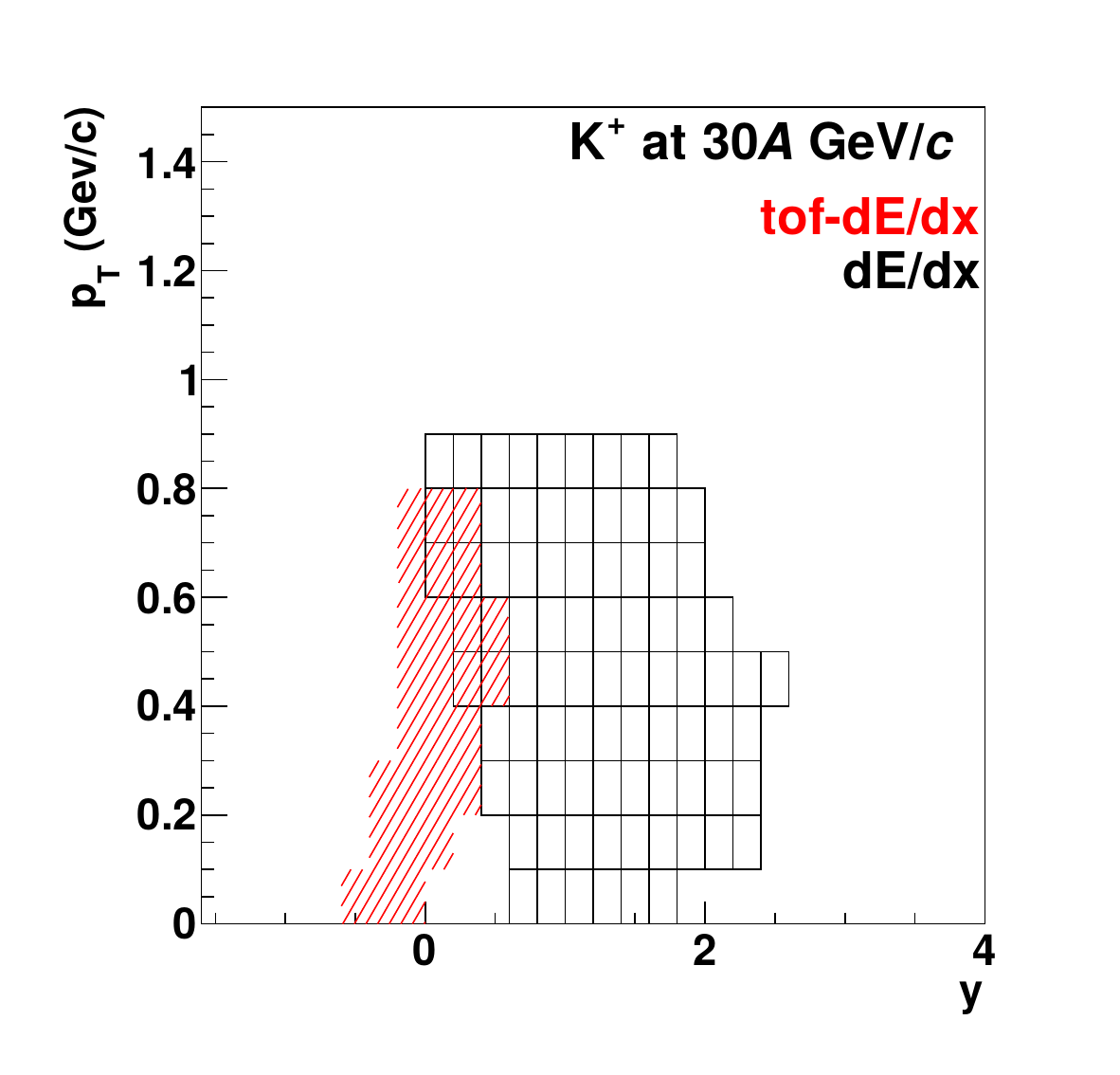}
        \includegraphics[width=0.3\textwidth]{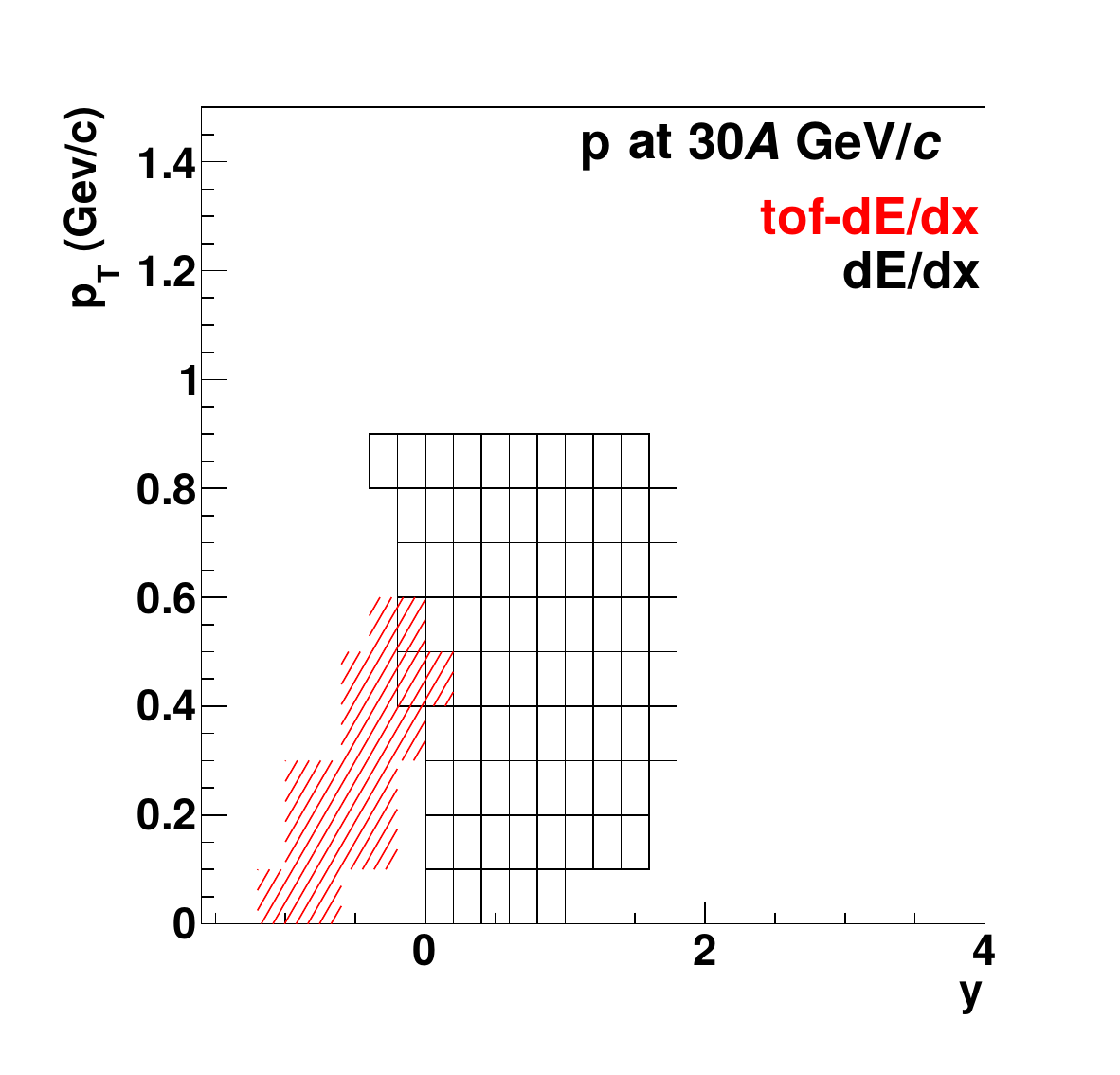}
        \end{center}
        \caption{Acceptance of the $tof$-\dEdx and \dEdx methods for identification of 
                 pions, kaons and protons in the 20\% most \textit{central} Be+Be interactions at 30\AGeVc.
                }

        \label{fig:methodacc19}
\end{figure}

\begin{figure}[!ht]
       \begin{center}
       \includegraphics[width=0.3\textwidth]{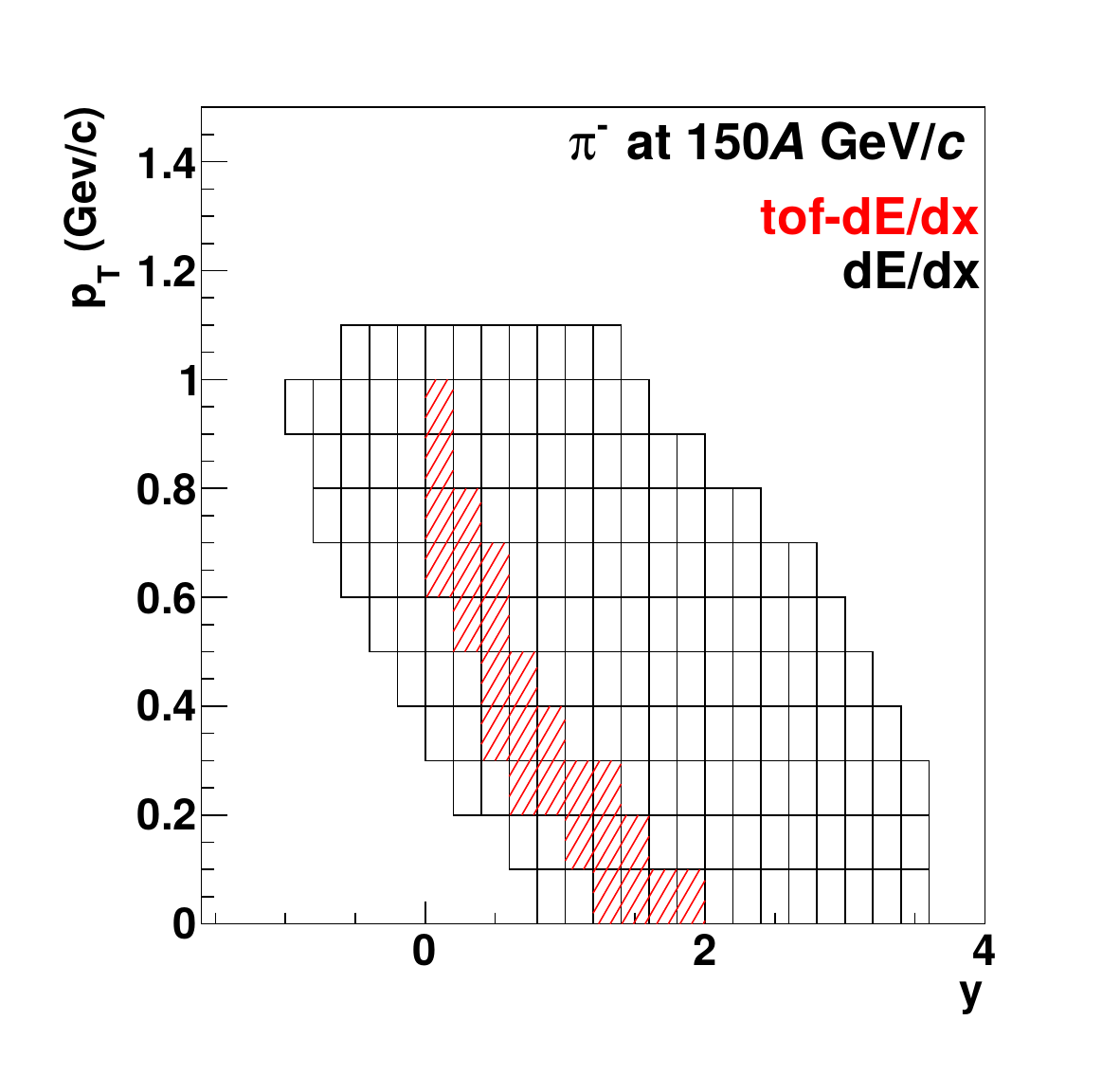}
       \includegraphics[width=0.3\textwidth]{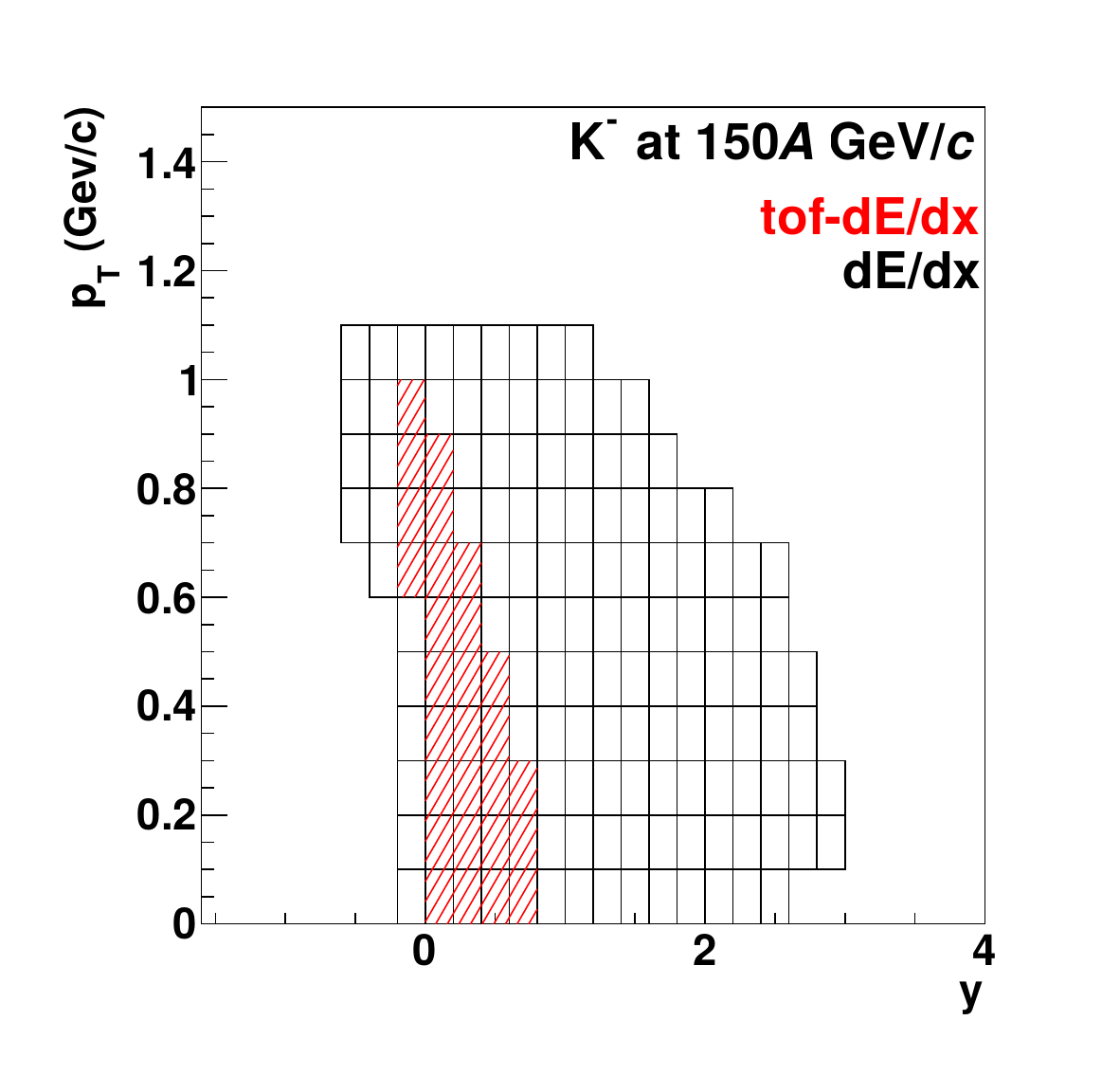}
       \includegraphics[width=0.3\textwidth]{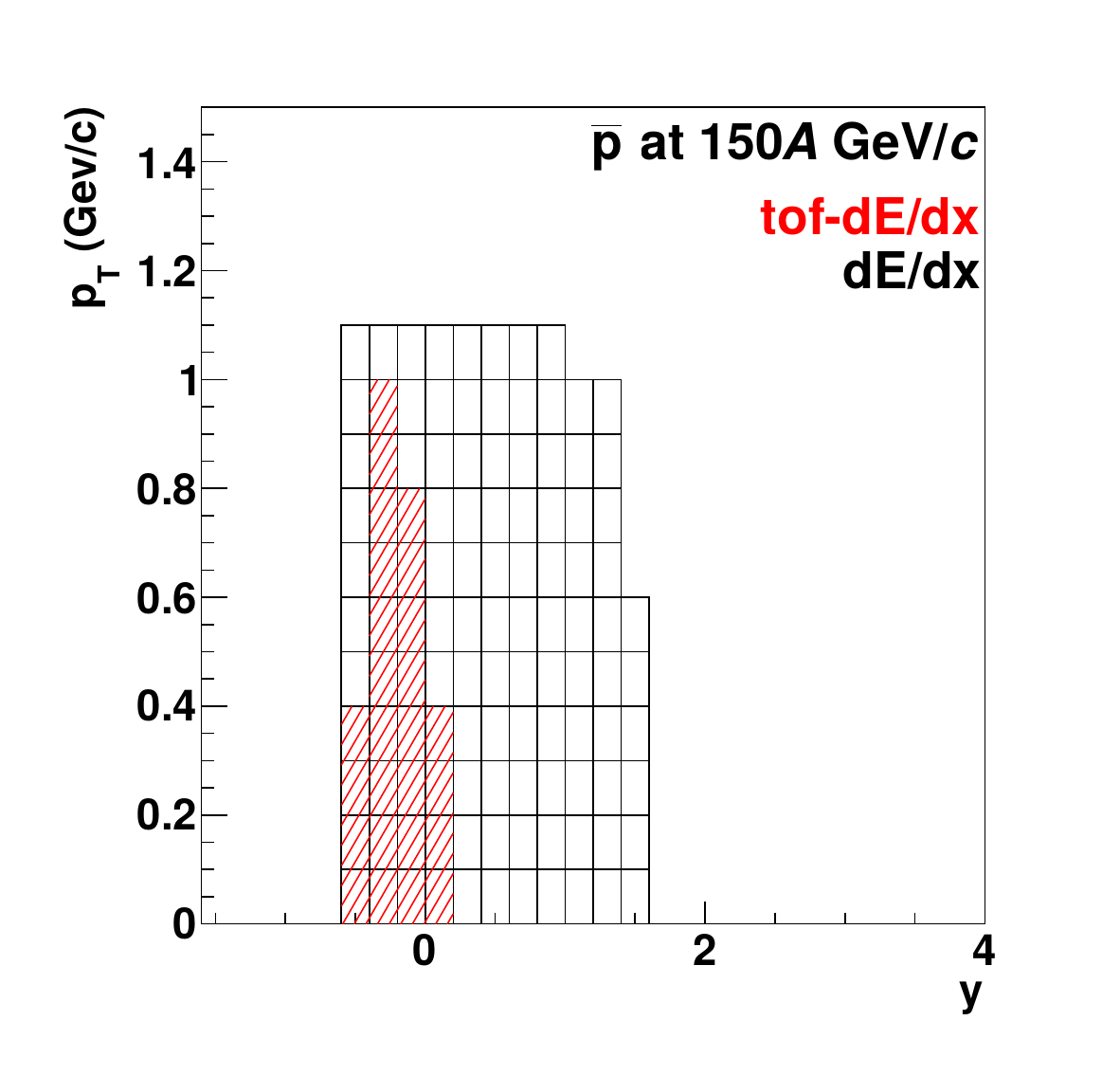}\\
       \includegraphics[width=0.3\textwidth]{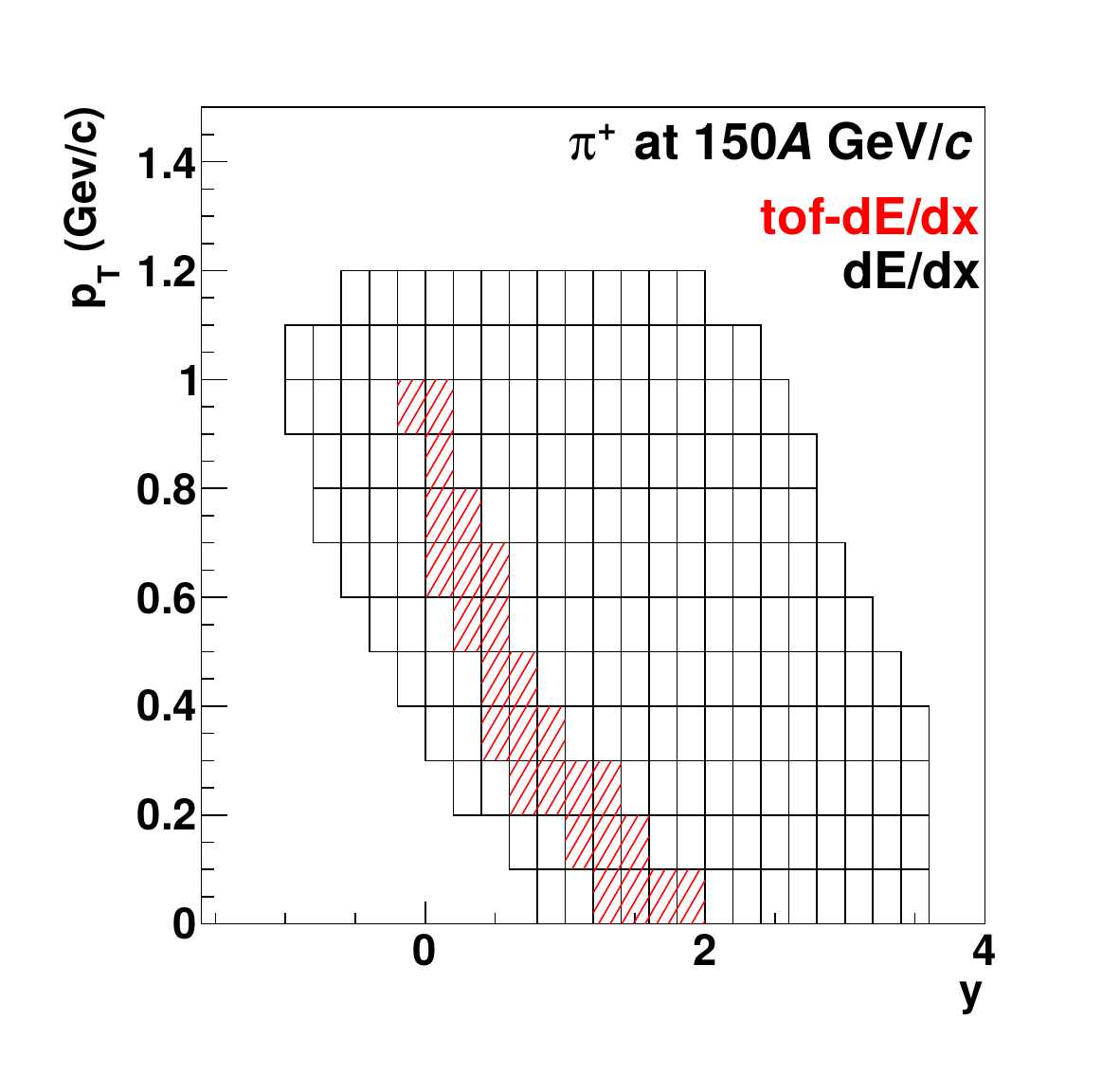}
       \includegraphics[width=0.3\textwidth]{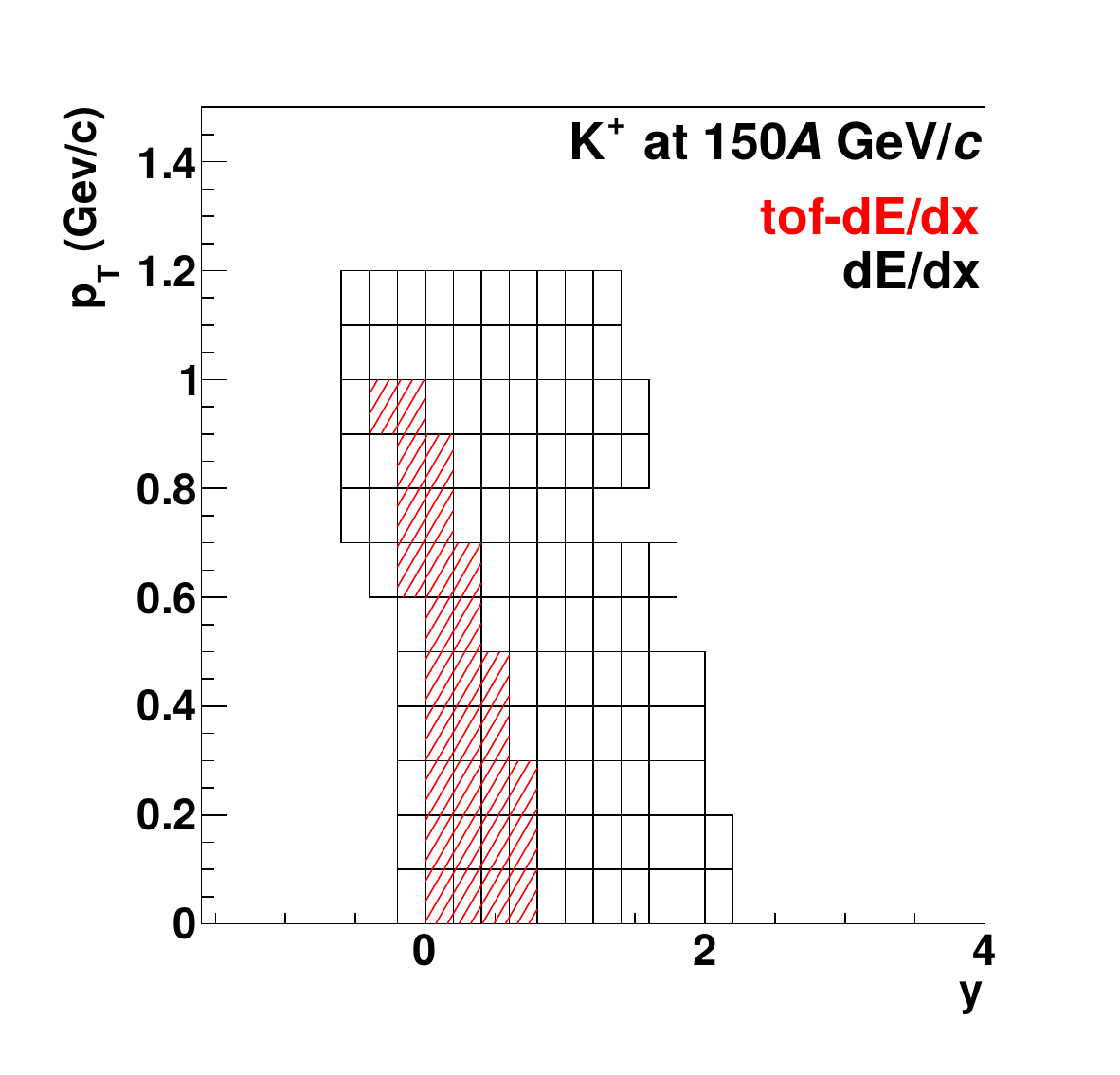}
       \includegraphics[width=0.3\textwidth]{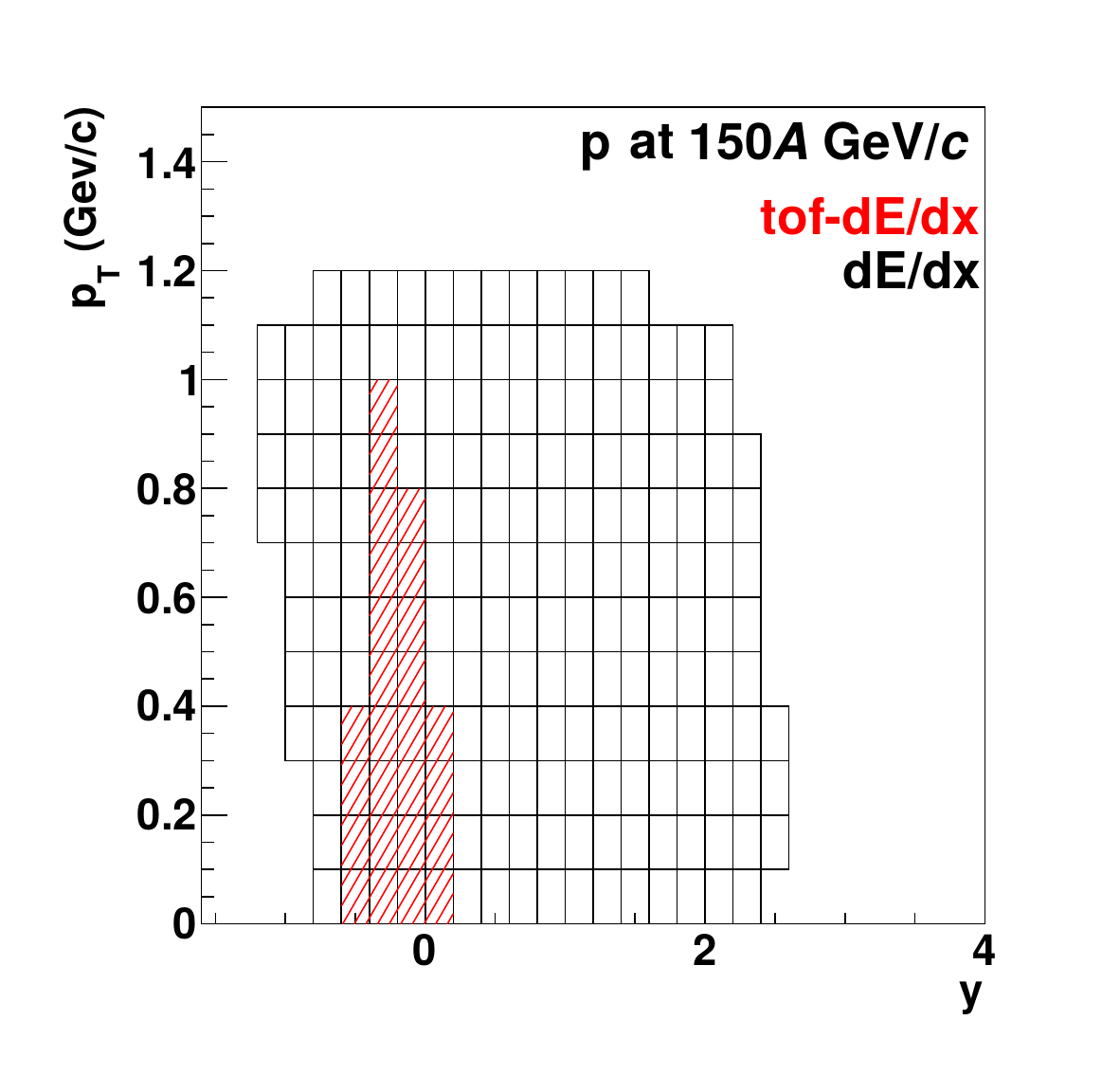}
       \end{center}
       \caption{Acceptance of the $tof$-\dEdx and \dEdx methods for identification of
                pions, kaons and protons in the 20\% most \textit{central} Be+Be interactions at 150\AGeVc.
               }

       \label{fig:methodacc150}
\end{figure}

Charged particle identification in the \NASixtyOne experiment is based on 
the ionization energy loss, \dEdx, in the gas of the TPCs and the time of flight, $tof$, obtained
from the ToF-L and ToF-R walls. In the region of the relativistic rise of the ionization at 
large momenta the measurement of \dEdx alone allows identification. At lower momenta the
\dEdx bands for different particle species overlap and additional measurement of $tof$ is required
to remove the ambiguity. These two methods allow to cover most of the phase space in rapidity 
and transverse momentum which is of interest for the strong interaction programme of \NASixtyOne. 
The acceptance of the two methods is shown in Figs.~\ref{fig:methodacc19} and~\ref{fig:methodacc150}
for the 20\% most \textit{central} Be+Be interactions at 30 and 150\AGeVc, respectively.
At low beam energies the $tof$-\dEdx method extends the identification acceptance, 
while at top SPS energy it overlaps with the \dEdx method (for more details see Ref.~\cite{Kuich:2019}).

\subsubsection{Identification based on energy loss measurement~(\dEdx)}
\label{sec:dedx_id}

\begin{figure}[!ht]
        \begin{center}
       \includegraphics[width=0.9\textwidth]{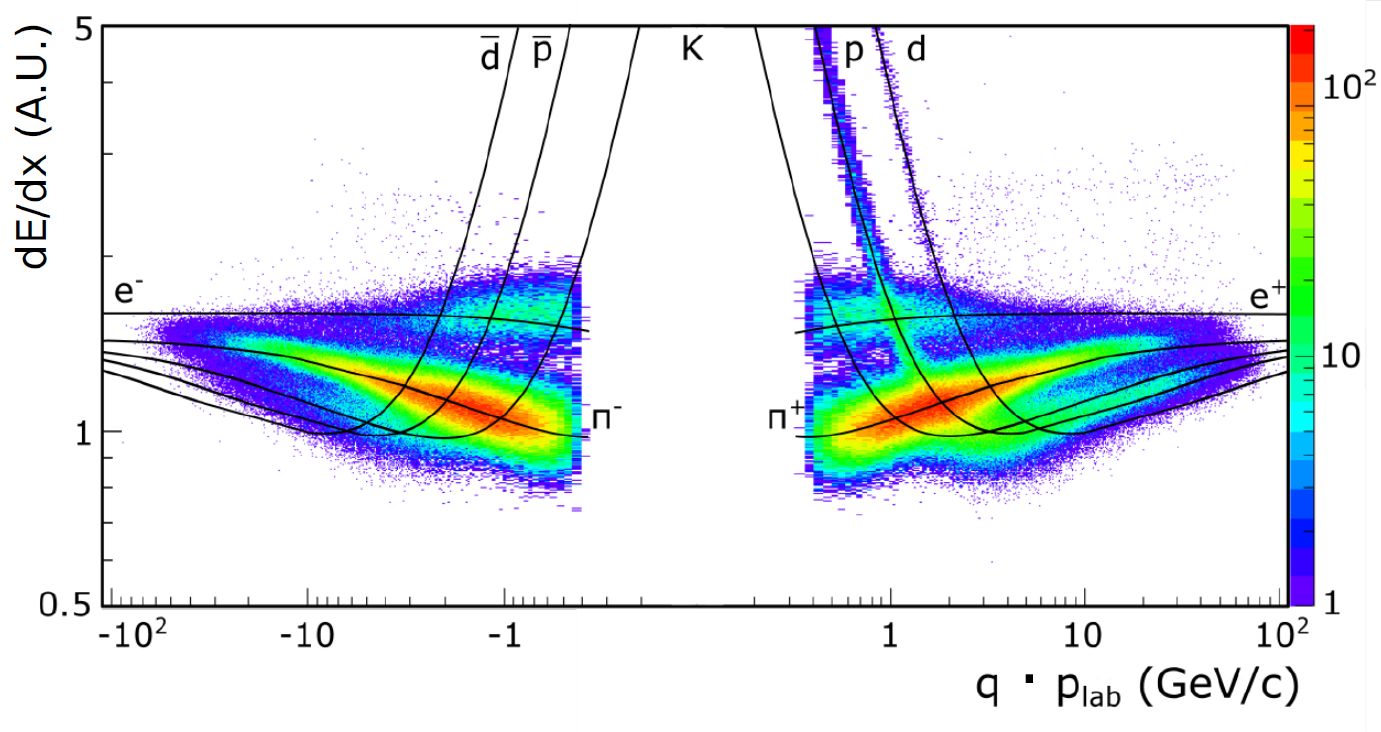}
        \end{center}
        \caption{
        Distribution of charged particles in the \dEdx -- $q\cdot$\plab~plane.
        The energy loss in the TPCs for different charged particles for events and tracks
        selected for the analysis of Be+Be collisions at 75\AGeVc
        (the target inserted configuration).
        Expectations for the dependence of the mean \dEdx on ~\plab for the considered particle
        types are shown by the curves calculated based on the Bethe-Bloch function.
}
        \label{fig:dedx}
\end{figure}

Time projection chambers provide measurements of energy loss \dEdx of charged particles 
in the chamber gas along their trajectories. Simultaneous measurements of \dEdx and $p_\text{lab}$~ 
allow to extract information on particle mass. The mass assignment follows the procedure which was developed
for the analysis of \textit{p+p} reactions as described in Ref.~\cite{Aduszkiewicz:2017sei}.
Values of \dEdx are calculated as the truncated mean (smallest 50\%)
of ionisation energy loss measurements along the track trajectory. As an example, \dEdx measured in
the 20\% most central Be+Be interactions at 75\AGeVc~is presented in Fig.~\ref{fig:dedx}, for positively and negatively charged particles, as a function of $q \cdot \plab$.

The contributions of $e^{+}$, $e^{-}$, $\pi^{+}$, $\pi^{-}$, $K^{+}$, $K^{-}$, $p$ and $\bar{p}$ are
obtained by fitting the \dEdx distributions separately for positively and negatively charged particles
in bins of  \plab~ and \pt~ with a sum of four functions~\cite{vanLeeuwen:2003ru,note_MvL}
each corresponding to the expected \dEdx distribution for the corresponding particle type. 
The small contribution of light (anti-)nuclei was neglected.

In order to ensure similar particle multiplicities in each bin, 20 logarithmic bins
are chosen in \plab~ in the  range $1-100$~\GeVc to cover the full detector acceptance.
Furthermore, the data are binned in 20 equal \pt~intervals in the range 0-2~\GeVc.

The distribution of \dEdx for tracks
of a given particle type  $i$ is parameterised as the sum of Gaussians with widths $\sigma_{i,l}$
depending on the particle type $i$ and the number of points $l$ measured in the TPCs.
Simplifying the notation in the fit formulae, the peak position of the \dEdx distribution
for particle type $i$ is denoted as $x_{i}$. The contribution of a reconstructed particle track
to the fit function reads:
\begin{equation}
\rho(x)=\sum_{i}\rho_{i}(x)=\sum\limits_{i=\pi,p,K,e} A_{i} \frac{1}{\sum\limits_{l} n_{l}} \sum\limits_{l} \frac{n_{l}}{\sqrt{2\pi}\sigma_{l}}exp \left [-\frac{1}{2}\left (\frac{x-x_{i}}{\sigma_{l}} \right ) ^{2} \right ]~,
\label{Eq:AsymGaus}
\end{equation}
where $x$ is the \dEdx of the particle, $n_{l}$ is the number of tracks with number of points $l$
in the sample and $A_{i}$ is the amplitude of the contribution of particles of type $i$.
The second sum is the weighted average of the line-shapes from the different numbers of measured points
(proportional to track-length) in the sample. The quantity $\sigma_{l}$ is written as:
\begin{equation}
\sigma_{l}=\sigma_{0}\left ( \frac{x_{i}}{x_{\pi}}\right )^{0.625} / \sqrt{n_{l}},
\label{Eq:sigma}
\end{equation}
where the width parameter $\sigma_{0}$ is assumed to be common for all particle types and bins.
A $1/\sqrt{l}$ dependence on number of points is assumed. The Gaussian peaks could in principle
be asymmetric if the tail of the Landau distribution persists to some extent even 
after truncation (for detail see \cite{Afanasev:1999iu}). However, no significant effect was found.

The fit function has 9 parameters (4 amplitudes, 4 peak positions and width) which 
are difficult to fit in each bin independently. Therefore the following constraints on the fitting parameters 
were adopted:
\begin{enumerate}[(i)]
\item positions of electrons, kaons and protons relative to pions were assumed to be $p_{T}$-independent,
\item the fitted amplitudes were required to be greater than or equal to 0,
\item the electron amplitude was set to zero for total momentum \plab~above 23.4~\GeVc (i.e. starting from 
      the 13$^{th}$ bin), as the electron contribution vanishes at high \plab,
\item if possible, the relative position of the positively charged kaon peak was taken to be the same as that of negatively charged kaons determined from the negatively charged particles in the bin of the same 
      \plab~ and \pt. This procedure helps to overcome the problem of the large overlap 
      between $K^+$ and protons in the \dEdx distributions.
\end{enumerate}
The constraints reduce the number of independently fitted parameters in each bin from 9 to 6, 
i.e. the amplitudes of the four particle types, the pion peak position and the width parameter $\sigma_{0}$.

\begin{figure}[ht]
        \begin{center}
        \includegraphics[width=0.45\textwidth]{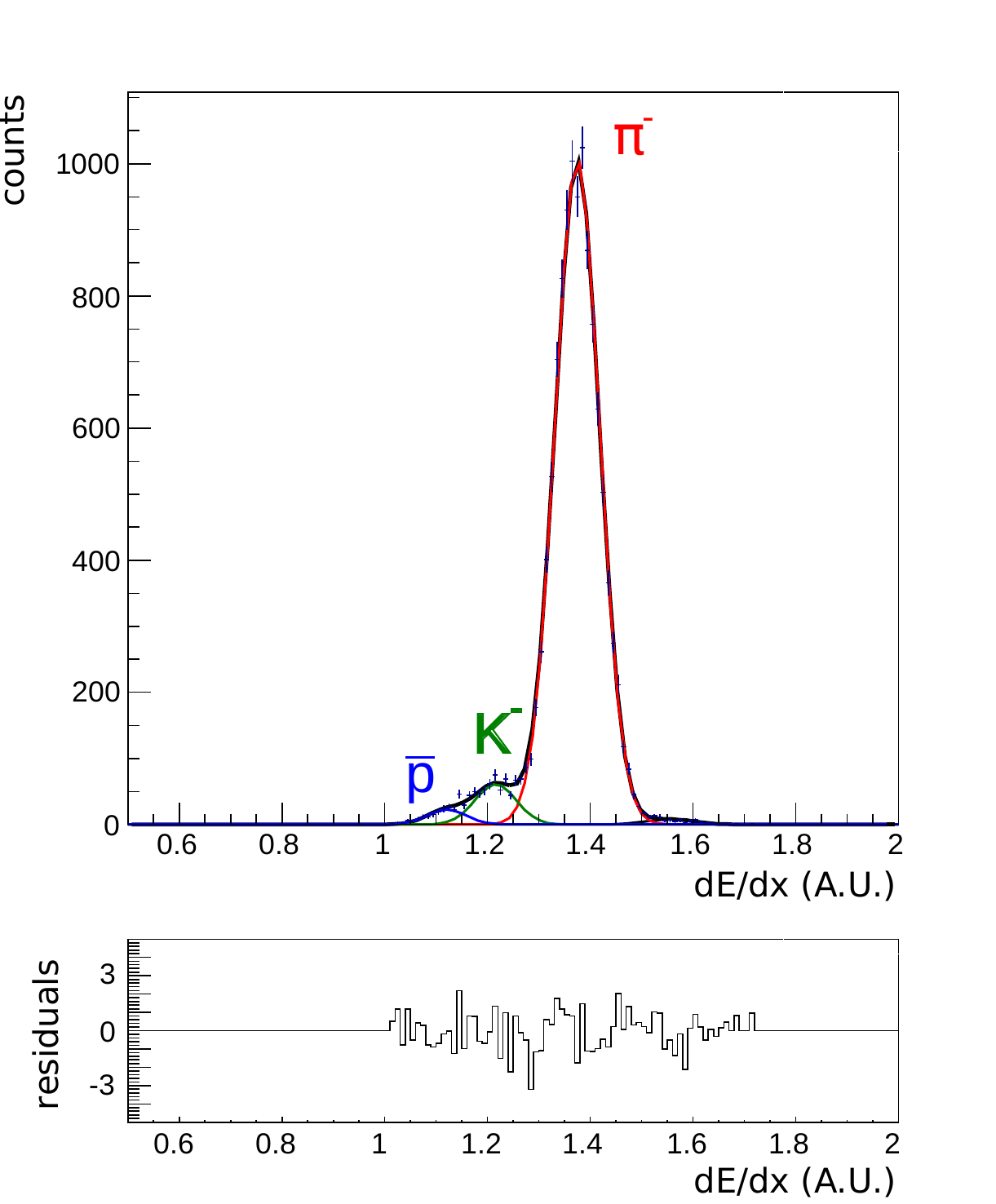}
        \includegraphics[width=0.45\textwidth]{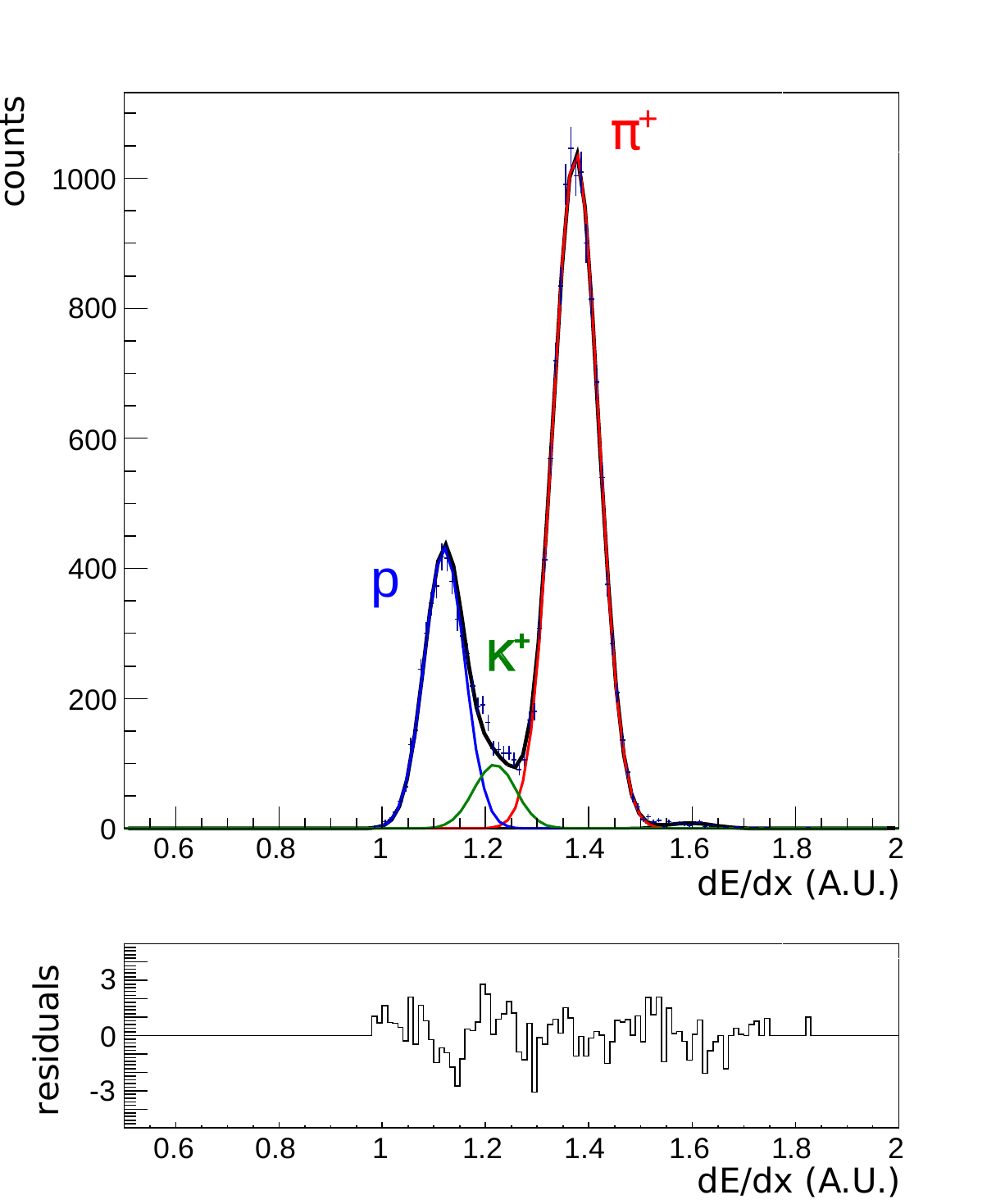}
        \end{center}
        \caption{
        The \dEdx distributions for negatively (\textit{top, left}) and positively (\textit{top, right}) charged
        particles in the bin 12.6 $\leq$~\plab~$\leq$ 15.8~\GeVc and 0.2 $\leq$~\pt~$\leq$ 0.3~\GeVc
        produced in PSD selected Be+Be collisions at 75\AGeVc.
        The fit by a sum of contributions from different particle types is shown by solid lines.
        The corresponding residuals (the difference between the data and fit divided
        by the statistical uncertainty of the data) is shown in the bottom plots.   }
        \label{fig:exfit}
\end{figure}

\begin{figure}[!ht]
        \begin{center}
        \includegraphics[width=0.45\textwidth]{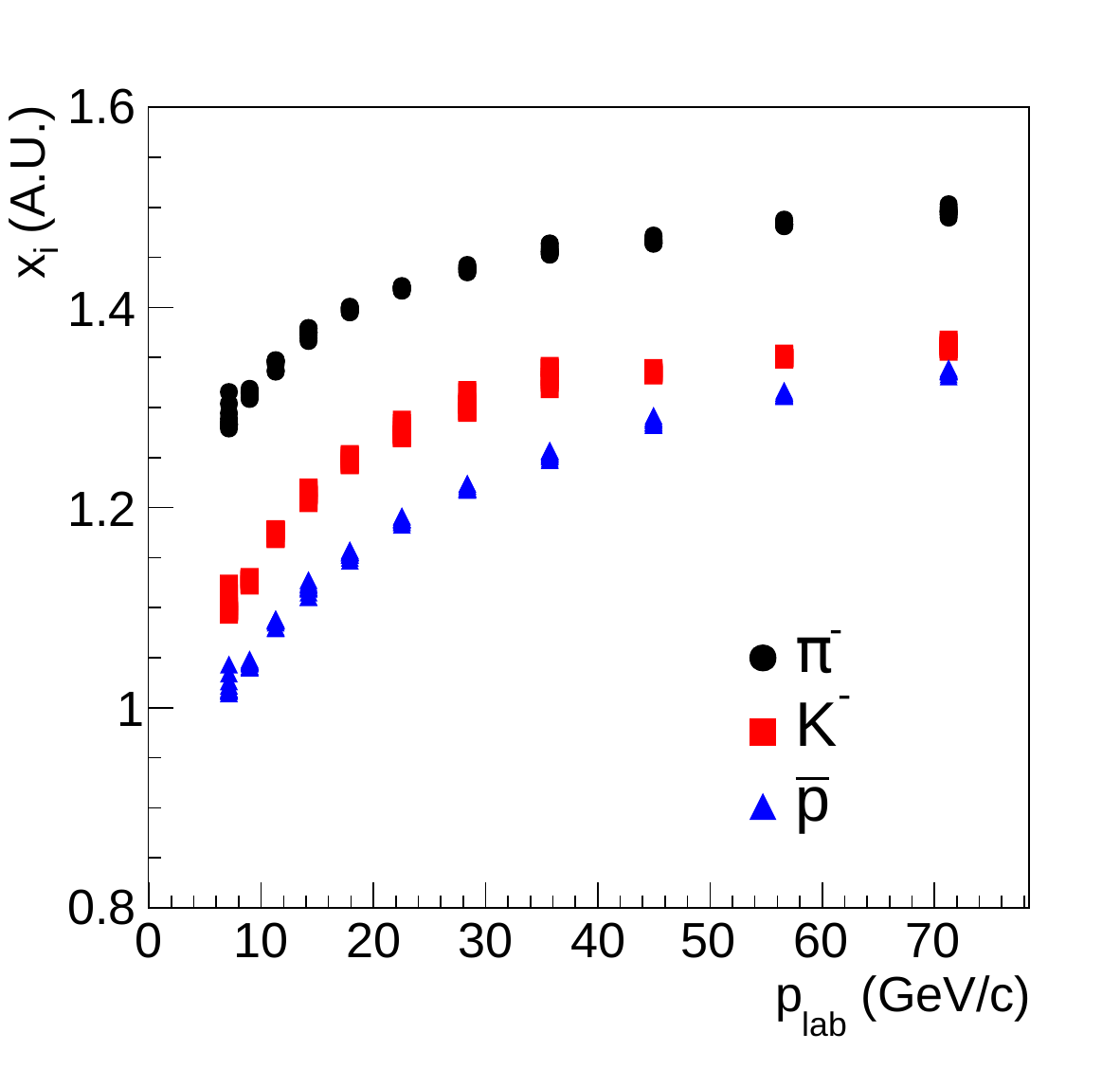}
        \includegraphics[width=0.45\textwidth]{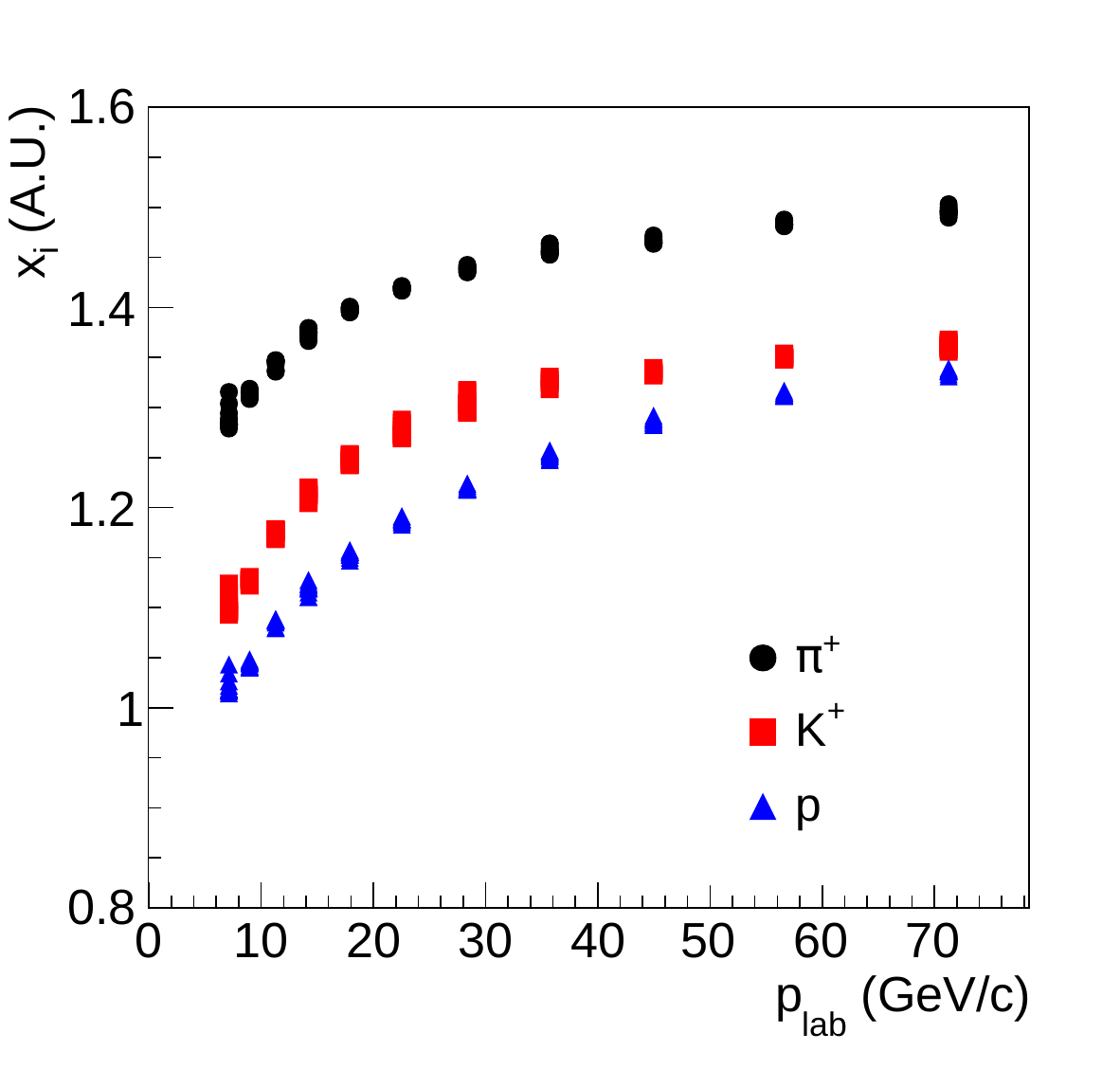}
        \end{center}
        \caption{
        Fitted peak positions in selected Be+Be interactions at 150~\AGeVc for different particles as a function 
        of \plab. The $x_i$ for different \pt at each \plab show little variation and  mostly overlap.}
        \label{fig:fitpos}
\end{figure}

Examples of fits are shown in Fig.~\ref{fig:exfit} and the values of the fitted peak positions $x_i$ are plotted in
Fig.~\ref{fig:fitpos} versus momentum for different particle types $i$ in selected Be+Be interactions at 150~\AGeVc. As expected, the values of $x_i$ increase with \plab~but do not depend on \pt.

In order to ensure good fit quality, only bins with total number of tracks grater than 1000 (500 in the 19\AGeVc sample) are used 
for further analysis. The Bethe-Bloch curves for different particle types cross each other at low values of 
the total momentum. Thus, the technique is not sufficient for particle identification at low \plab~and
bins with \plab~< 3.98~\GeVc (bins 1-5) are excluded from the analysis based solely on \dEdx measurements.

\subsubsection{Identification based on time of flight and energy loss measurements
($tof$-\dEdx)}
\label{sec:tof_id}
Identification of $\pi^{+}$, $\pi^{-}$, $K^{+}$, $K^{-}$, $p$ and $\bar{p}$
 at low momenta (from 2-8~\GeVc)
is possible when measurement of \dEdx  is combined with time of flight information $tof$.
Timing signals from the constant-fraction discriminators and signal amplitude information are
recorded for each tile of the ToF-L/R walls.
Only hits which satisfy quality criteria~(see Ref.~\cite{Anticic:2011ny} for detail)
are selected for the analysis. The coordinates of the track intersection with the front face  
are used to match the track to tiles with valid $tof$ hits. The position of
the extrapolation point on the scintillator tile is used to correct the measured value of $tof$ for
the propagation time of the light signal. The distribution of the difference between the corrected
$tof$ measurement and the value calculated from the extrapolated track trajectory length 
with the assumed mass hypothesis can be well described by a Gaussian with standard deviation 
of 80 ps for ToF-R and 100 ps for ToF-L. 
These values represent the $tof$ resolution including all detector effects.

Momentum phase space is subdivided into bins of 1 \GeVc in \plab~and 0.1 \GeVc~in \pt. Only bins with more than
200 (500 in the 150\AGeVc sample) entries were used for extracting yields with the $tof$-\dEdx method.

\begin{figure}[!ht]
        \begin{center}
        \includegraphics[width=0.45\textwidth]{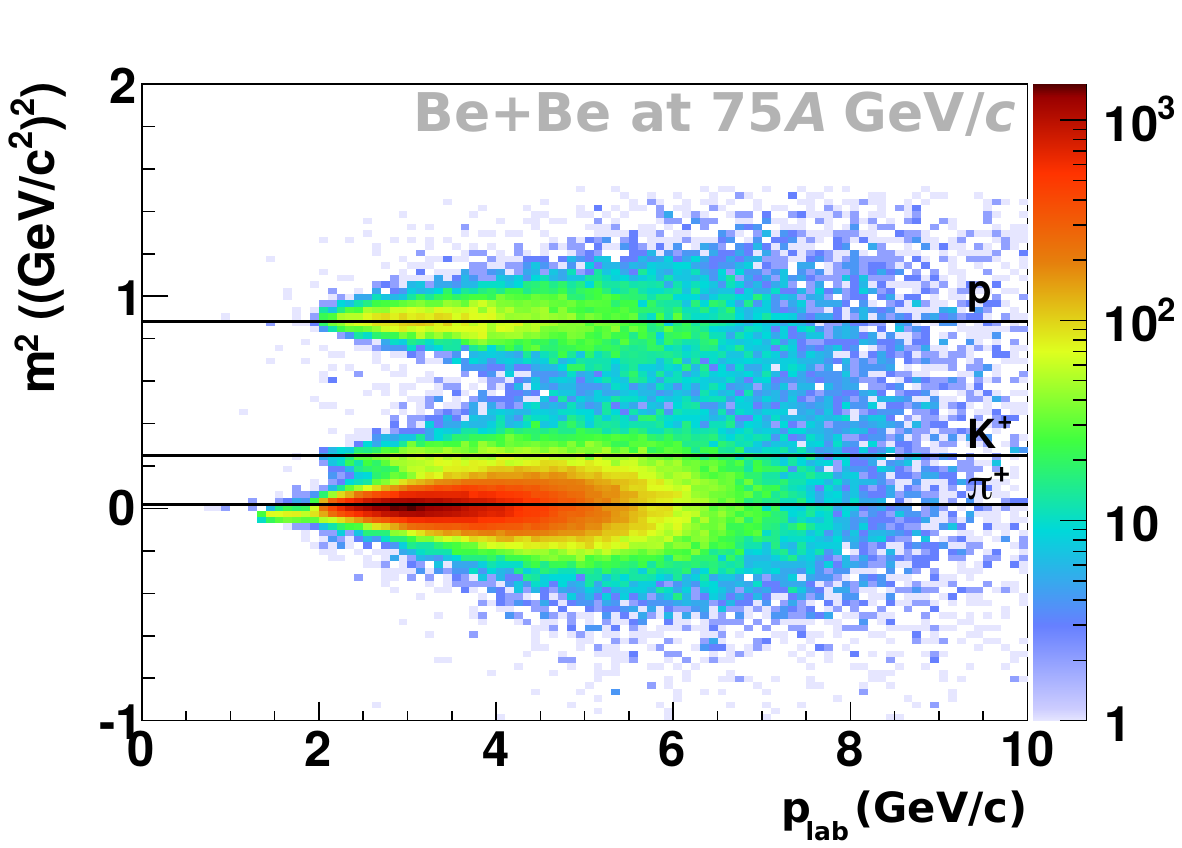}
        \includegraphics[width=0.45\textwidth]{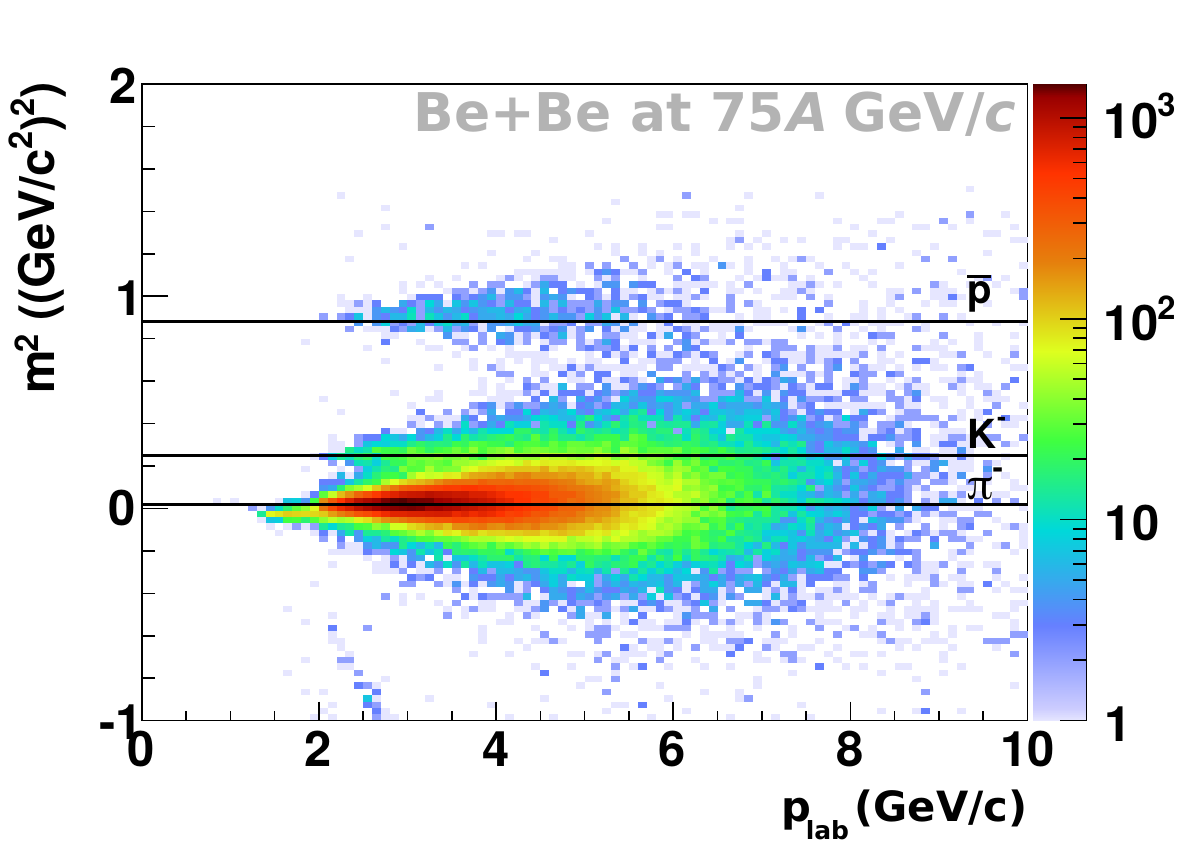}
        \end{center}
        \caption{Mass squared derived using time-of-flight measured by ToF-R ({\em right}) and
             ToF-L ({\em left}) versus laboratory momentum for particles produced in PSD selected Be+Be collisions
             at 75\AGeVc (target inserted configuration). The lines show the expected mass squared values for different hadrons species.}
        \label{fig:m2vsp}
\end{figure}
The square of the particle mass $m^2$ is obtained from $tof$, the momentum \p and
the fitted trajectory length $l$:
\begin{equation} 
m^2=(cp)^2
\left(\frac{c^2~tof^2}{l^2}-1 \right)~.
\label{eq:m2}
\end{equation}
For illustration distributions of $m^2$ versus \plab~are plotted in Fig.~\ref{fig:m2vsp}
for positively ({\em left}) and negatively ({\em right}) charged hadrons produced in Be+Be interactions at 75\AGeVc.
Bands which correspond to different particle types are visible. Separation between pions and kaons is possible 
up to momenta of about 5~\GeVc, between pions and protons up to about 8~\GeVc.

\begin{figure}[!ht]
        \begin{center}
        \includegraphics[width=0.45\textwidth]{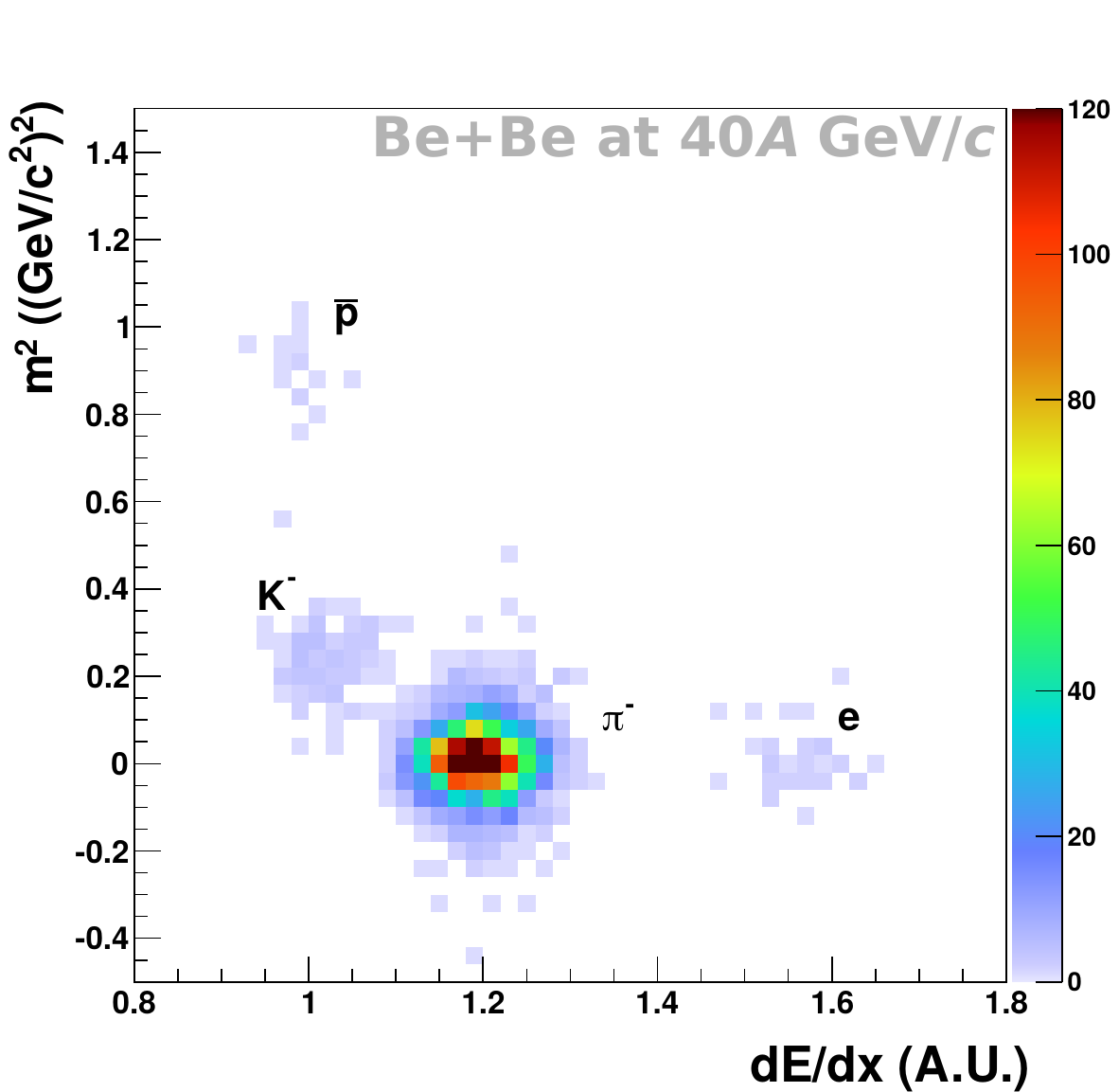}
        \includegraphics[width=0.45\textwidth]{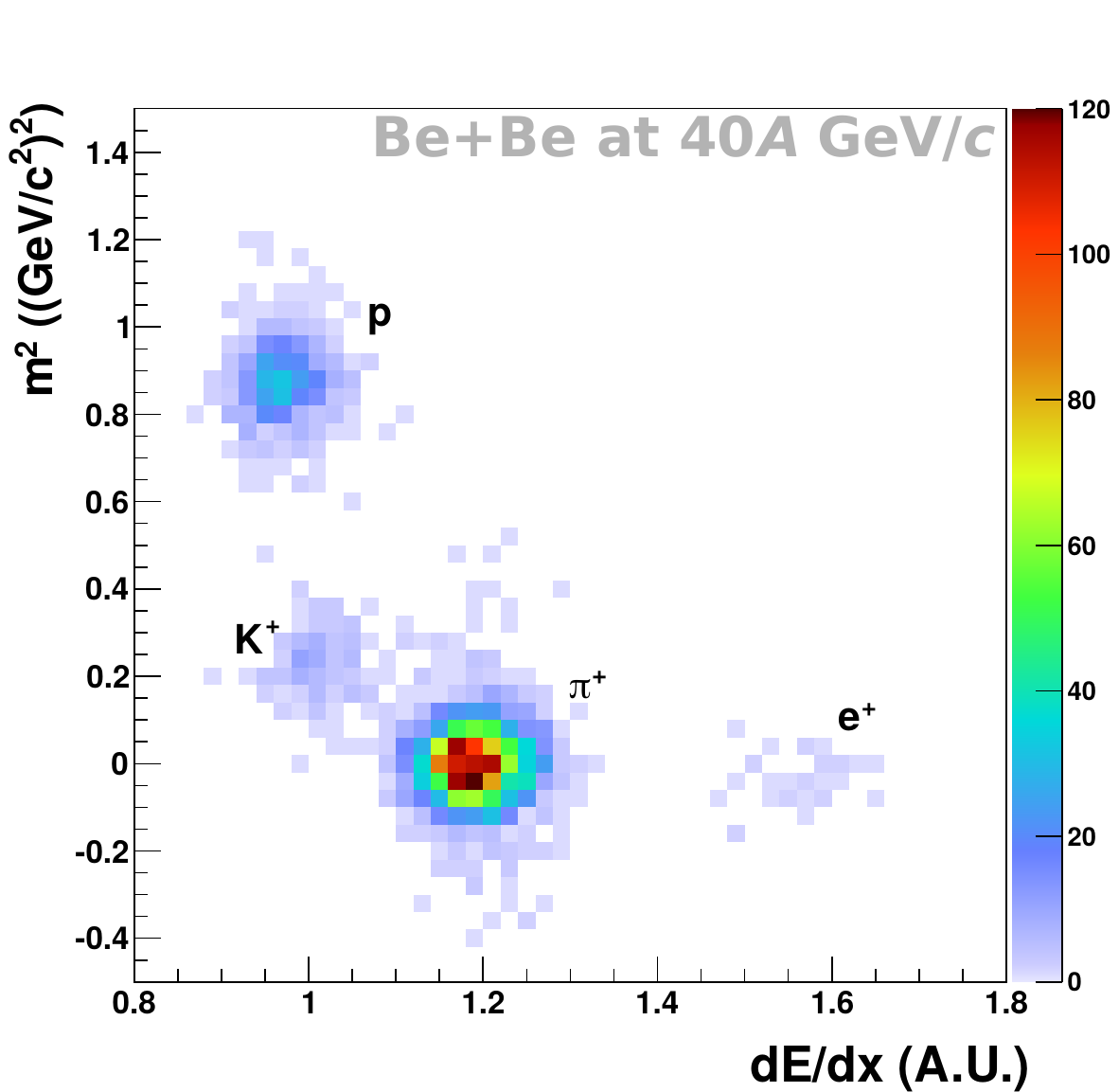}
        \end{center}
        \caption{
                 Particle number distribution in the $m^2$-\dEdx plane for negatively ({\em left}) 
                 and positively ({\em right}) charged particles with momenta 
                 3 $<$ \plab $<$ 4~\GeVc and 0.3 $<$ \pt $<$ 0.4~\GeVc) for 
                 PSD selected Be+Be collisions at 40\AGeVc. 
                }
        \label{fig:tofdedx}
\end{figure}

Example distributions of particles in the $m^2$--\dEdx plane for the selected Be+Be interactions 
at 40\AGeVc are presented in Fig.~\ref{fig:tofdedx}. Simultaneous \dEdx and $tof$ measurements lead to 
improved separation between different hadron types. In this case a simple Gaussian parametrization of 
the \dEdx distribution for a given hadron type can be used.

The $tof$-\dEdx identification method proceeds by fitting the 2-dimensional distribution of particles 
in the \dEdx-$m^2$ plane. Fits were performed in the momentum range from 1-8~\GeVc and 
transverse momentum range 0-1~\GeVc. For positively charged particles the fit function
included contributions of $p$, $K^+$, $\pi^+$ and $e^+$, and for negatively charged particles 
the corresponding anti-particles were considered. The fit function for a given particle type was assumed 
to be a product of a Gauss function in \dEdx and a sum of two Gauss functions in $m^2$ 
(in order to describe the broadening of the $m^2$ distributions with momentum ). 
In order to simplify the notation in the fit formulae, the peak positions of the \dEdx and $m^2$ Gaussians
for particle type $j$ are denoted as $x_{j}$ and $y_{j}$, respectively. The contribution of a reconstructed 
particle track to the fit function reads:

\begin{equation}
\begin{aligned}
\rho(x,y) &= \sum _{j = p,\pi,K,e}\rho_j(x, y)\\ \noalign{\vskip20pt}
&= \sum _{j}\ \frac{N_j}{2\pi\ \sigma_{x}}\ exp \left[ {-\frac{(x-x_j)^2}{2 \sigma^2 _{x}}} \right] \Bigg(
          \frac{f}{\sigma_{y_1}}\ exp \left [{\frac{(y-y_j )^2}{2 \sigma^2 _{y_1}}} \right] 
        + \frac{(1-f)}{\sigma_{y_2}} exp \left[ {\frac{(y-y_j)^2}{2 \sigma^2 _{y_2}}} \right] \Bigg) ,
        \end{aligned}
\label{eq:2dgaus}
\end{equation}

\vspace{20pt}
where $N_{j}$  and $f$ are amplitude parameters, $x_j$, $\sigma_{x}$ are means and width of the \dEdx
Gaussians and $y_j$, $\sigma_{y1}$, $\sigma_{y2}$ are mean and width of the $m^2$ Gaussians, respectively.
The total number of parameters in Eq.~\ref{eq:2dgaus} is 16. Imposing the constraint of normalisation to
the total number of tracks $N$ in the kinematic bin 
\begin{equation}
N=\sum _{i} N_i
\end{equation}
the number of parameters is reduced to 15. Two additional assumptions were adopted:
\begin{enumerate}[(i)]
\item the fitted amplitudes were required to be greater than or equal to 0,
\item $\sigma_{y1} < \sigma_{y2} $ and $f > 0.7 $, the ''core'' distribution dominates the
$m^2$ fit.
\end{enumerate}

\begin{figure}[!ht]
        \begin{center}
        \includegraphics[width=0.55\textwidth]{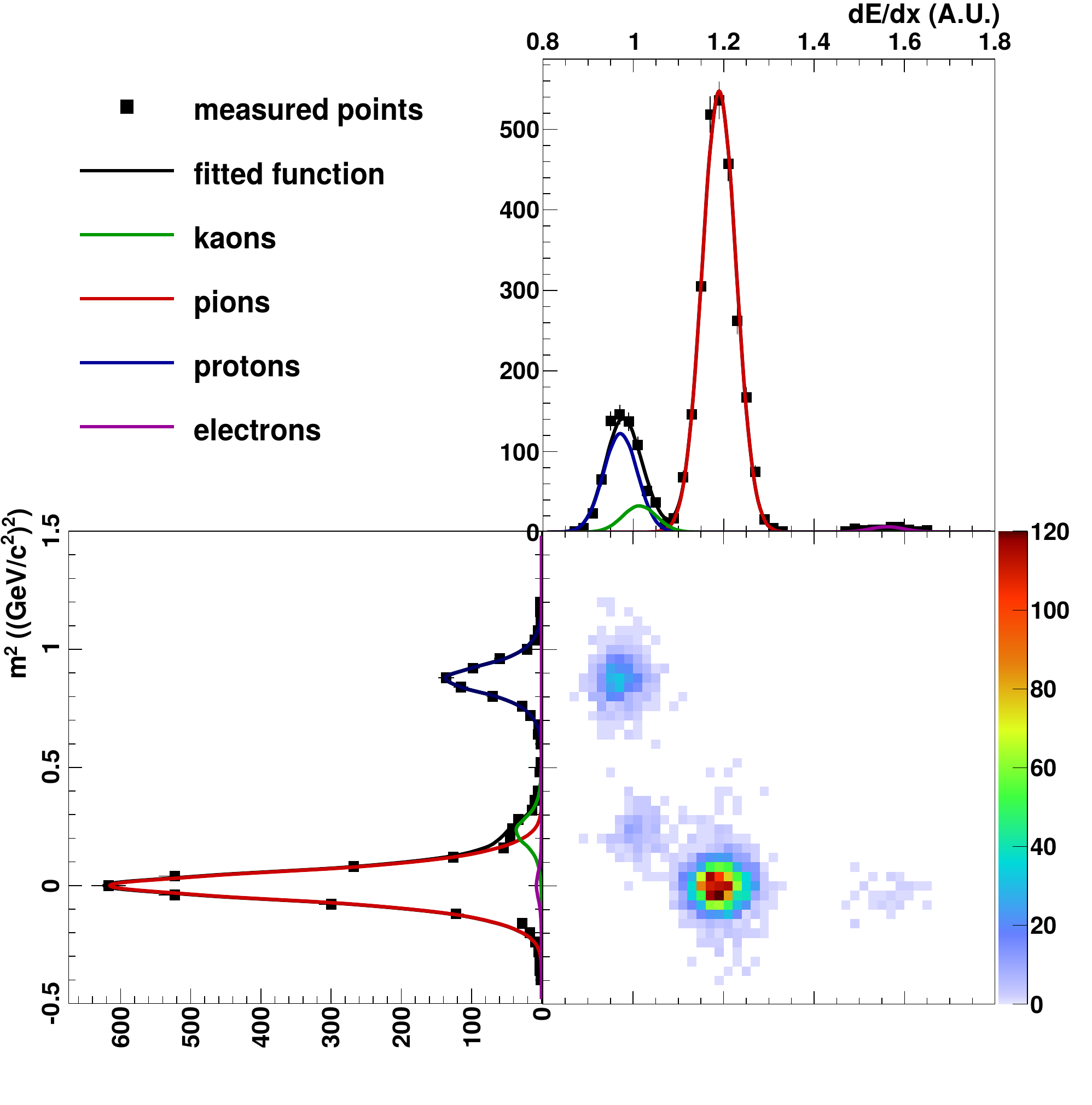}
        \includegraphics[width=0.34\textwidth]{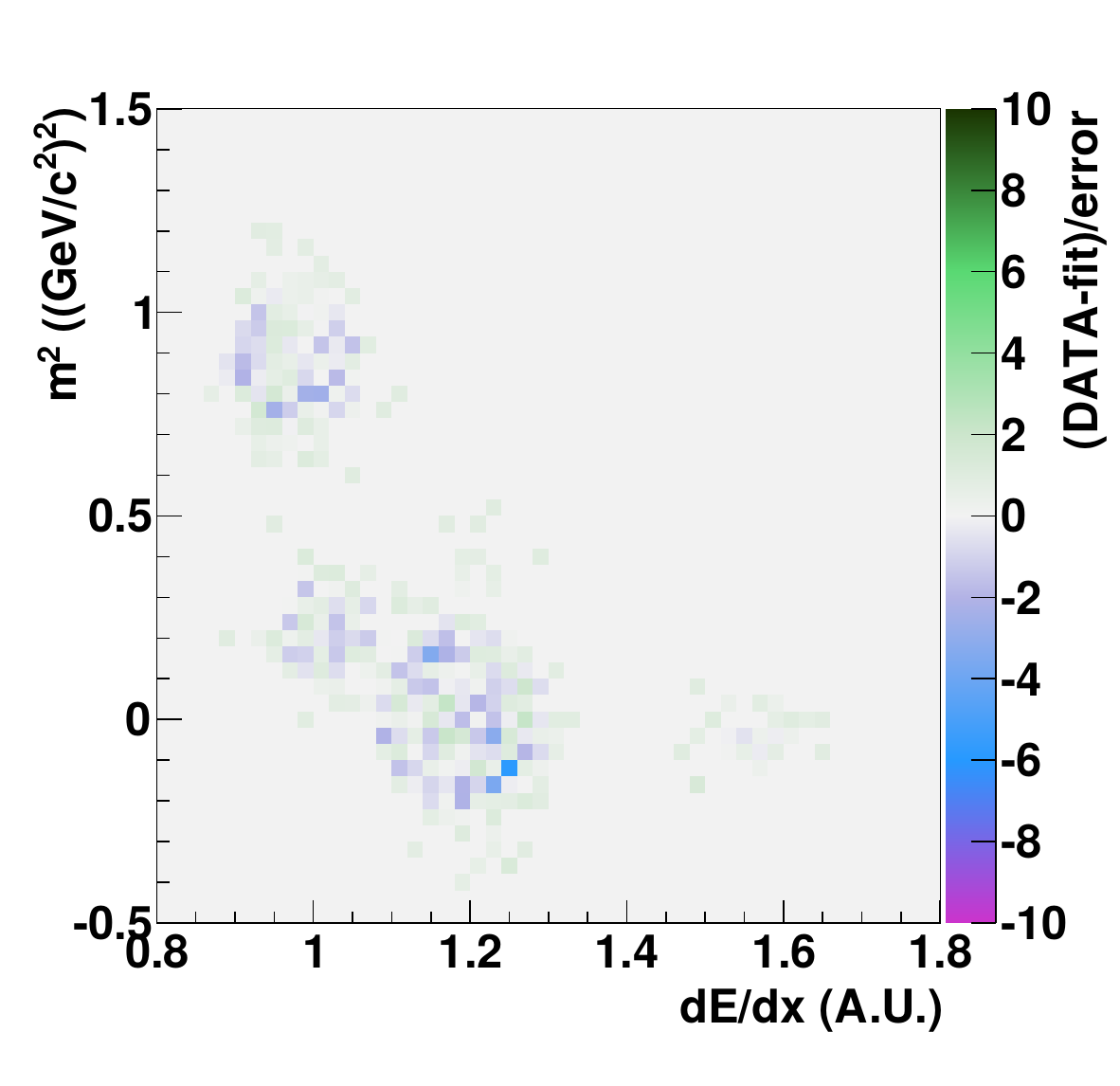}
        \end{center}
        \caption{
                 Example of the $tof$-\dEdx fit (Eq.~\protect{\ref{eq:2dgaus}}) obtained in a single phase-space bin 
                 (3 $<$ \plab $<$ 4~\GeVc and 0.3 $<$ \pt $<$ 0.4~\GeVc) for positively charged particles in PSD
                 selected Be+Be collisions at 40\AGeVc. Lines show projections of the fits for pions (red), 
                 kaons (green), protons (blue) and electrons (magenta). Bottom right panel shows the fit residuals.
                }
        \label{fig:exmtof}
\end{figure}

An example of the $tof$-\dEdx fit obtained in a single phase-space bin for positively charged particles 
in PSD selected Be+Be collisions at 40\AGeVc is shown in Fig.~\ref{fig:exmtof}.

The $tof$-\dEdx method allows to fit the kaon yield close to mid-rapidity. This
is not possible using the \dEdx method. Moreover, the kinematic domain in which pion and proton yields 
can be fitted is enlarged. The results from both methods partly overlap
at the highest beam momenta. In these regions the results from the \dEdx method were selected since
they have smaller uncertainties.

\subsubsection{Probability method}
\label{sec:propability}

The fit results allow to calculate the probability $P_i$ that a measured particle is
of a given type $i$ = $\pi, K, p, e$. For the \dEdx fits
(see Eq.~\ref{Eq:AsymGaus}) one gets:
\begin{equation}
P_{i}^{dE/dx}(\plab,\pt)=\frac{\rho_{i}^{dE/dx}(\plab,\pt)}
{\sum\limits_{i=\pi, K, p, e}^{} \rho_{i}^{dE/dx}(\plab, \pt)},
\label{eq:propdedx}
\end{equation}
where $\rho_{i}$ is the value of the fitted function in a given (\plab, \pt) bin
calculated for \dEdx  of the particle.

\begin{figure}[!ht]
        \begin{center}
        \includegraphics[width=0.45\textwidth]{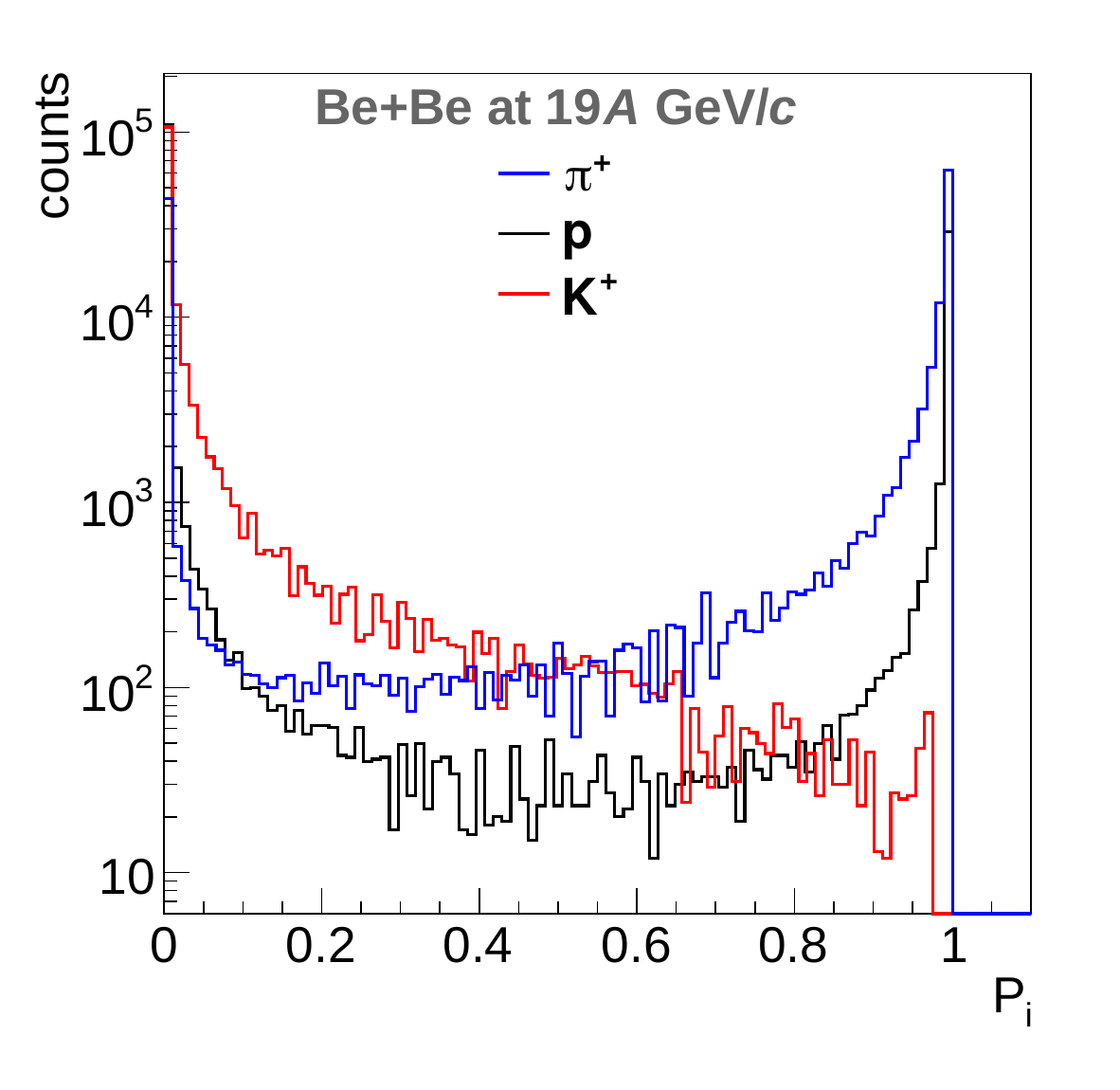}
        \includegraphics[width=0.45\textwidth]{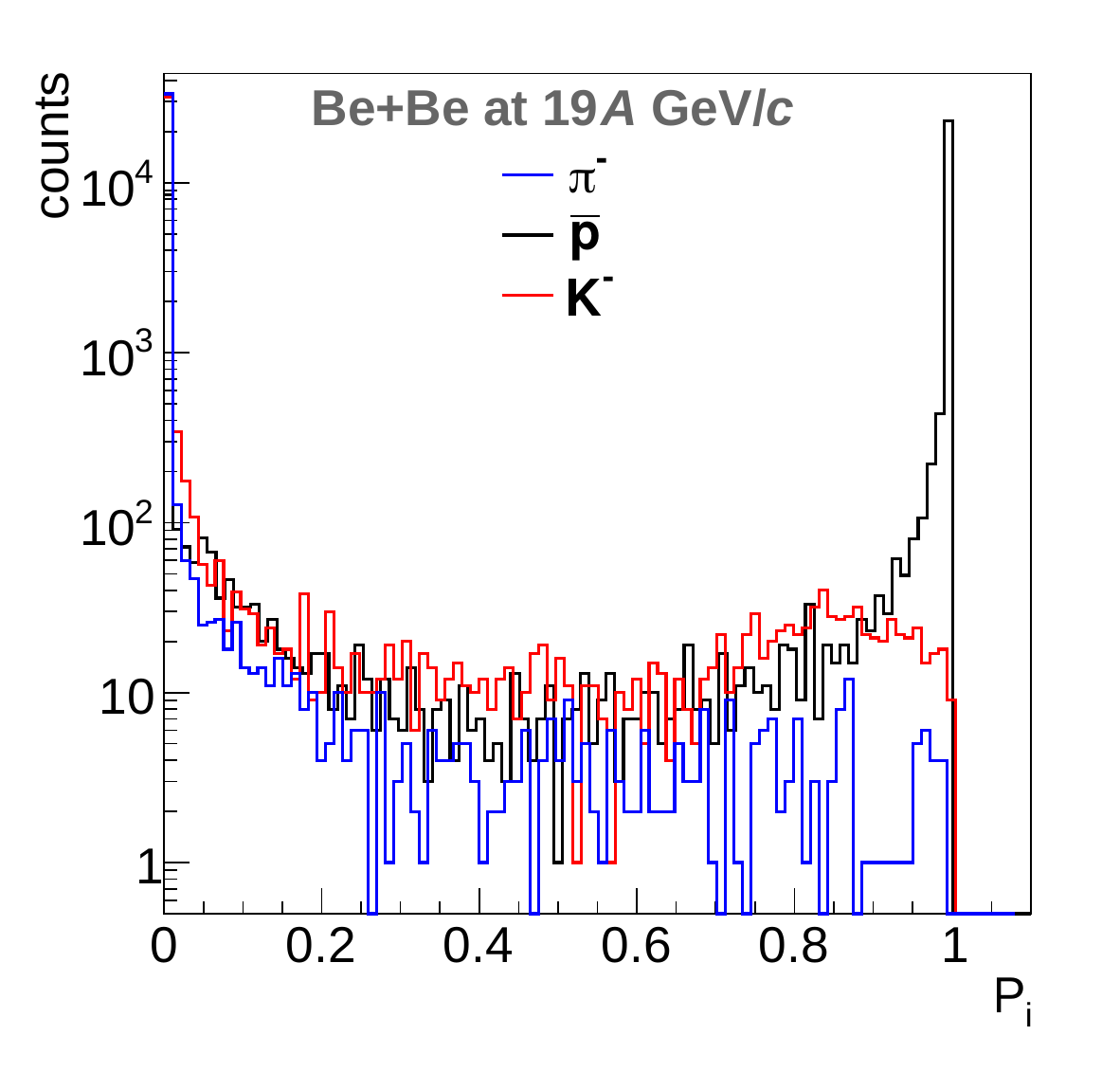}\\
        \includegraphics[width=0.45\textwidth]{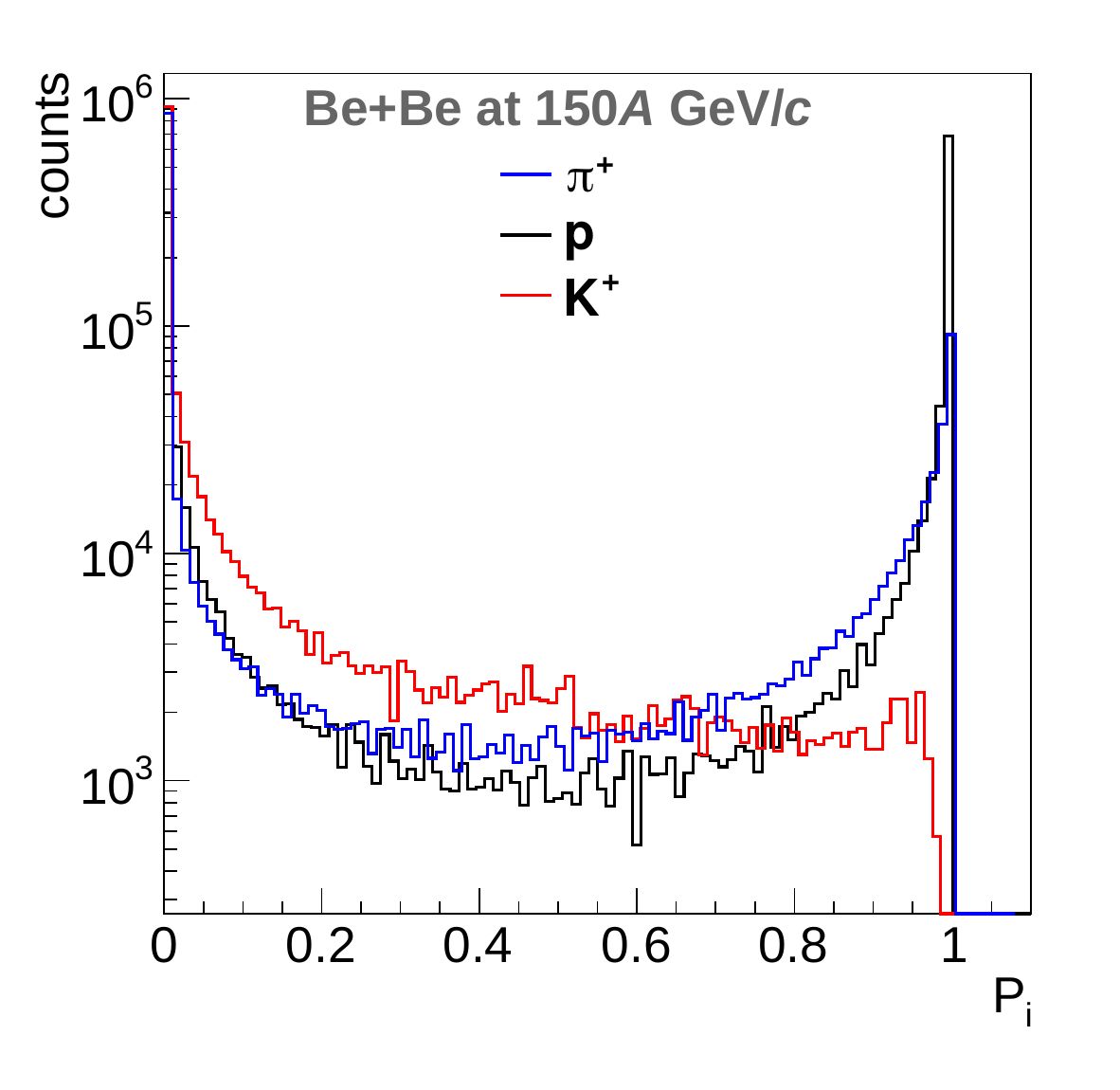}
        \includegraphics[width=0.45\textwidth]{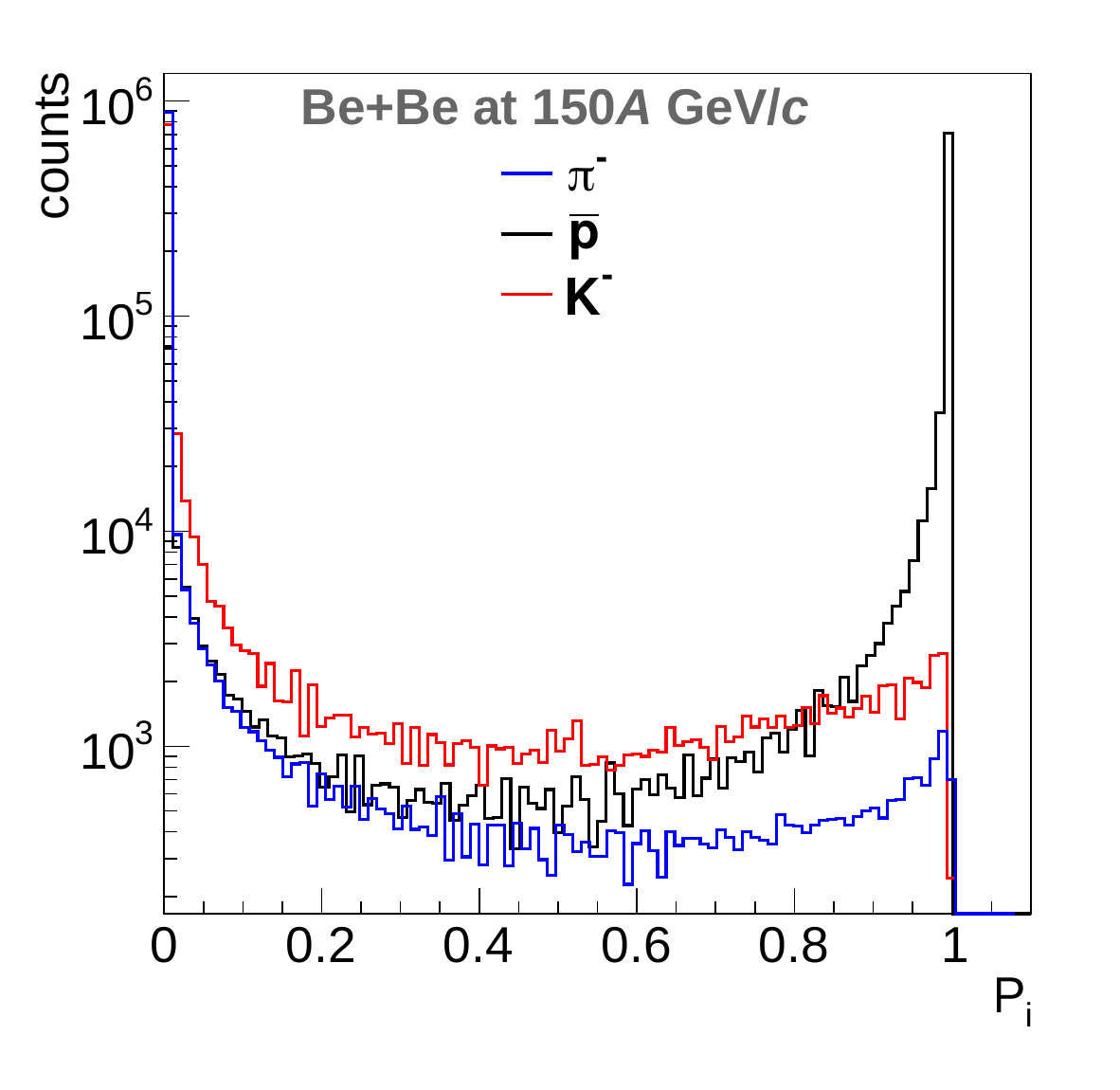}
        \end{center}
        \caption{
                 Probability of a track being a pion, kaon, proton for positively \textit{(left)} 
                 and negatively \textit{(right)} charged particles from \dEdx measurements in PSD selected Be+Be collisions 
                 at 19$A$ and 150\AGeVc.
                 }
        \label{fig:propdedx}
\end{figure}

Similarly the $tof$-\dEdx fits (see Eq.~\ref{eq:2dgaus}) give the particle type probability as
\begin{equation}
P_{i}^{dE/dx, m^2}(\plab,\pt) = \frac{\rho_{i}^{dE/dx,m^2}(\plab,\pt)}
{\sum\limits_{i=\pi, K, p,e}^{} \rho_{i}^{dE/dx,m^2}(\plab, \pt)}.
\label{eq:proptof}
\end{equation}

\begin{figure}[!ht]
        \begin{center}
        \includegraphics[width=0.45\textwidth, trim={0.5cm 0.5cm 0.5cm 0.5cm}, clip]{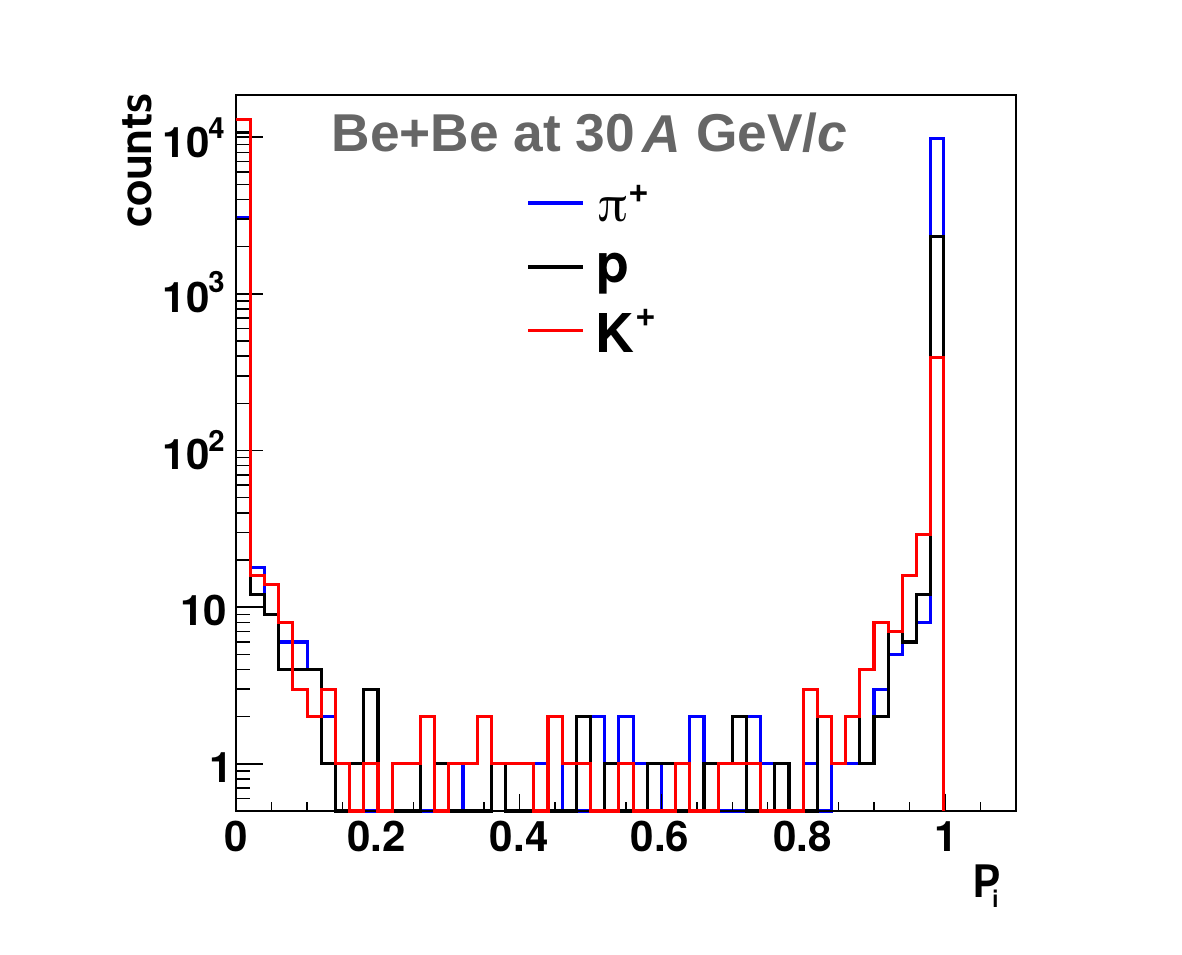}
        \includegraphics[width=0.45\textwidth, trim={0.5cm 0.5cm 0.5cm 0.5cm}, clip]{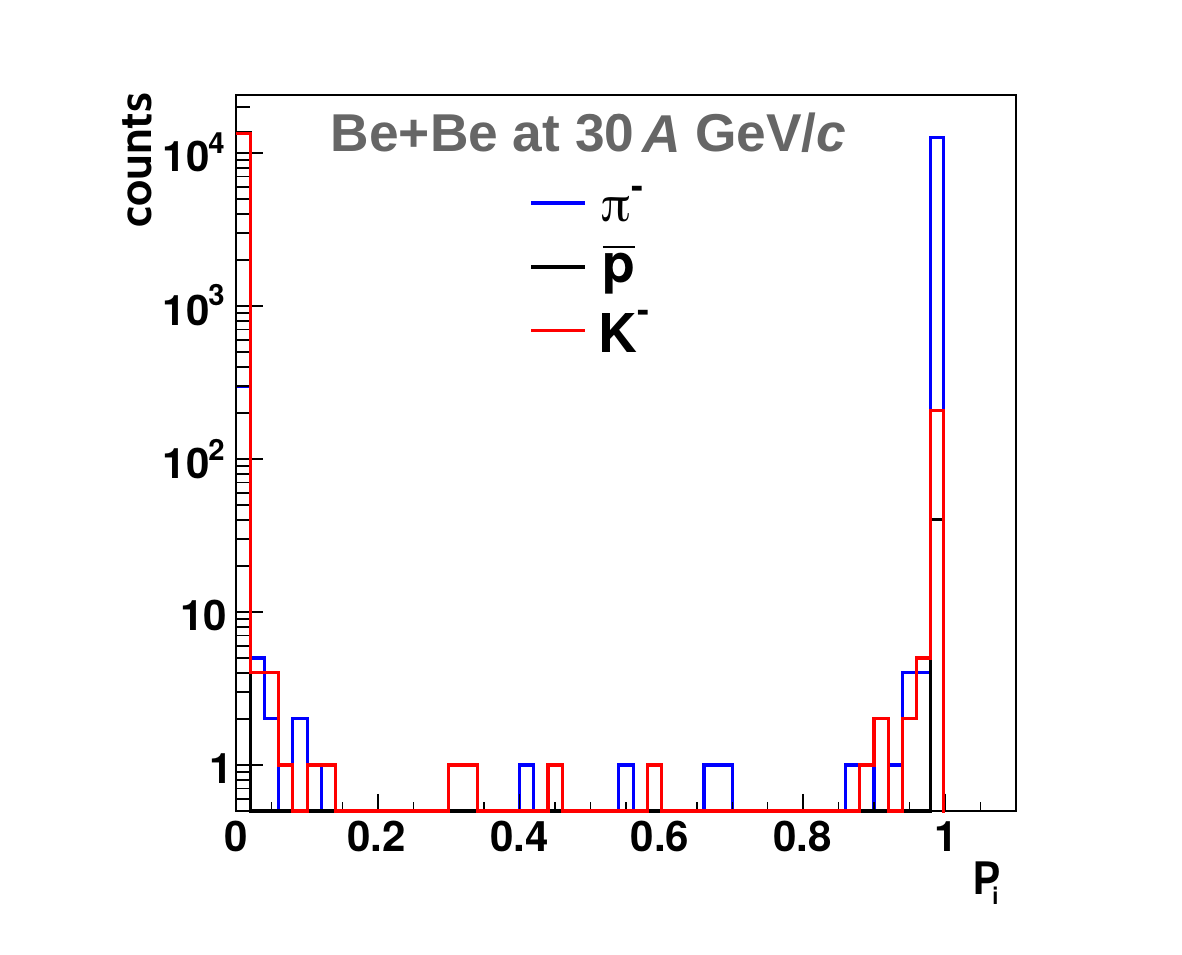}\\
        \includegraphics[width=0.45\textwidth, trim={0.5cm 0.5cm 0.5cm 0.5cm}, clip]{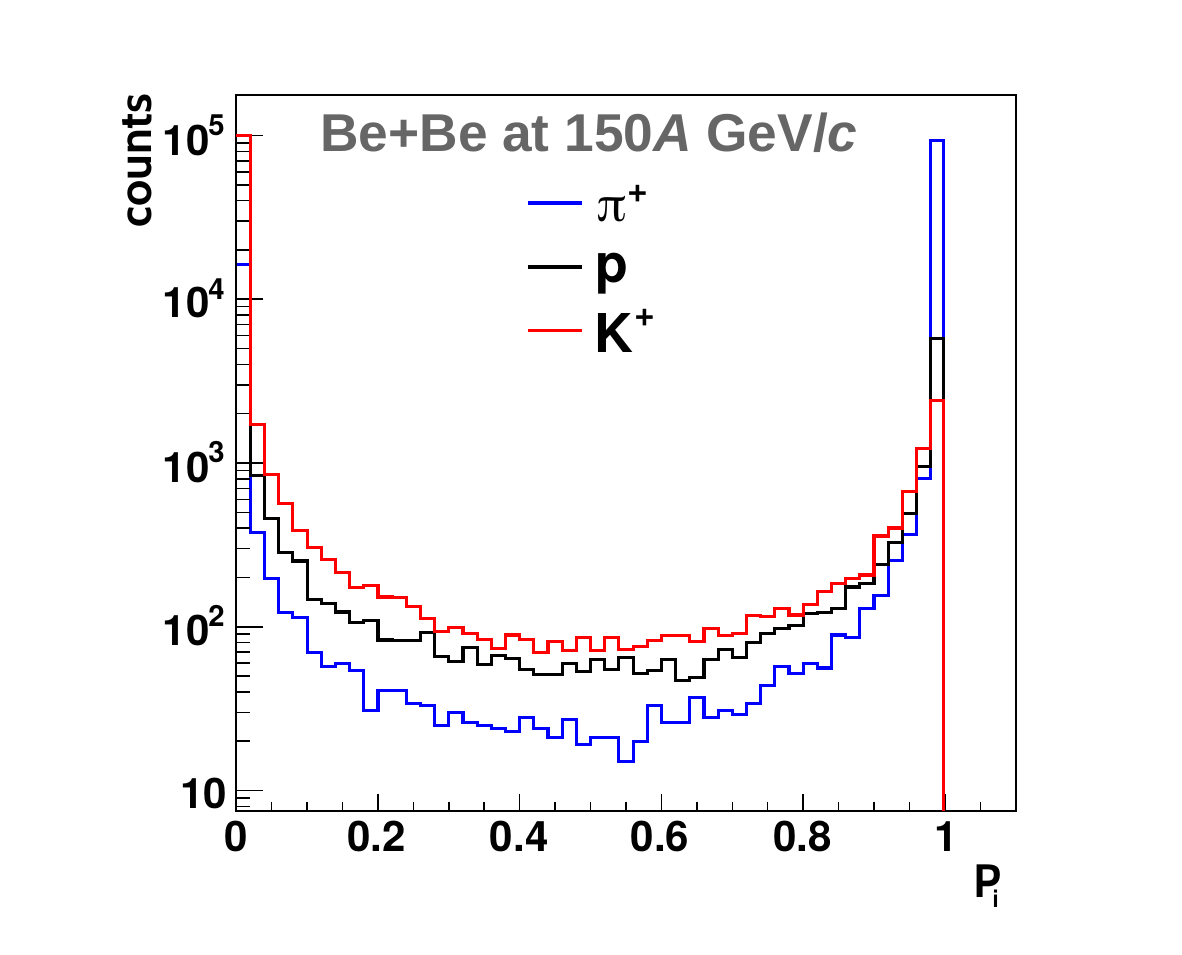}
        \includegraphics[width=0.45\textwidth, trim={0.5cm 0.5cm 0.5cm 0.5cm}, clip]{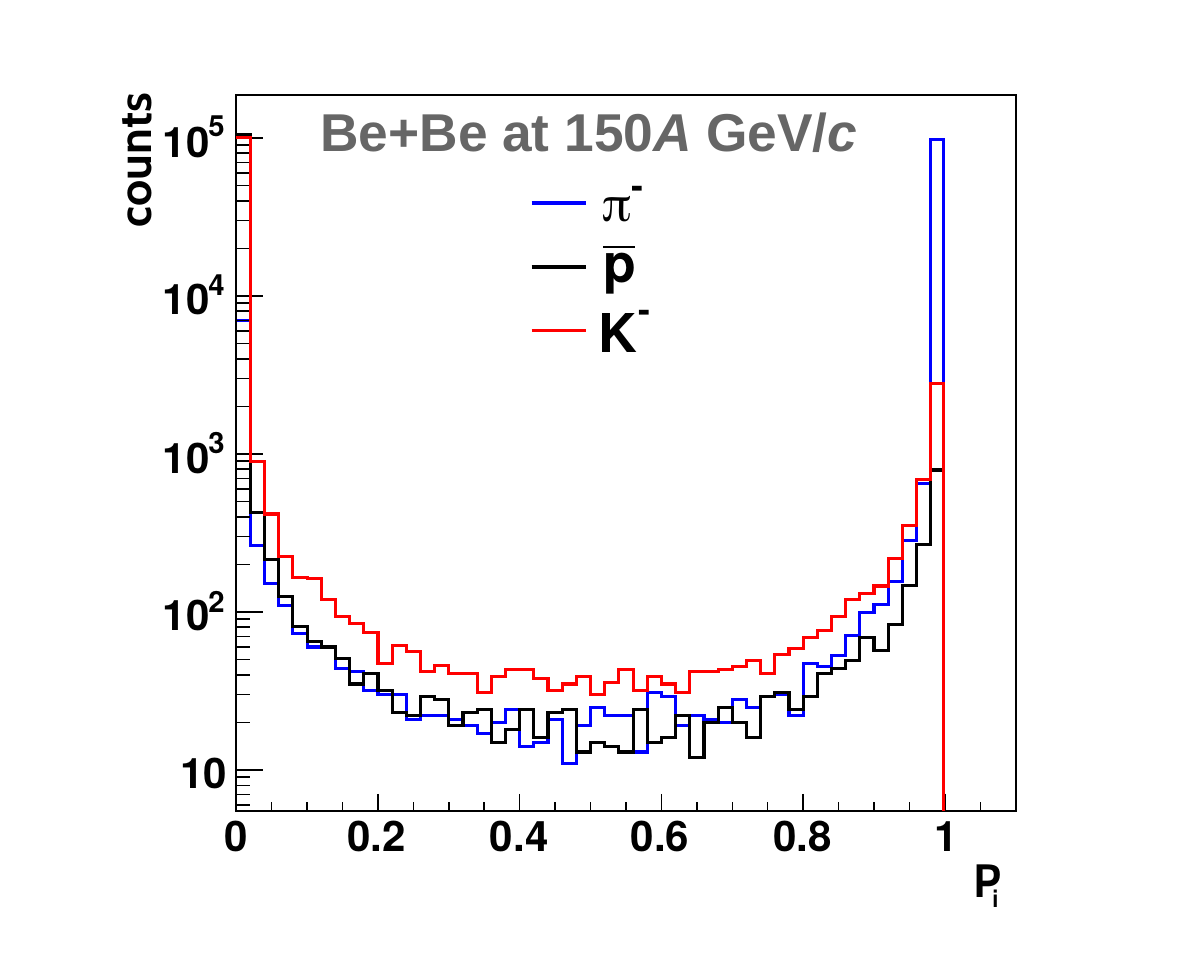}
        \end{center}
        \caption{
                 Probability of a track being a pion, kaon, proton for positively (\textit{left}) 
                 and negatively (\textit{right}) charged particles from $tof$-\dEdx measurements 
                 in PSD selected Be+Be collisions at 30$A$ (\textit{top}) and 150\AGeVc (\textit{bottom)}.}
        \label{fig:proptof}
\end{figure}

For illustration, particle type probability distributions for positively and negatively charged particles
produced in PSD selected Be+Be collisions at 19$A$ and 150\AGeVc are presented in Fig.~\ref{fig:propdedx}
for the \dEdx fits and in Fig.~\ref{fig:proptof} for the $tof$-\dEdx fits. In the case of 
perfect particle type identification the probability distributions in Figs.~\ref{fig:propdedx} 
and \ref{fig:proptof} will show entries at 0 and 1 only. In the case of incomplete particle identification (overlapping \dEdx or $tof$-\dEdx distributions) values between these extremes will also be populated.

The probability method allows to transform fit results performed in
($p_\text{lab}$, $p_{T}$) bins to results in (\y, \pt) bins.
Hence, for the probability method the mean number of identified particles in a given kinematical bin
(e.g. (\plab, \pt)) is given by~\cite{Rustamov:2012bx}:
\begin{equation}
n[i]^{raw}_{dEdx}(\y,\pt)=\frac{1}{N_{ev}}\sum_{j=1}^{N_{trk}}P_{i}^{dEdx}(\plab,\pt) ,
\label{eq:ntrkdEdx}
\end{equation}
for the \dEdx identification method and:
\begin{equation}
n[i]^{raw}_{dEdx,m^2}(\y,\pt)=\frac{1}{N_{ev}}\sum_{j=1}^{N_{trk}}P_{i}^{dEdx,m^2}(\plab,\pt) ,
\label{eq:ntrktof}
\end{equation}
for the $tof$-\dEdx procedure,
where $P_i$ is the probability of particle type $i$ given by
Eqs.~\ref{eq:propdedx} and~\ref{eq:proptof}, $j$ the summation index running over all entries $N_{trk}$ in the bin, $N_{ev}$ is the number of selected events.

Statistical uncertainties of multiplicities calculated with probability method were derived from the variance of the distribution of $P_i$ in the (\plab,\pt) bin:

\begin{equation}
\sigma^{2}_{n[i]^{raw}}(\y, \pt) = \frac{1}{N_{ev}}( \sum_{j=1}^{N_{trk}} P_{i}^j(\plab,\pt)^2 - (\sum_{k=1}^{N_{trk}} P_{i}^j(\plab,\pt)) / N_{trk} )
\label{eq:sigmaraw}
\end{equation}

\FloatBarrier
\subsection{Corrections and uncertainties}
\label{sec:Corrections}

In order to estimate the true number of each type of identified particle produced in Be+Be interactions 
a set of corrections was applied to the extracted raw results.
These were obtained from a simulation of the \NASixtyOne detector followed by event reconstruction
using the standard reconstruction chain. Only inelastic Be+Be interactions 
were simulated in the target material. 
The \EposLong model (version CRMC 1.5.3)~\cite{Werner:2005jf} was selected to generate primary inelastic 
interactions as it best describes the \NASixtyOne measurements. A \GeantThree based program chain
was used to track particles through the spectrometer, generate decays and secondary interactions and
simulate the detector response (for details see Ref.~\cite{Abgrall:2013pp_pim}). Simulated events were 
then processed using the standard \NASixtyOne reconstruction chain and reconstructed tracks were
matched to the simulated particles based on the cluster positions. 
The event selection was based on a dedicated simulation of the energy 
recorded by the PSD~(see Sec.~\ref{sec:centrality}).
Corrections depend on the particle identification technique 
(i.e. \dEdx or $tof$-\dEdx).
Hadrons which were not produced in the primary interaction can amount to a significant fraction of the selected
tracks. Thus a special effort was undertaken to evaluate and subtract this contribution.
The correction factors were calculated in the same bins of \y and \pt as the particle spectra.
Bins with a correction factor lower than 1.5 and higher than 4
are caused by low acceptance or high contamination of non primary particles 
and were rejected.

\subsubsection{Corrections for the \dEdx method}

The correction factor $c_i^{dEdx}$(\y,\pt) for biasing effects listed in Sec.~\ref{sec:event_selection} items
(b) - (g) was calculated as:
\begin{equation}
    c_i^{dEdx}(\y,\pt)=\frac{n[i]^{MC}_{gen}(\y, \pt)}{n[i]^{MC}_{sel} (\y, \pt)},
\label{eq:cdedx}
\end{equation}
where $n[i]^{MC}_{gen}$ is the number \textit{primary} particles in the (\y,\pt) bin for simulated events and $n[i]^{MC}_{sel}$ 
the number of reconstructed tracks passing all event and track selection cuts.
The uncertainty of $c_i^{dEdx}$(\y, \pt) was calculated assuming that the denominator 
$n[i]^{MC}_{sel}(\y, \pt)$ is a subset of the numerator $n[i]^{MC}_{gen}(\y, \pt)$ and thus 
has a binomial distribution. The uncertainty of $c_i^{dEdx}$(\y, \pt) is thus given by:
\begin{equation}
    \sigma_{c_i^{dEdx}(\y,\pt)}=c_i^{dEdx}(\y,\pt) \sqrt{\frac{n[i]^{MC}_{gen}(\y, \pt)-n[i]^{MC}_{sel}(\y, \pt)}{n[i]^{MC}_{gen} \cdot n[i]^{MC}_{sel}}},
\label{eq:sigmacdedx}
\end{equation}
The mean multiplicity for particle type $i$ in the PSD selected events in a (\y,\pt) is calculated as:
\begin{equation}
    n[i]^{dEdx}(\y,\pt) = c_i^{dEdx}(\y,\pt) \cdot n[i]^{raw}_{dEdx}(\y,\pt).
\end{equation}
\label{eq:ndedx}

\subsubsection{Corrections for the $tof$-\dEdx method}


The corrections for the $tof$-\dEdx method were calculated based on simulation and data. The ToF tile efficiency $\epsilon_{pixel}(p,\pt)$ was calculated from the data in the following way. Each reconstructed track was extrapolated to the ToF walls and if it crossed one of the ToF tiles it was classified as a hit
and summed in $n[i]_{tof}$ . Moreover, it was accepted as a valid ToF hit, summed in $n[i]_{hit}$, if the signal satisfied quality criteria given in Ref.~\cite{Anticic:2011ny}. Finally, the ToF tile efficiency $\epsilon_{pixel}(p,\pt)$ was calculated as the ratio of the number of tracks $n[i]_{tof}$ crossing a particular tile to the number of tracks $n[i]_{hit}$ with valid ToF hits. The corresponding efficiency factor $\epsilon_{pixel}(p,\pt)$ is given by: 
\begin{equation}
\epsilon_{pixel}(p,\pt) = \frac{n[i]_{hit}}{n[i]_{tof}}.
\label{eq:tile}
\end{equation}
This ToF pixel efficiency factor was used in the MC simulation by weighting each reconstructed MC track passing all event and track selection cuts by the efficiency factor of the corresponding tile.
Then, the number of selected MC tracks $n[i]^{MC}_{sel}$ becomes a sum of weights:
\begin{equation}
n[i]^{MC}_{sel} = \sum_{j=1}^{N_{trk}}
\epsilon_{pixel}^{j}(p,\pt).
\label{eq:mcseltof}
\end{equation}
Only hits in working tiles with efficiency higher than 50\% were taken into account in the identification and correction procedures.
This results in the following correction factor for biasing effects listed in Sec.~\ref{sec:event_selection} items (b)-(g) as well as ToF efficiency, $tof$-\dEdx method acceptance, secondary interactions and contribution of particles other than \textit{primary}:
\begin{equation}
    c_i^{dEdx, m^2}(\y,\pt)=\frac{n[i]^{MC}_{gen}}{n[i]^{MC}_{sel}},
\label{eq:ctof}
\end{equation}
where $n[i]^{MC}_{gen}$ is  the  number  \textit{primary} particles  in  the  (\y,\pt)  bin  for  simulated  events  and $n[i]^{MC}_{sel}$ the  number  of reconstructed tracks passing all event and track selection cuts weighted by the tile efficiency factor $\epsilon_{pixel}(p,\pt)$ given by Eg. \ref{eq:mcseltof}.

The uncertainty of $c_i^{dEdx,m^2}(\y,\pt)$ was calculated assuming that the denominator $n[i]^{MC}_{sel}(y,pT)$ is a subset of the nominator $n[i]^{MC}_{gen}(y,pT)$ and thus has a binomial distribution. The uncertainty of $c_i^{dEdx, m^2}$(\y, \pt) was calculated as follows:
\begin{equation}
     \sigma_{c_i^{dEdx,m^2}}(\y,\pt)=c_i^{dEdx, m^2}(\y,\pt) \sqrt{\frac{n[i]^{MC}_{gen}(\y, \pt)-n[i]^{MC}_{sel}(\y, \pt)}{n[i]^{MC}_{gen} \cdot n[i]^{MC}_{sel}}},
\label{eq:sigmactof}
\end{equation}

The mean multiplicity for particle type $i$ in the PSD selected events in a (\y,\pt) bin for the $tof$-\dEdx method is defined by:
\begin{equation}
\label{eq:ntof}
    n[i]^{dEdx,m^2}(\y,\pt) =  c_i^{dEdx, m^2}(\y,\pt) \cdot n[i]^{raw}_{dEdx,m^2} ,
\end{equation}

%
%
%
%
%
%

\FloatBarrier
\subsection{Corrected spectra}

Final spectra of different types of hadrons produced in Be+Be interactions are calculated as:
\begin{equation}
\label{finalresdEdx}
\frac{d^{2}n}{dydp_T} =  \frac{1}{\Delta\y\cdot\Delta\pt \cdot N_{ev}} \cdot c_i(\y,\pt) \cdot n[i]^{raw}(\y,\pt),
\end{equation}
where $\Delta\y$ and $\Delta\pt$ are the bin sizes, $N_{ev}$ is the total number of accepted events, $n[i]^{raw}$ represents the mean multiplicity for particle type $i$ in the \y,\pt bin obtained as $n[i]^{dEdx}$ and $n[i]^{dEdx,m^2}$ for the \dEdx and $tof$-\dEdx identification method, respectively and finally the correction factor $c_i$ stands for $c_i^{dEdx}(\y,\pt)$~(Eq.~\ref{eq:cdedx}) or $c_i^{dEdx,m^2}(\y,\pt)$~(Eq.~\ref{eq:ctof})
for the \dEdx and $tof$-\dEdx identification method, respectively.

\begin{figure*}
\begin{center}
\newcolumntype{S}{>{\centering} m{0.01\textwidth} }
\newcolumntype{A}{>{\centering} m{0.21\textwidth} }
\hspace{-10mm}
\begin{tabular}{S A A A A A A}
& \hspace{-12mm}\small19\AGeVc & \hspace{-22mm}\small30\AGeVc & \hspace{-32mm}\small40\AGeVc & \hspace{-42mm}\small75\AGeVc  & \hspace{-53mm}\small150\AGeVc
\tabularnewline
\hspace{-7mm}$\pi^{-}$ &
\hspace{-15mm}
\includegraphics[width=0.2\textwidth]{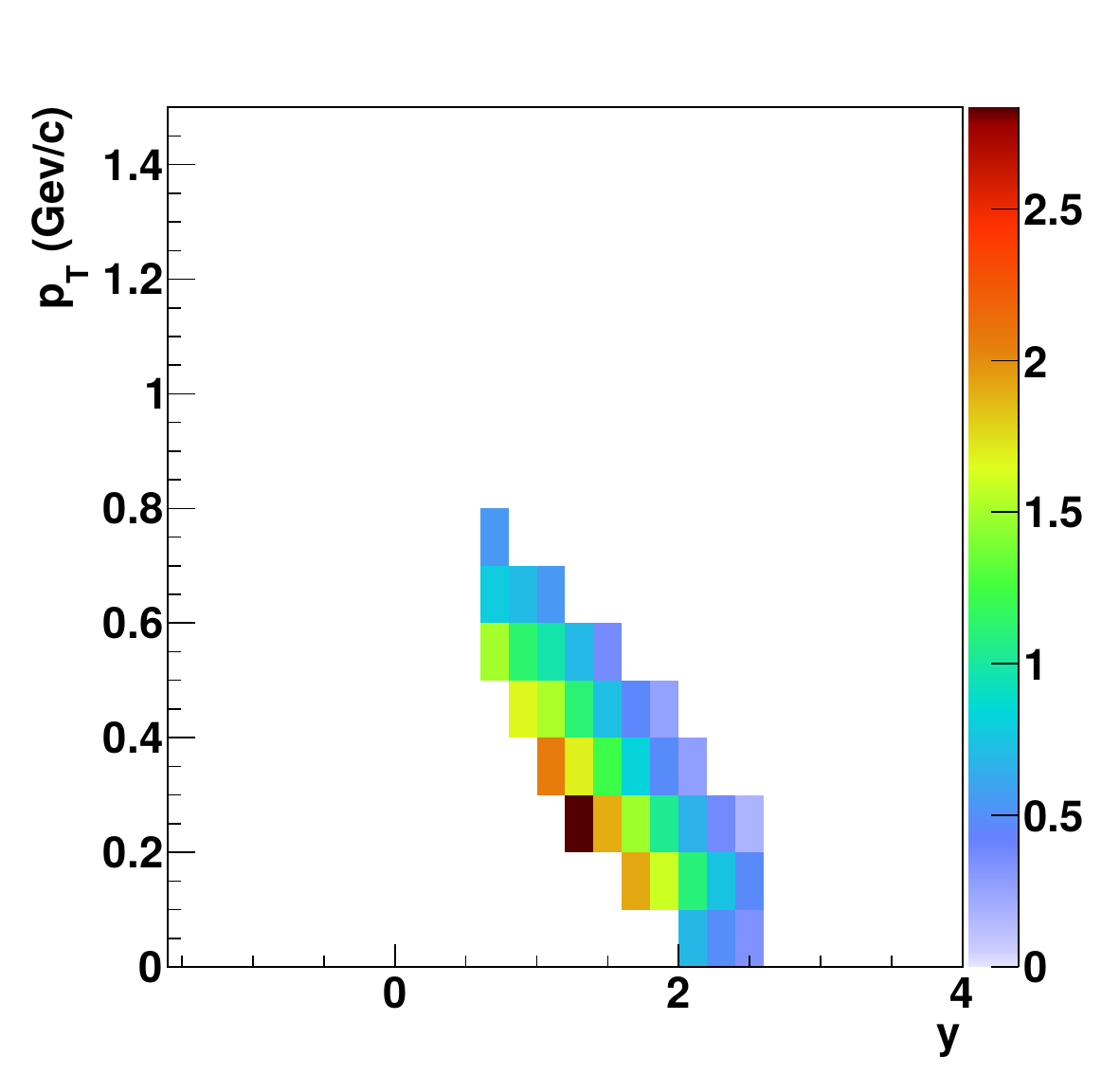} &
\hspace{-25mm}
\includegraphics[width=0.2\textwidth]{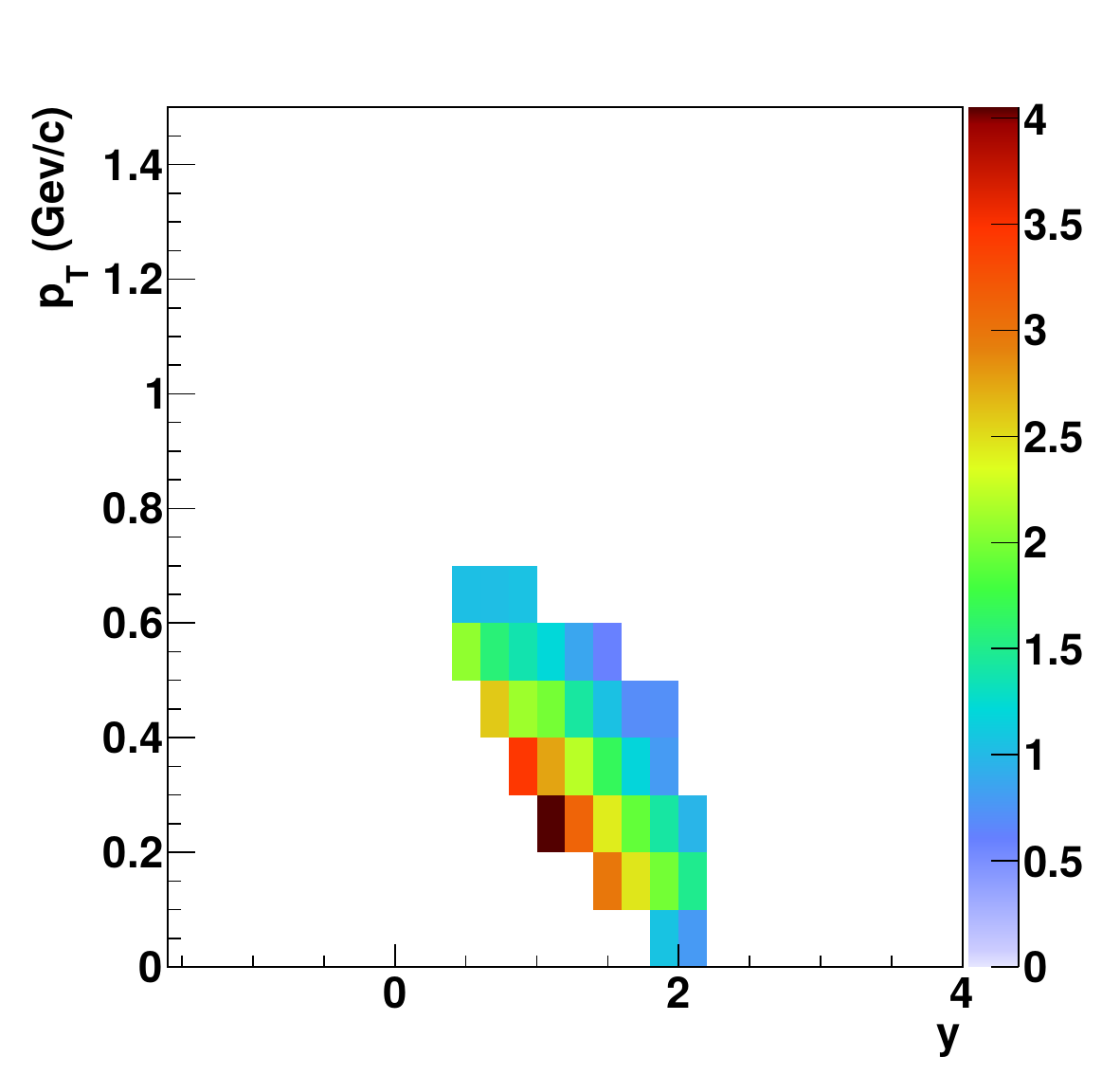} &
\hspace{-35mm}
\includegraphics[width=0.2\textwidth]{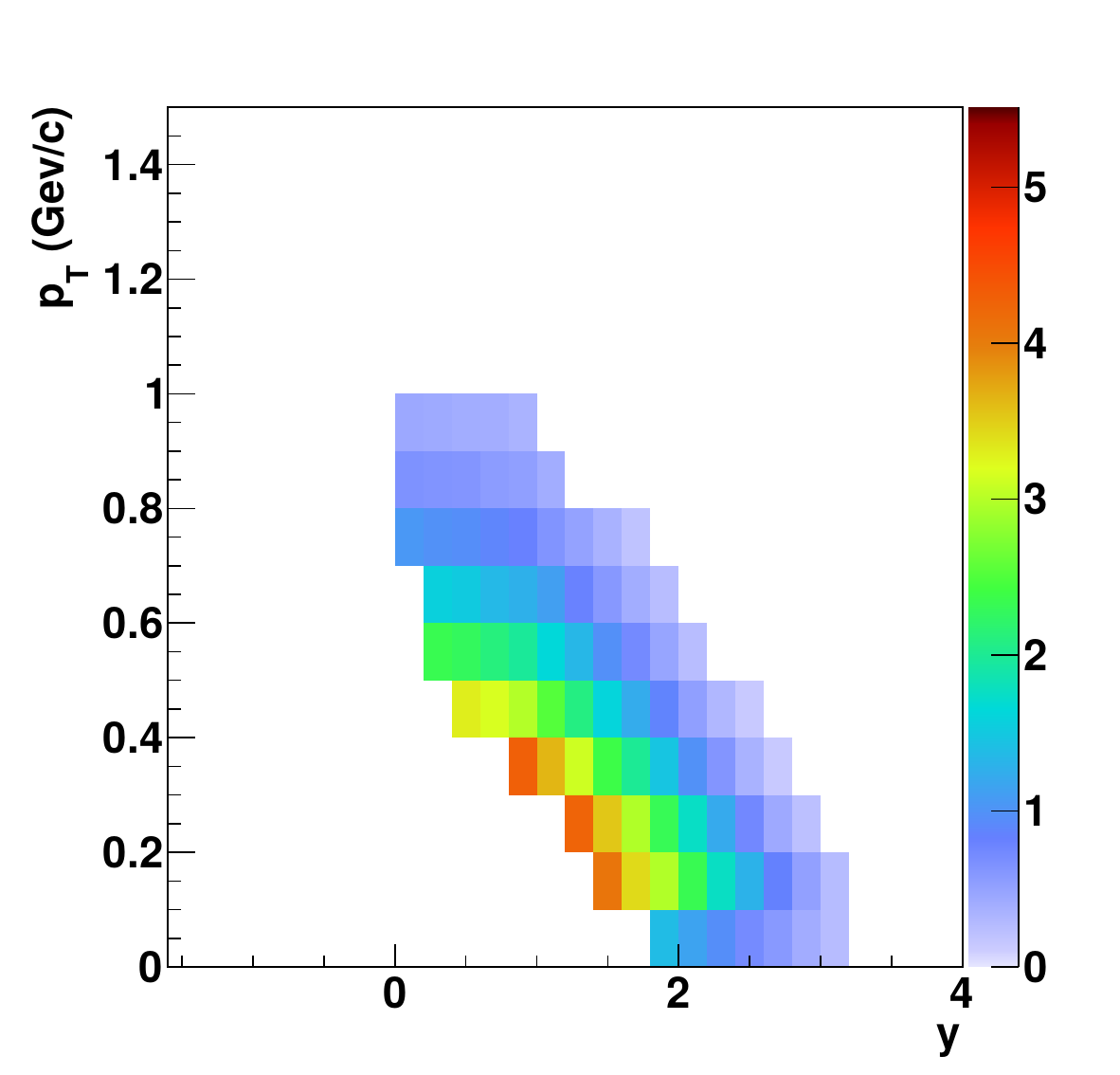} &
\hspace{-45mm}
\includegraphics[width=0.2\textwidth]{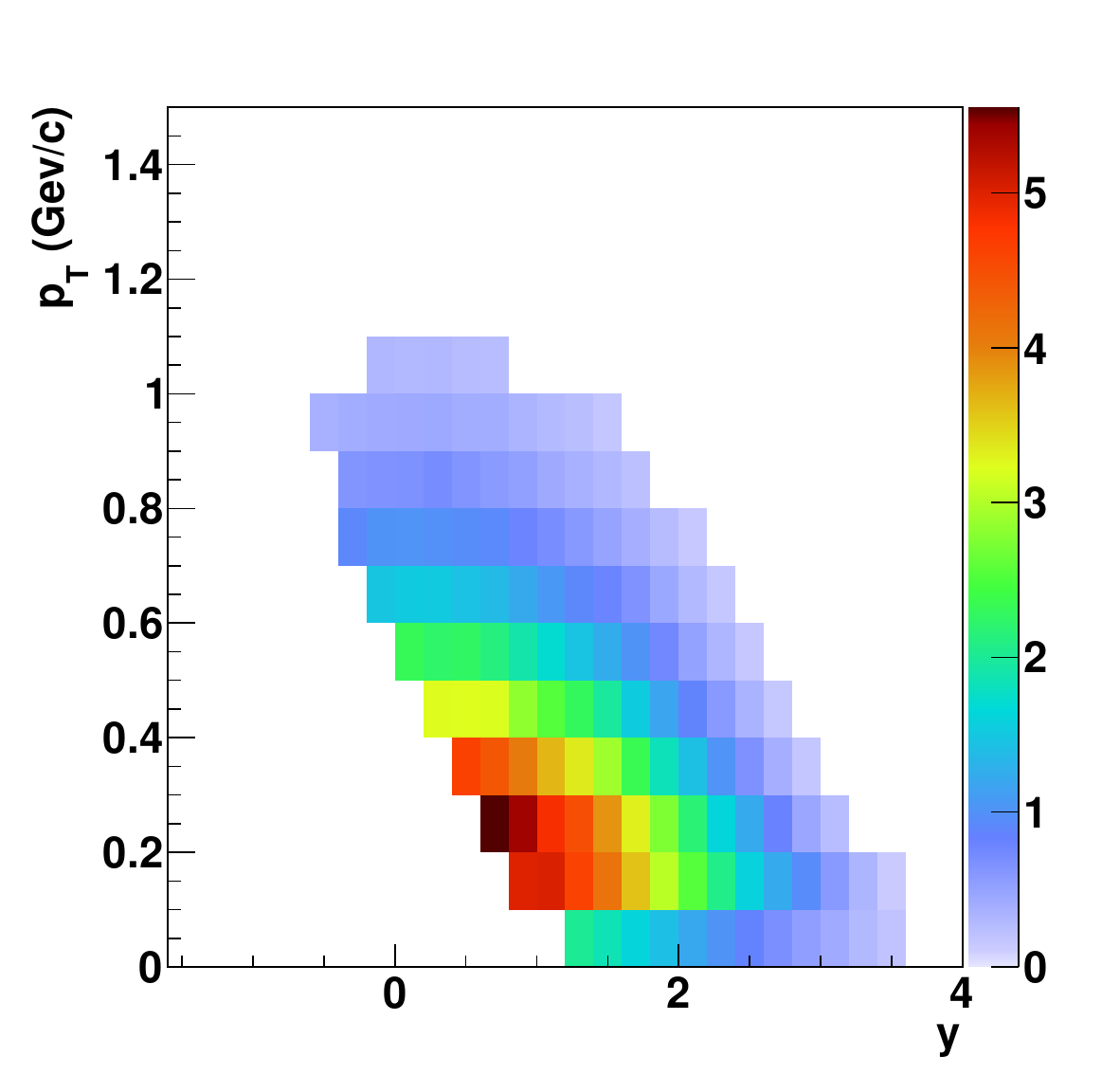} &
\hspace{-55mm}
\includegraphics[width=0.2\textwidth]{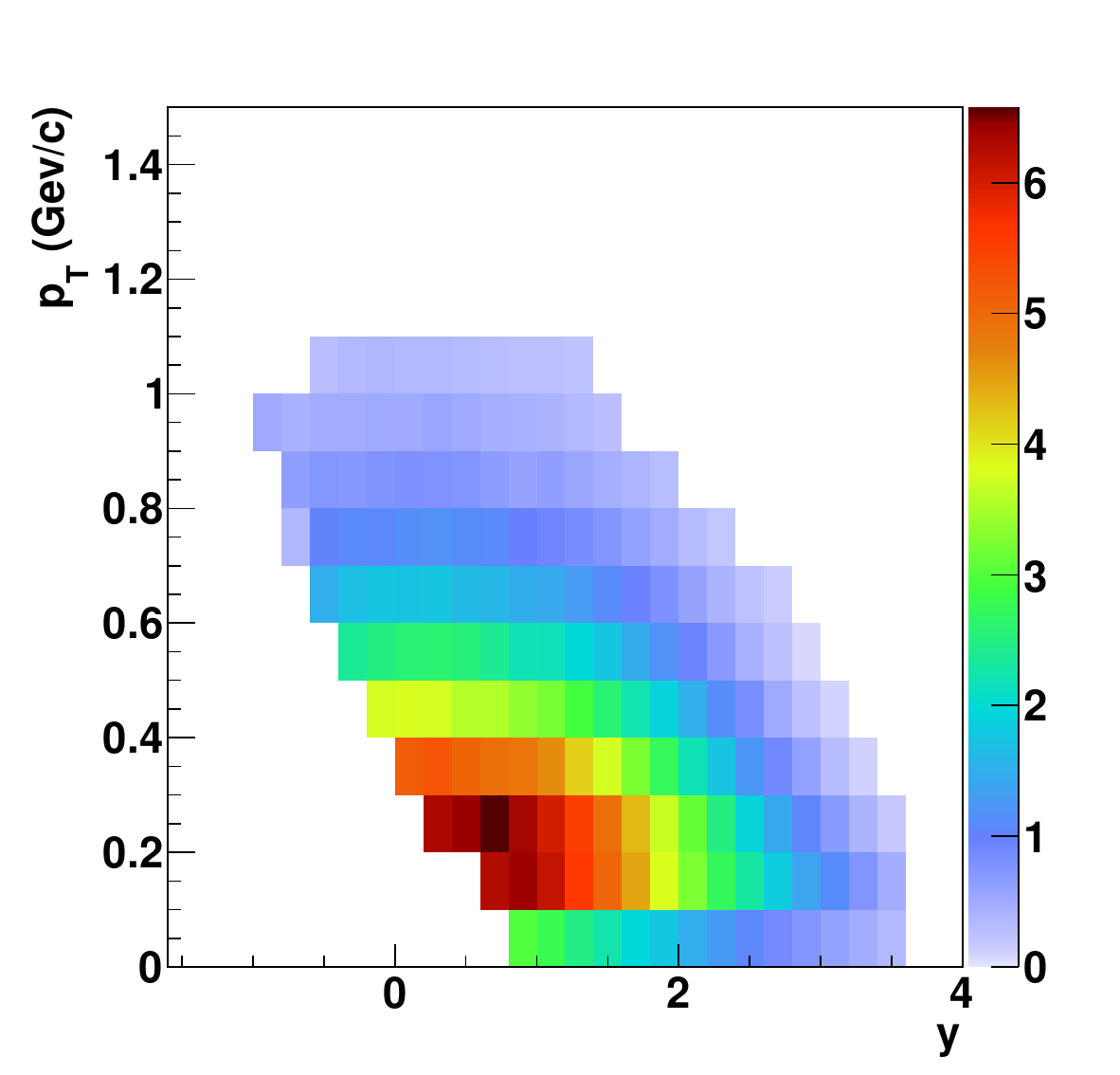}
\tabularnewline
\hspace{-7mm}$\pi^{+}$ &
\hspace{-15mm}
\includegraphics[width=0.2\textwidth]{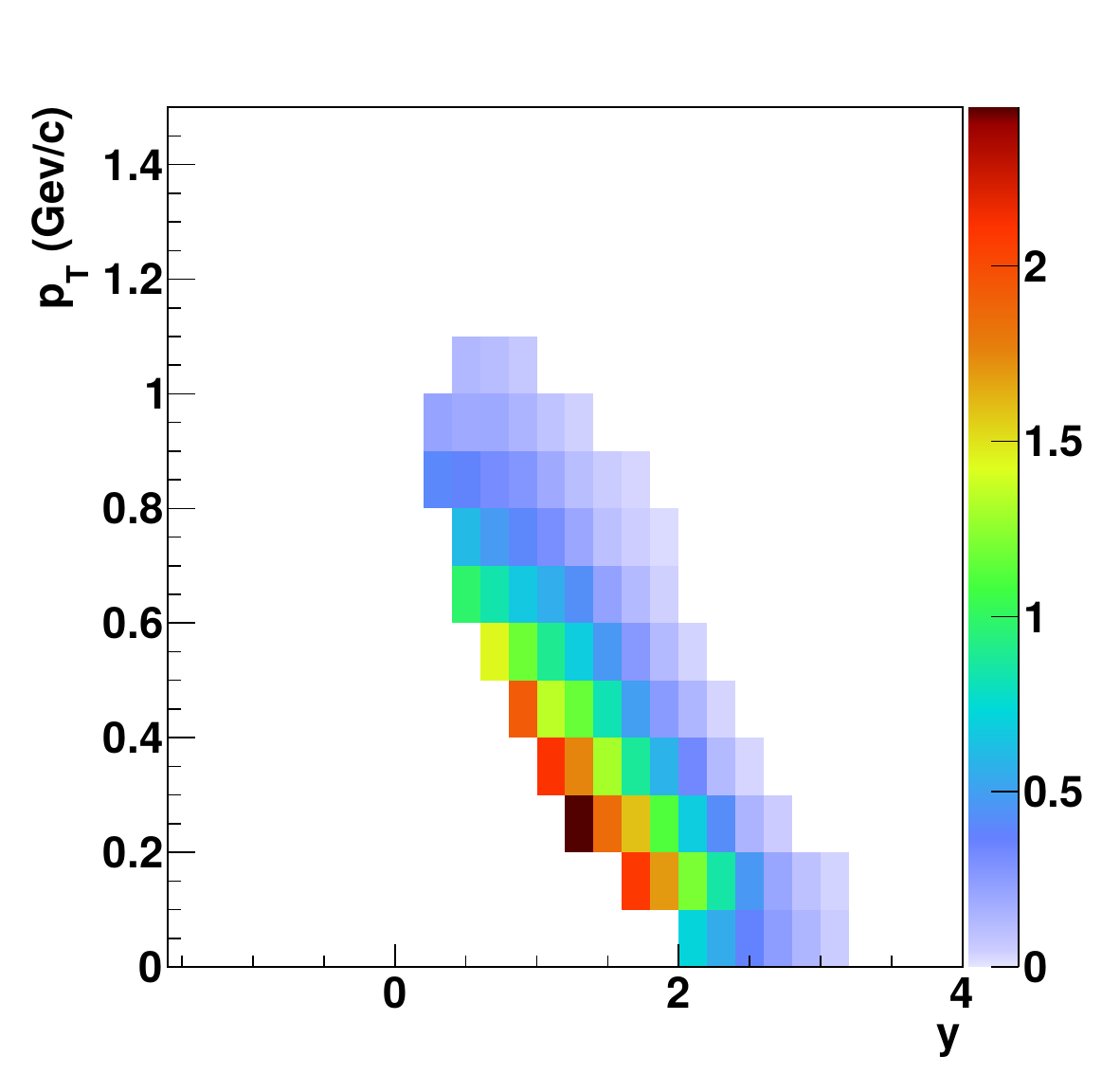} &
\hspace{-25mm}
\includegraphics[width=0.2\textwidth]{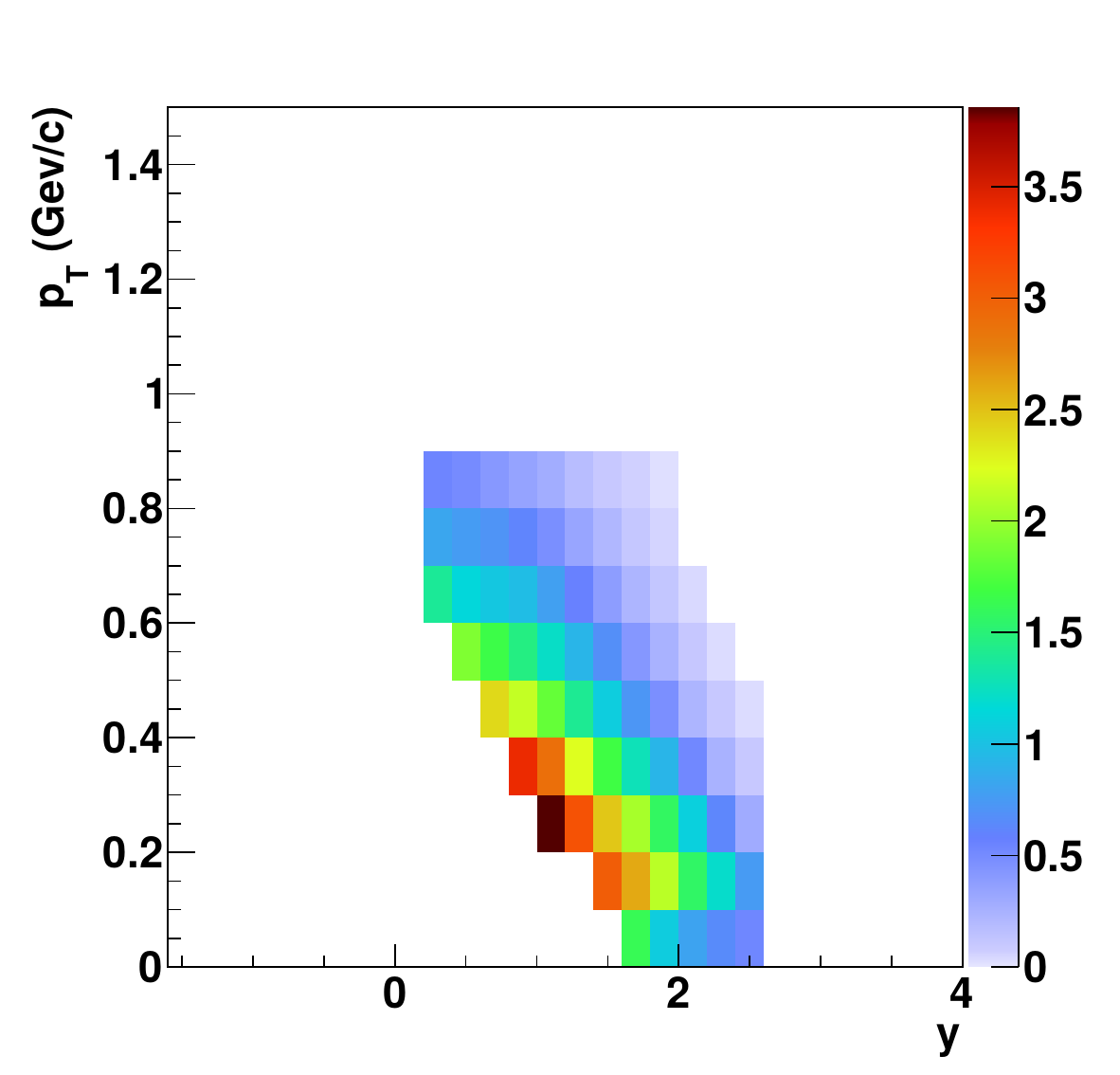} &
\hspace{-35mm}
\includegraphics[width=0.2\textwidth]{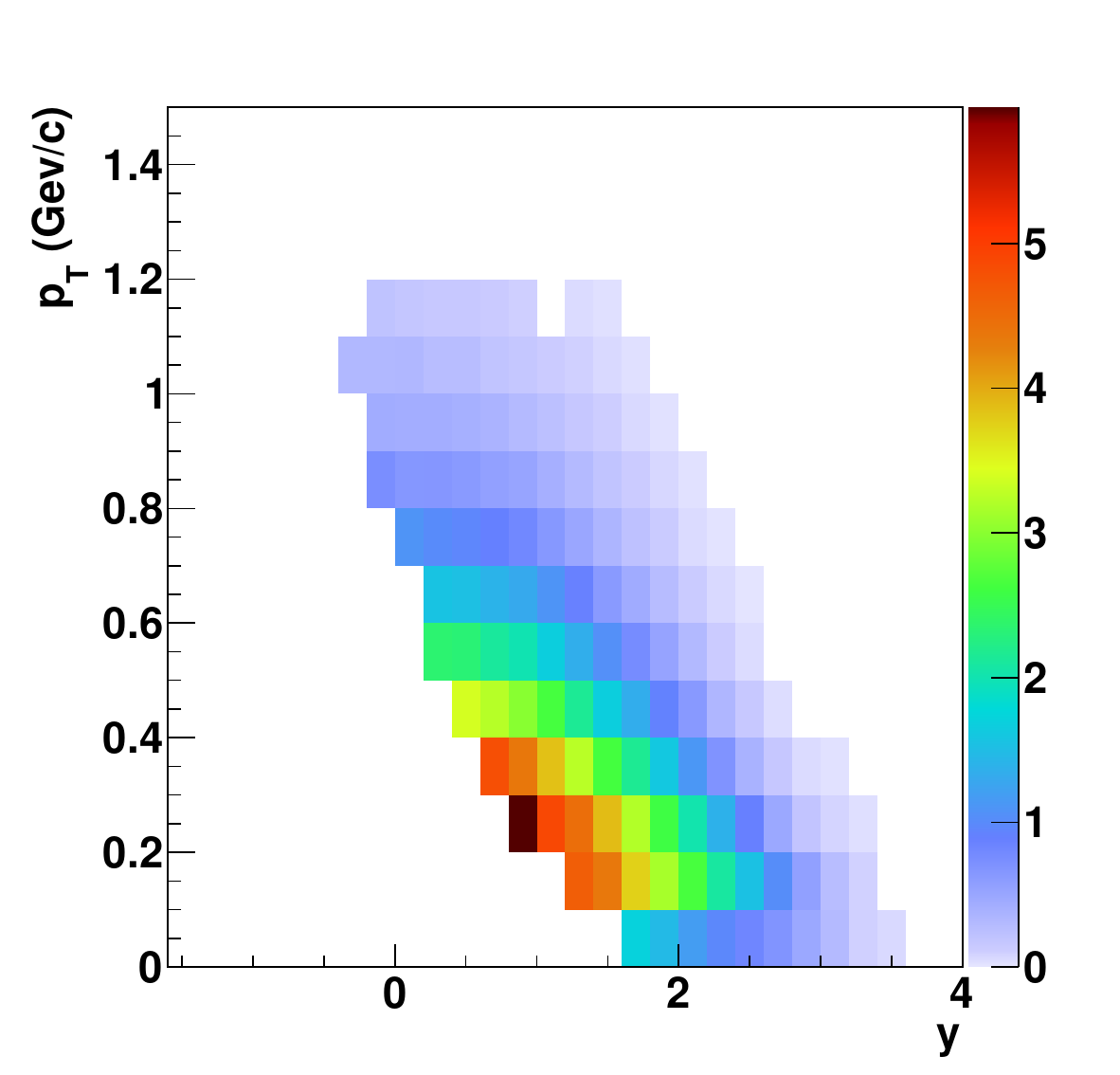} &
\hspace{-45mm}
\includegraphics[width=0.2\textwidth]{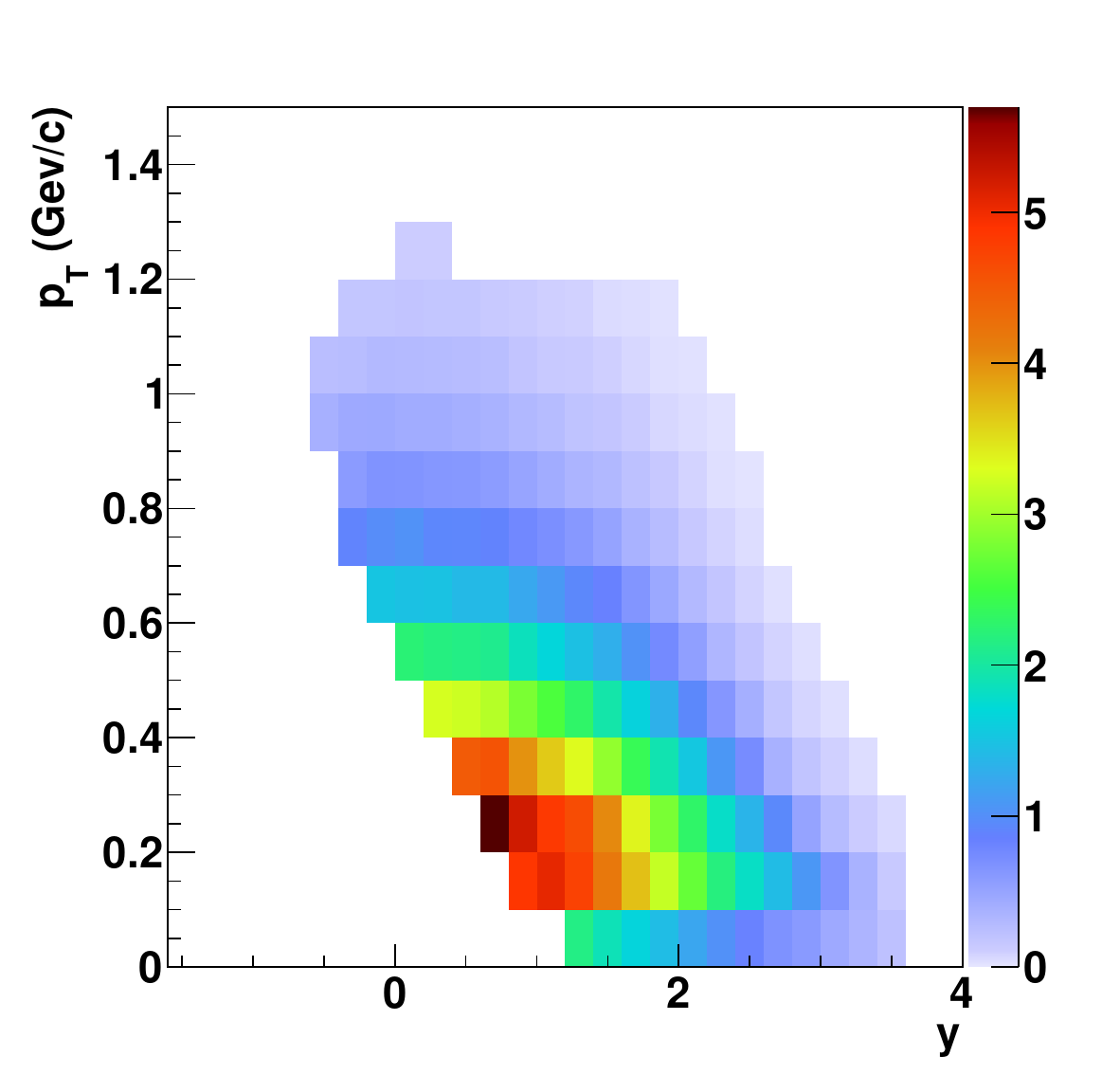} &
\hspace{-55mm}
\includegraphics[width=0.2\textwidth]{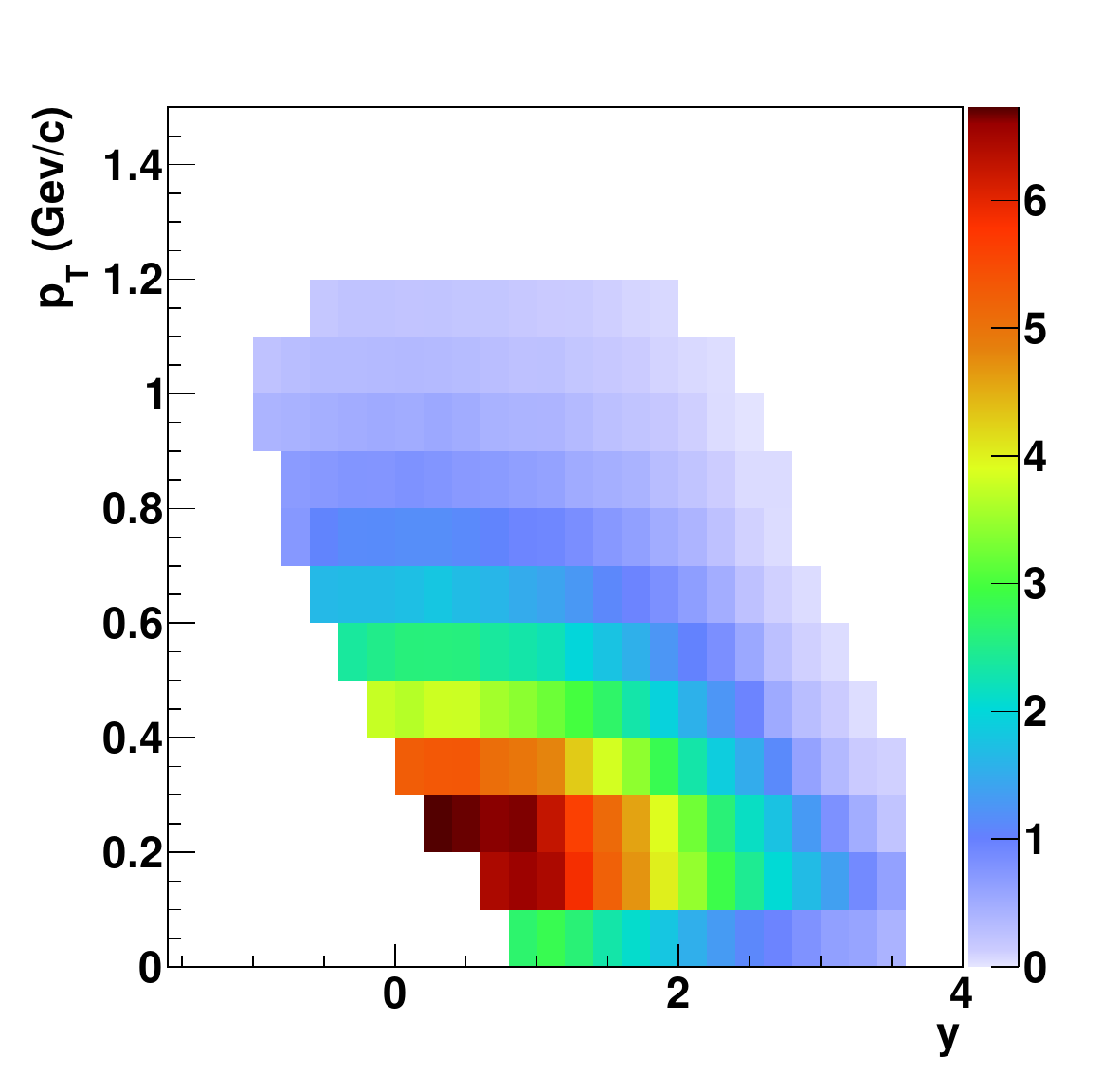}
\tabularnewline
\hspace{-7mm}$K^{-}$ &
\hspace{-15mm}
\includegraphics[width=0.2\textwidth]{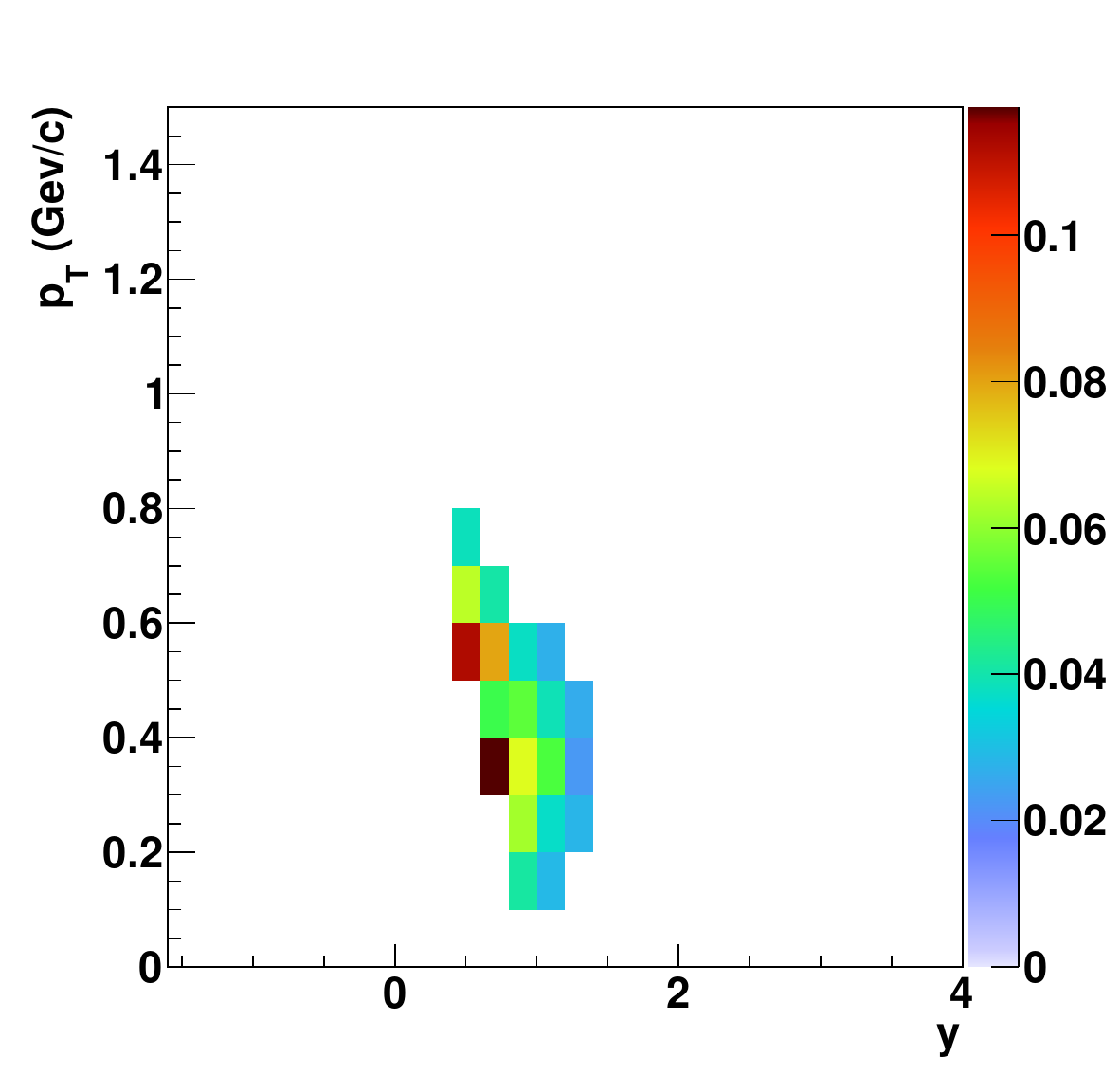} &
\hspace{-25mm}
\includegraphics[width=0.2\textwidth]{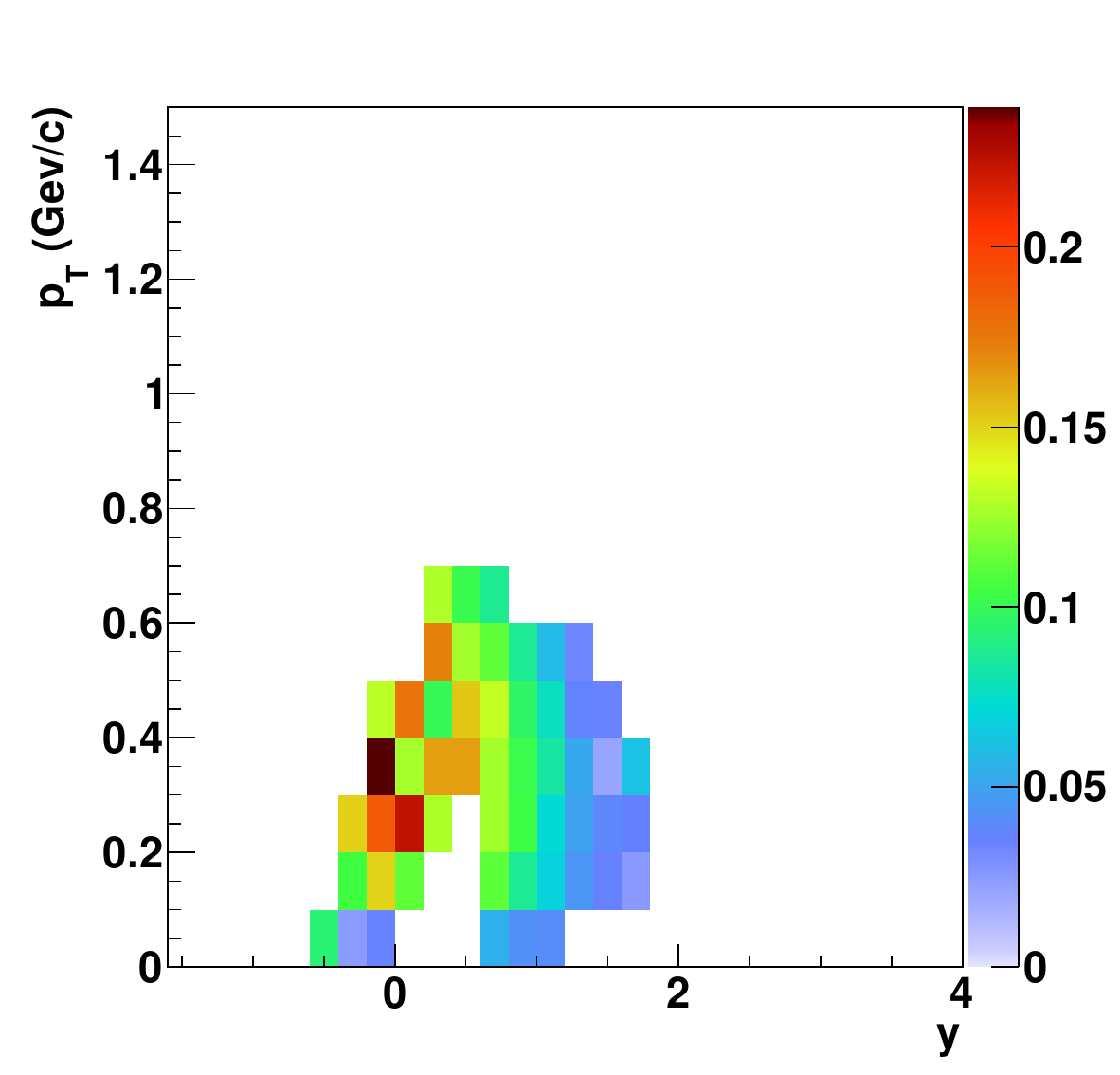} &
\hspace{-35mm}
\includegraphics[width=0.2\textwidth]{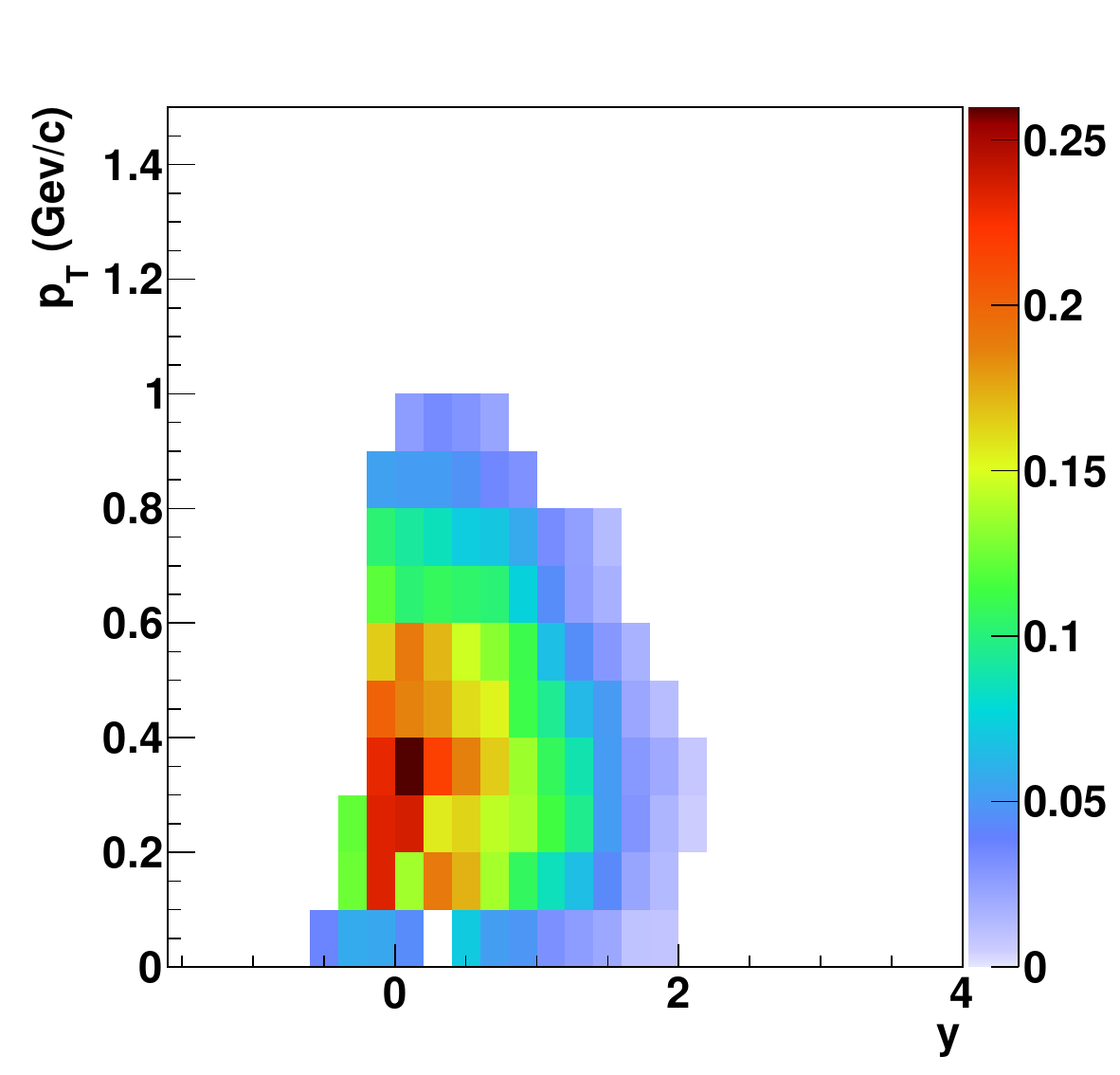} &
\hspace{-45mm}
\includegraphics[width=0.2\textwidth]{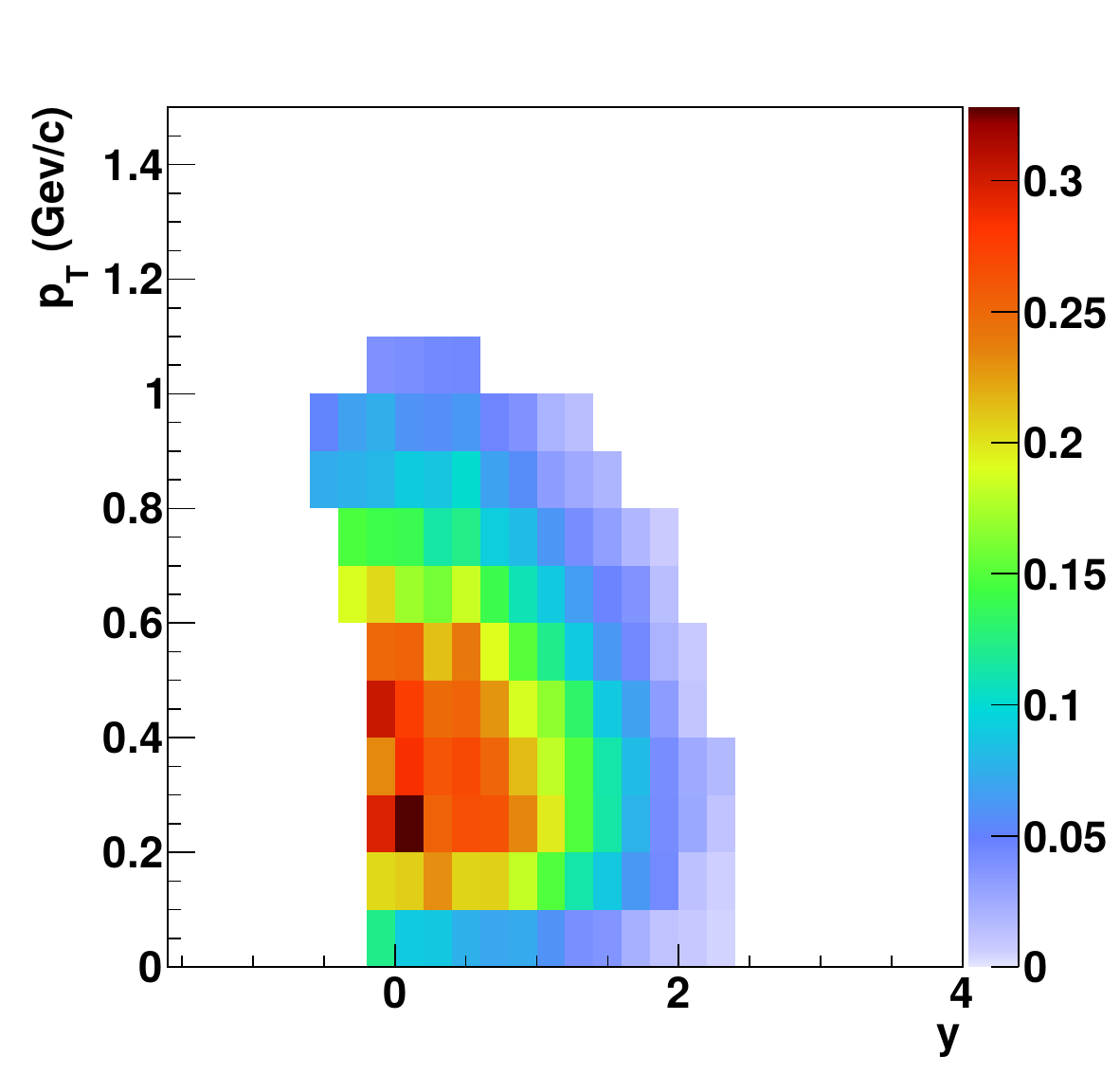} &
\hspace{-55mm}
\includegraphics[width=0.2\textwidth]{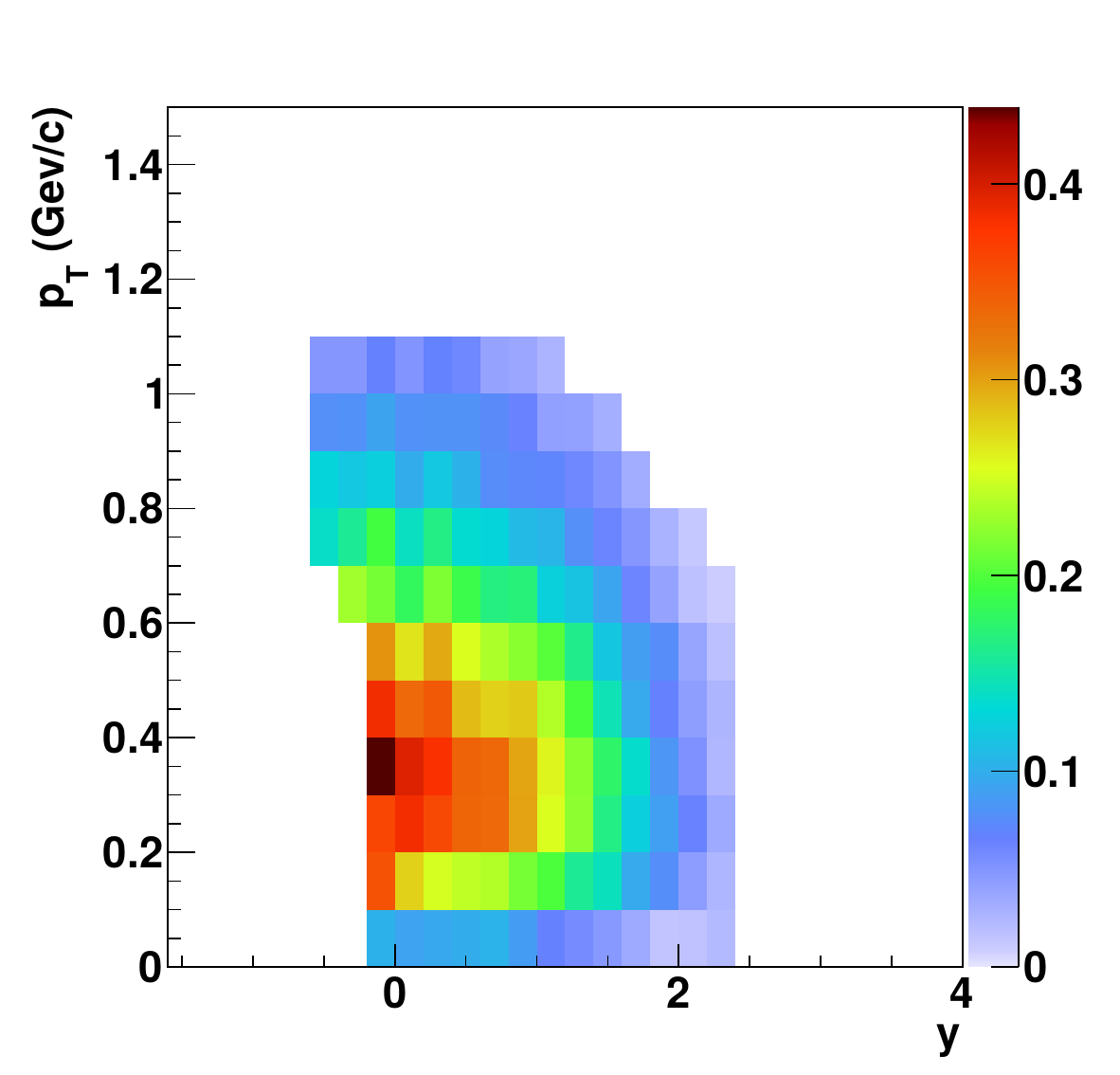}
\tabularnewline
\hspace{-7mm}$K^{+}$ &
\hspace{-15mm}
\includegraphics[width=0.2\textwidth]{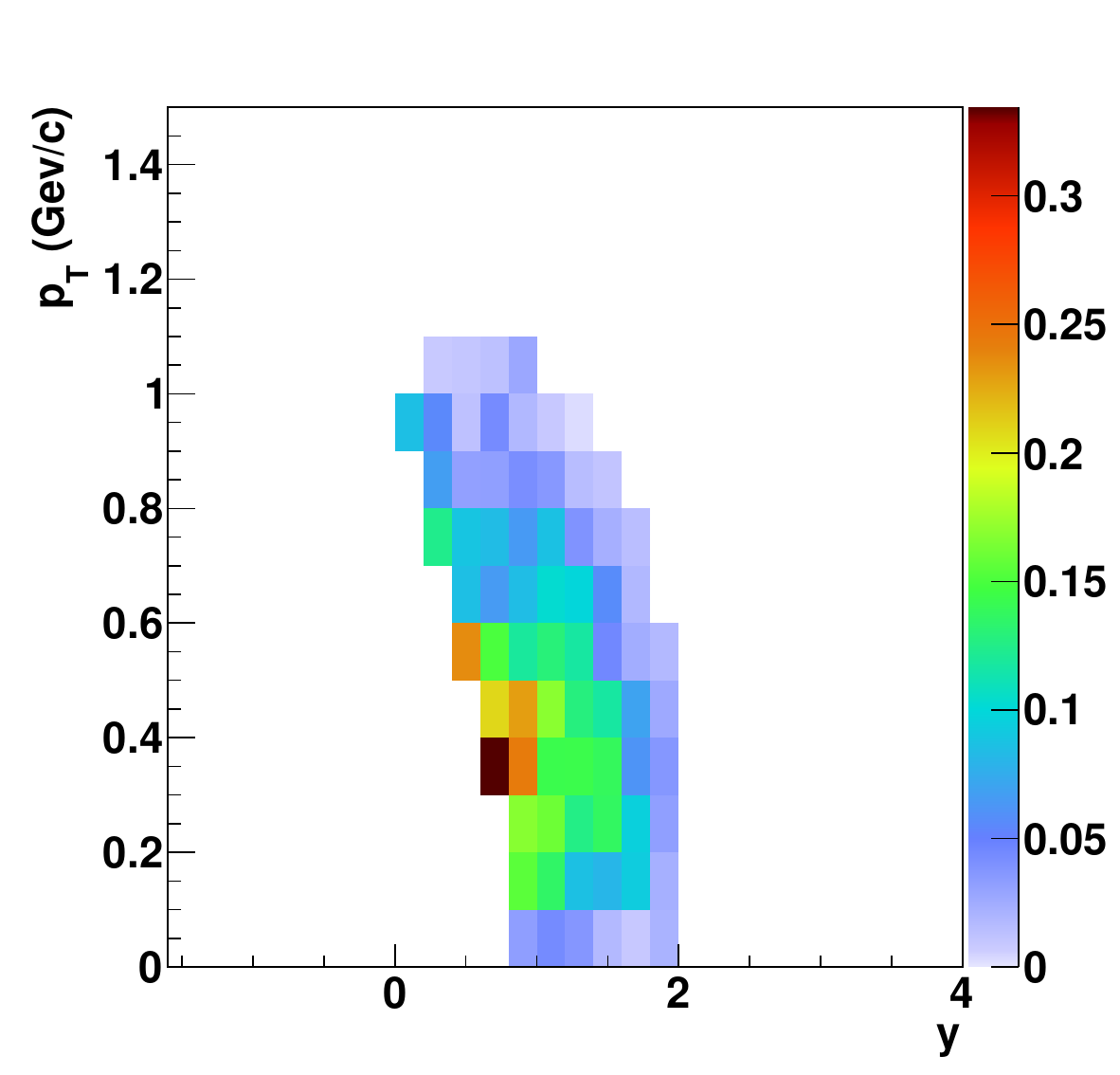} &
\hspace{-25mm}
\includegraphics[width=0.2\textwidth]{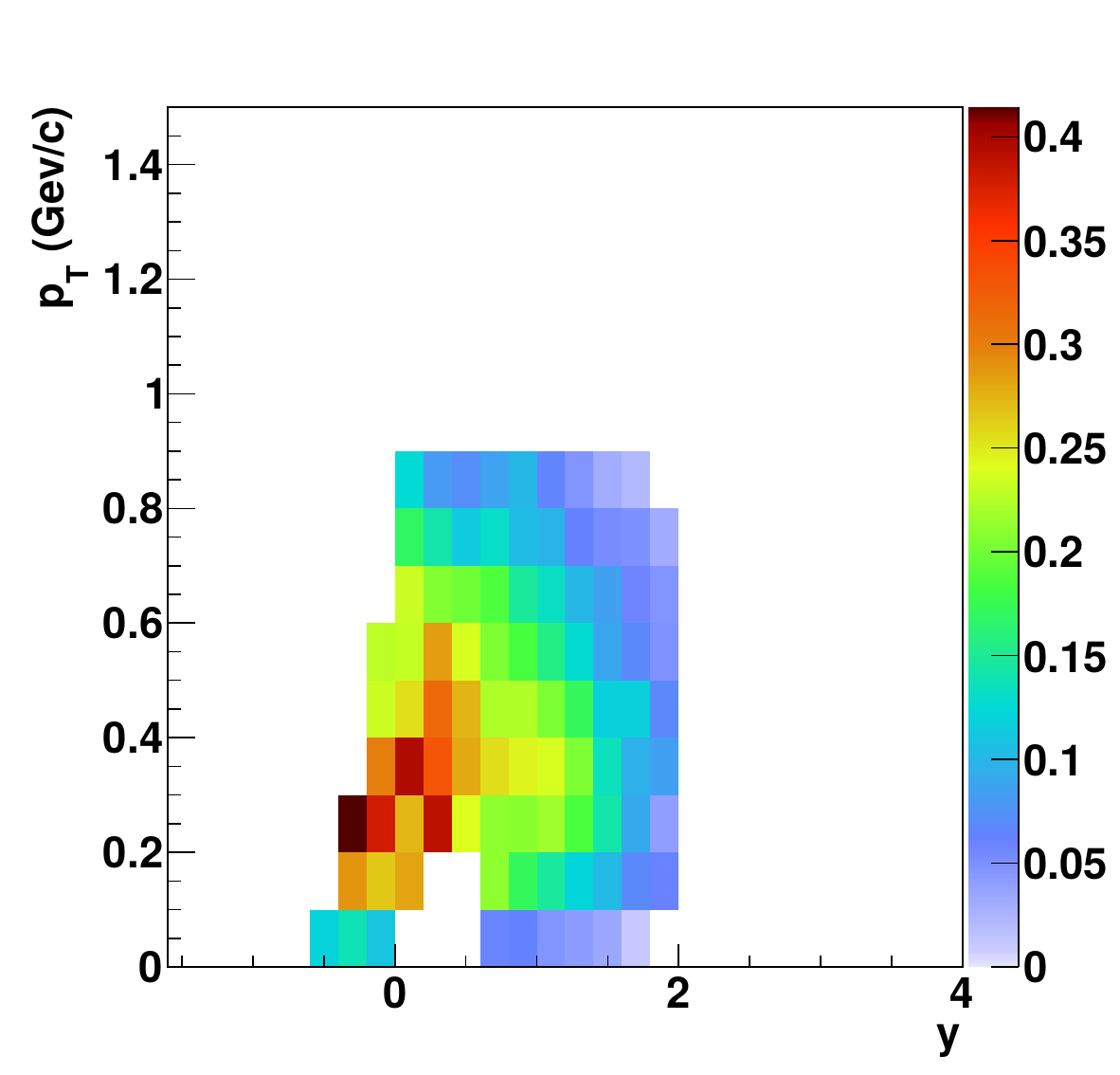} &
\hspace{-35mm}
\includegraphics[width=0.2\textwidth]{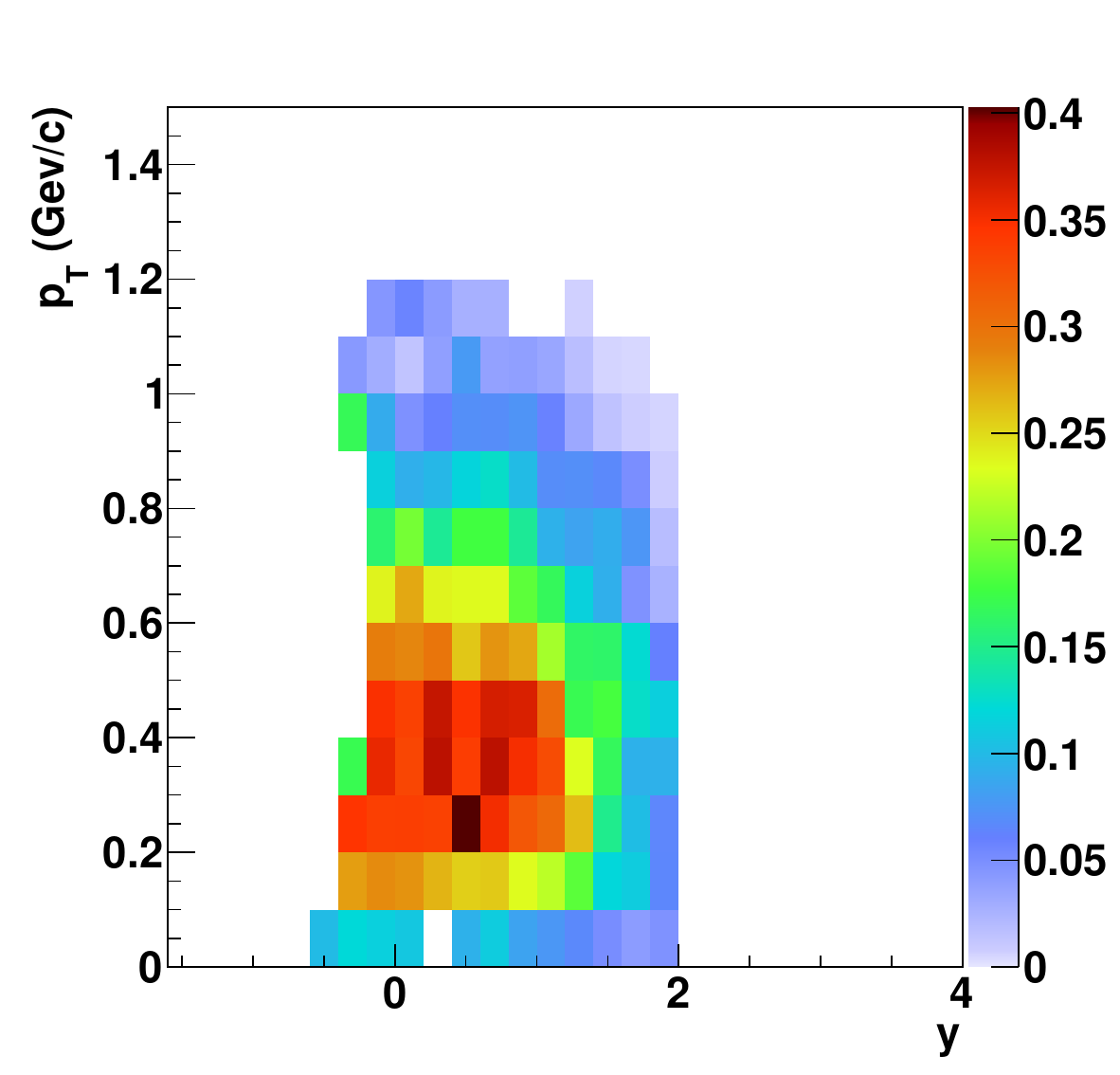} &
\hspace{-45mm}
\includegraphics[width=0.2\textwidth]{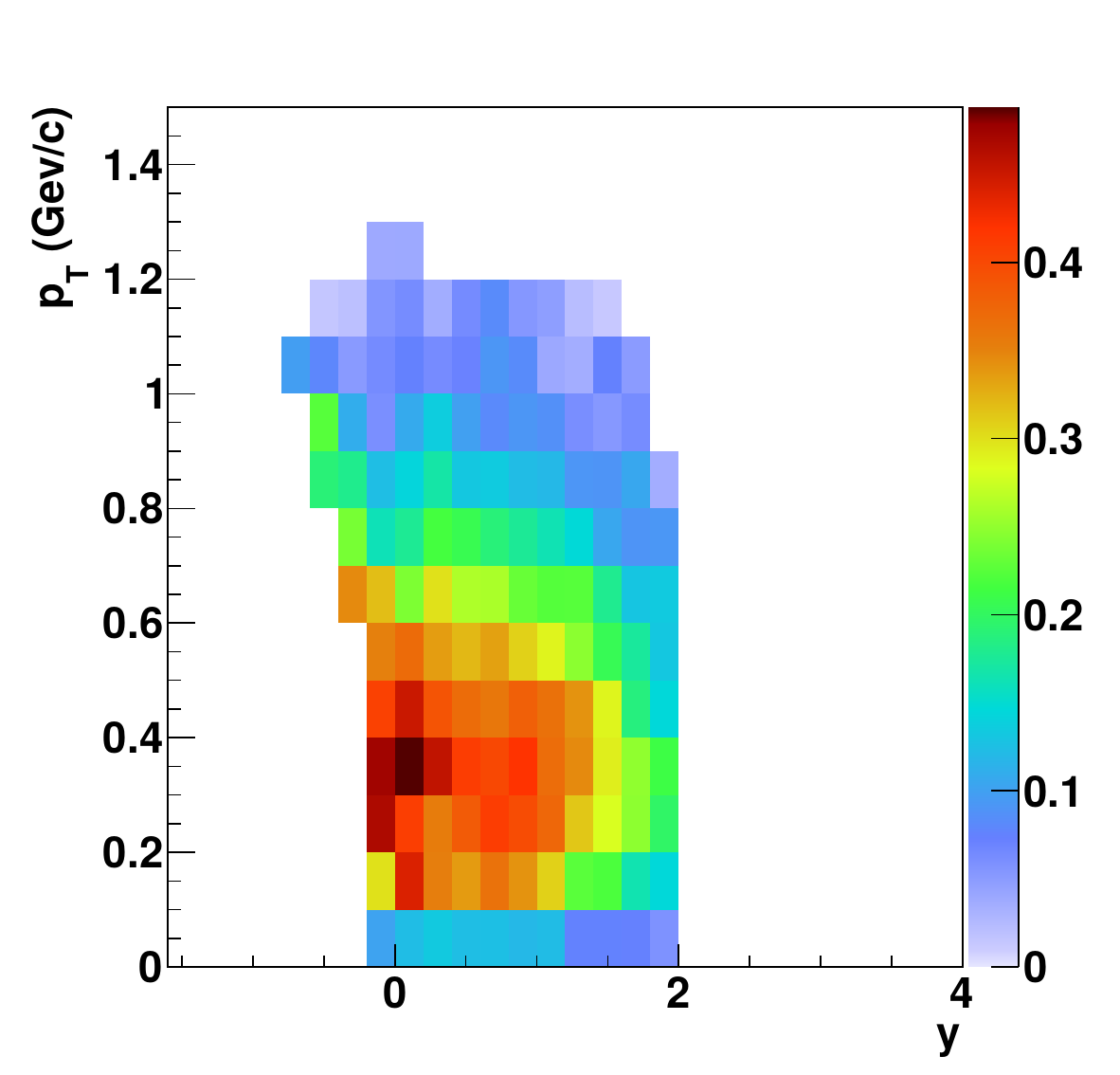} &
\hspace{-55mm}
\includegraphics[width=0.2\textwidth]{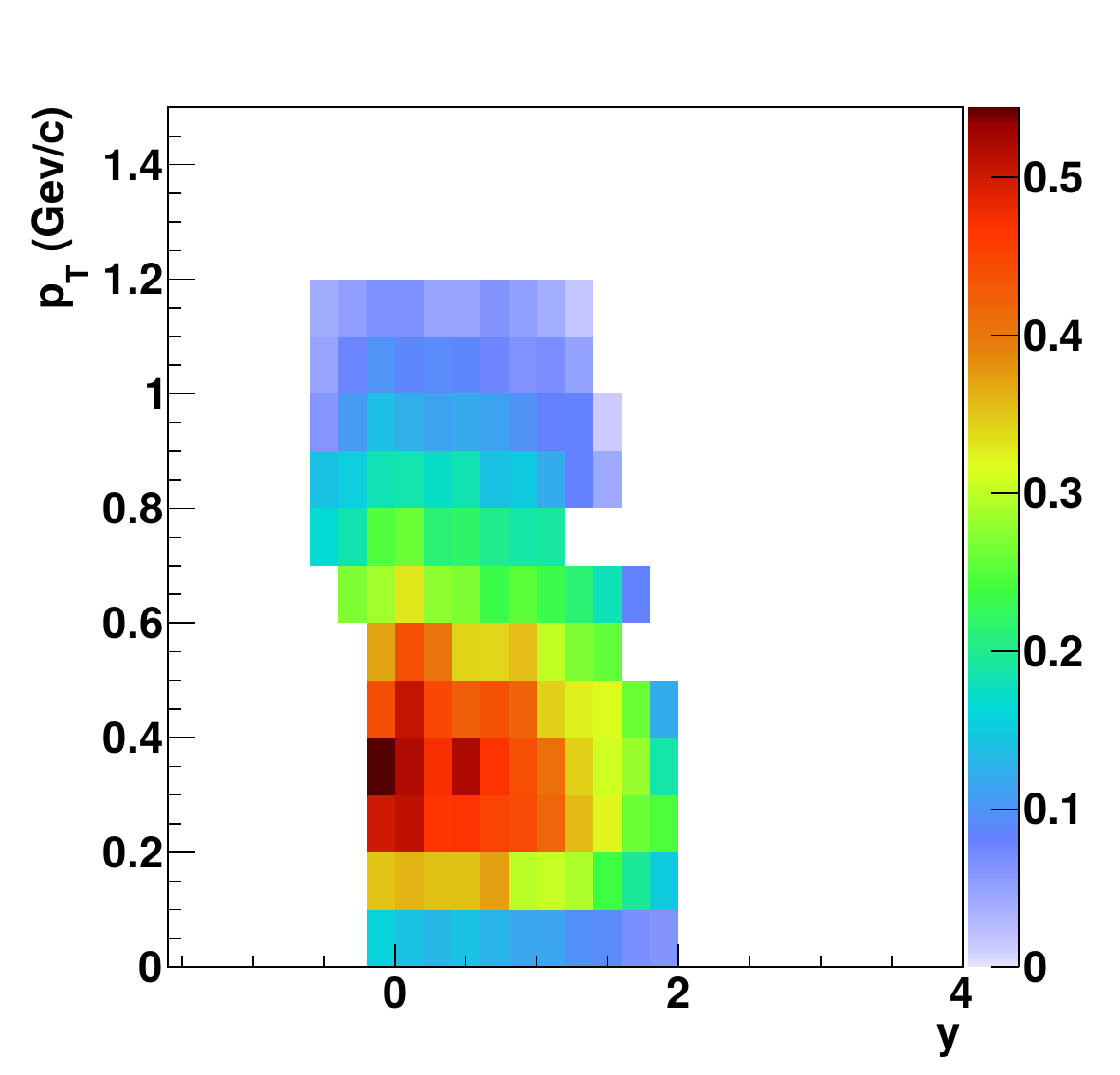}
\tabularnewline
\hspace{-7mm}$p$ &
\hspace{-15mm}
\includegraphics[width=0.2\textwidth]{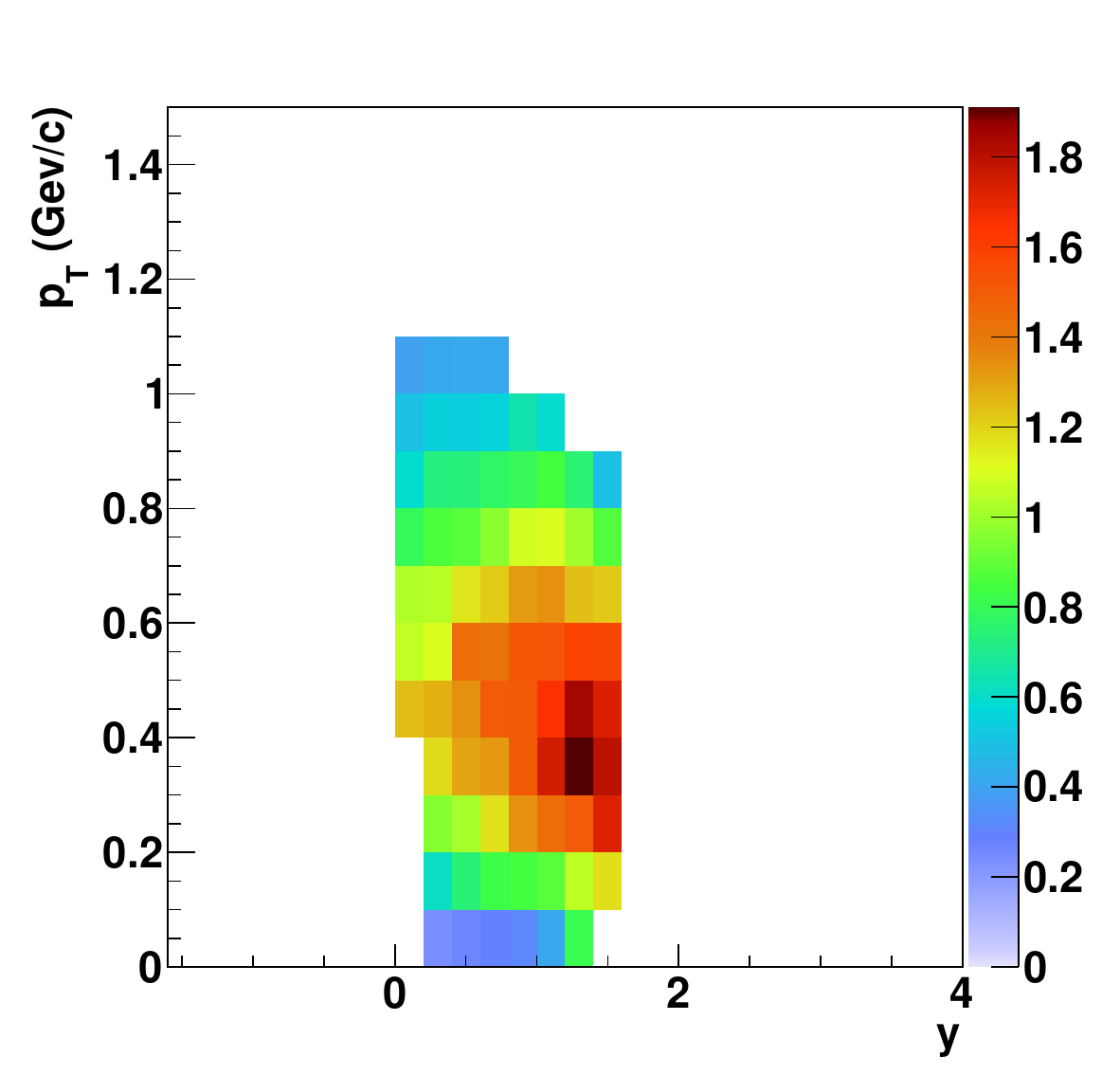} &
\hspace{-25mm}
\includegraphics[width=0.2\textwidth]{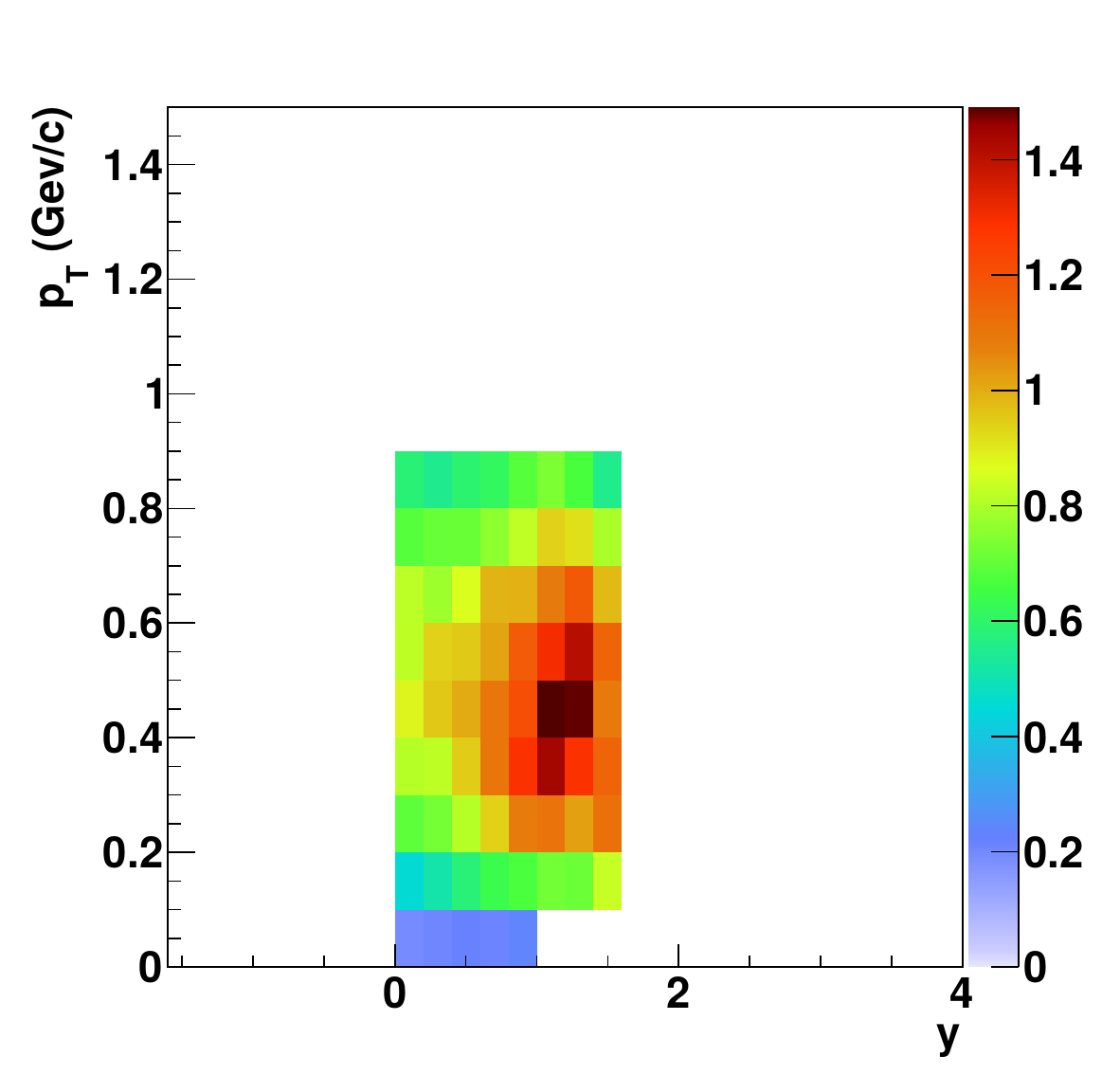} &
\hspace{-35mm}
\includegraphics[width=0.2\textwidth]{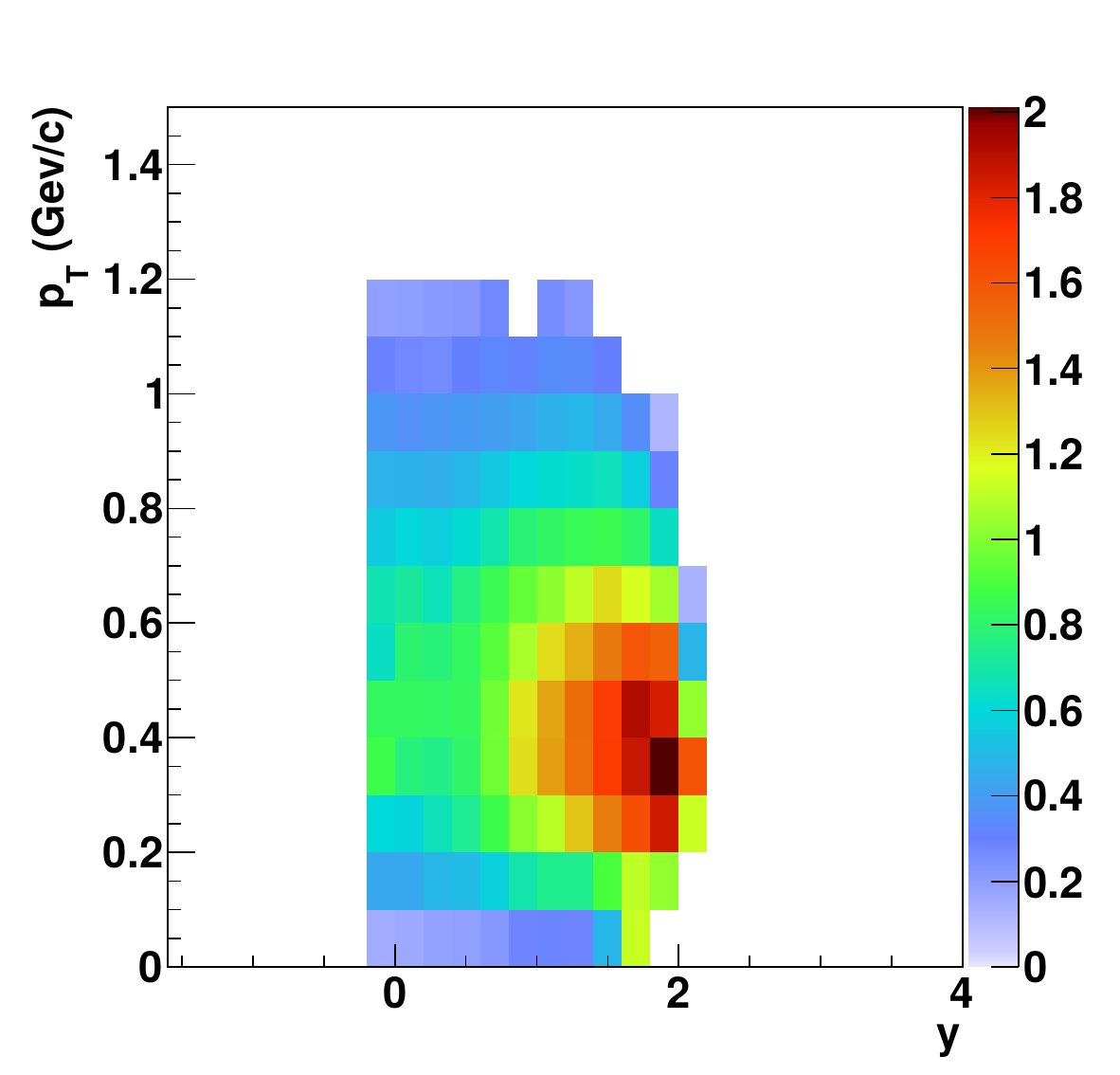} &
\hspace{-45mm}
\includegraphics[width=0.2\textwidth]{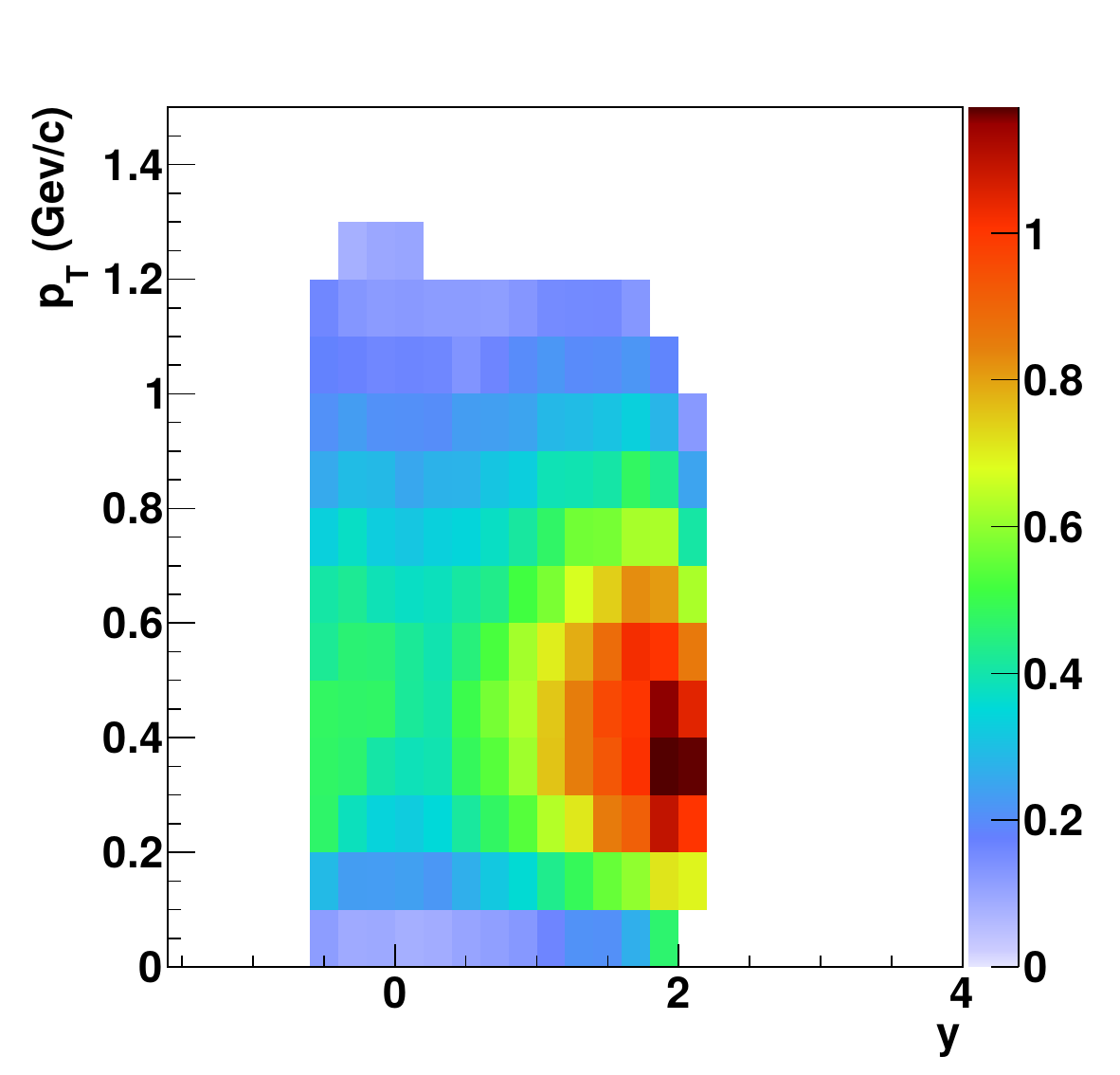} &
\hspace{-55mm}
\includegraphics[width=0.2\textwidth]{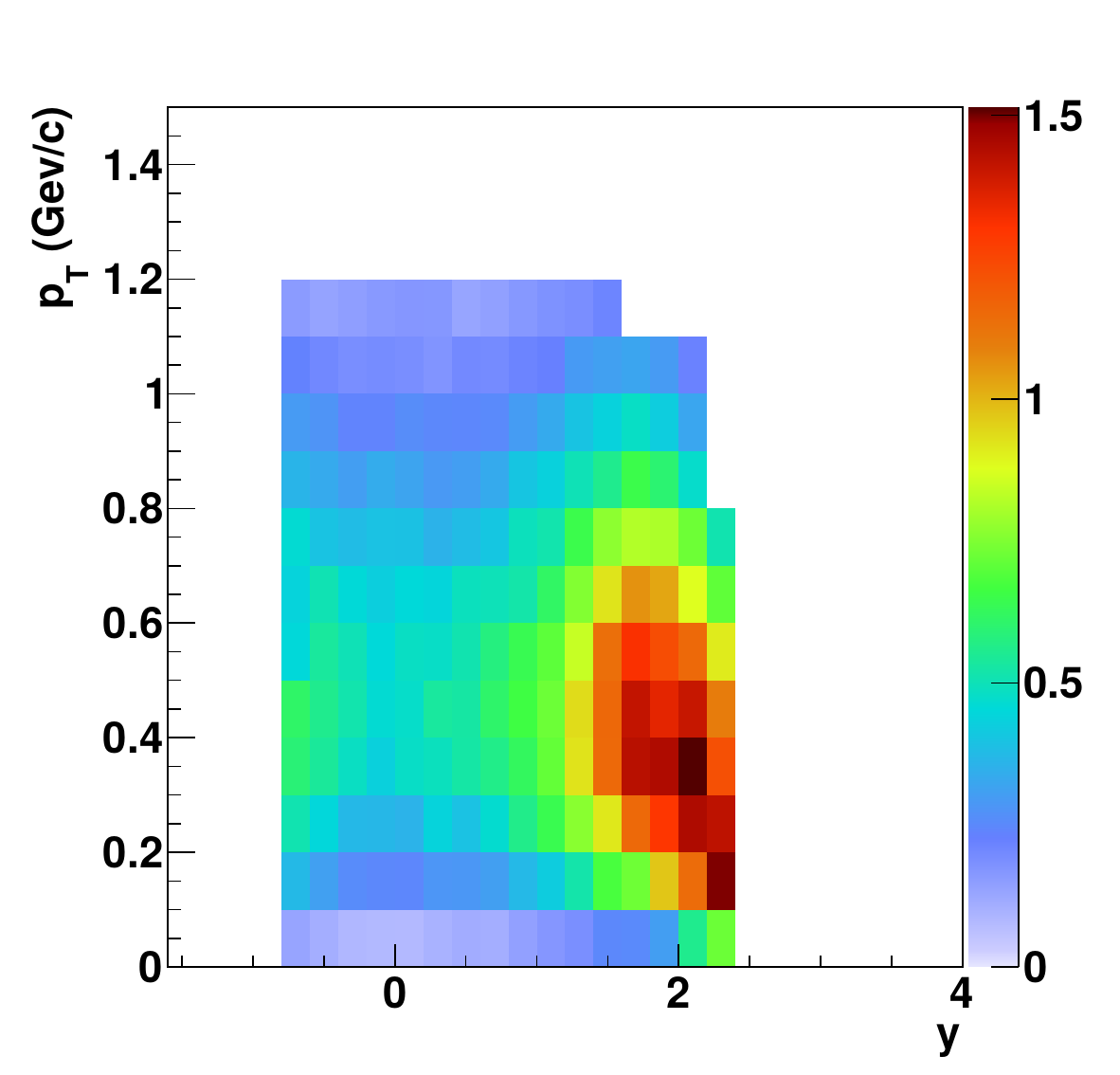}
\tabularnewline
\hspace{-7mm}$\bar{p}$ &
\hspace{-15mm}
&
\hspace{-25mm}
&
\hspace{-35mm}
\includegraphics[width=0.2\textwidth]{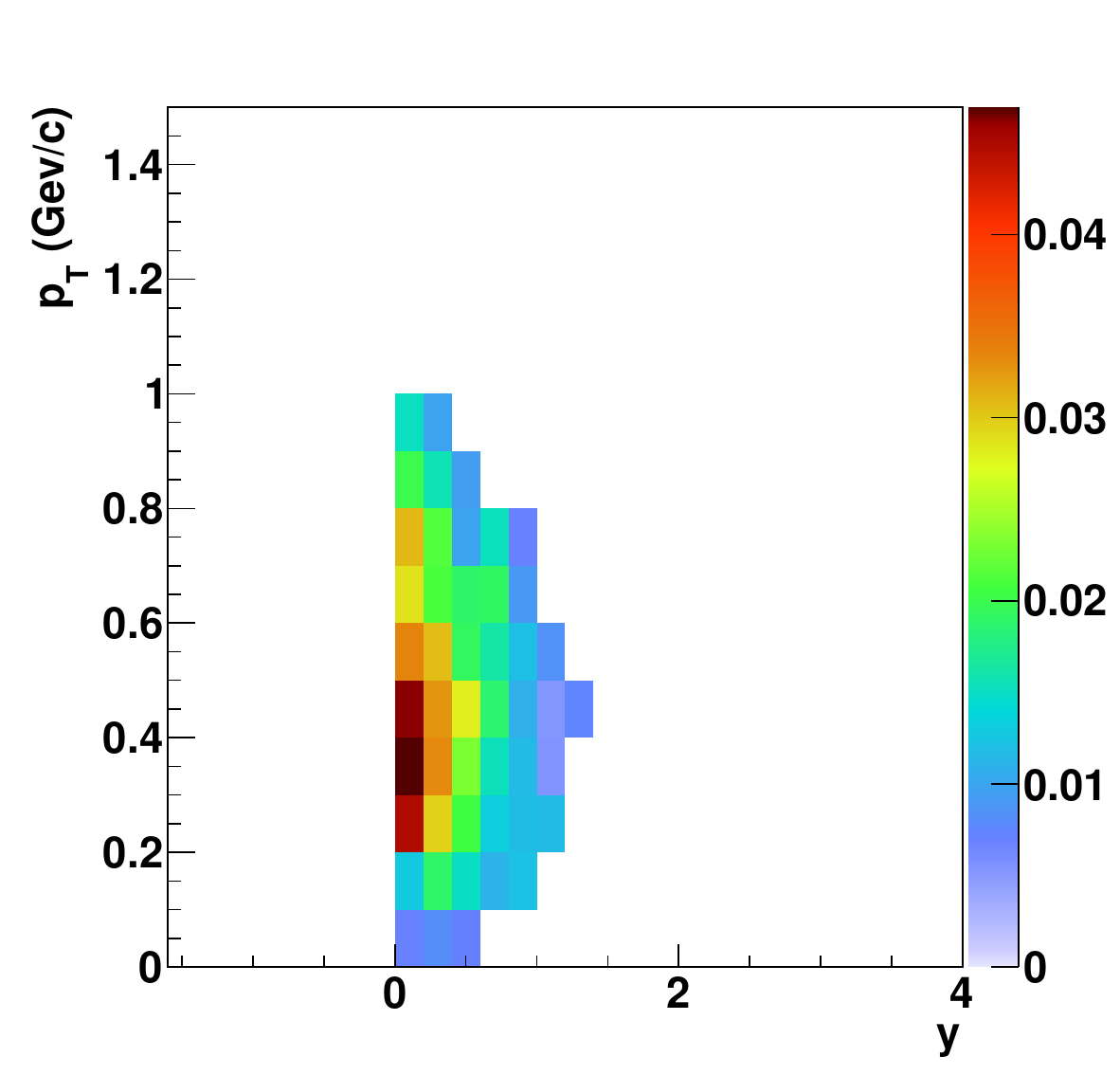} &
\hspace{-45mm}
\includegraphics[width=0.2\textwidth]{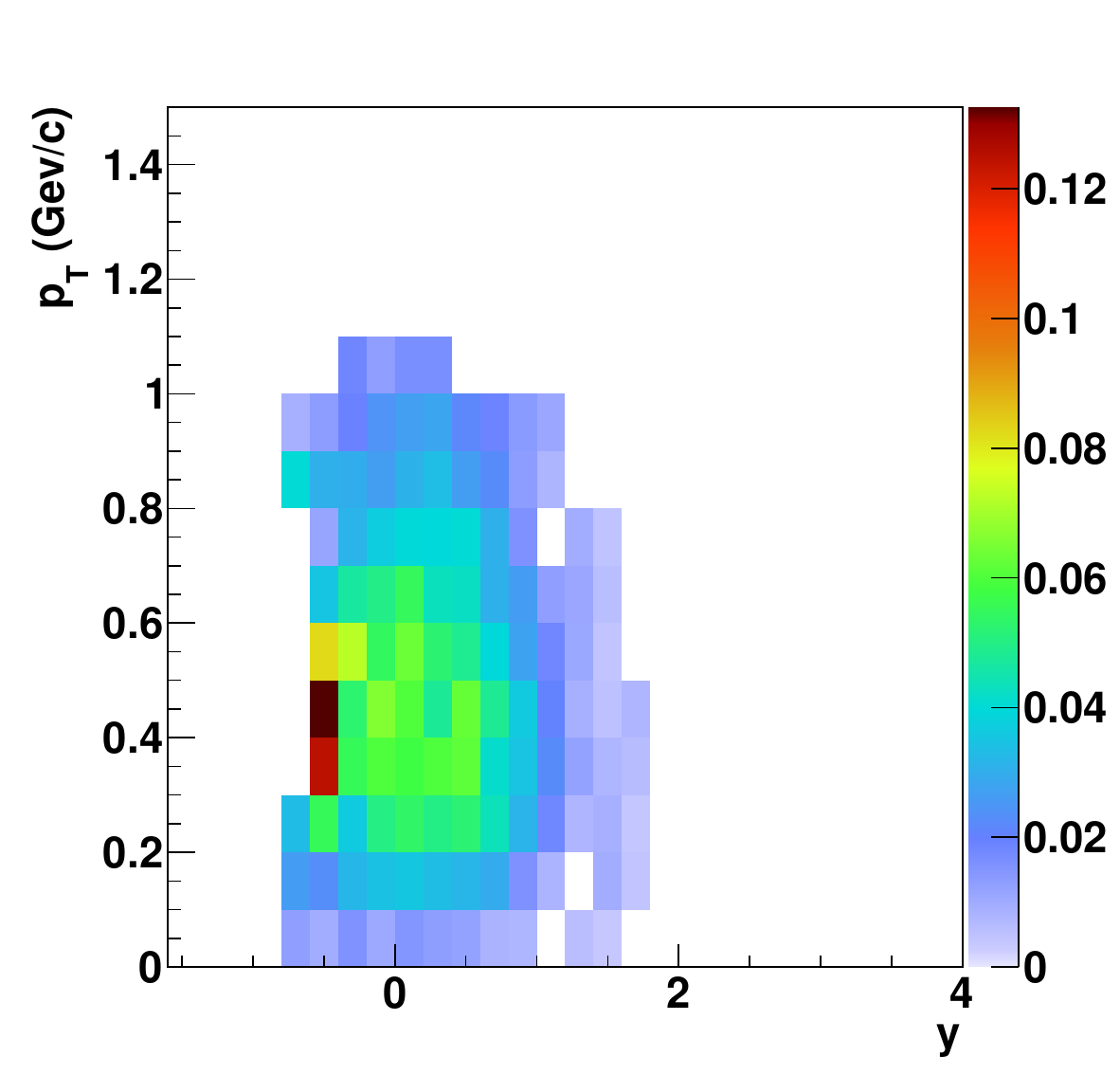} &
\hspace{-55mm}
\includegraphics[width=0.2\textwidth]{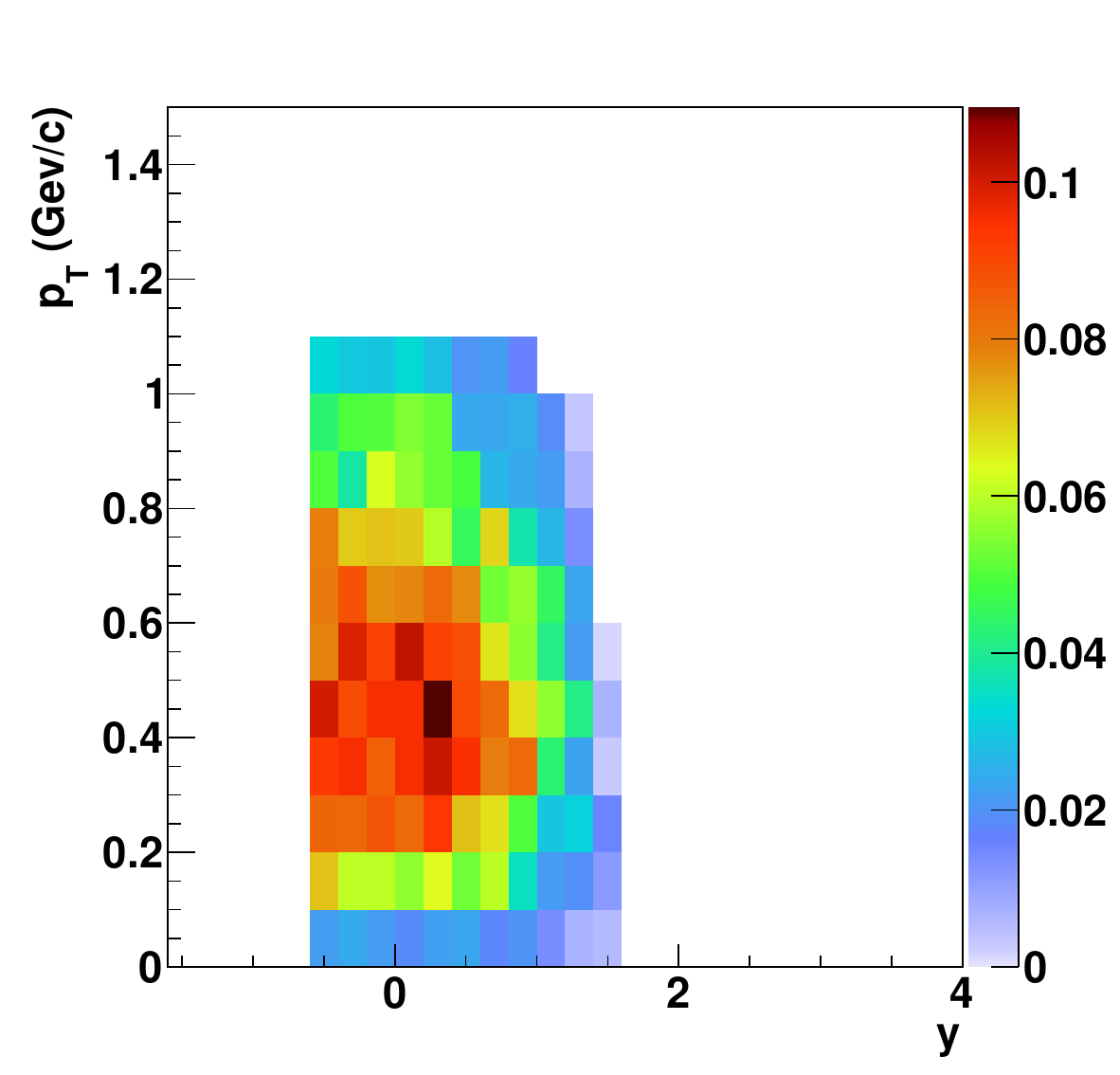}
\tabularnewline
\end{tabular}
\end{center}
\caption{Two-dimensional distributions (\y vs. $p_{T}$) of yields
         of $\pi^{-}$, $\pi^{+}$, $K^{-}$, $K^{+}$, 
         $p$ and $\bar{p}$ produced in the 20\% most \textit{central} Be+Be interactions at 
         19$A$, 30$A$, 40$A$, 75$A$ and 150\AGeVc.}
\label{fig:final2D}
\end{figure*}

Resulting two-dimensional distributions $\frac{d^{2}n}{dydp_T}$ of $\pi^{-}, \pi^{+}, K^{-}, K^{+}, 
p$ and $\bar{p}$ produced in the 20\% most \textit{central} Be+Be collisions at different SPS energies 
are presented in Fig.~\ref{fig:final2D}.

\subsubsection{Statistical uncertainties}
\label{sec:statistics}

Statistical uncertainties of yields (Eq.~\ref{finalserrdEdx}) were calculated as:


\begin{equation}
\label{finalserrdEdx}
\sigma_{stat}^2 =\frac{1}{\Delta\y^{2}\cdot\Delta\pt^{2}\cdot{N_{ev}^{2}}}
\left( \left(c_i(\y,\pt) \sigma_{n[i]^{raw}}(\y, \pt)\right)^2  + \left(\sigma_{c_i(y, p_T)}n[i]^{raw} \right)^2\right)   ,
\end{equation}
where $\sigma_{n[i]^{raw}}$ (Eq.~\ref{eq:sigmaraw}) is the uncertainty of the uncorrected particle multiplicity
$n[i]^{raw}$ (Eqs.~\ref{eq:ntrkdEdx} and \ref{eq:ntrktof}), $\sigma_{c_i(y, p_T)}$ (Eqs.~\ref{eq:sigmacdedx} 
and \ref{eq:sigmactof}) denotes the uncertainty of the correction factors $c^{dEdx}_i(y,pT)$
(Eqs.~\ref{eq:cdedx}) or $c^{dEdx, m^2}_i(y,pT)$ (Eq.~\ref{eq:ctof})  for the \dEdx or $tof$-\dEdx
identification method, $i$ is the particle type and $N_{ev}$ the total number of accepted events.


\subsubsection{Systematic uncertainties}
\label{sec:systematics}

\begin{figure}[!ht]
\begin{center}
\includegraphics[width=0.32\textwidth, trim={2.5cm 1.5cm 1cm 2cm}, clip]{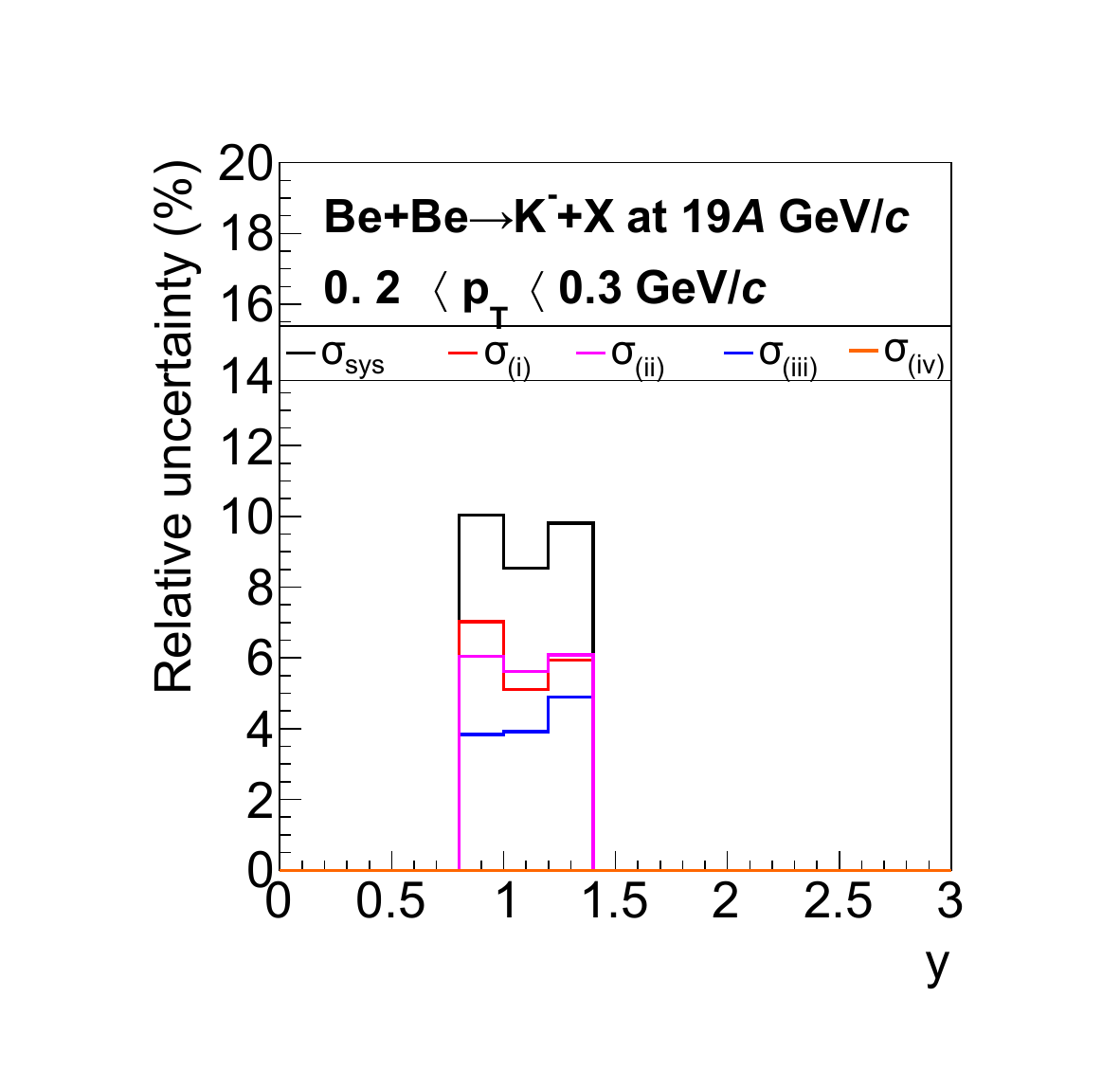}
\includegraphics[width=0.32\textwidth, trim={2.5cm 1.5cm 1cm 2cm}, clip]{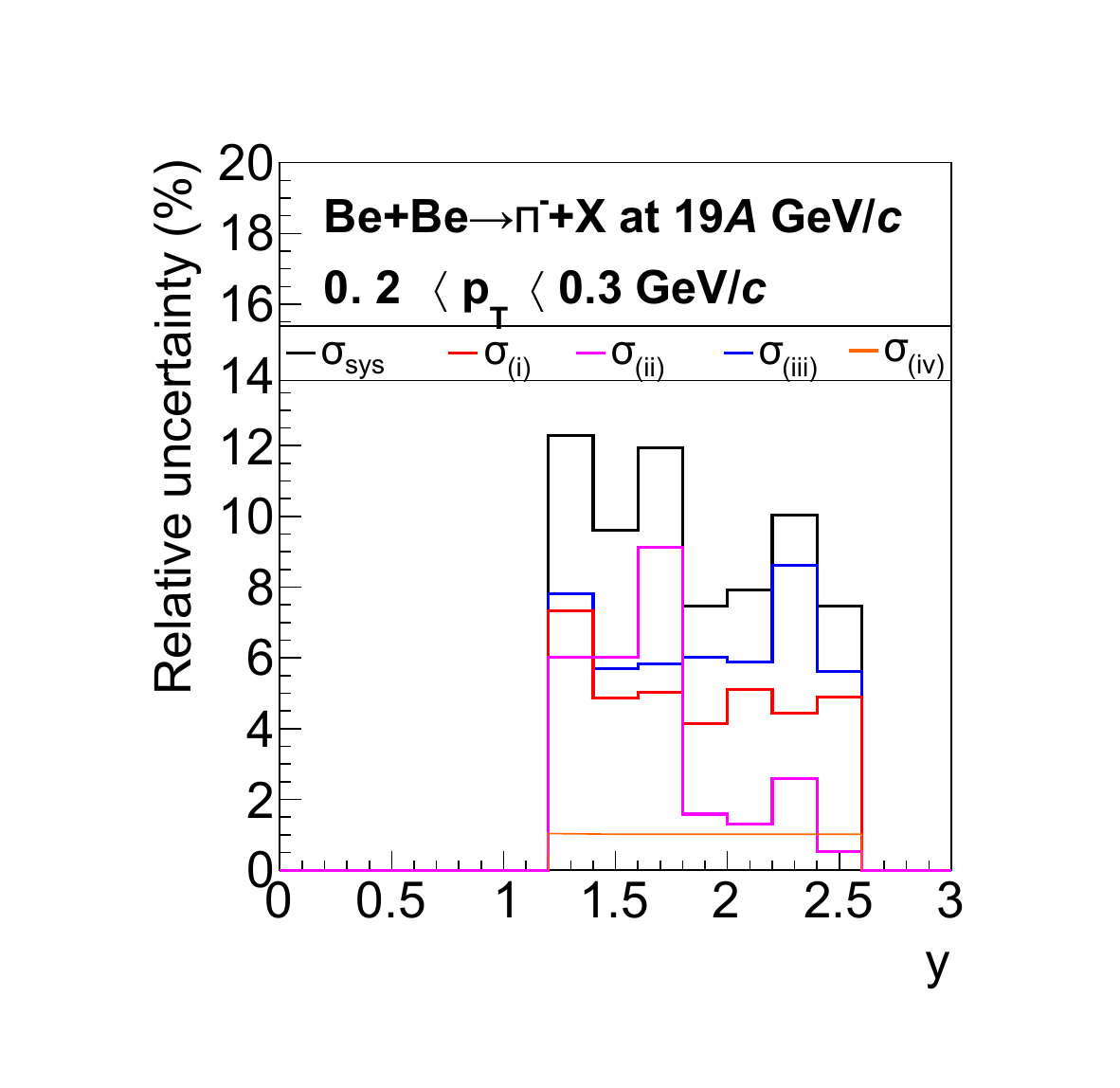}\\
\includegraphics[width=0.32\textwidth, trim={2.5cm 1.5cm 1cm 2cm}, clip]{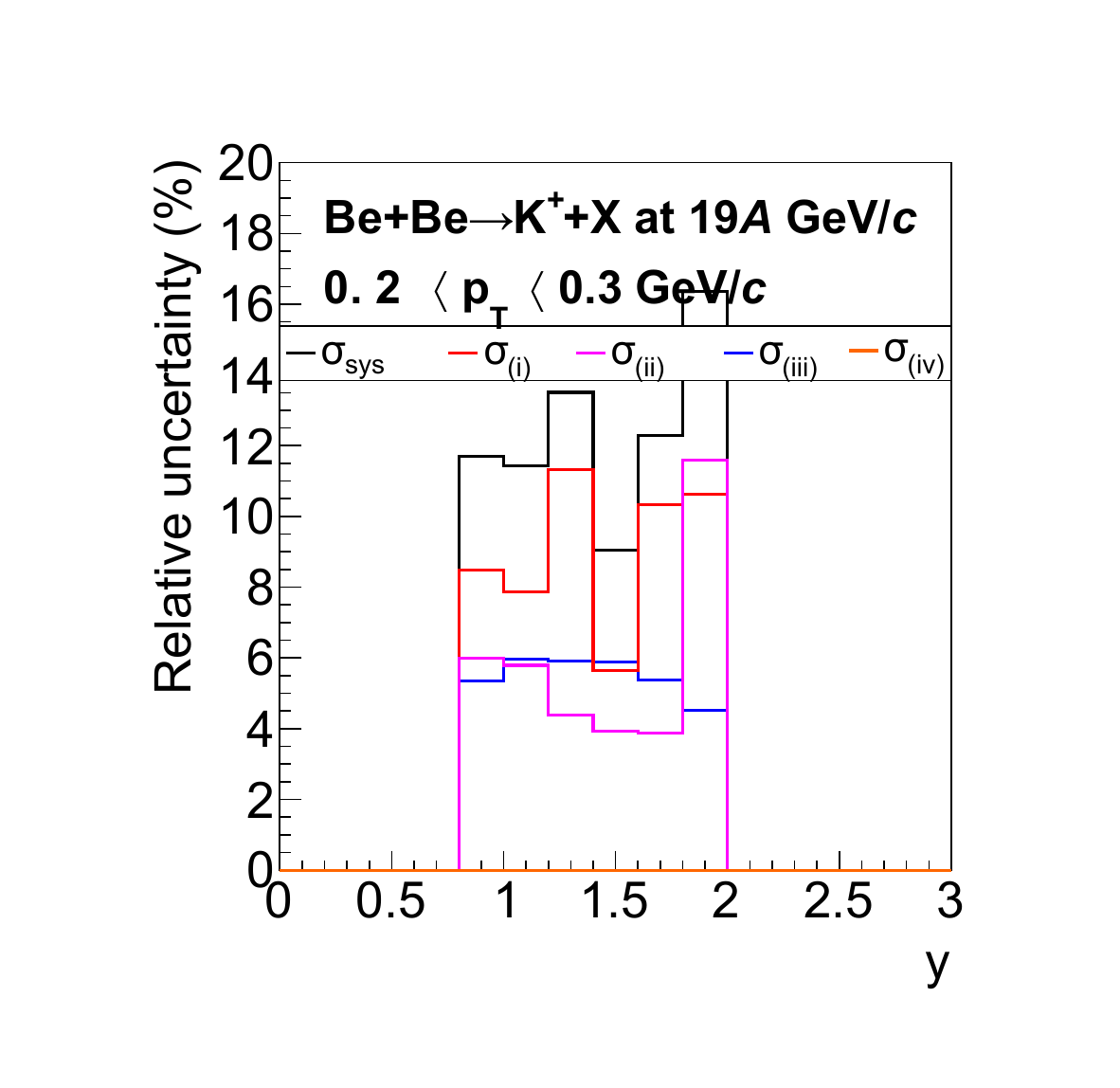}
\includegraphics[width=0.32\textwidth, trim={2.5cm 1.5cm 1cm 2cm}, clip]{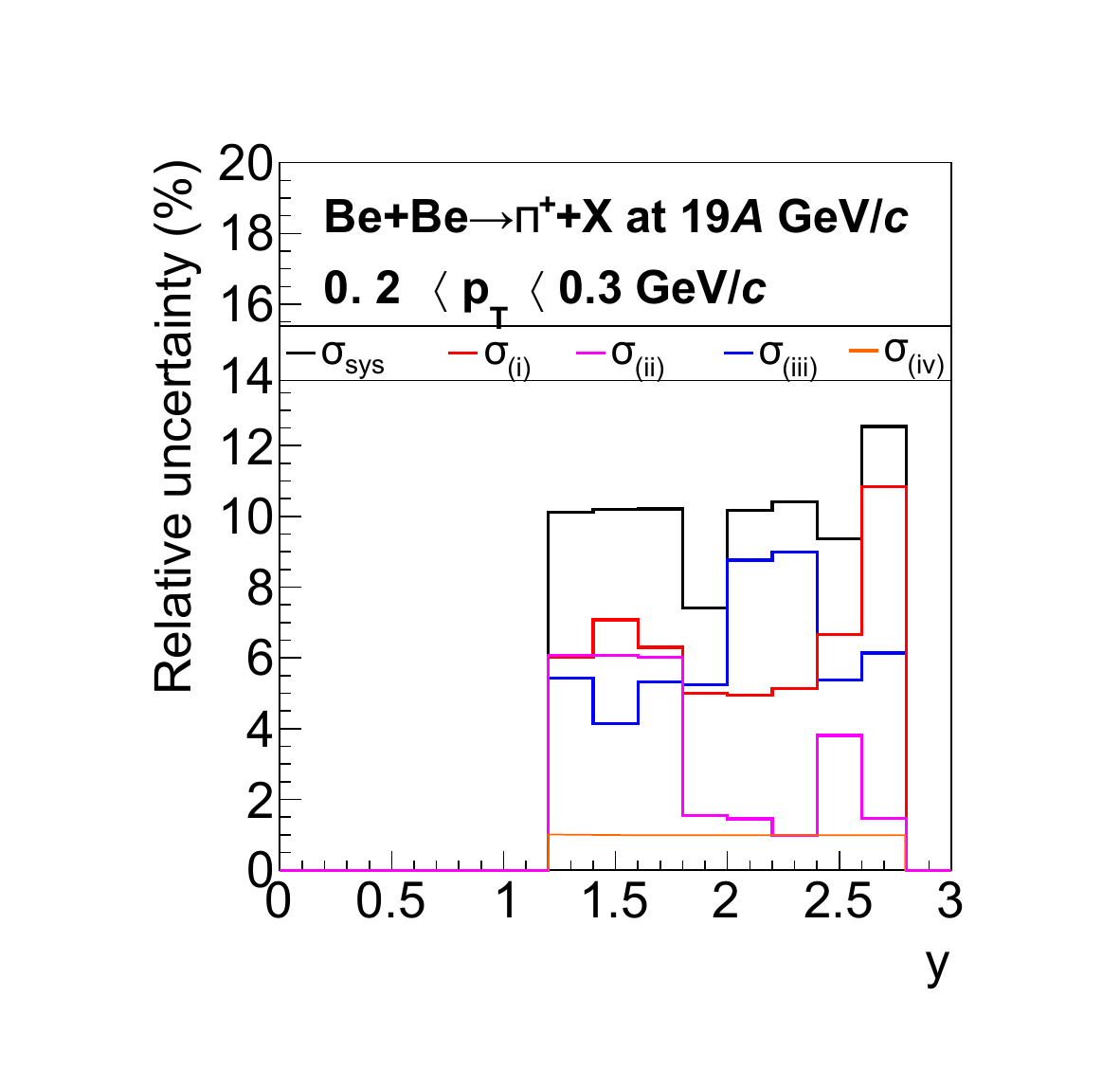}
\includegraphics[width=0.32\textwidth, trim={2.5cm 1.5cm 1cm 2cm}, clip]{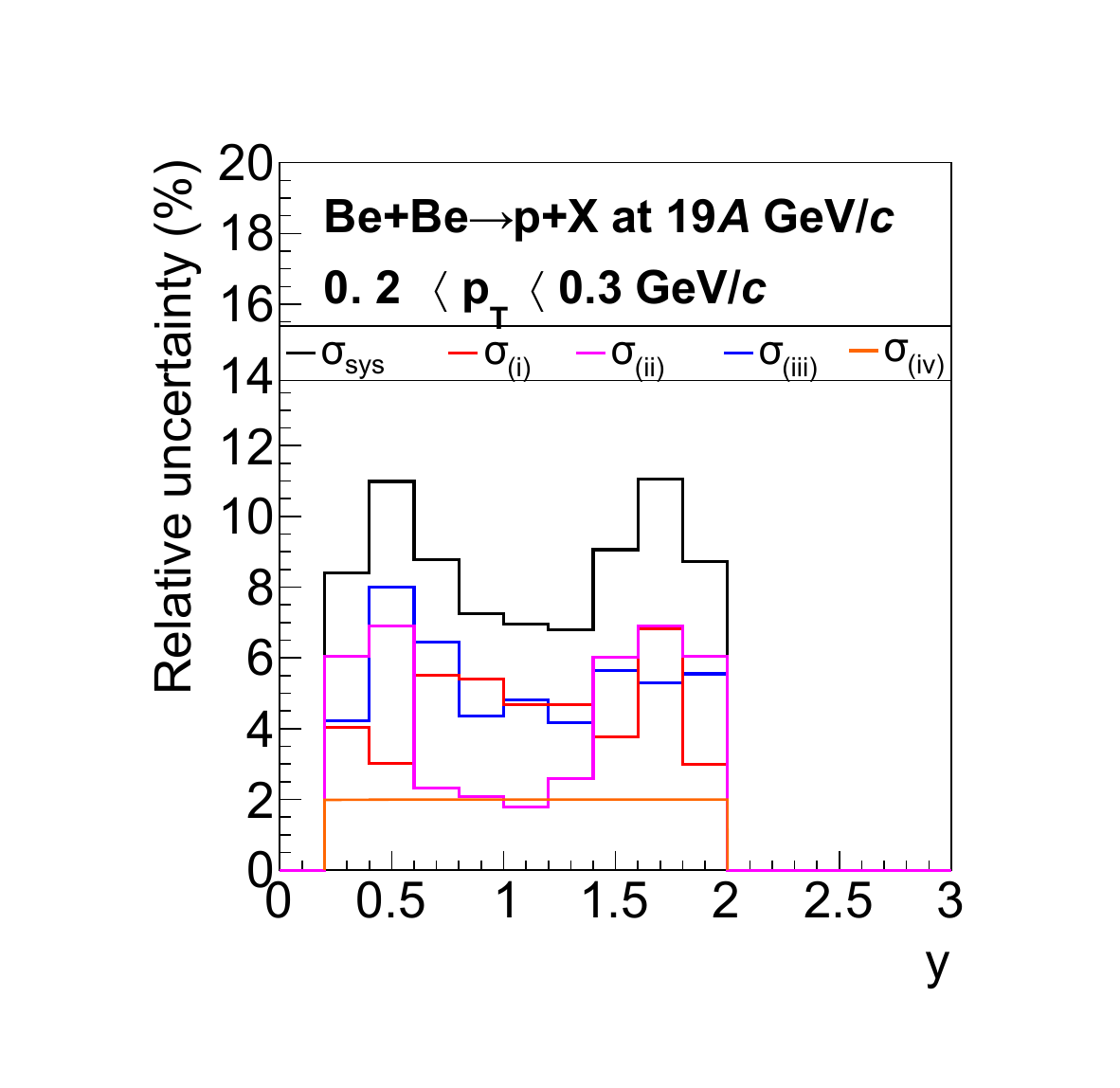}
\end{center}
\caption{Contributions to the systematic uncertainty of particle spectra obtained from the \dEdx method 
         at 19\AGeVc as a function of rapidity for the transverse momentum interval 
         between $0.2 - 0.3$~\GeVc. $\sigma_{\textrm{(i)}}$ (red lines) refers to event selection 
         $\sigma_{\textrm{(ii)}}$ (magenta lines) to the track selection procedure, 
         $\sigma_{\textrm{(iii)}}$ (blue lines) to the identification technique and $\sigma_{\textrm{(iv)}}$ (orange lines) to the
         contamination by feeddown from weak decays of strange particles. 
         Black lines, $\sigma_{\textrm{(sys)}}$, show the 
         total systematic uncertainty calculated as the square root of the sum of squares of the components.}
\label{fig:sysexa30}
\end{figure}

\begin{figure}[!ht]
\begin{center}
\includegraphics[width=0.3\textwidth, trim={2.5cm 1.5cm 1cm 2cm}, clip]{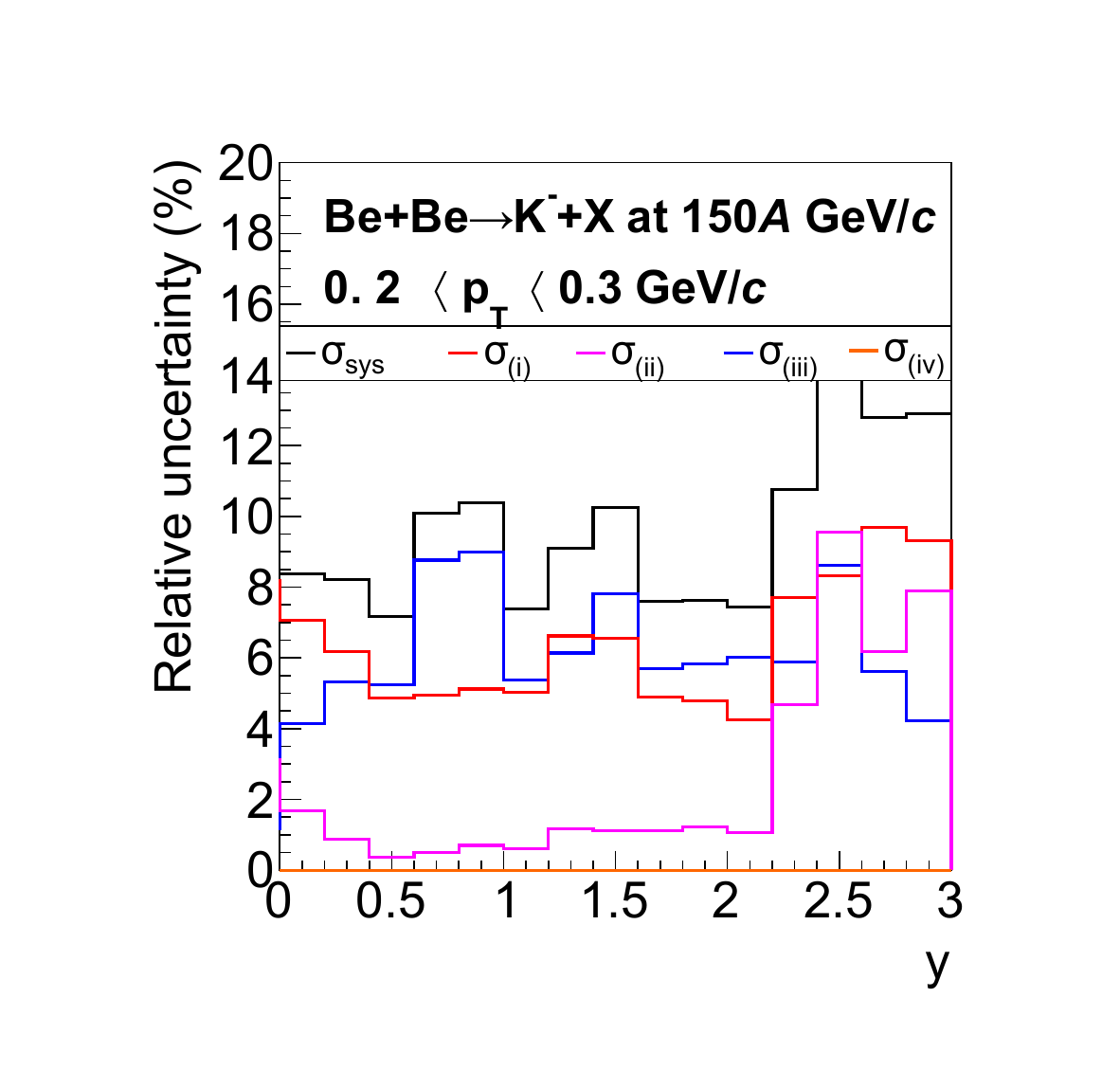}
\includegraphics[width=0.3\textwidth, trim={2.5cm 1.5cm 1cm 2cm}, clip]{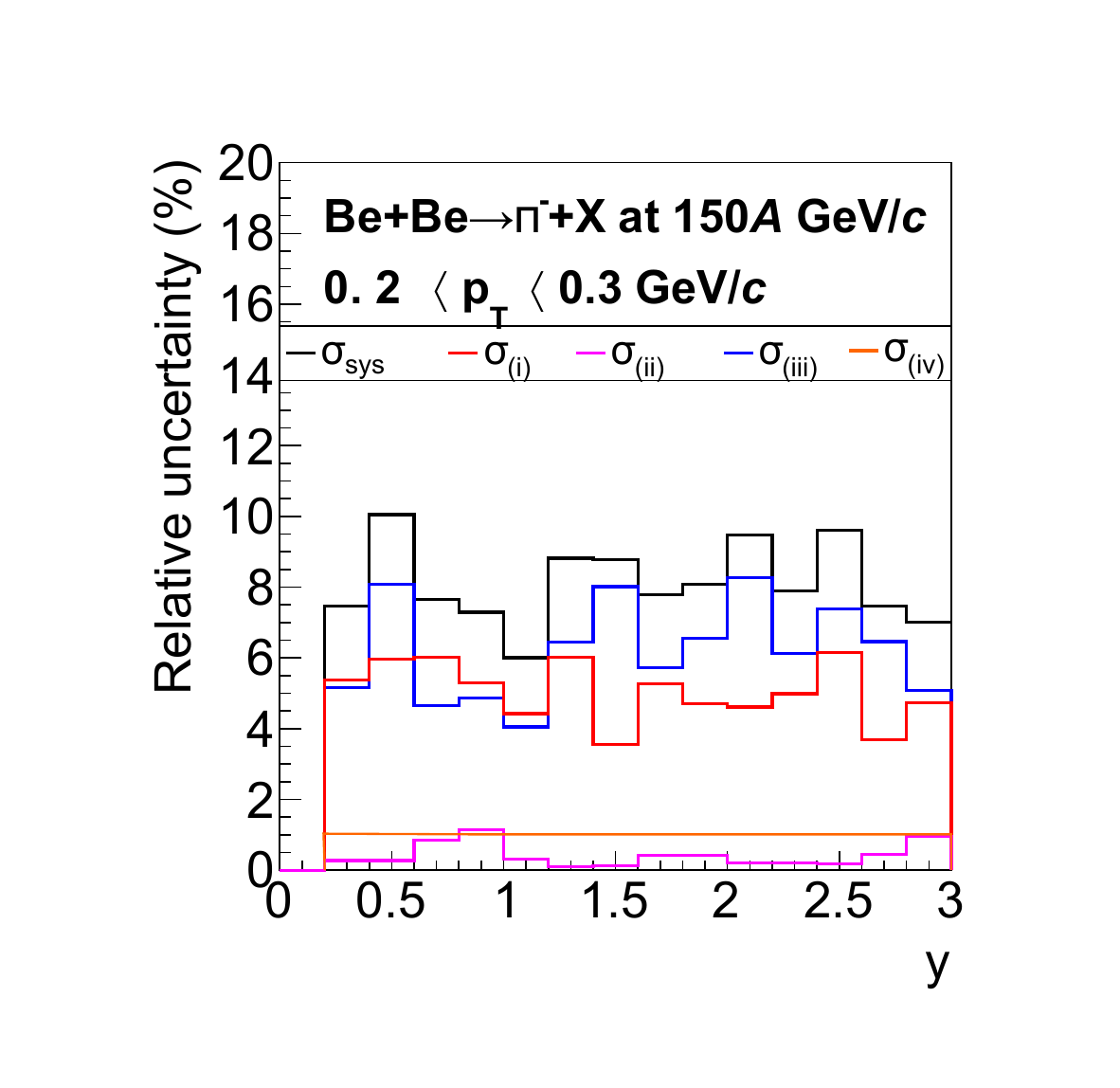}
\includegraphics[width=0.3\textwidth, trim={2.5cm 1.5cm 1cm 2cm}, clip]{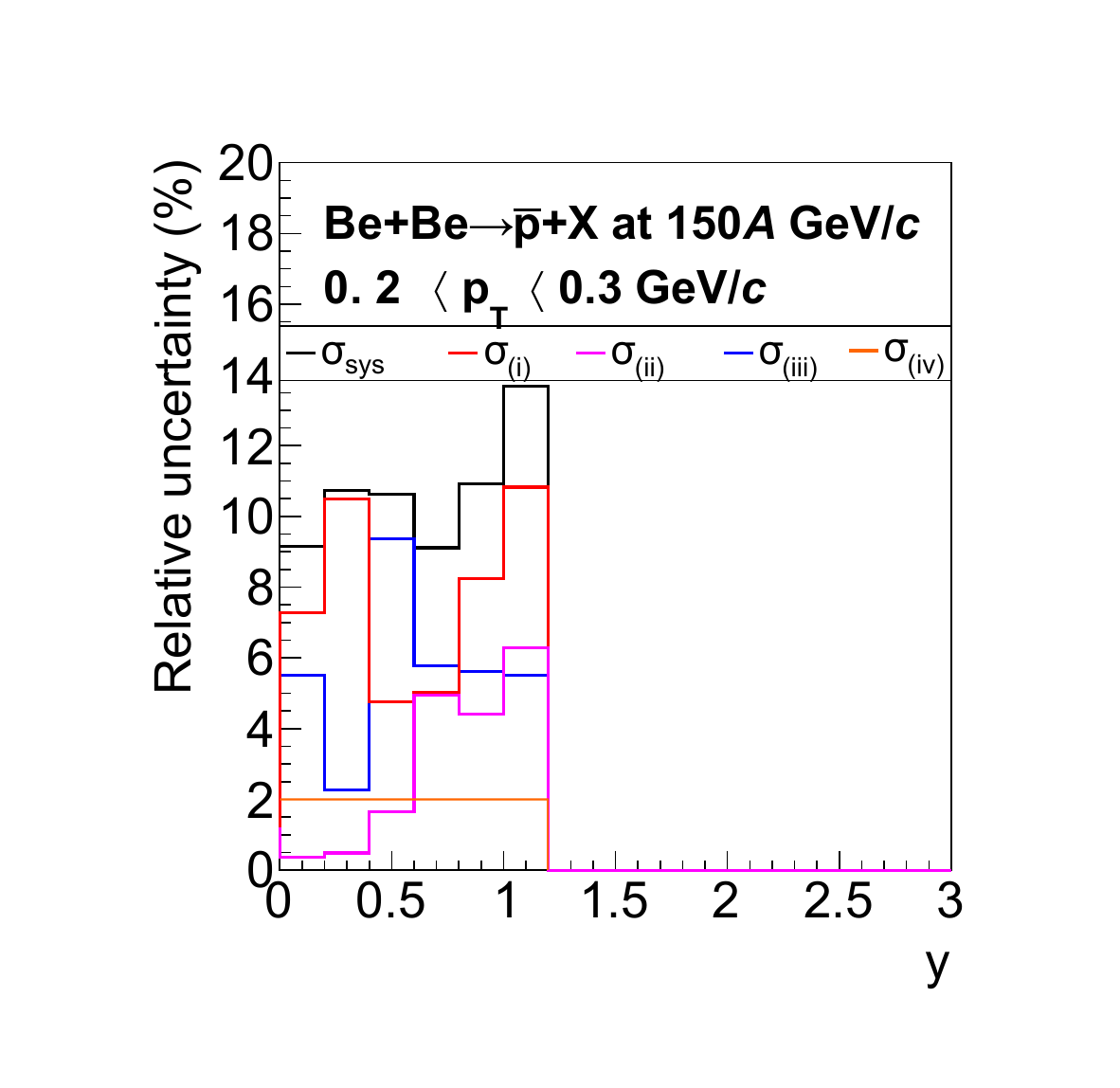}\\
\includegraphics[width=0.3\textwidth, trim={2.5cm 1.5cm 1cm 2cm}, clip]{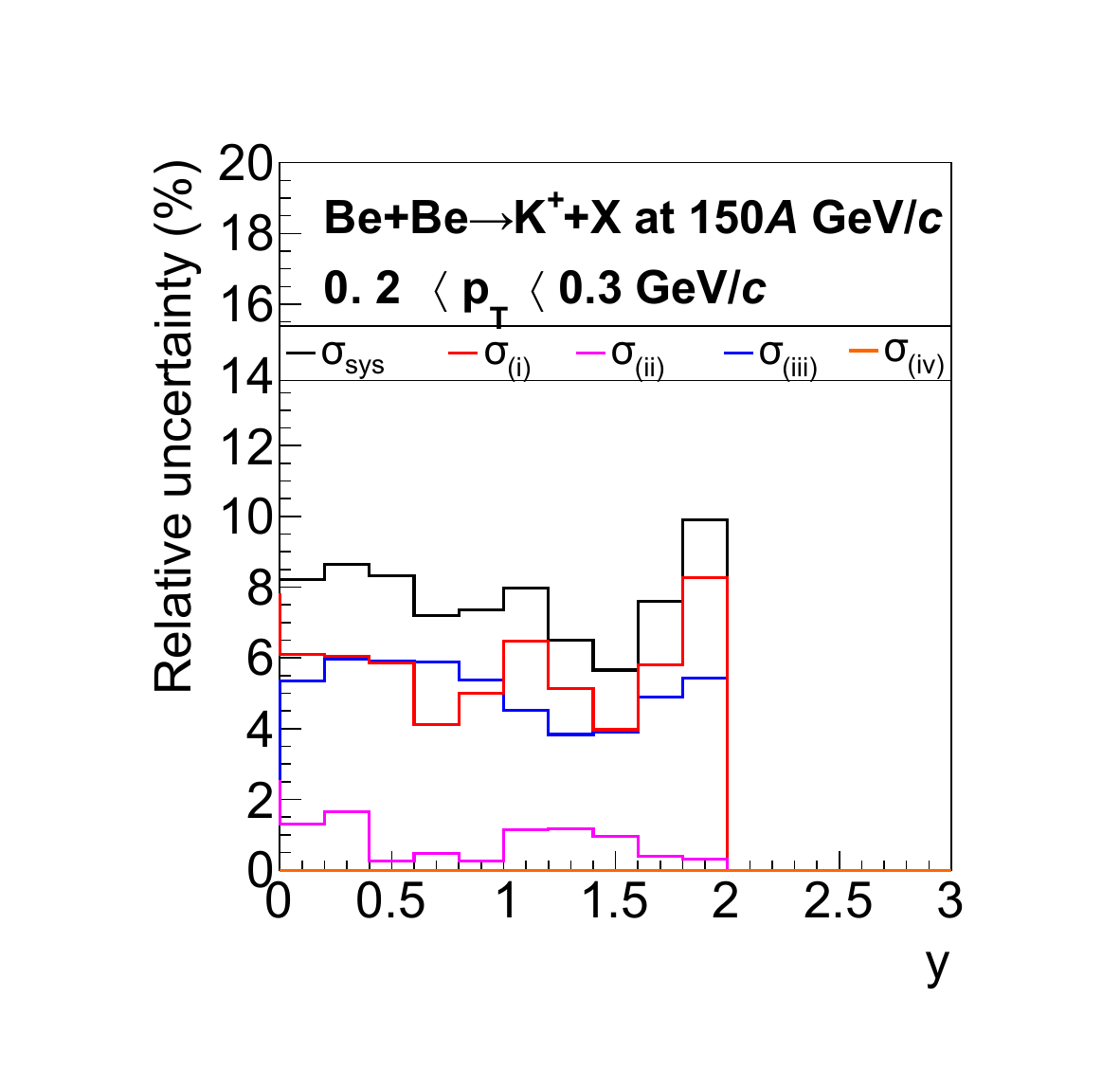}
\includegraphics[width=0.3\textwidth, trim={2.5cm 1.5cm 1cm 2cm}, clip]{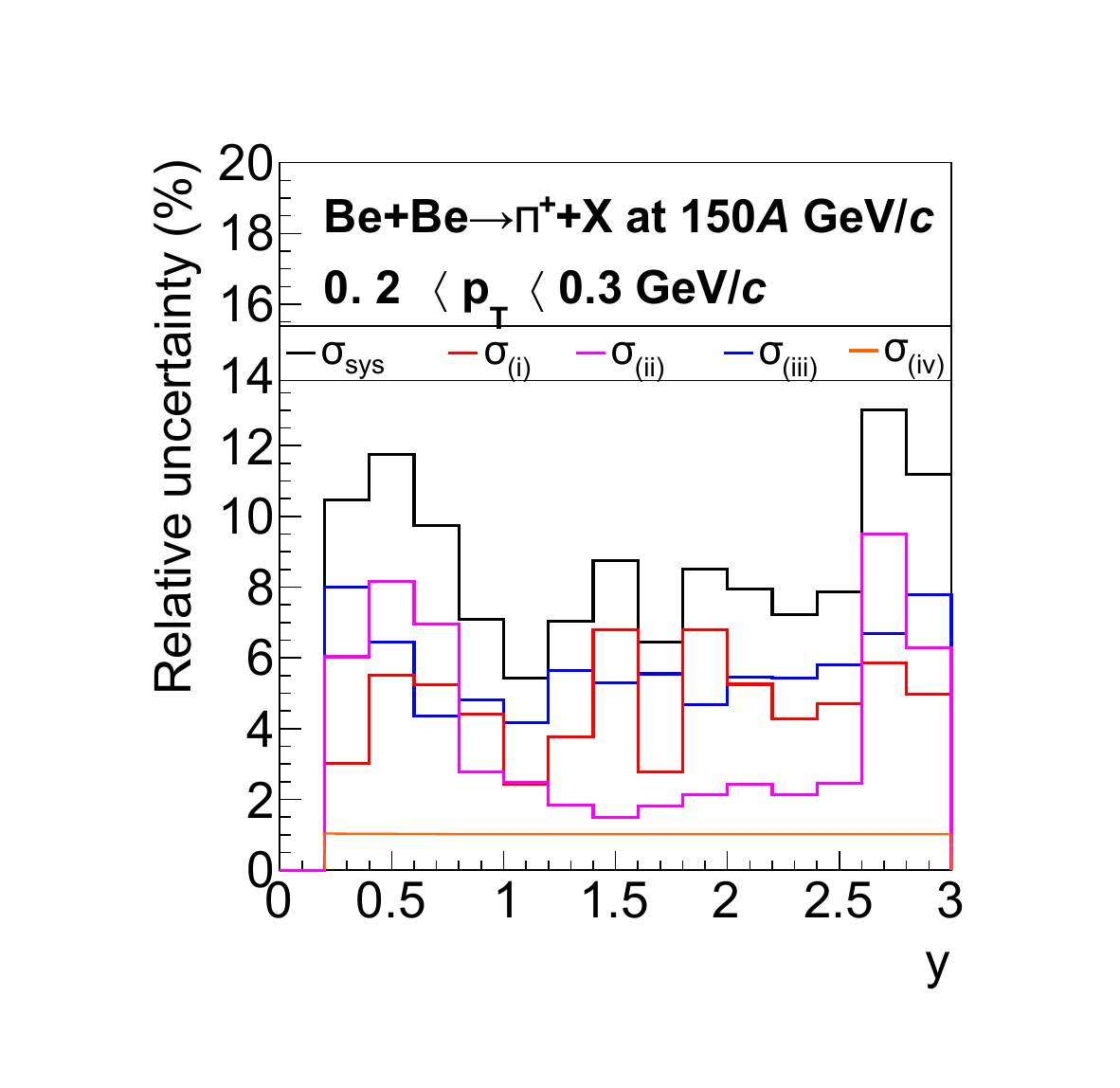}
\includegraphics[width=0.3\textwidth, trim={2.5cm 1.5cm 1cm 2cm}, clip]{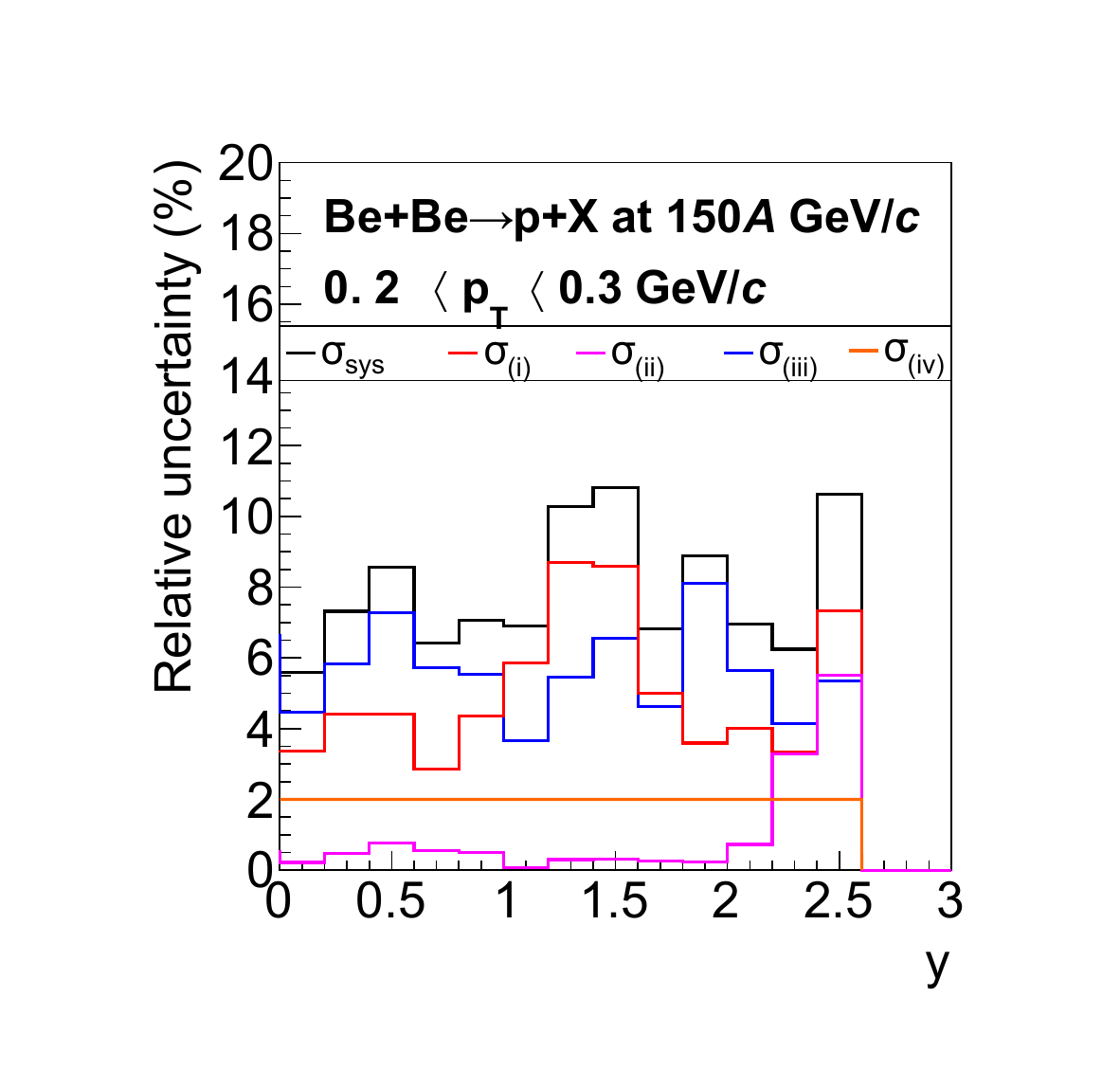}
\end{center}
\caption{Contributions to the systematic uncertainty of particle spectra obtained from the \dEdx method 
         at 150\AGeVc as a function of rapidity for the transverse momentum interval 
         between $0.2 - 0.3$~\GeVc. $\sigma_{\textrm{(i)}}$ (red lines) refers to event selection 
         $\sigma_{\textrm{(ii)}}$ (magenta lines) to the track selection procedure, 
         $\sigma_{\textrm{(iii)}}$ (blue lines) to the identification technique and $\sigma_{\textrm{(iv)}}$ (orange lines) to the
         contamination by feeddown from weak decays of strange particles.
         Black lines, $\sigma_{\textrm{(sys)}}$, show the 
         total systematic uncertainty calculated as the square root of the sum of squares of the components.}
\label{fig:sysexa150}
\end{figure}

\begin{figure}[!ht]
\begin{center}
\includegraphics[width=0.32\textwidth, trim={2cm 0.5cm 1.5cm 1cm}, clip]{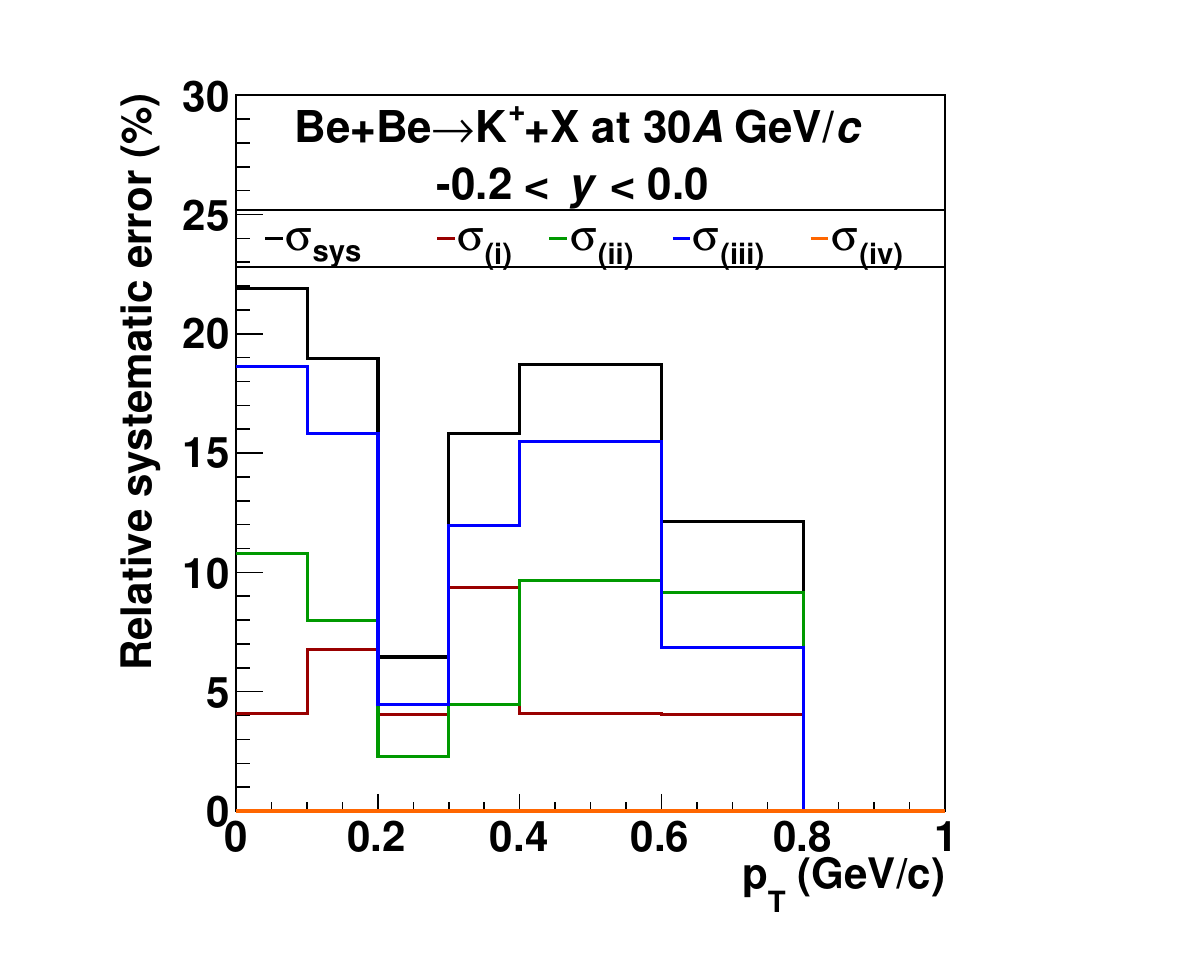}
\includegraphics[width=0.32\textwidth,trim={2cm 0.5cm 1.5cm 1cm}, clip]{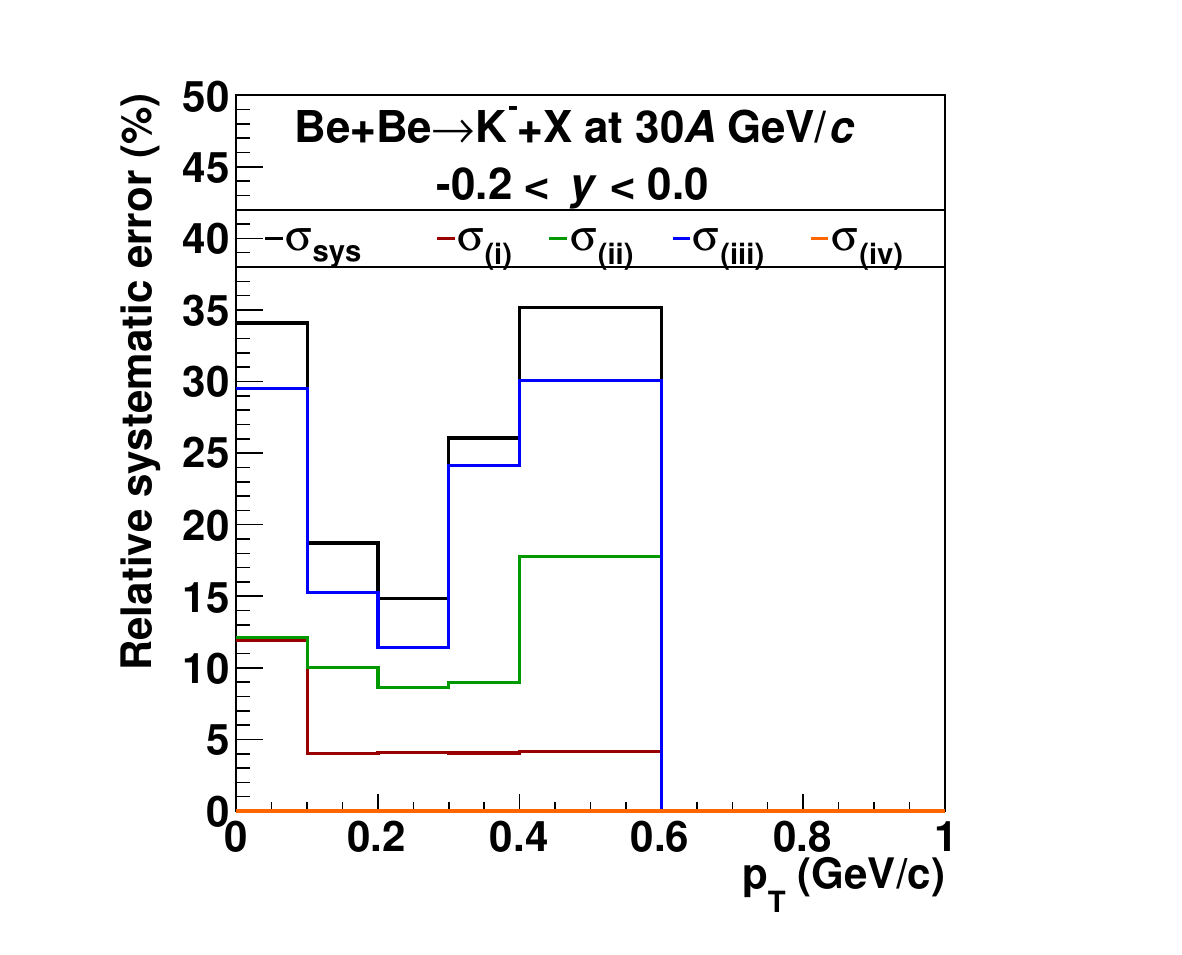}
\includegraphics[width=0.32\textwidth,trim={2cm 0.5cm 1.5cm 1cm}, clip]{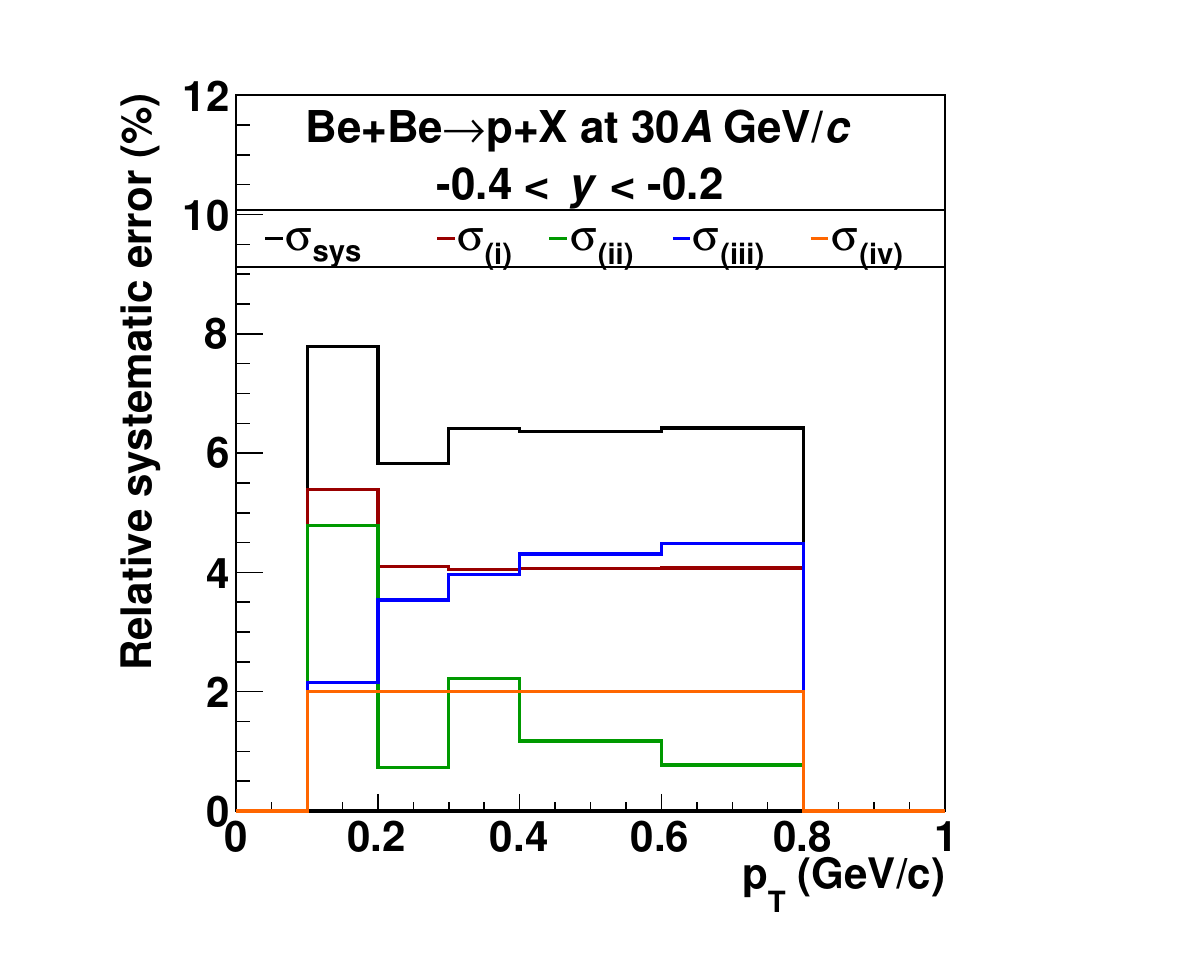}
\end{center}
\caption{Contributions to the systematic uncertainty of particle spectra obtained from the $tof$-\dEdx method at 
         30\AGeVc as a function of \pt for the rapidity interval -0.2 to 0.0~\GeVc 
         ($K^+$ and $K^-$) and -0.4 to -0.2~\GeVc (protons).
         $\sigma_{\textrm{(i)}}$ (red lines) refers to event selection 
         $\sigma_{\textrm{(ii)}}$ (green lines) to the track selection procedure, 
         $\sigma_{\textrm{(iii)}}$ (blue lines) to the identification technique and $\sigma_{\textrm{(iv)}}$ (orange lines) to the
         contamination by feeddown from weak decays of strange particles.
         Black lines, $\sigma_{\textrm{(sys)}}$, show the total systematic uncertainty calculated as the square root 
         of the sum of squares of the components.}
\label{fig:sysexa30t}
\end{figure}

\begin{figure}[!ht]
\begin{center}
\includegraphics[width=0.3\textwidth,trim={2cm 0.5cm 1.5cm 1cm}, clip]{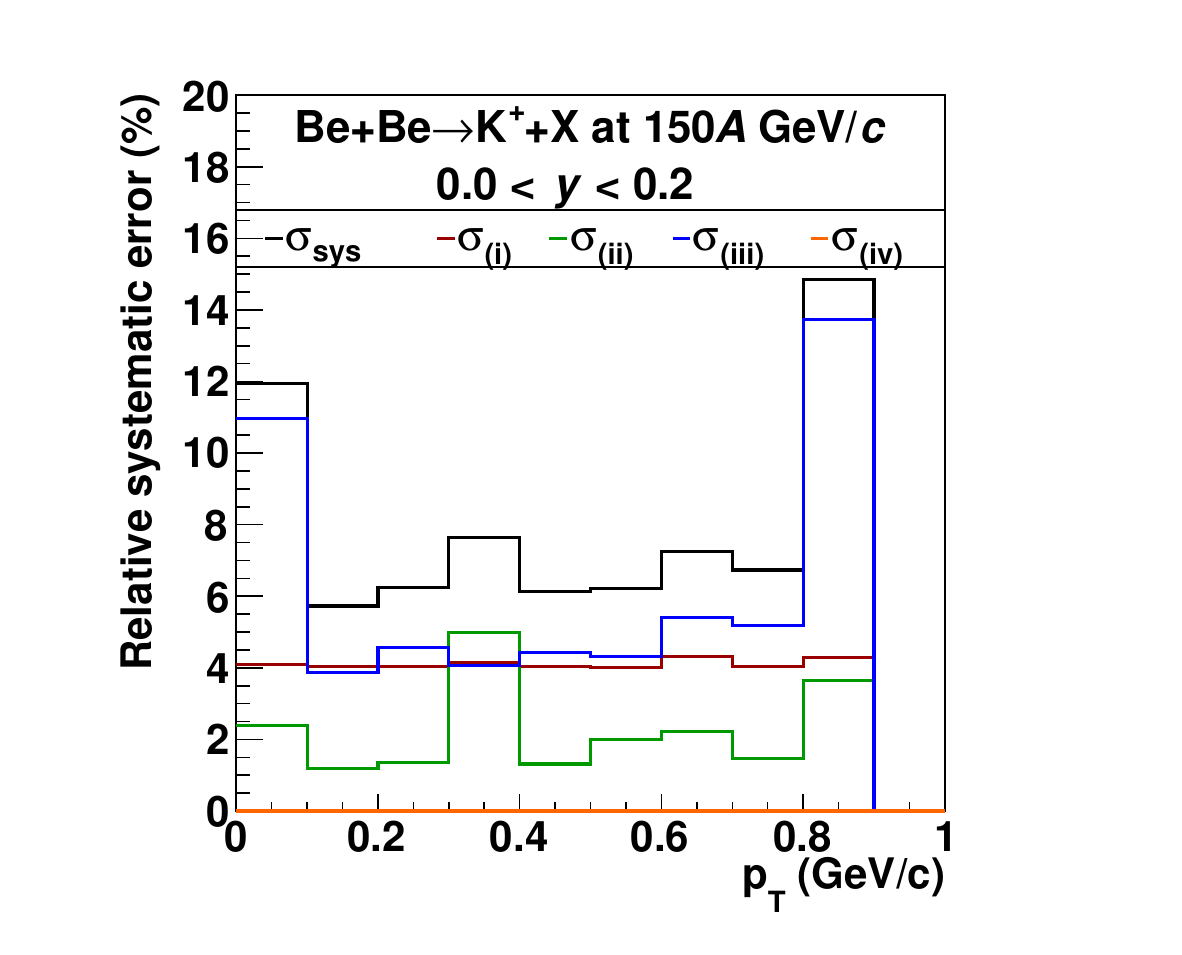}
\includegraphics[width=0.3\textwidth,trim={2cm 0.5cm 1.5cm 1cm}, clip]{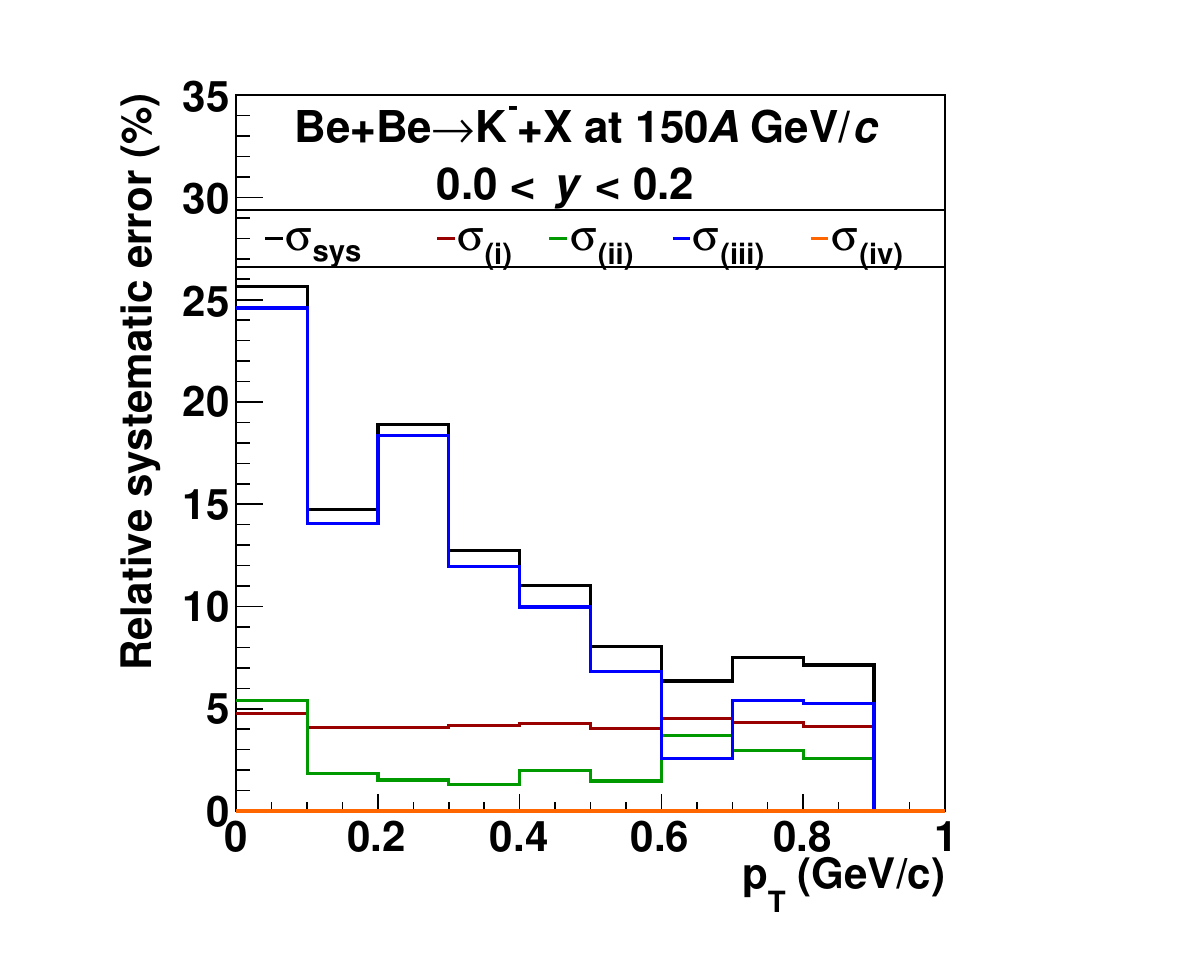}
\includegraphics[width=0.3\textwidth,trim={2cm 0.5cm 1.5cm 1cm}, clip]{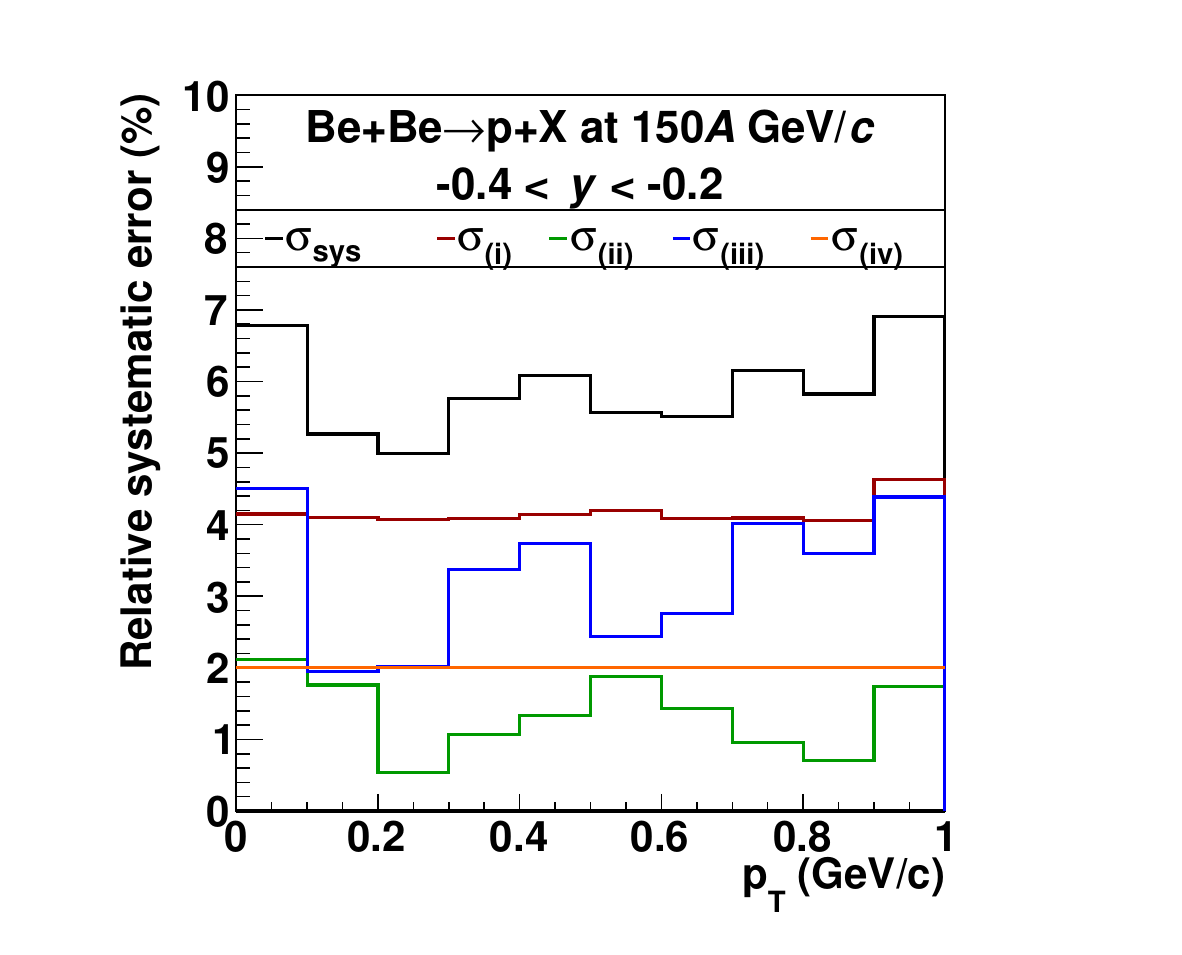}
\end{center}
\caption{Contributions to the systematic uncertainty of particle spectra obtained from the $tof$-\dEdx method 
         at 150\AGeVc as a function of \pt for the rapidity interval 0.0~to~0.2~\GeVc
         ($K^+$ and $K^-$) and -0.4 to -0.2~\GeVc (protons). 
         $\sigma_{\textrm{(i)}}$ (red lines) refers to event selection 
         $\sigma_{\textrm{(ii)}}$ (green lines) to the track selection procedure, 
         $\sigma_{\textrm{(iii)}}$ (blue lines) to the identification technique and $\sigma_{\textrm{(iv)}}$ (orange lines) to the
         contamination by feeddown from weak decays of strange particles.
         Black lines, $\sigma_{\textrm{(sys)}}$, show the total systematic uncertainty calculated as the square root 
         of the sum of squares of the components.}
\label{fig:sysexa150t}
\end{figure}

The contributions to the systematic uncertainty for the \dEdx and $tof$-\dEdx methods in 
selected \pt intervals at beam momenta of 19$A$ (30$A$) and 150~\AGeVc are presented in 
Figs.~\ref{fig:sysexa30}-\ref{fig:sysexa150t}.
Assuming they are uncorrelated, the total systematic uncertainty, also shown in the plots, was calculated 
as the square root of the sum of squares of the described components.

The considered contributions to systematic uncertainties are listed below:

\begin{enumerate}[(i)]

\item Event selection:

Systematic uncertainty of final particle yields due to the slightly different procedures of event selection
used for data ($E_{PSD}$) and simulated events ($E_F$) (see Sec.~\ref{sec:centrality}).

Systematic uncertainty related to the rejection of events with additional tracks from off-time particles 
was estimated by changing the width of the time window in which no second beam particle is allowed by $\pm~1~\mu s$ 
with respect to the nominal value of $\pm$4.5$~\mu$s. The maximal difference of the results was assigned as the 
systematic uncertainty of the selection. This contribution does not affect the results of the $tof-$\dEdx identification method.

Systematic uncertainty due to the choice of selection window for the \coordinate{z}-position of the fitted vertex was estimated
by varying the selection criteria for the data and the \EposLong model in the range of $\pm$25~\cm around the nominal position of the target.

\item Track selection:

Systematic uncertainty from the value of the cut on the track impact parameter at the primary vertex was estimated 
by varying the cut for the data and the \EposLong model by $\pm$50\% around the nominal value.

Systematic uncertainty originating from the requirement on the number of measured points in the magnetic field 
(minimum number of points in VTPCs and GTPC) was estimated by changing the nominal requirement 
on the number of measured points by $\pm~5$ (33\% of the standard selection) and $\pm~10$ (66\% of the standard selection) 
for the \dEdx and $tof-$\dEdx identification methods, respectively.

\item Particle identification:

Uncertainties of the \dEdx identification method were studied and estimated by a 10\% variation of the parameter constraints for
Eq.\ref{Eq:AsymGaus} fitted to the \dEdx spectra.

In case of $tof$-\dEdx identification, systematic uncertainties were estimated by shifting the
mean ($x_j$ and $y_j$) of the two-dimensional Gaussians (Eq.~\ref{eq:2dgaus}) fitted to the
$m^2-dE/dx$ distributions by $\pm1\%$.

An additional systematic uncertainty arises for the $tof-$\dEdx method from the quality requirements on the signals
registered in the ToF pixels. This systematic uncertainty was estimated by changing the nominal thresholds by $\pm$10\%.

\item Feeddown correction:

The determination of the feeddown correction is based on the \EposLong model which reasonably describes 
the available cross section data for strange particles in \pp collisions (see e.g. for $K^{+}$, $K^{-}$ Ref.~\cite{Aduszkiewicz:2017sei} and for $\Lambda$ at 158~\GeVc Ref.~\cite{Aduszkiewicz:2015dmr}). Systematic uncertainty comes from the lack of precise knowledge of the production cross section in Be+Be collisions 
of $K^{+}$, $K^{-}$, $\Lambda$, $\Sigma^{+}$, $\Sigma^{-}$, $K^{0}_{s}$ and $\bar{\Lambda}$ in case of pions, 
and in addition  of $\Sigma^{+}$ in case of protons, and $\bar{\Lambda}$ in case of $\bar{p}$. 


                            
\end{enumerate}

\FloatBarrier
\section{Results}

Two dimensional distributions $\frac{d^{2}n}{dydp_T}$ of $\pi^{-}$, $\pi^{+}$, $K^{-}$, $K^{+}$, 
$p$ and $\bar{p}$ produced in the 20\% most \textit{central} Be+Be interactions at different SPS energies
are presented in Fig.~\ref{fig:final2D}. Where available, results from the \dEdx method were used because of their smaller statistical uncertainties. Results from the $tof$-\dEdx method were taken to extend the 
momentum space coverage. Reflection symmetry around \y = 0 was used for $\frac{d^{2}n}{dydp_T}$ near 
mid-rapidity (see discussion in Sec.\ref{sec:yspectra}). Empty bins in momentum space (mostly for lower energies) are caused by insufficient acceptance for the identification methods used in the analysis. 

\subsection{Transverse momentum and transverse mass spectra}

\begin{figure*}
                \begin{center}
                \includegraphics[width=0.3\textwidth]{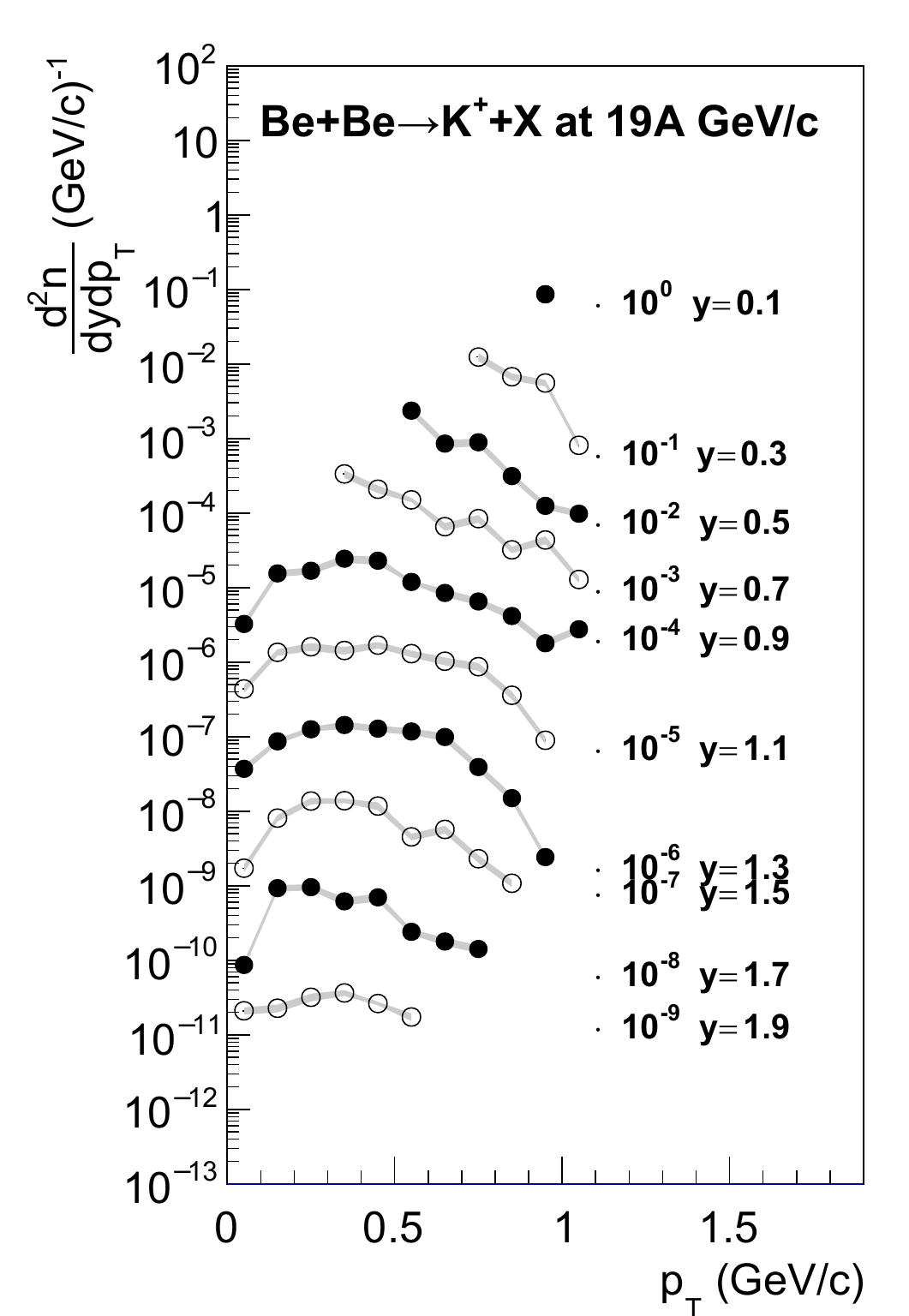}
                \includegraphics[width=0.3\textwidth]{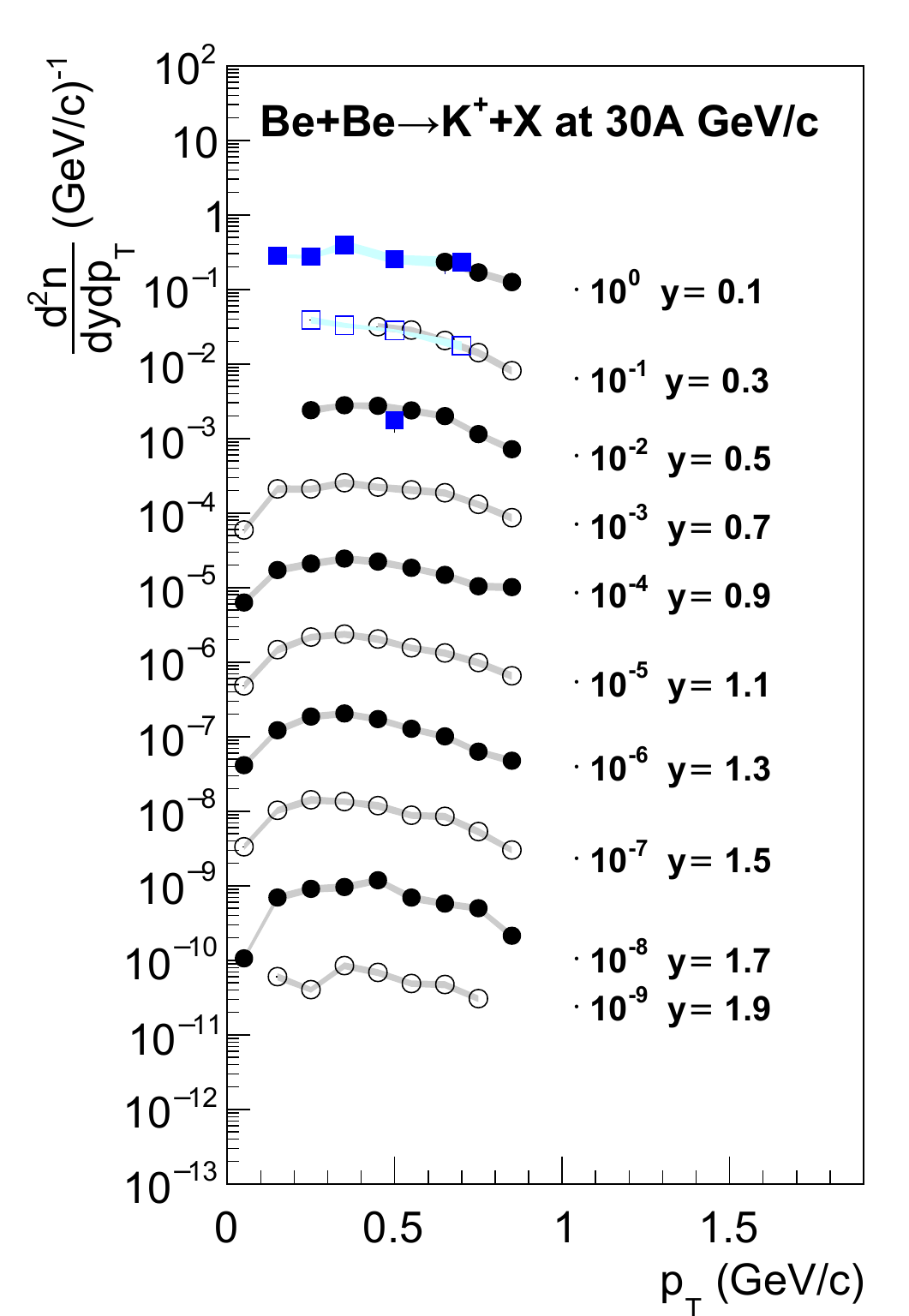}\\
                \includegraphics[width=0.3\textwidth]{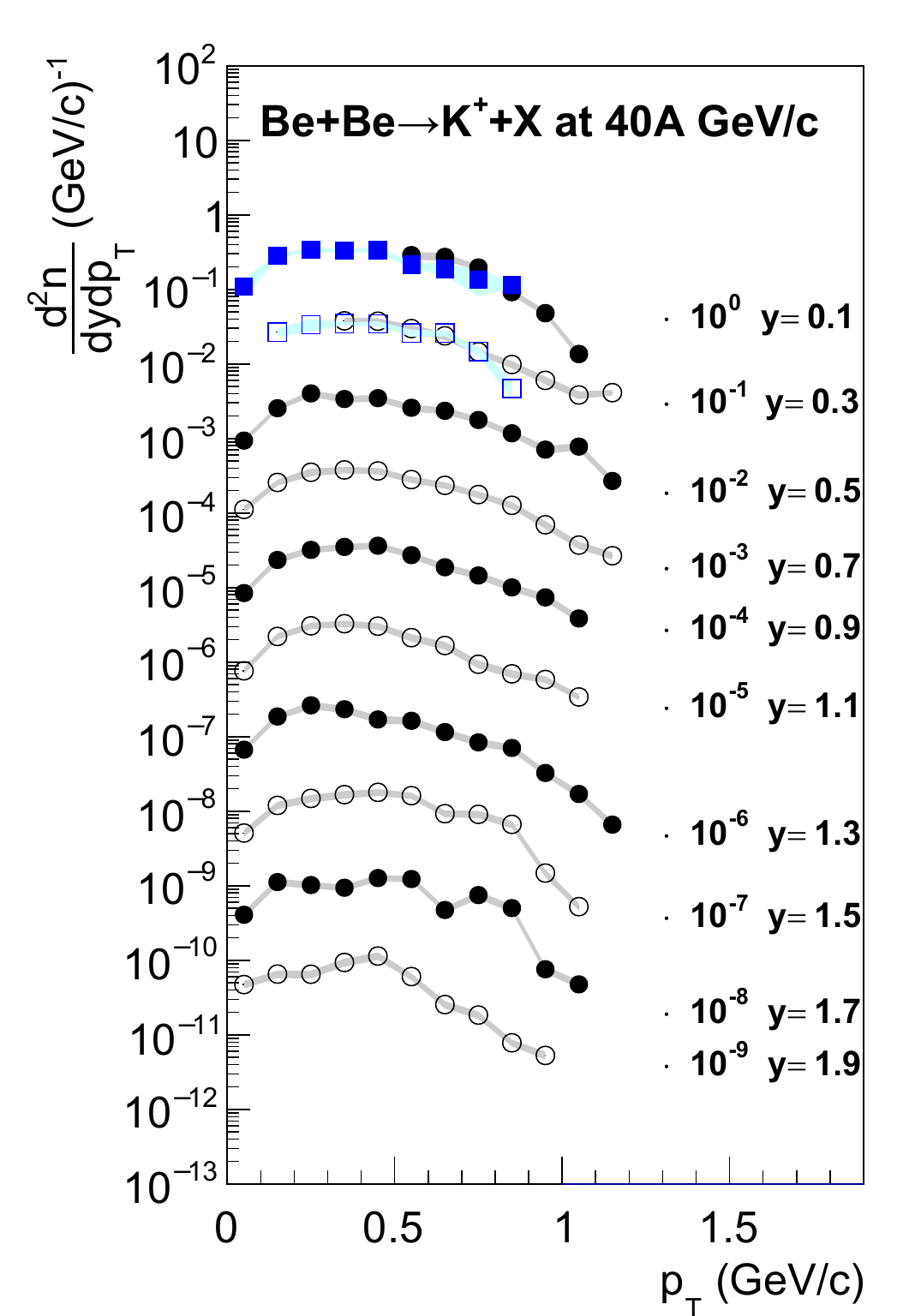}
                \includegraphics[width=0.3\textwidth]{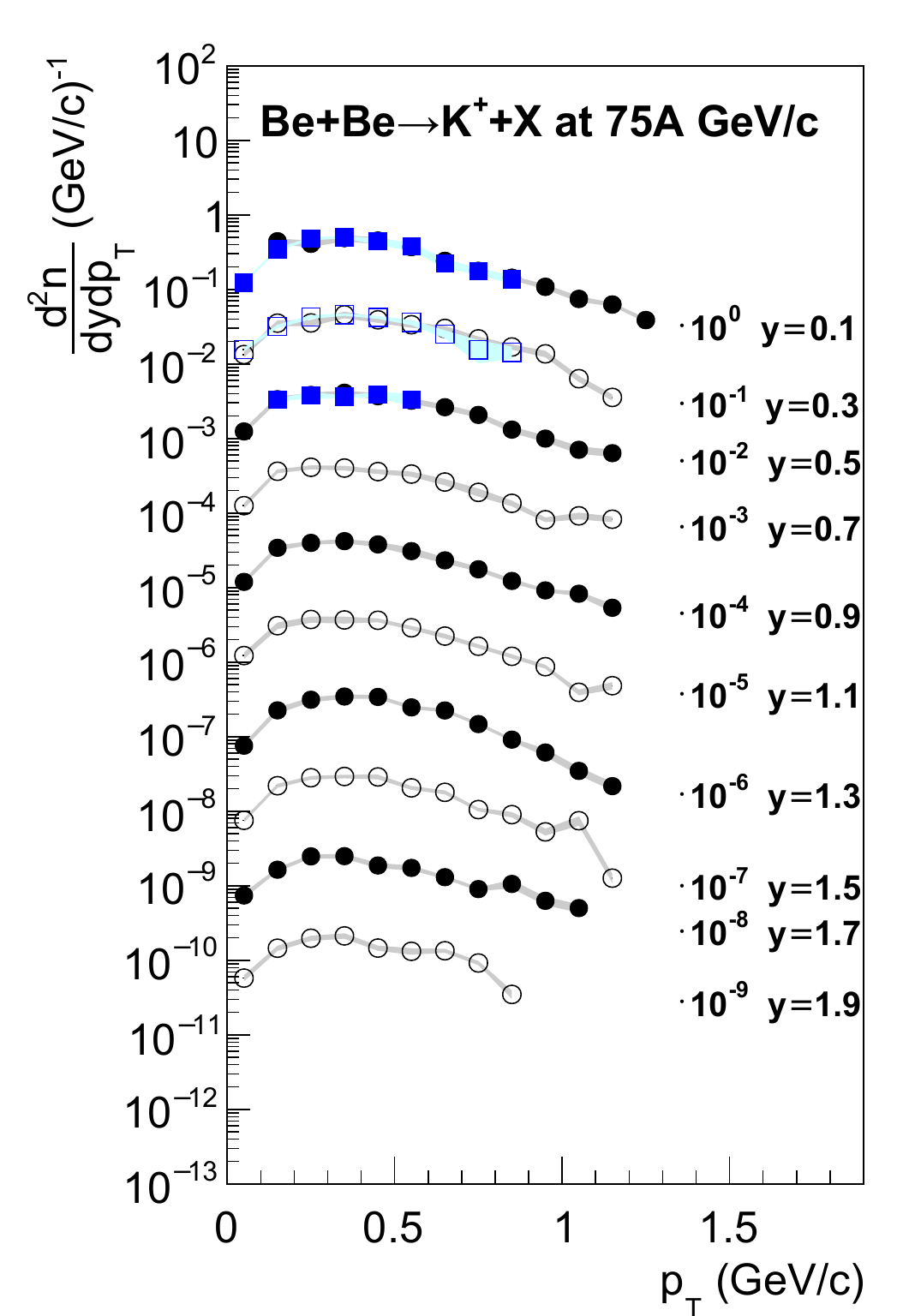}
                \includegraphics[width=0.3\textwidth]{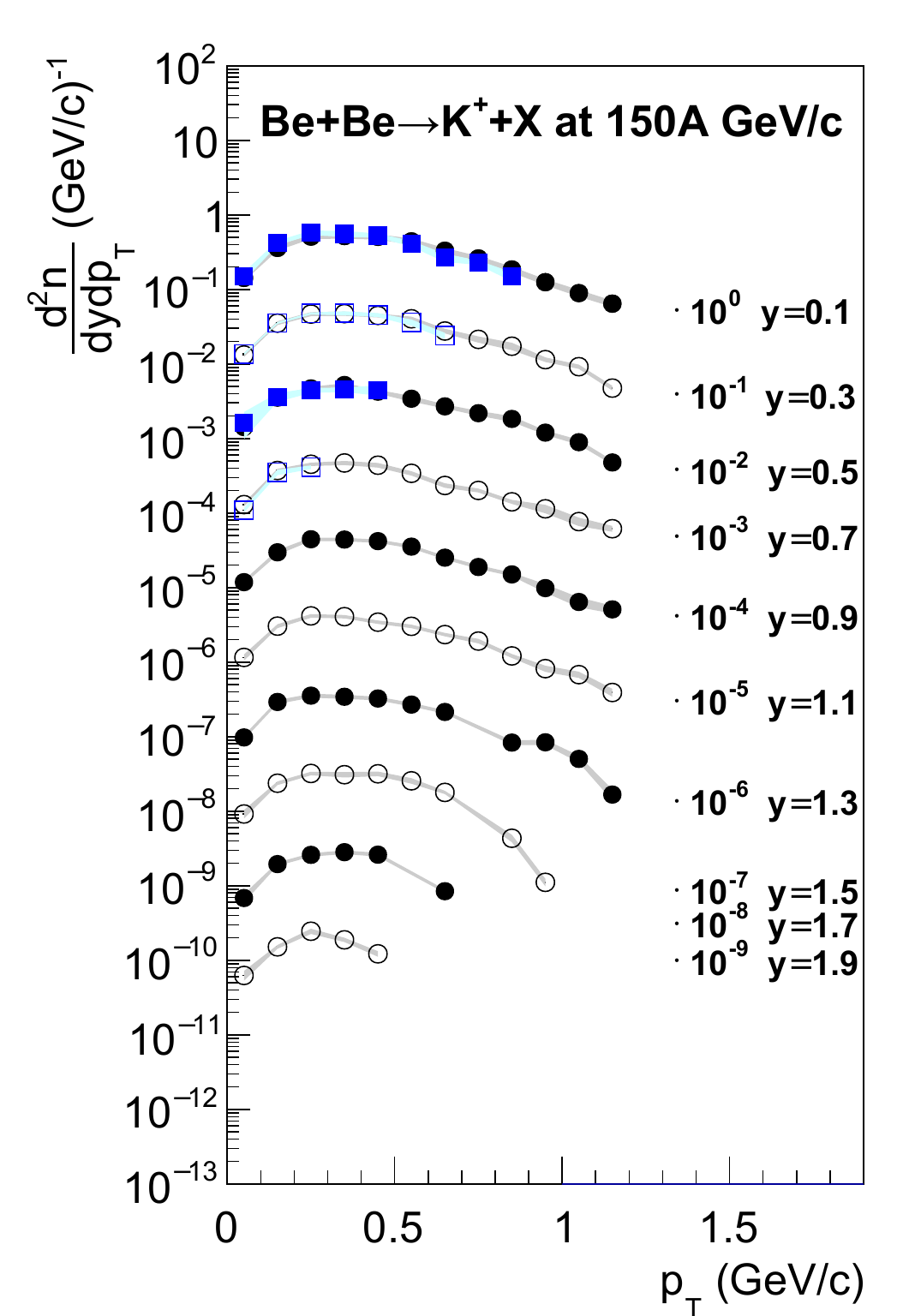}
                \end{center}
                \caption{Transverse momentum spectra in rapidity slices of $K^{+}$ produced
                         in the 20\% most \textit{central} Be+Be collisions. Rapidity values given in the
                         legends correspond to the middle of the corresponding interval. Black dots (blue squares)
                         show results of the \dEdx ($tof$-\dEdx) analysis, respectively. Shaded bands
                         show systematic uncertainties.}
                \label{fig:nptkpls}
\end{figure*}

\begin{figure*}
                \begin{center}
                \includegraphics[width=0.3\textwidth]{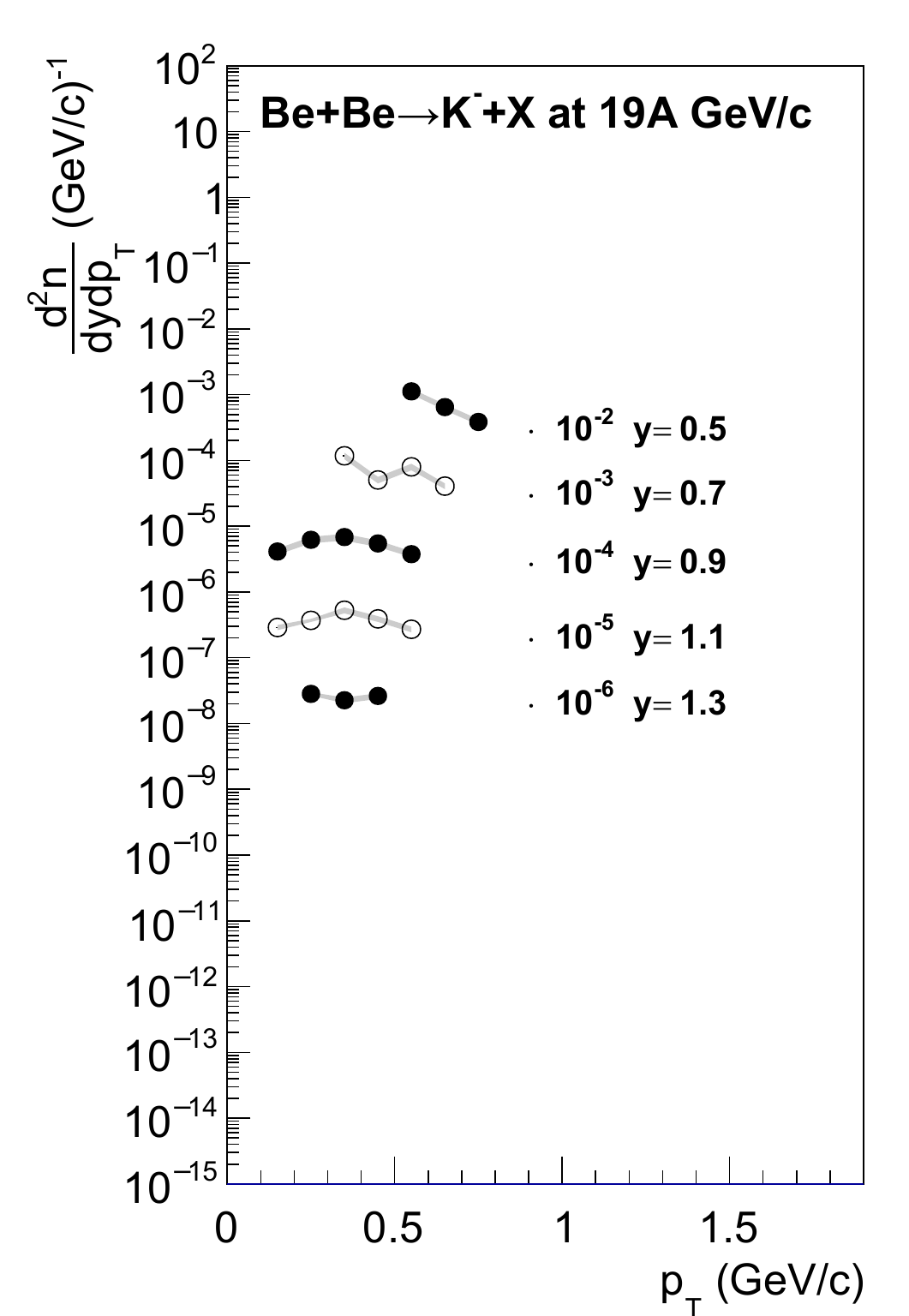}
                \includegraphics[width=0.3\textwidth]{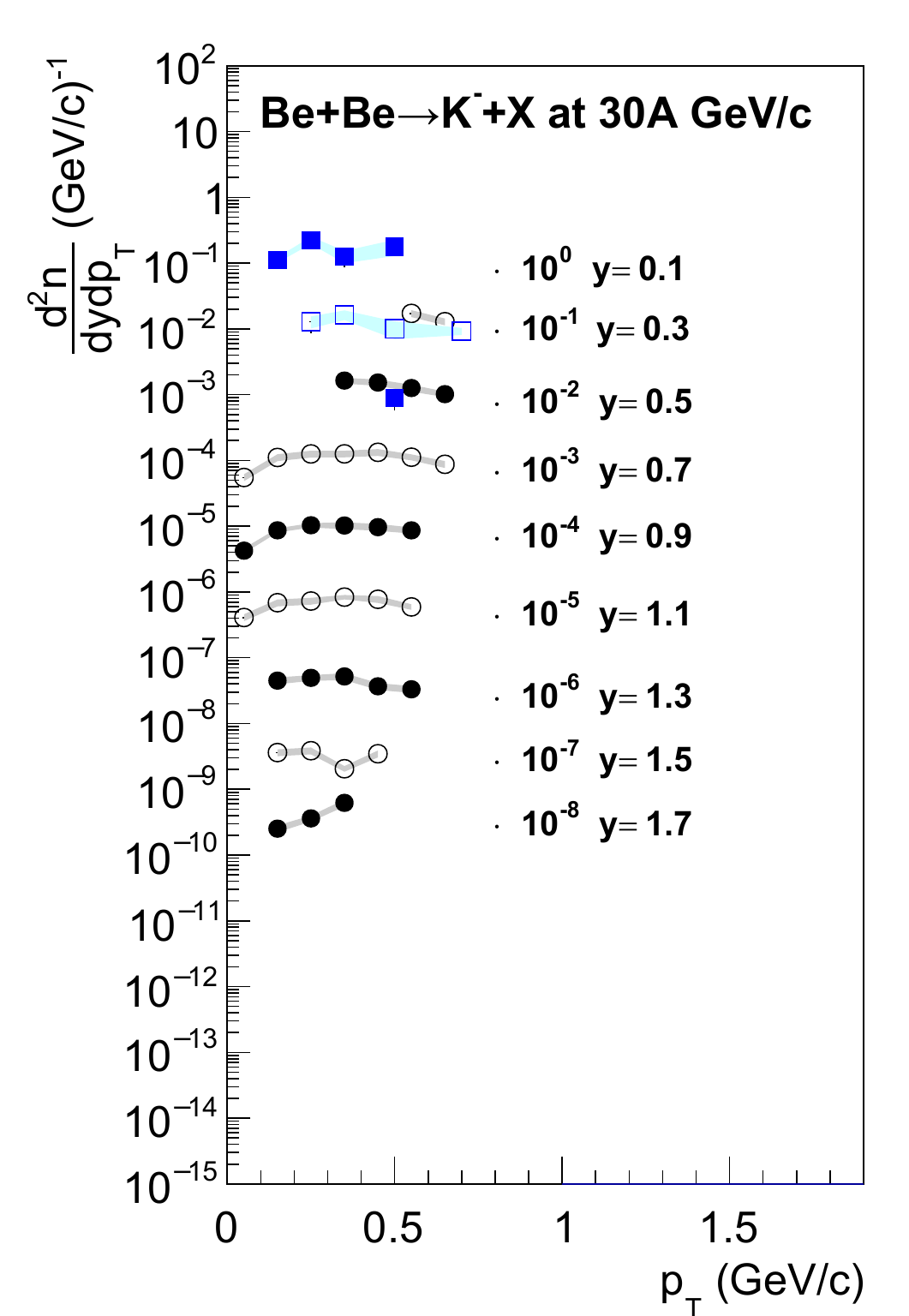}\\
                \includegraphics[width=0.3\textwidth]{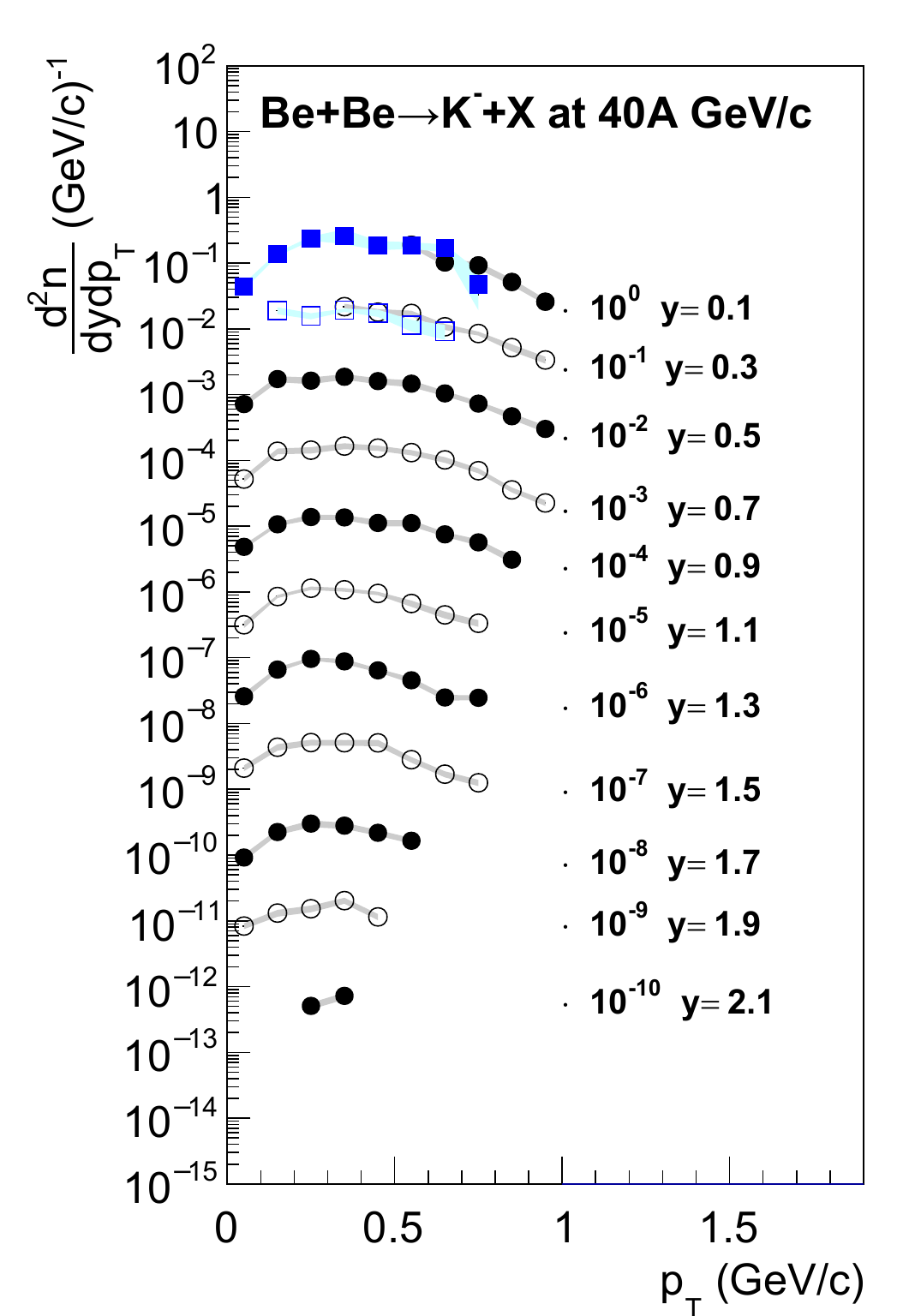}
                \includegraphics[width=0.3\textwidth]{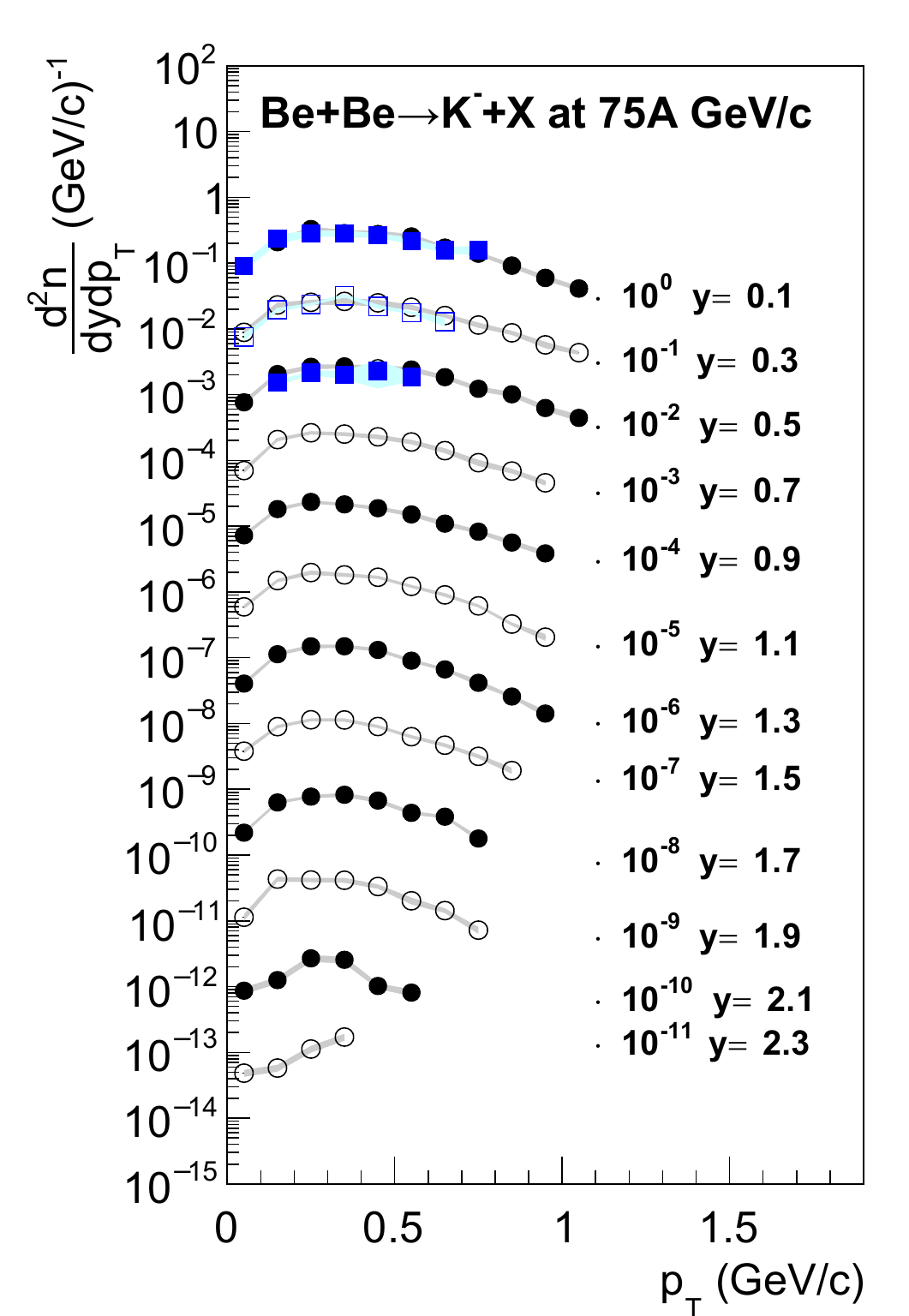}
                \includegraphics[width=0.3\textwidth]{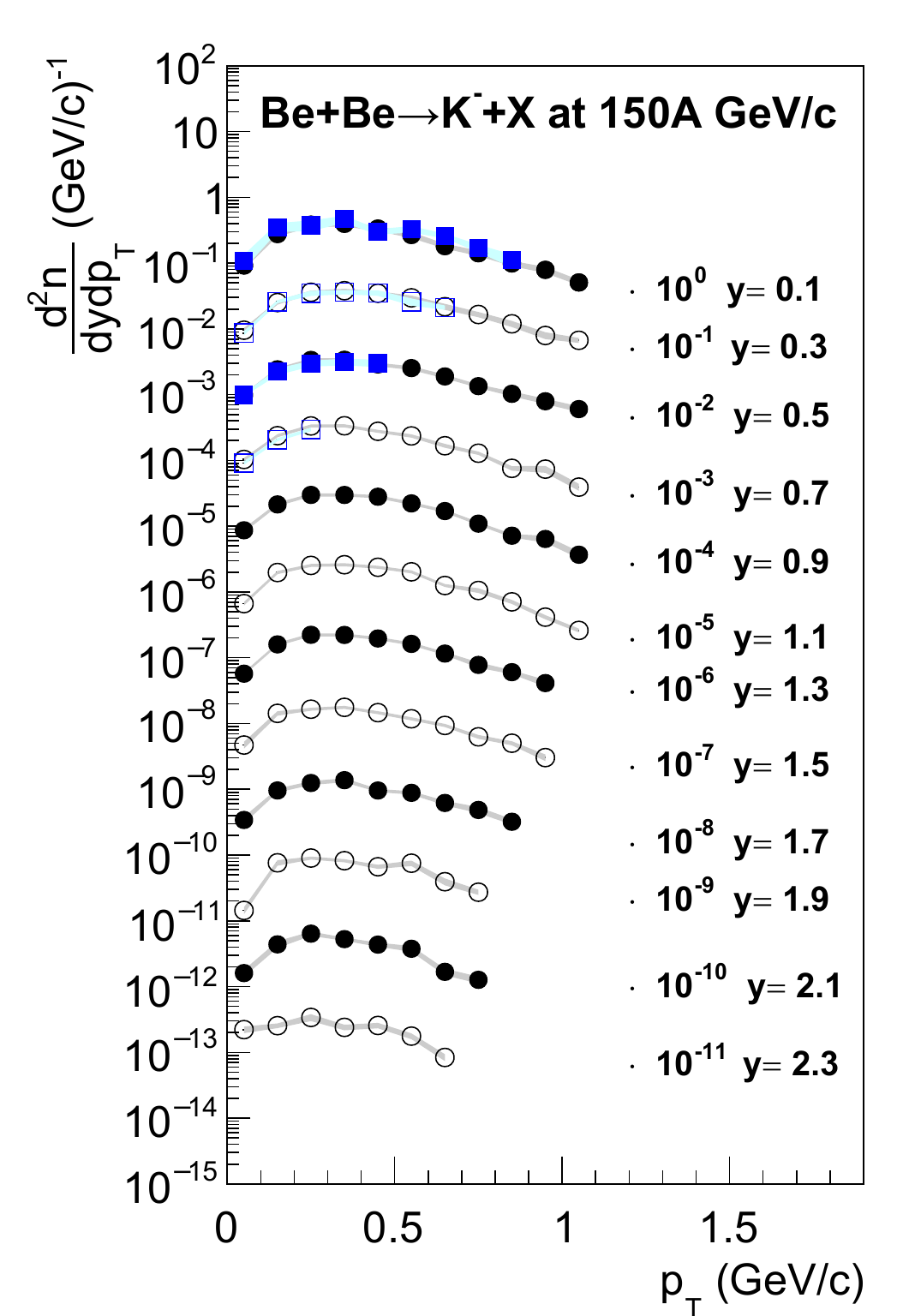}
                \end{center}
                \caption{Transverse momentum spectra in rapidity slices of $K^{-}$ produced 
                         in the 20\% most \textit{central} Be+Be collisions.
                         Rapidity values given in the 
                         legends correspond to the middle of the corresponding interval. Black dots (blue squares)
                         show results of the \dEdx ($tof$-\dEdx) analysis, respectively. Shaded bands 
                         show systematic uncertainties.}
                \label{fig:nptkmns}
\end{figure*}

\begin{figure*}
                \begin{center}
                \includegraphics[width=0.3\textwidth]{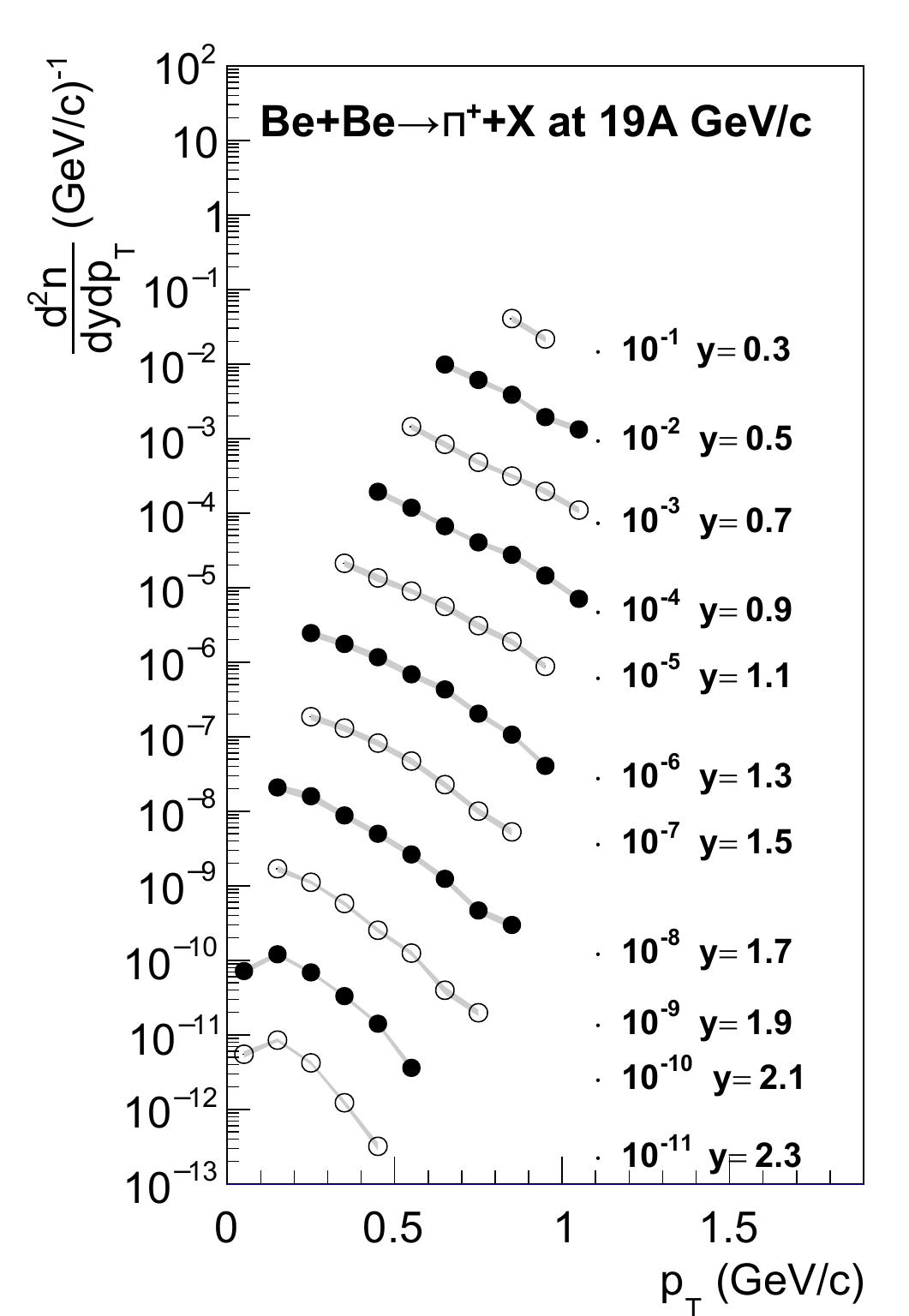}
                \includegraphics[width=0.3\textwidth]{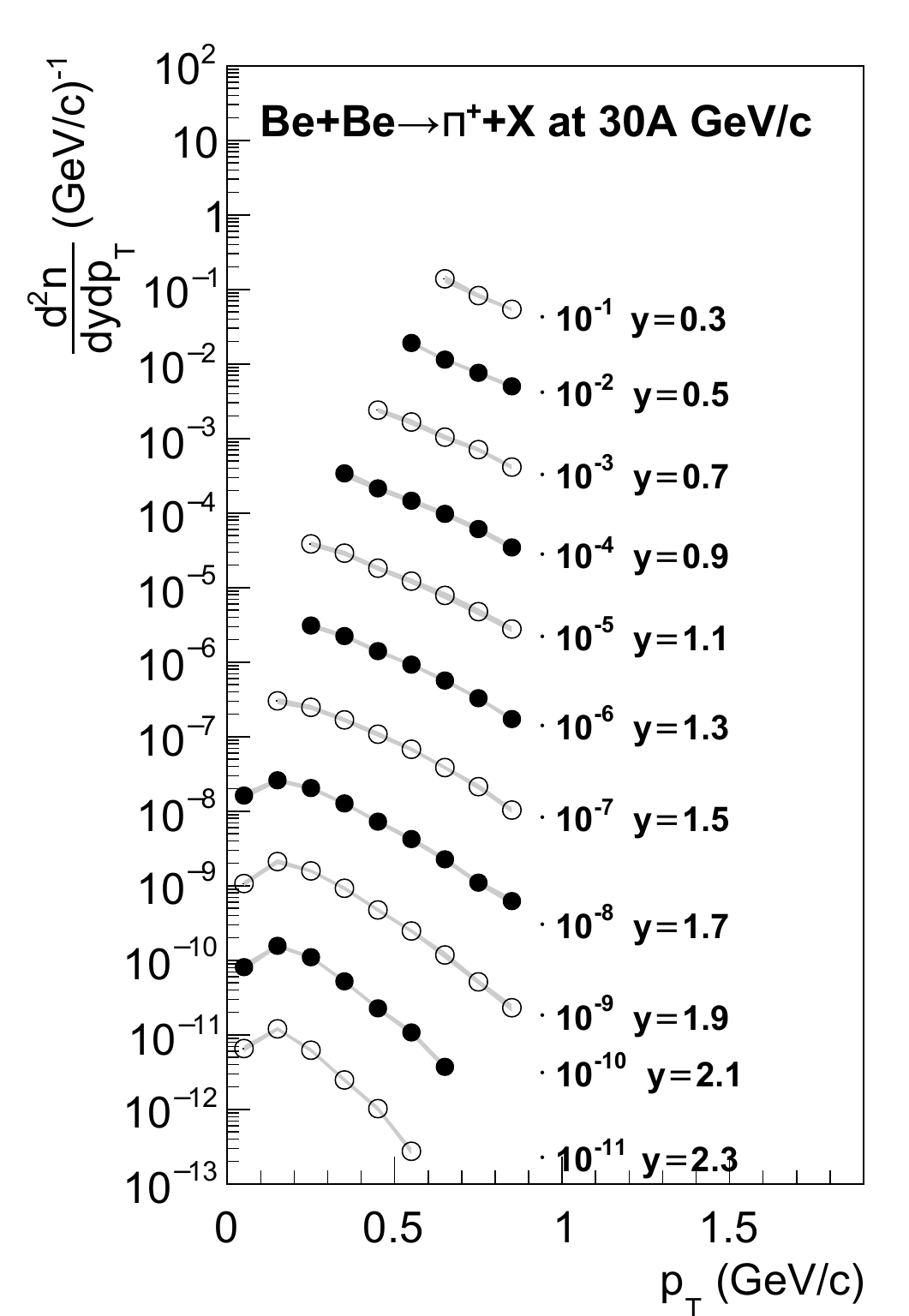}\\
                \includegraphics[width=0.3\textwidth]{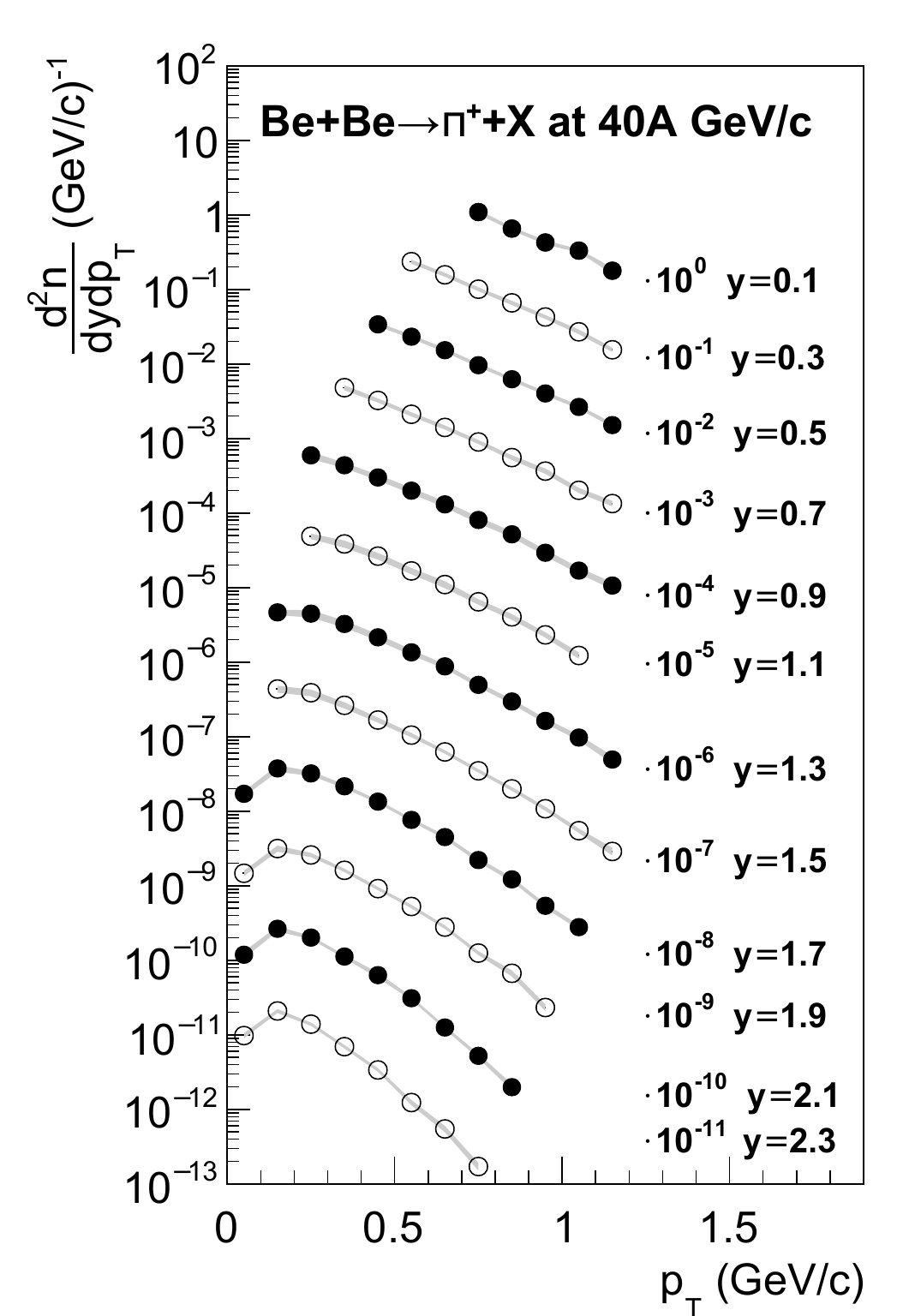}
                \includegraphics[width=0.3\textwidth]{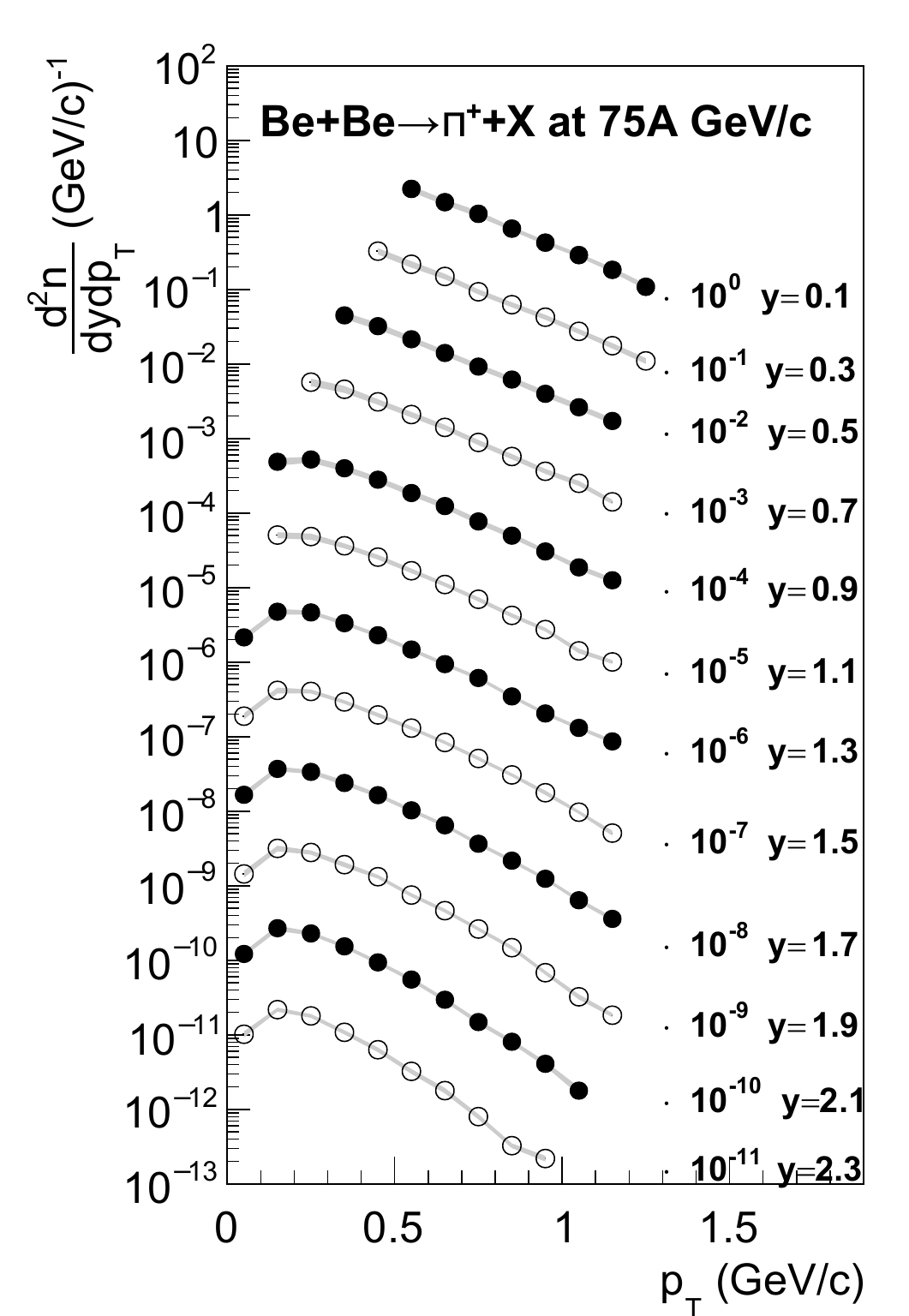}
                \includegraphics[width=0.3\textwidth]{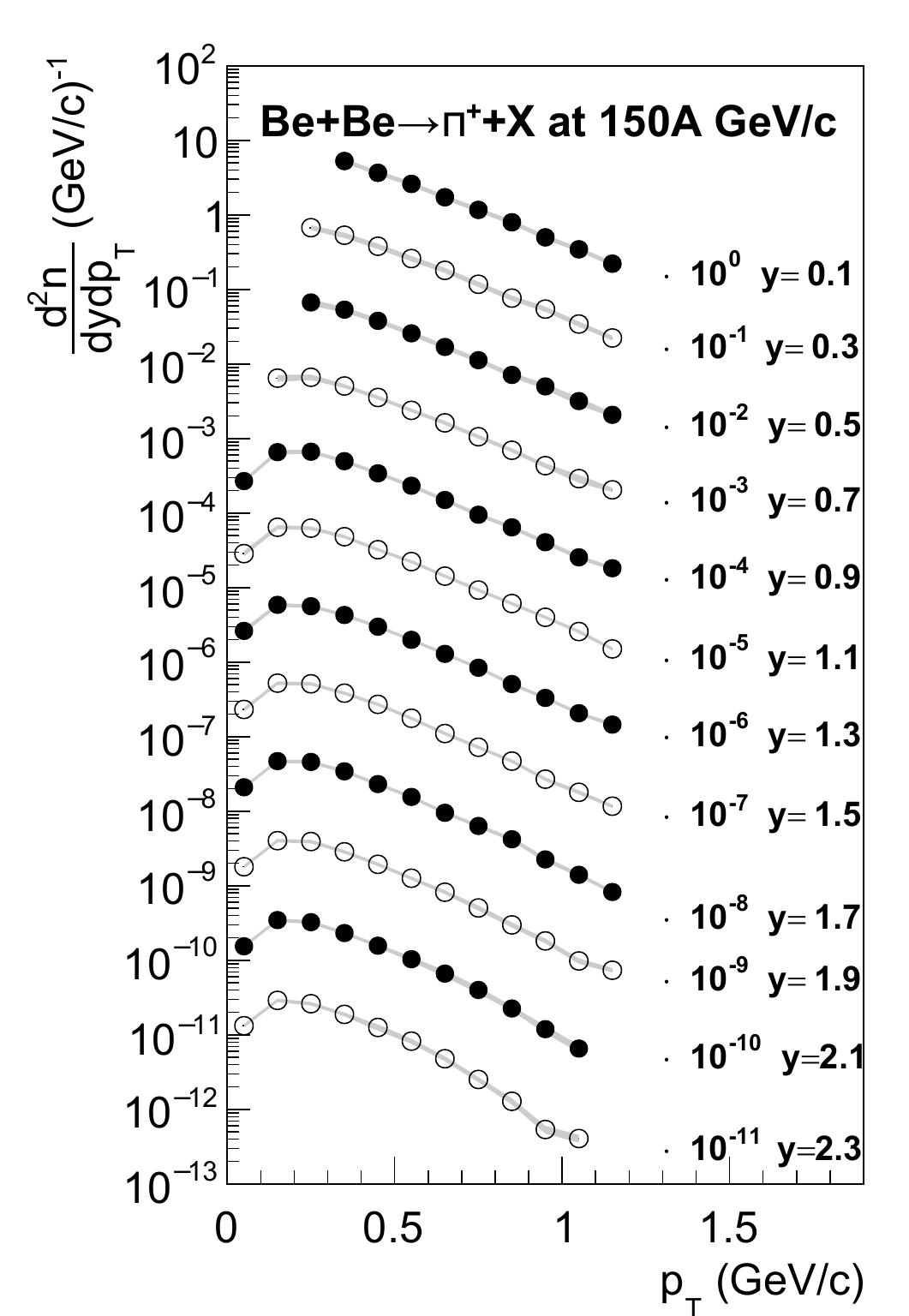}
                \end{center}
                \caption{Transverse momentum spectra in rapidity slices of $\pi^{+}$ produced
                         in the 20\% most \textit{central} Be+Be collisions. Rapidity values given in the
                         legends correspond to the middle of the corresponding interval. Presented results were obtained with the \dEdx analysis method. Shaded bands
                         show systematic uncertainties.}
                \label{fig:nptpipls}
\end{figure*}

\begin{figure*}
                \begin{center}
                \includegraphics[width=0.3\textwidth]{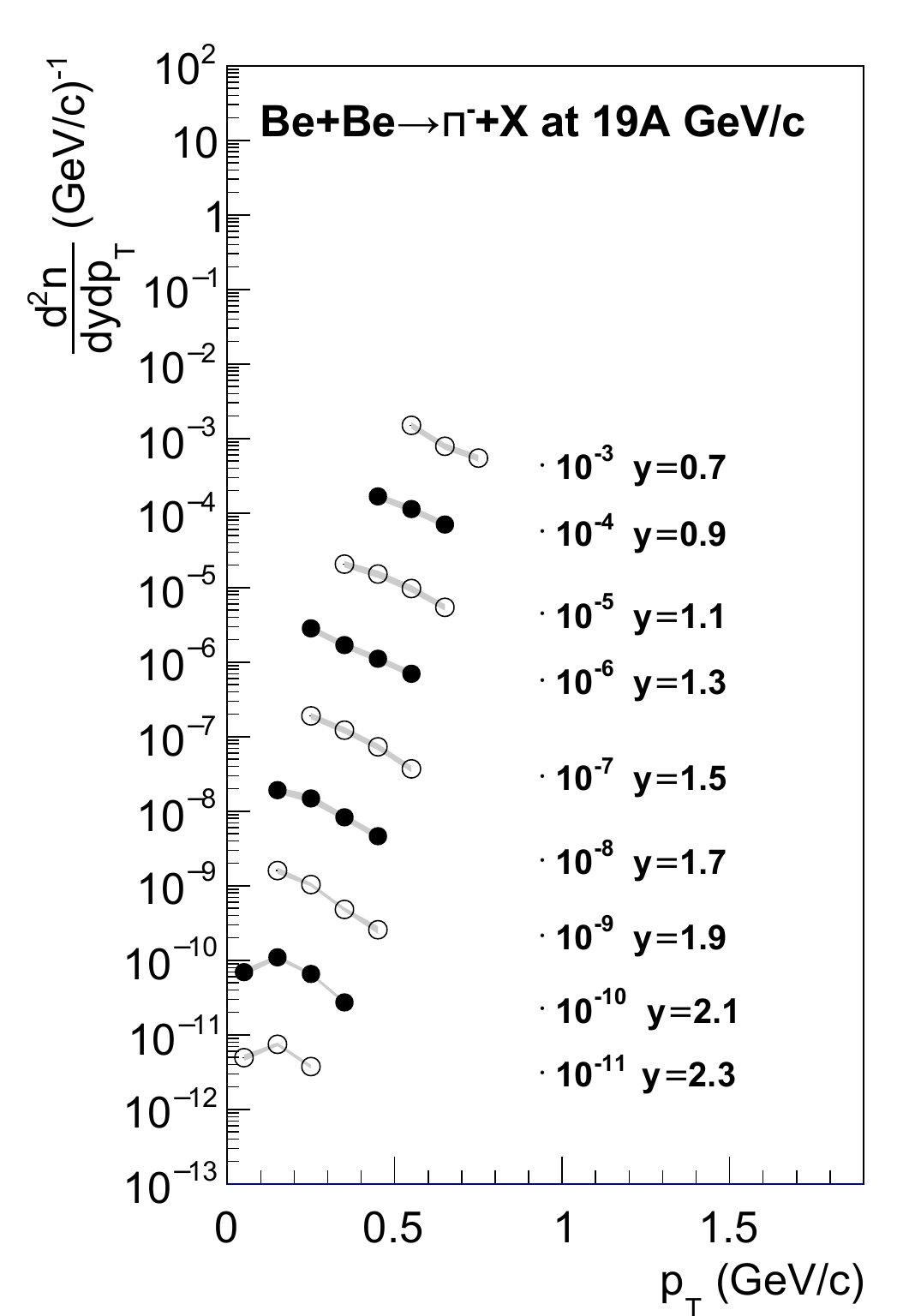}
                \includegraphics[width=0.3\textwidth]{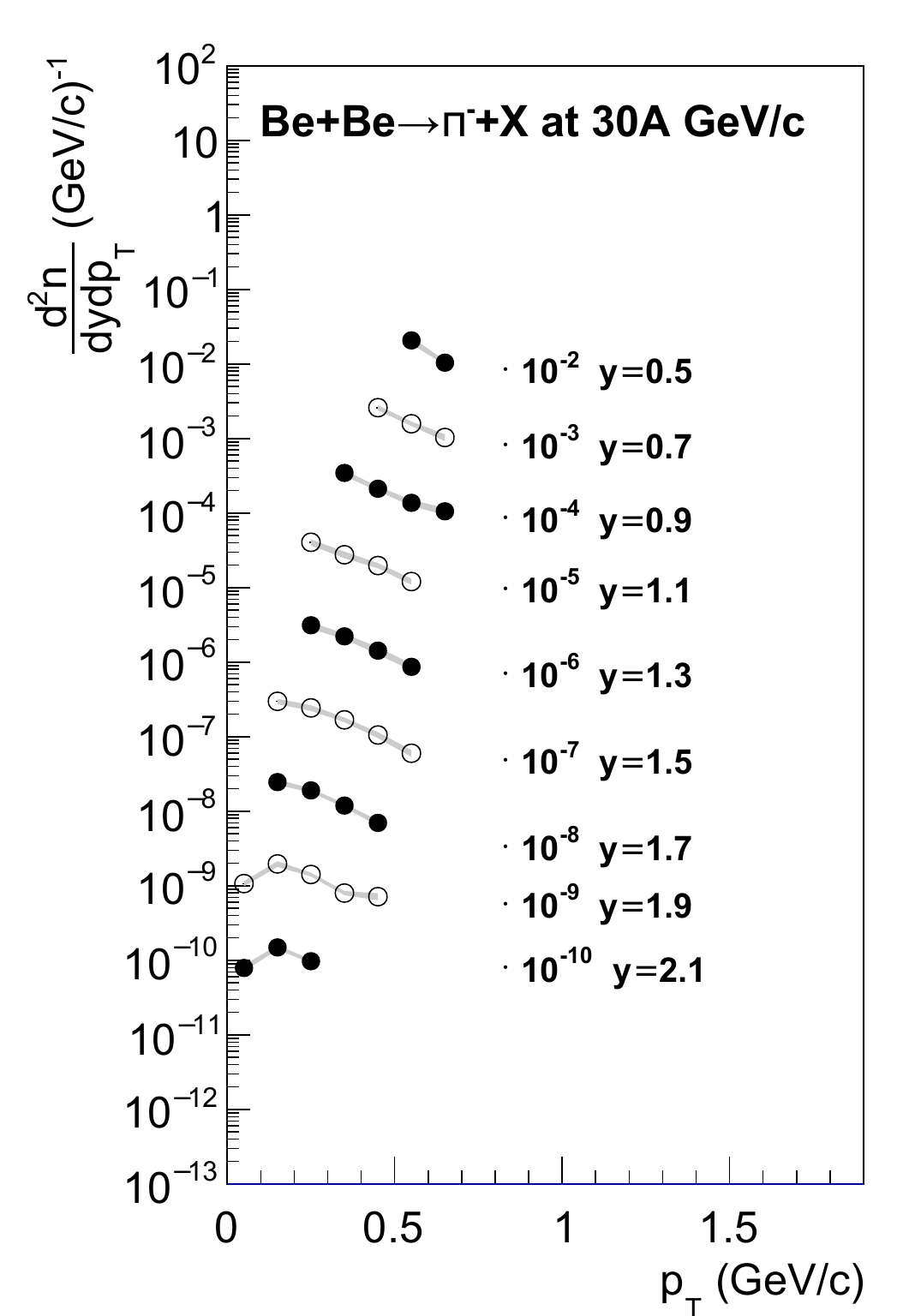}\\
                \includegraphics[width=0.3\textwidth]{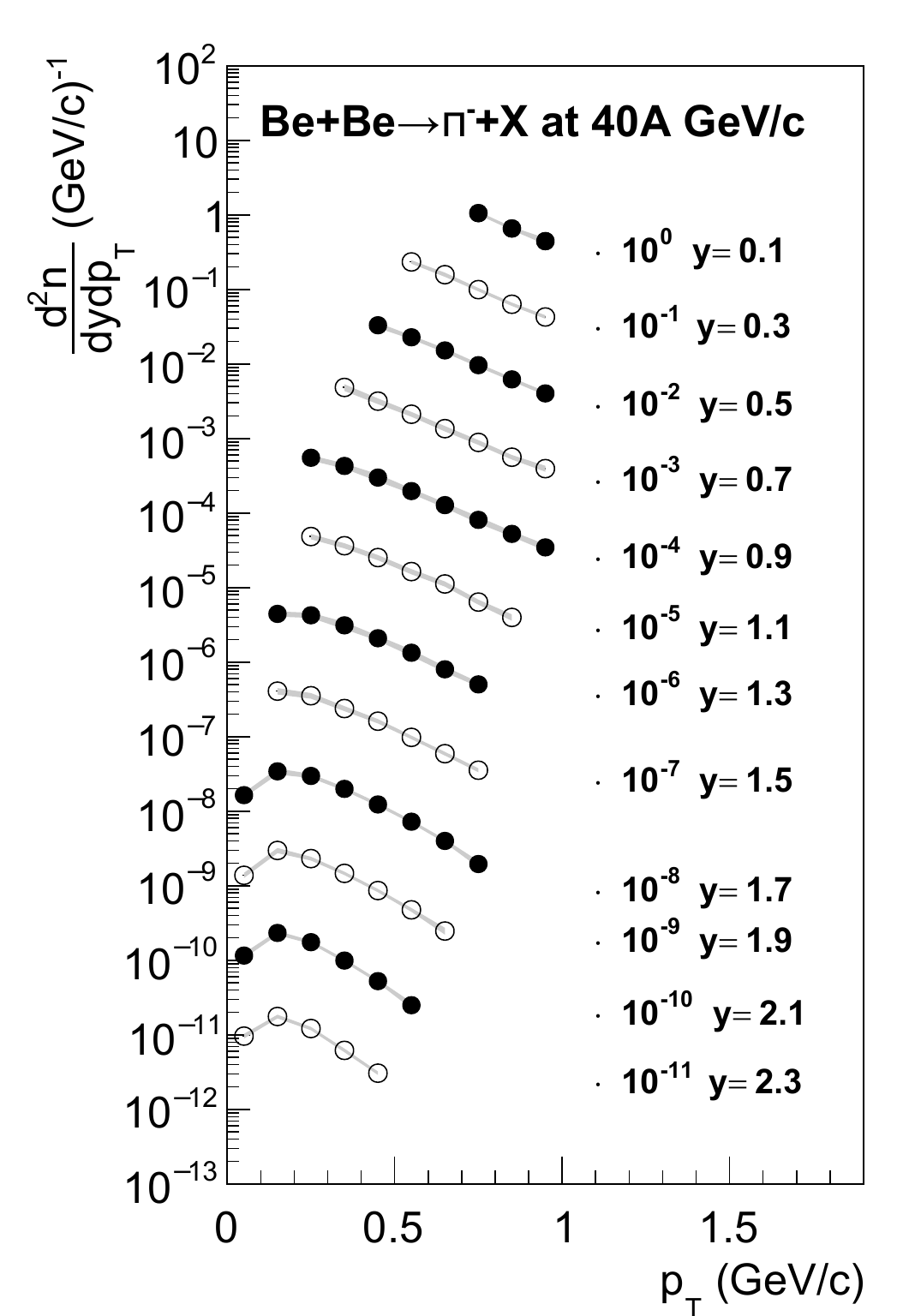}
                \includegraphics[width=0.3\textwidth]{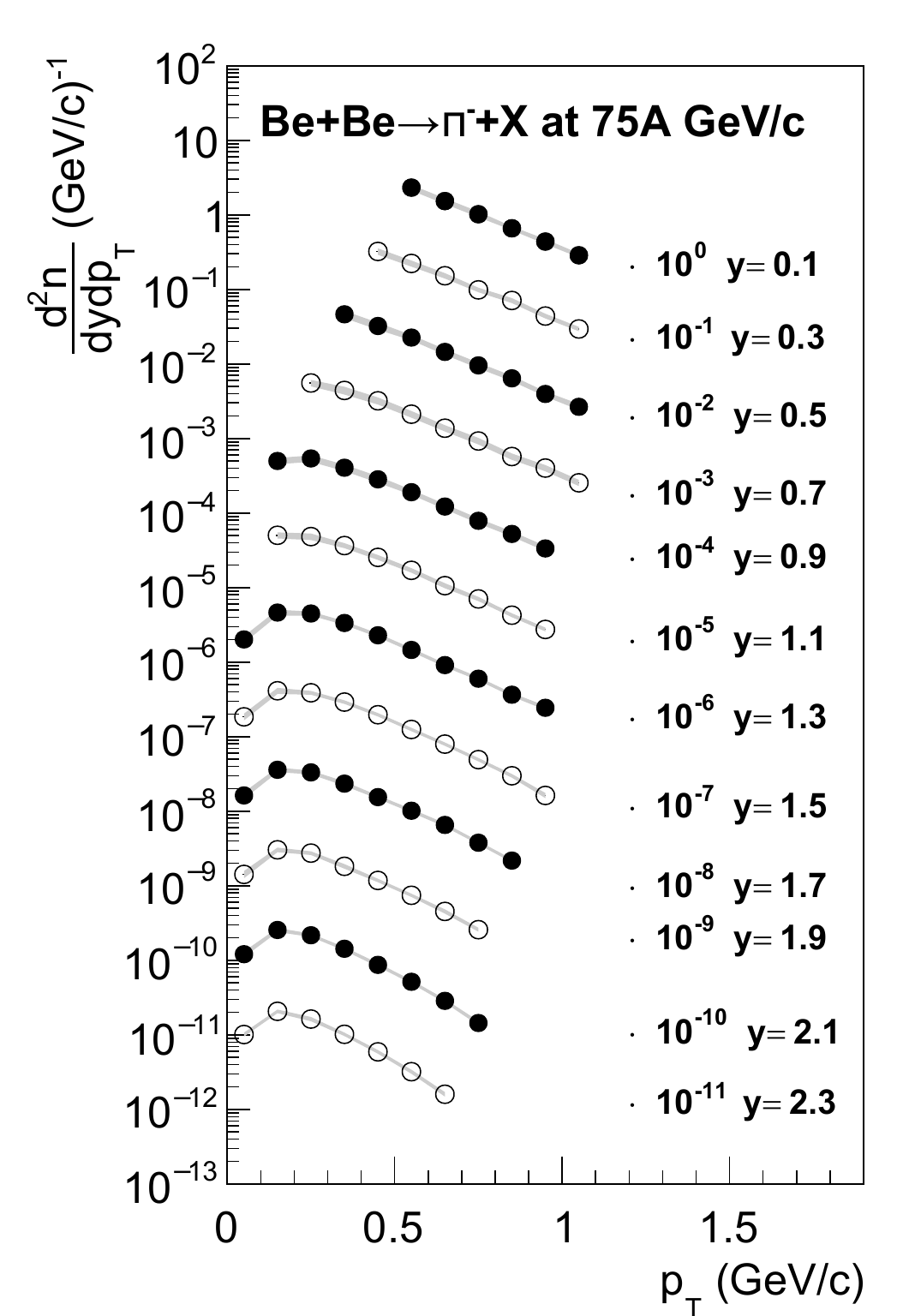}
                \includegraphics[width=0.3\textwidth]{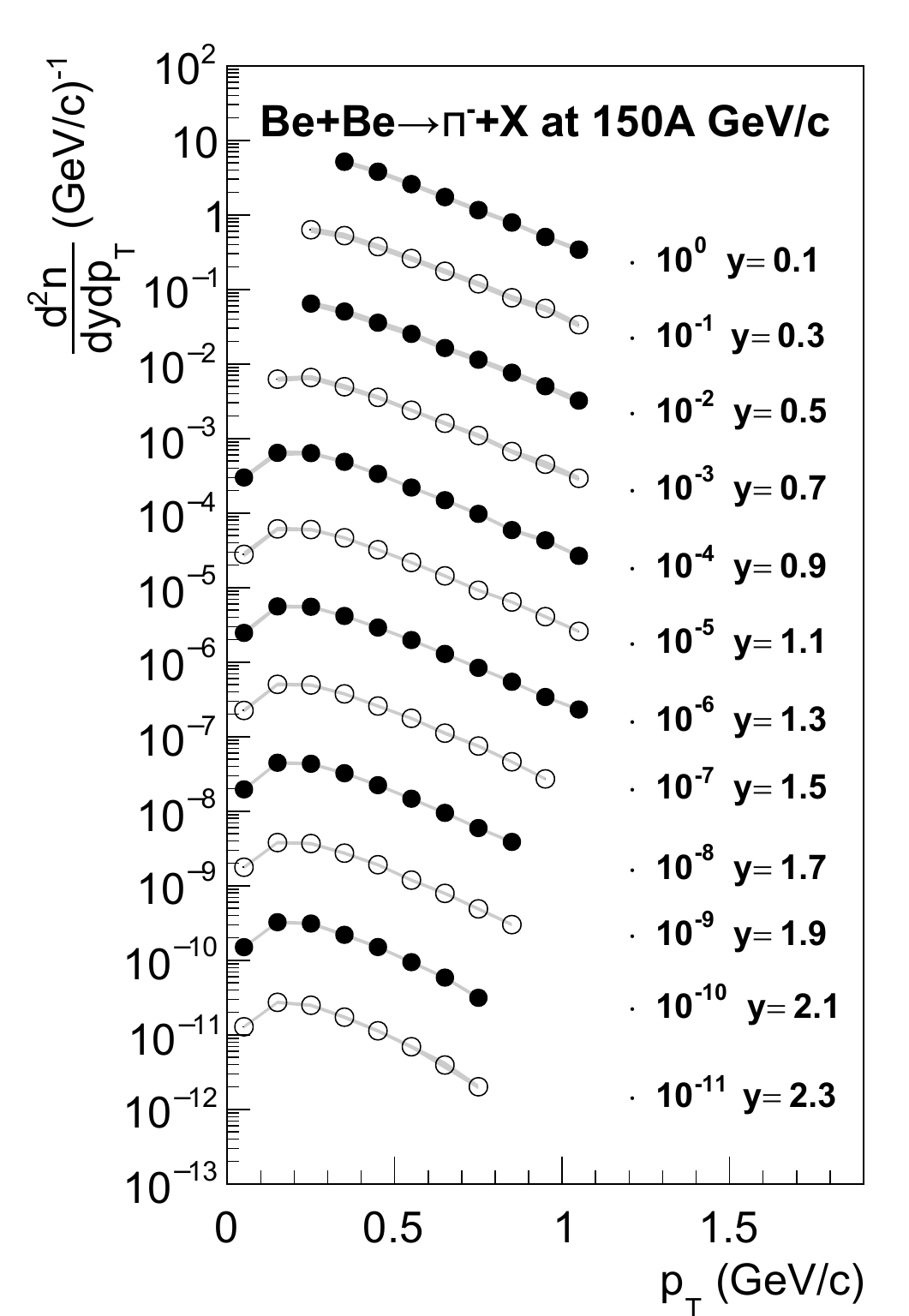}
                \end{center}
                \caption{Transverse momentum spectra in rapidity slices of $\pi^{-}$ produced
                         in the 20\% most \textit{central} Be+Be collisions. Rapidity values given in the
                         legends correspond to the middle of the corresponding interval. Presented results were obtained with the \dEdx analysis method. Shaded bands
                         show systematic uncertainties.}
                \label{fig:nptpimns}
\end{figure*}

\begin{figure*}
                \begin{center}
                \includegraphics[width=0.3\textwidth]{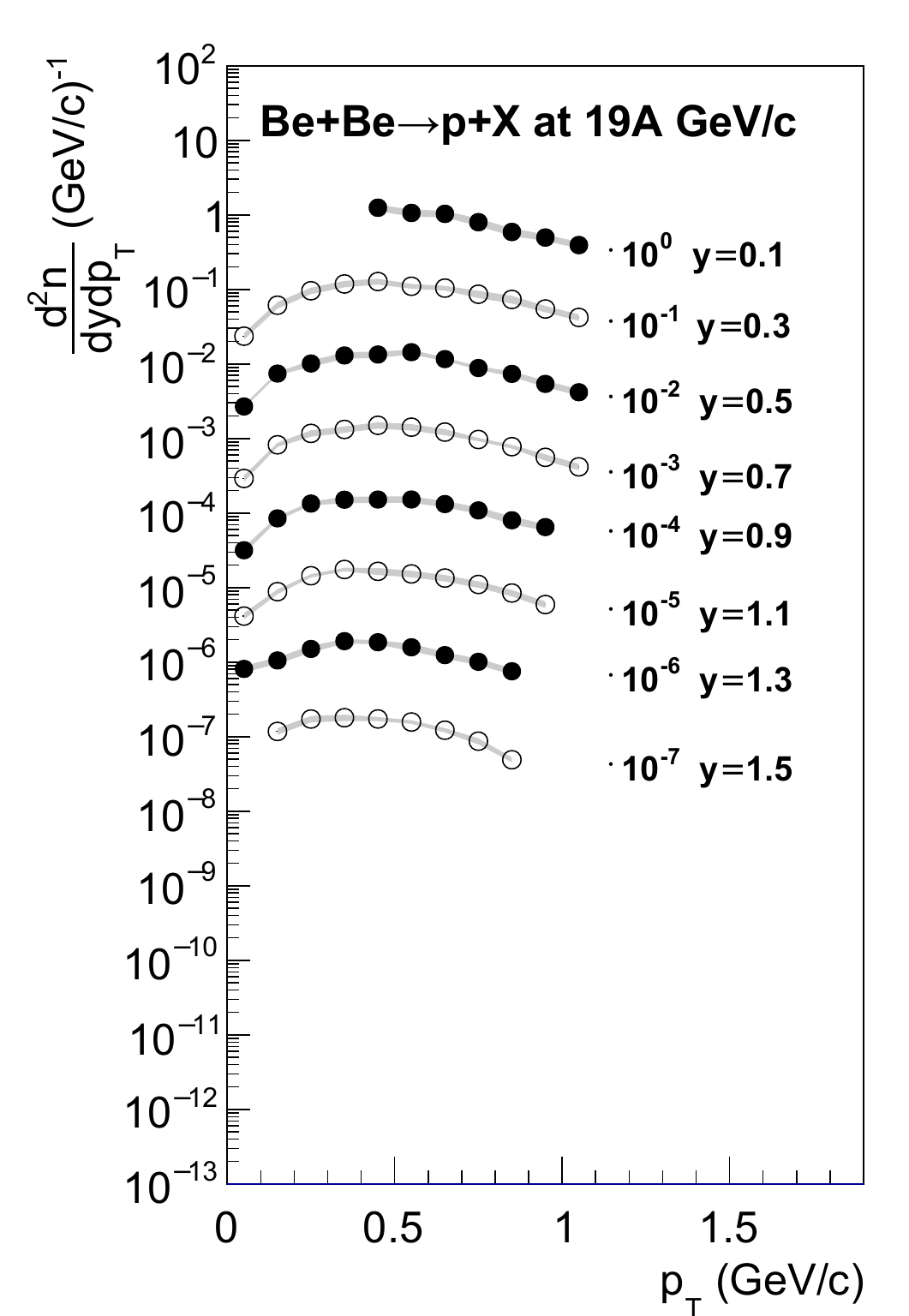}
                \includegraphics[width=0.3\textwidth]{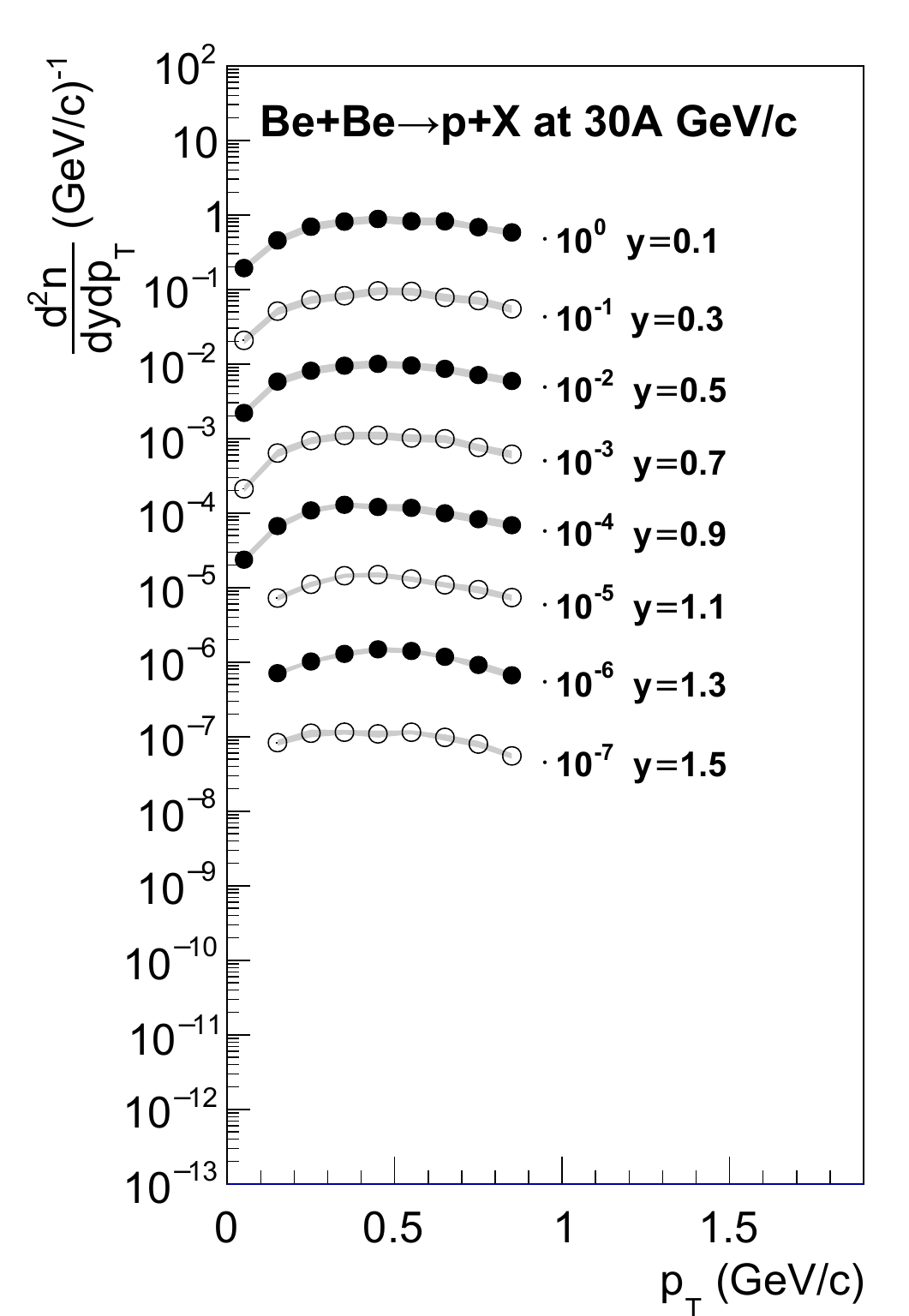}\\
                \includegraphics[width=0.3\textwidth]{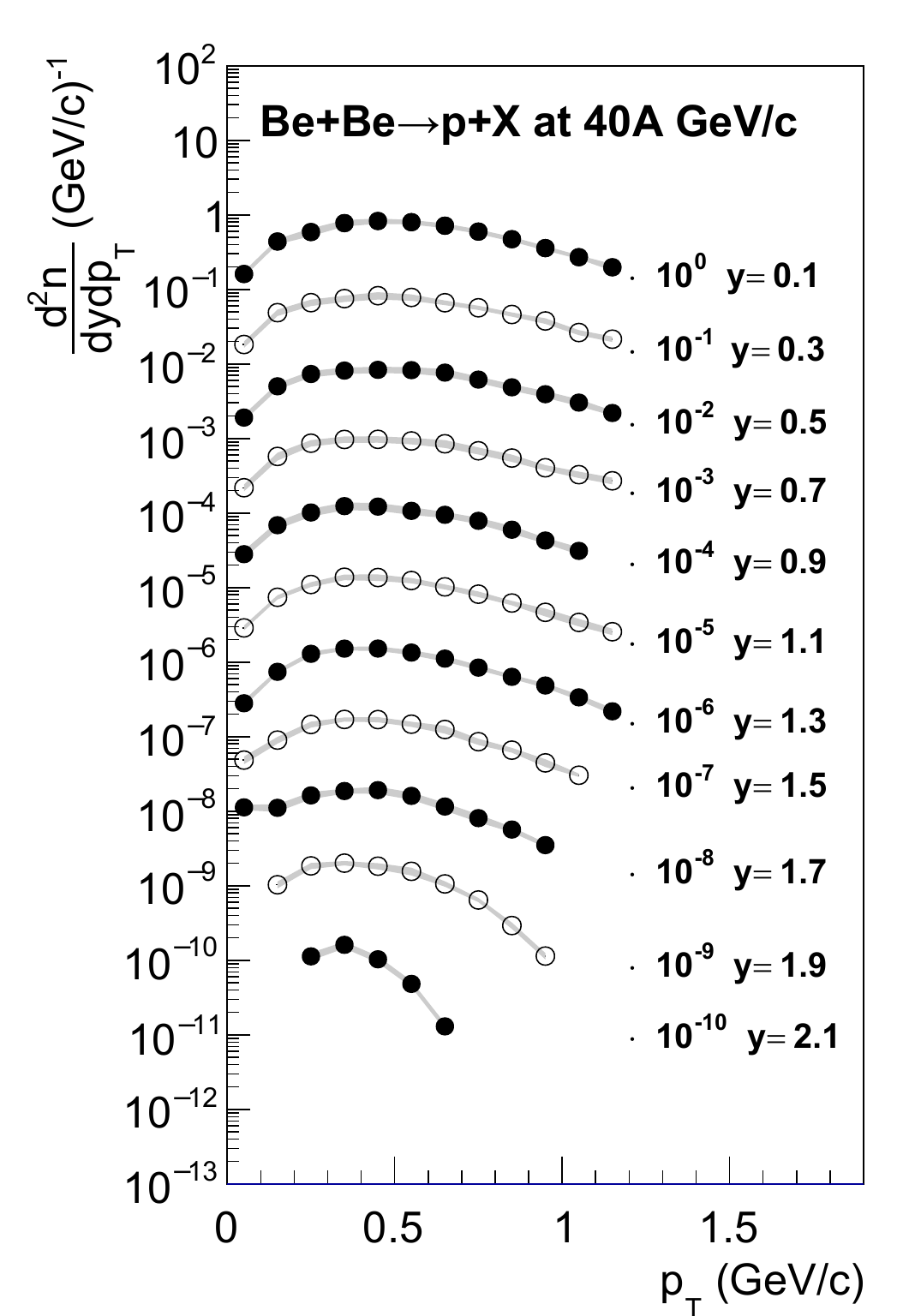}
                \includegraphics[width=0.3\textwidth]{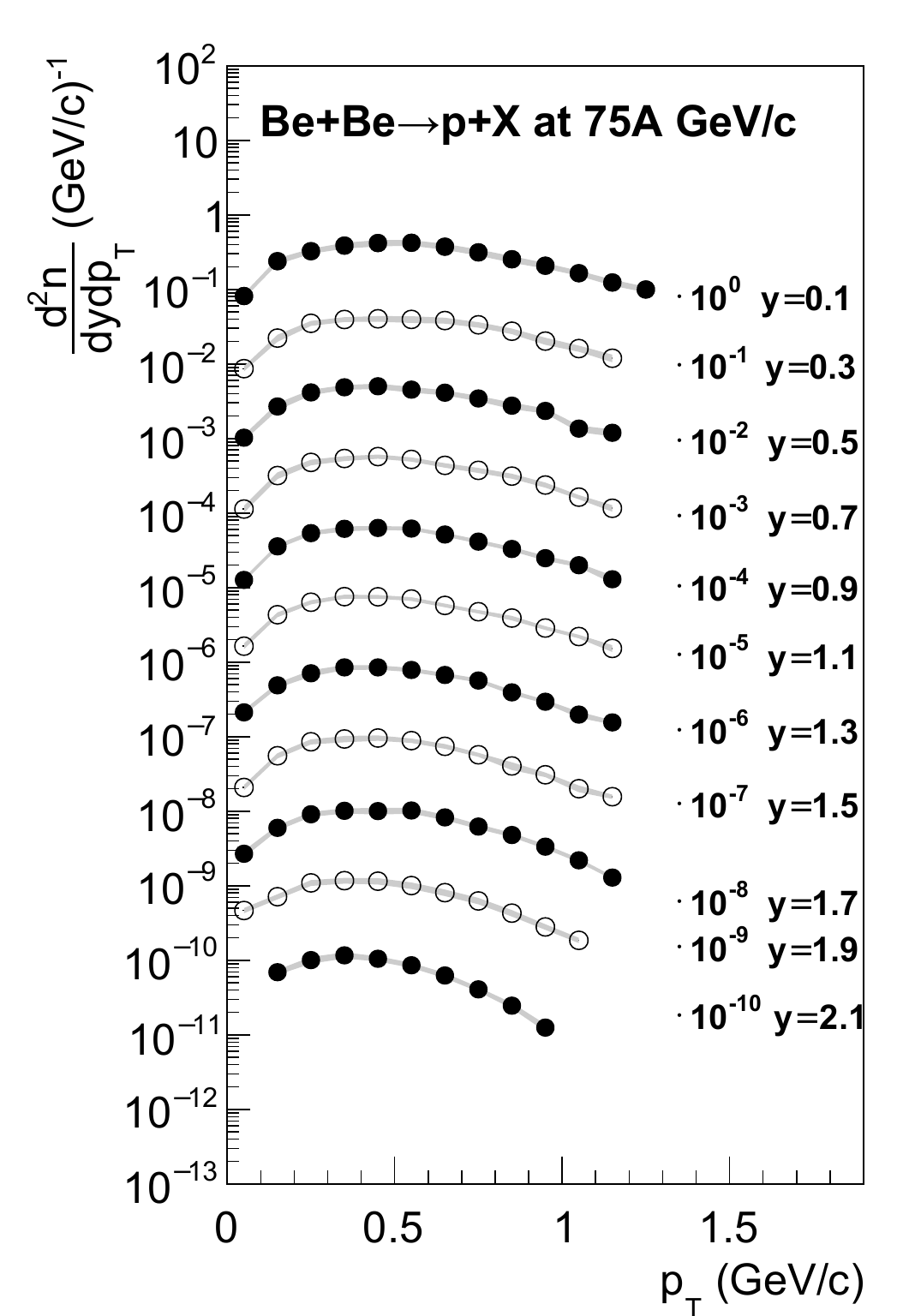}
                \includegraphics[width=0.3\textwidth]{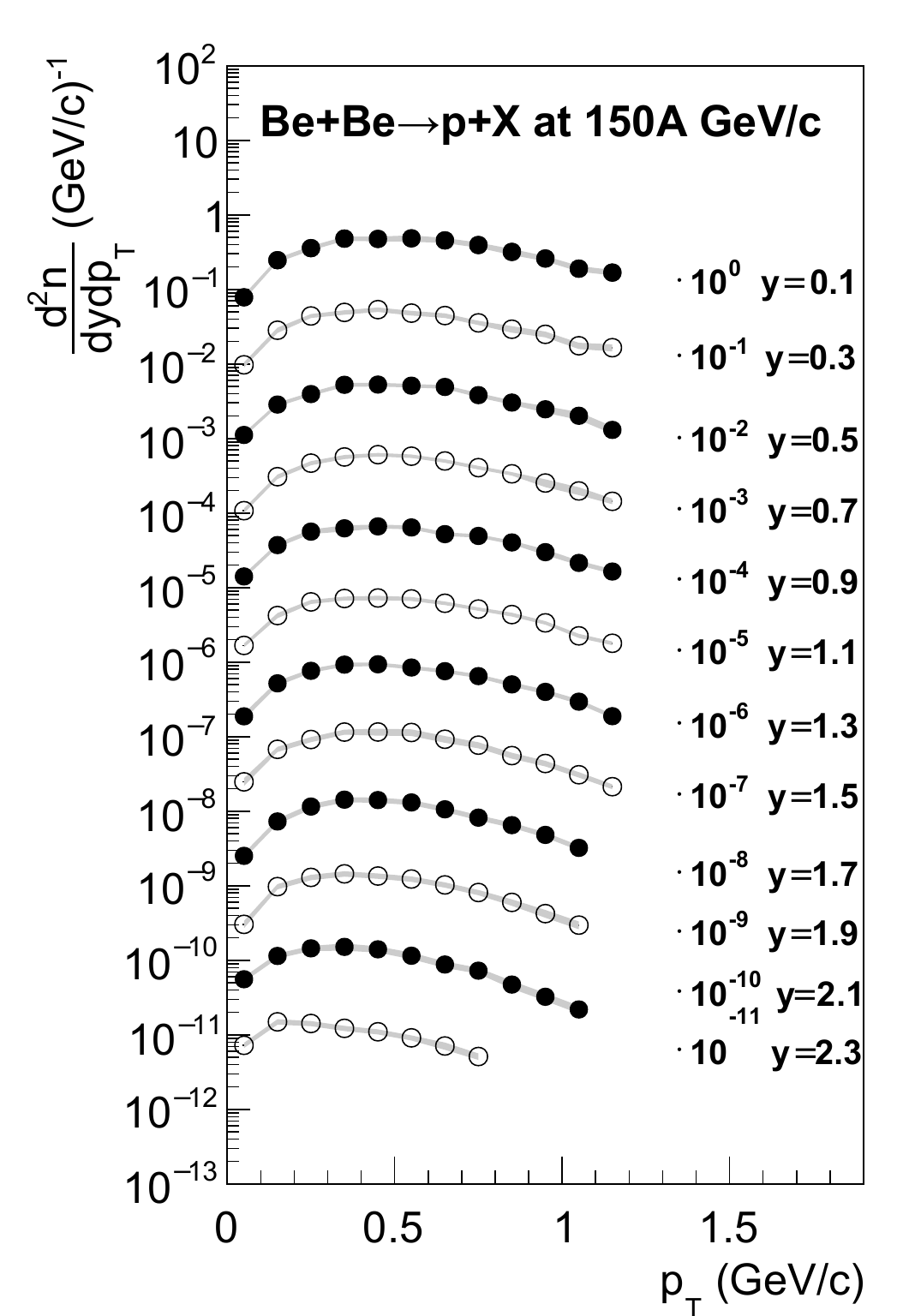}
                \end{center}
                \caption{Transverse momentum spectra in rapidity slices of protons produced
                         in the 20\% most \textit{central} Be+Be collisions. Rapidity values given in the
                         legends correspond to the middle of the corresponding interval. Presented results were obtained with the \dEdx analysis method. Shaded bands
                         show systematic uncertainties.}
                \label{fig:nptprot}
\end{figure*}

\begin{figure*}
                \begin{center}
                \includegraphics[width=0.3\textwidth]{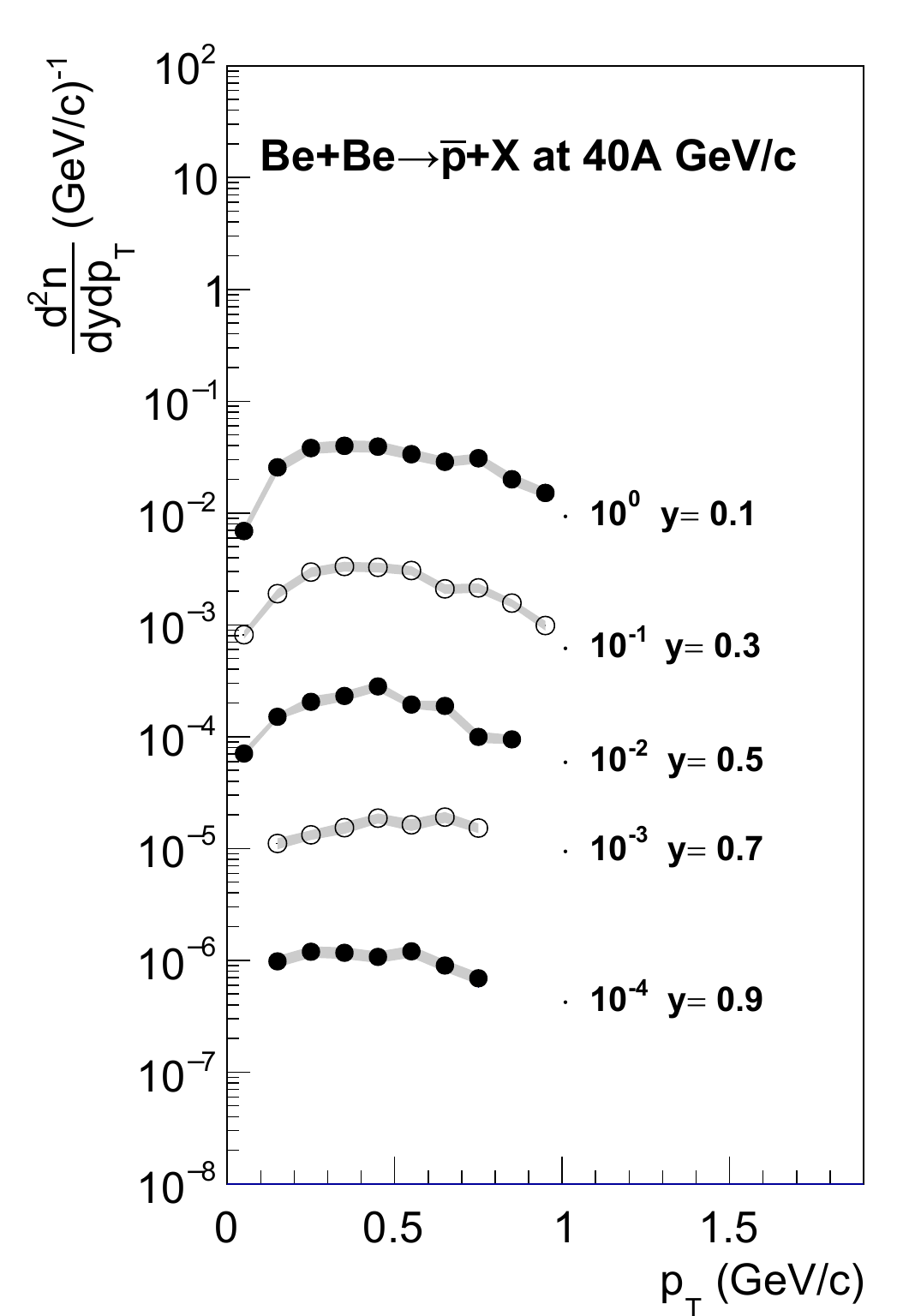}
                \includegraphics[width=0.3\textwidth]{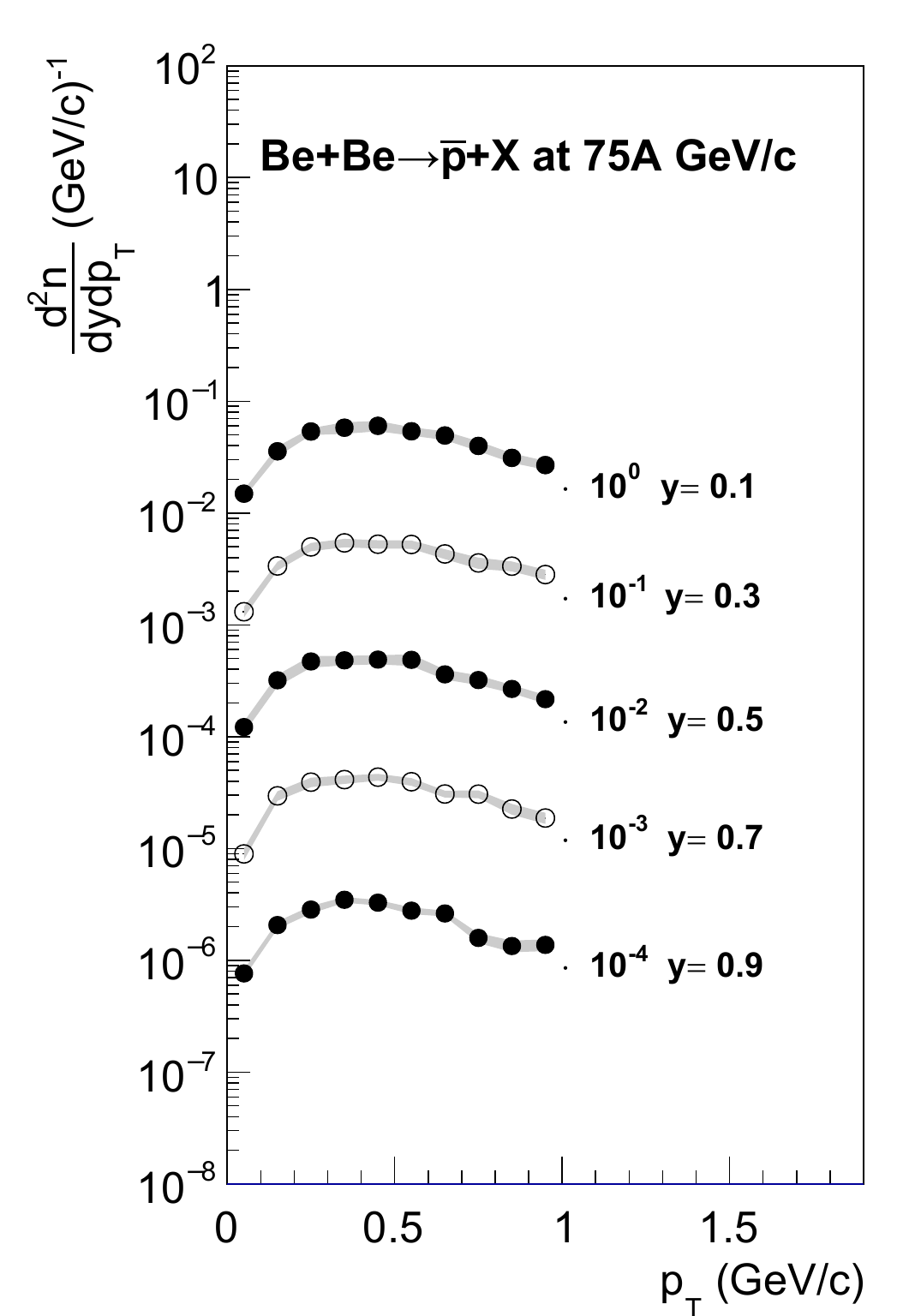}
                \includegraphics[width=0.3\textwidth]{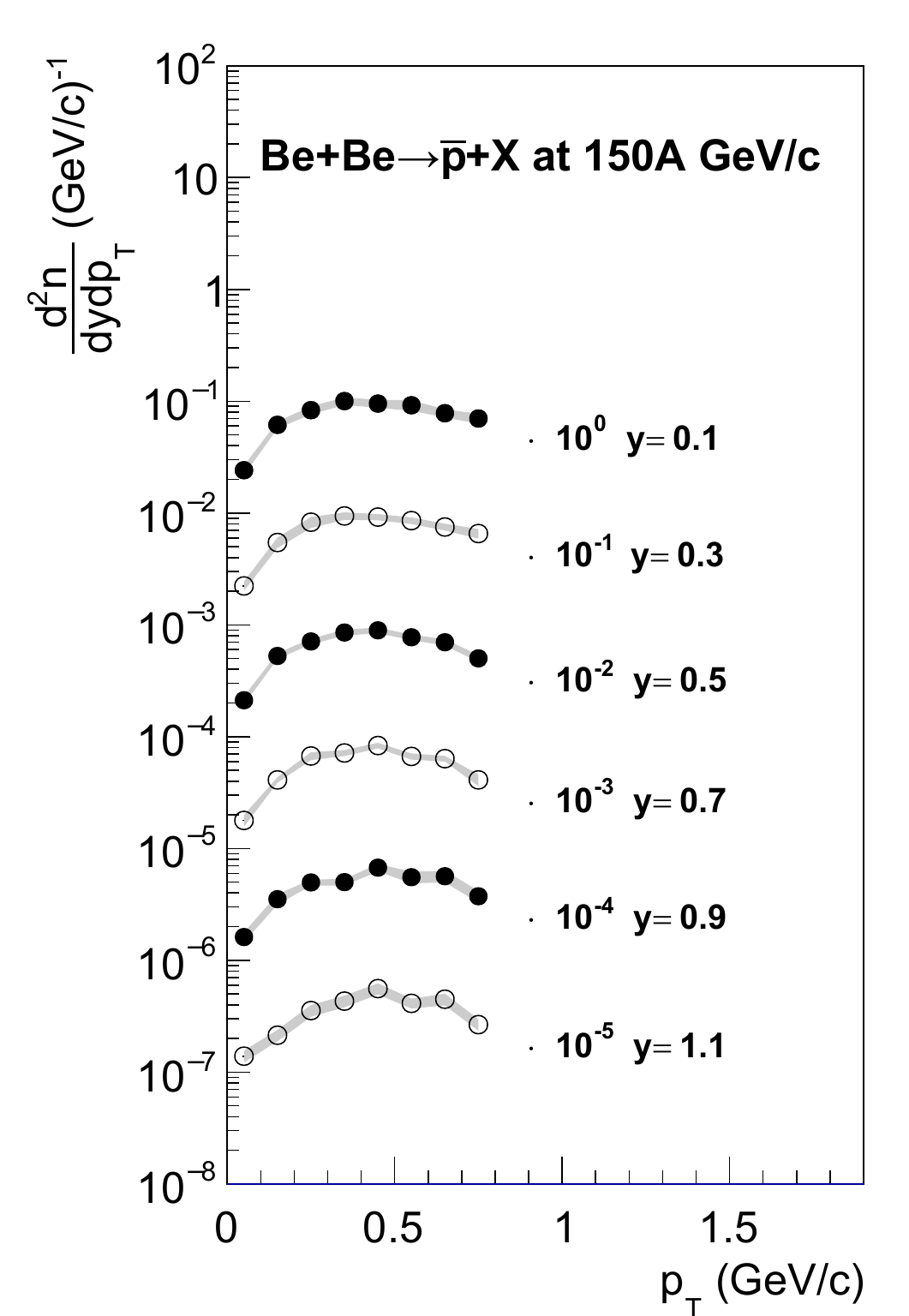}
                \end{center}
                \caption{Transverse momentum spectra in rapidity slices of antiprotons produced
                         in the 20\% most \textit{central} Be+Be collisions. Rapidity values given in the
                         legends correspond to the middle of the corresponding interval. Presented results were obtained with the \dEdx analysis method. Shaded bands
                         show systematic uncertainties.}
                \label{fig:nptaprot}
\end{figure*}
  
Resulting double differential spectra as a function of transverse momentum $p_{T}$ in intervals of 
rapidity \y of $K^{+}$, $K^{-}$, $\pi^{+}$, $\pi^{-}$, $p$ and $\bar{p}$ produced 
at 19A, 30A, 40A, 75A, 150\AGeVc beam momentum are shown in Figs.~\ref{fig:nptkpls}-\ref{fig:nptaprot}.
Entries in Fig.~\ref{fig:final2D} below mid-rapidity were reflected in order to fill the gaps 
in acceptance as far as possible. Spectra in successive rapidity intervals 
were scaled by appropriate factors for better visibility. Vertical bars on data points correspond 
to statistical, shaded bands to systematic uncertainties. Systematic uncertainties plotted in the logarithmic scale look small.

\begin{figure}[!ht]
\begin{center}
\includegraphics[width=0.45\textwidth]{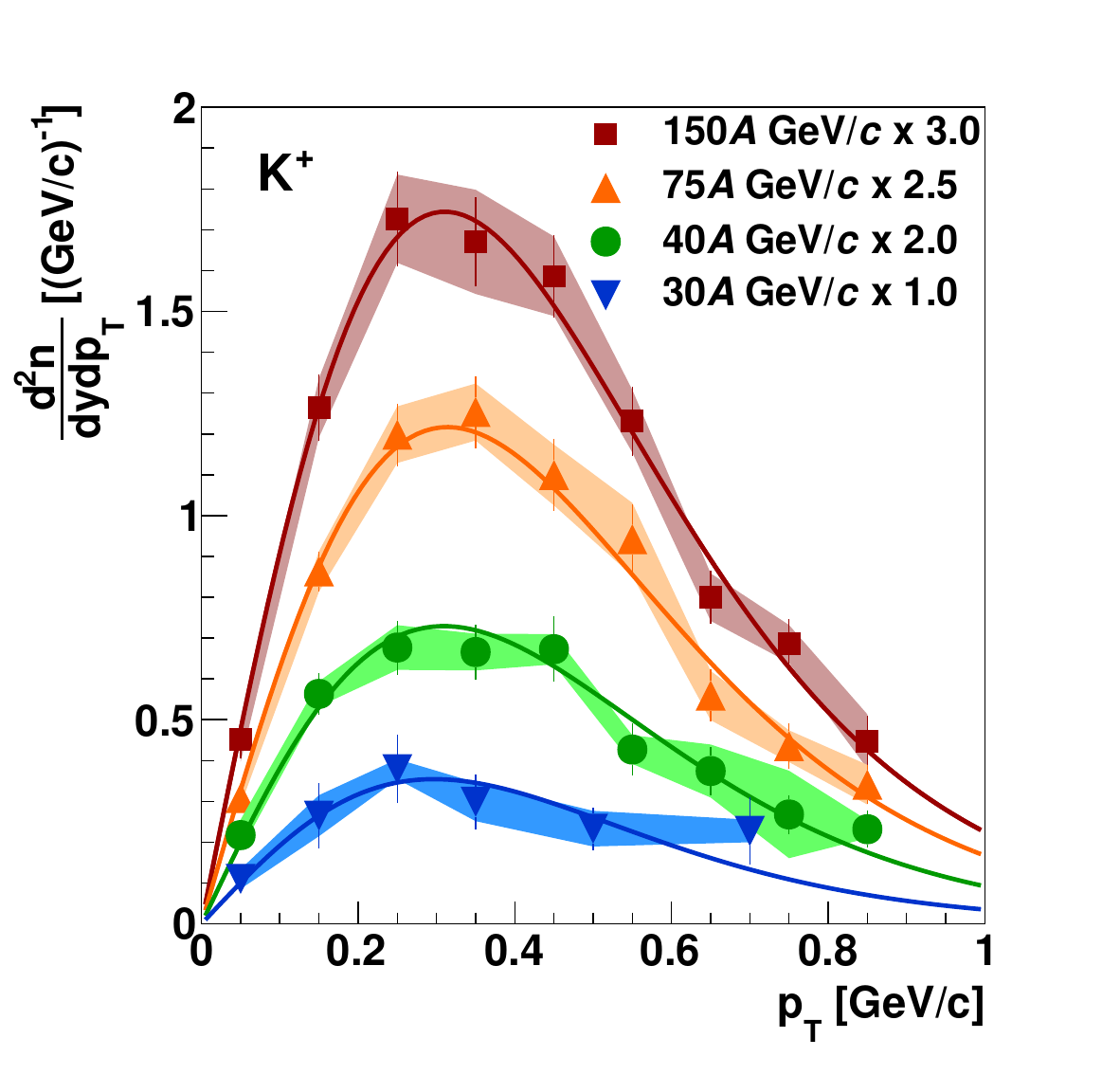}
\includegraphics[width=0.45\textwidth]{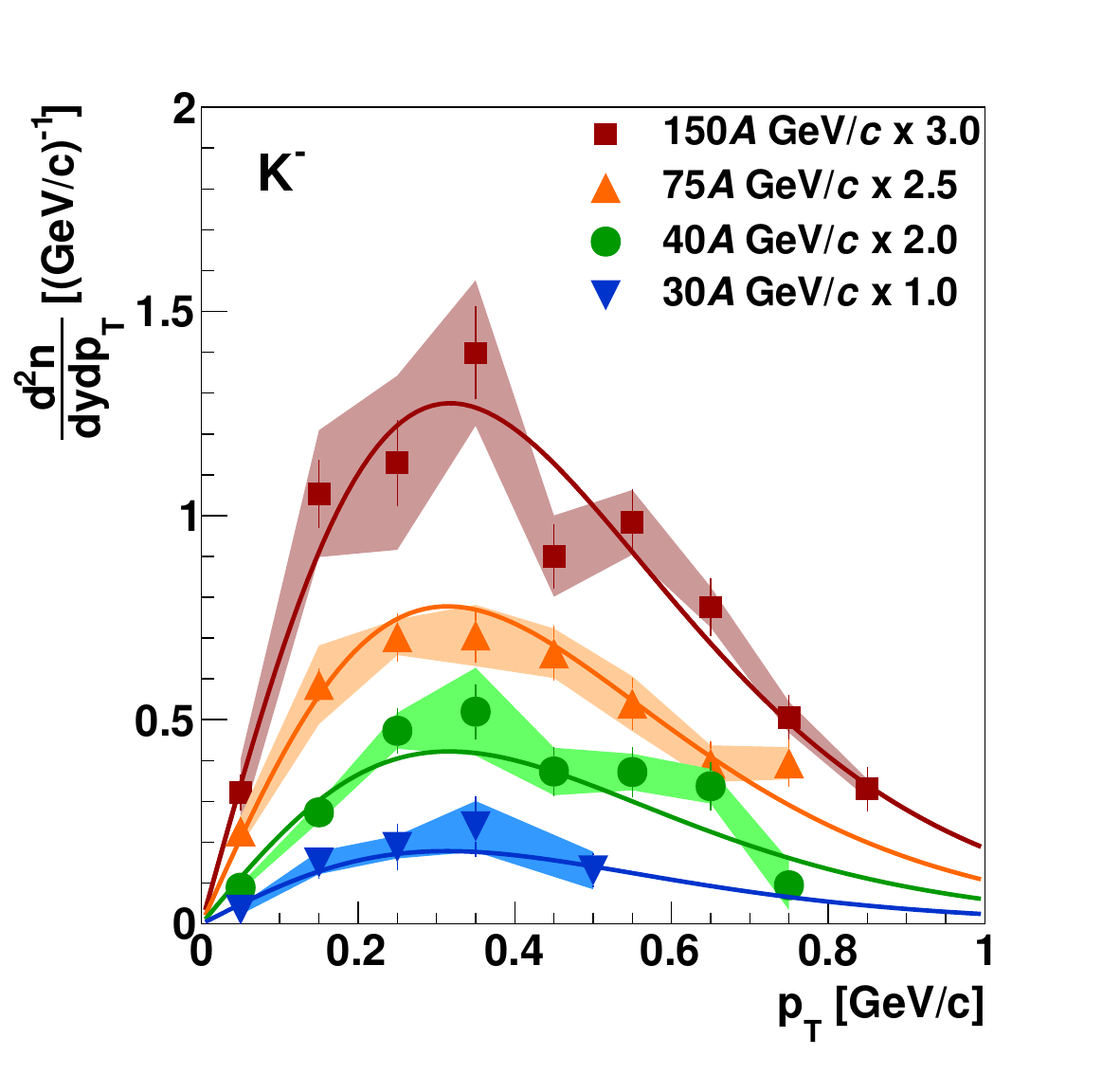}
\end{center}
\caption{Transverse momentum spectra of $K^{+}$ (\textit{left}) and $K^{-}$ (\textit{right}) mesons 
produced in 0 <\y< 0.2 (-0.2 <\y< 0.0 for 30\AGeVc) in the 20\% most \textit{central} Be+Be collisions. 
Colored lines represent the fitted function (Eq.~{\ref{eq:inverse}}).
}
\label{fig:kaonsptfinal}
\end{figure}

\begin{figure}[!ht]
\begin{center}
\includegraphics[width=0.45\textwidth]{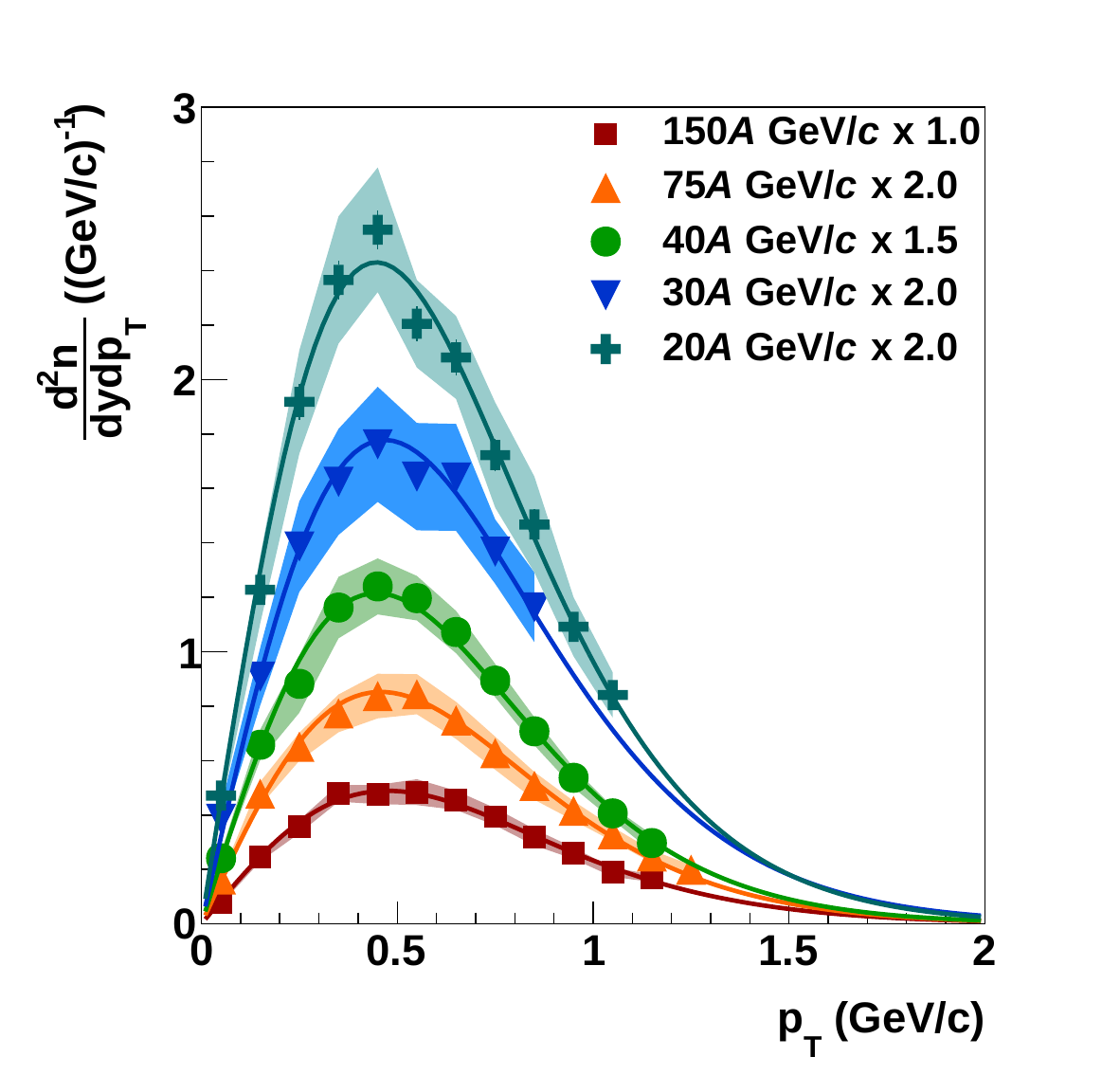}
\includegraphics[width=0.45\textwidth]{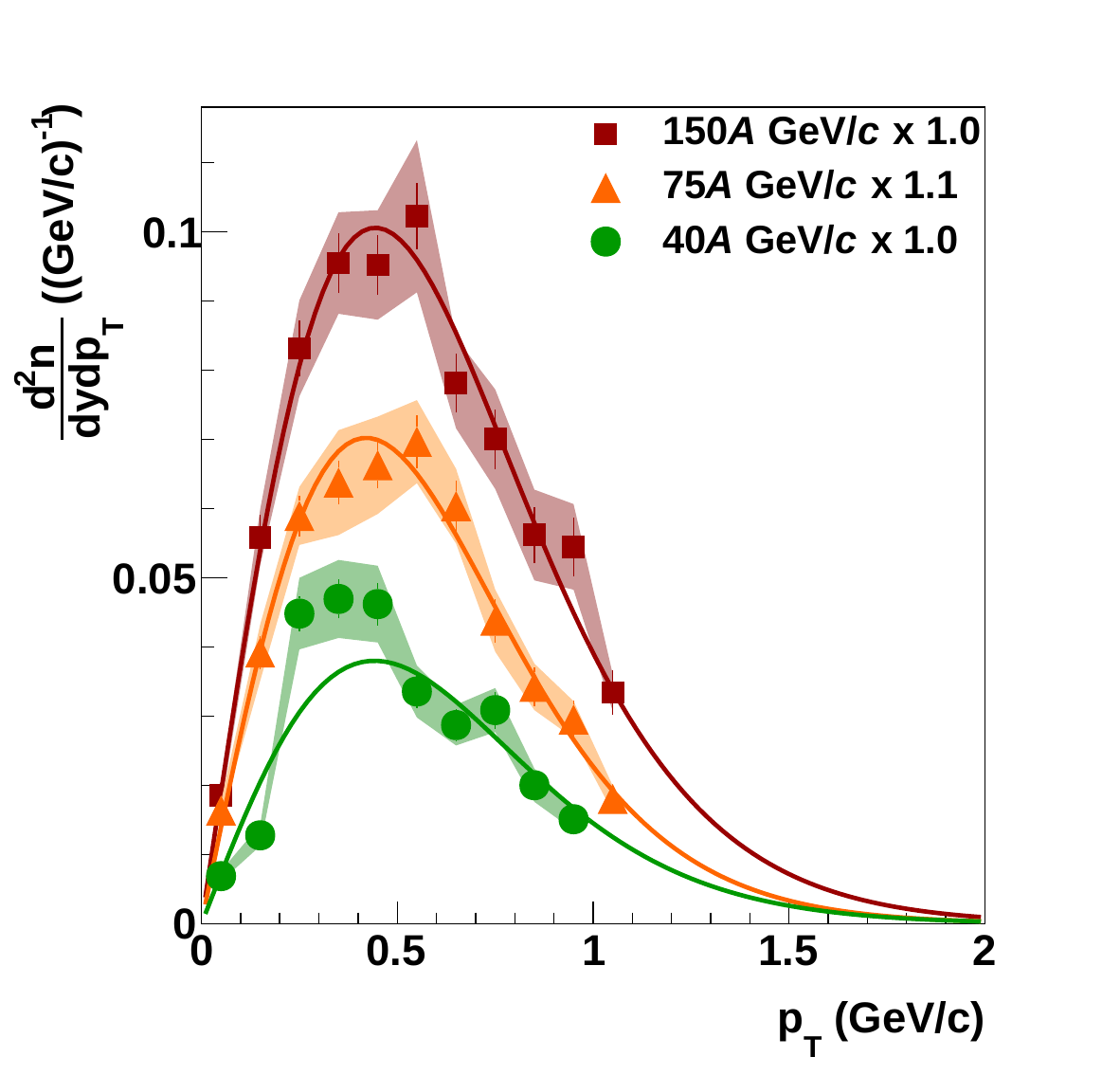}
\end{center}
\caption{Transverse momentum  spectra of protons (\textit{left)} and antiprotons (\textit{right}) 
produced in 0 <\y< 0.2 in the 20\% most \textit{central} Be+Be collisions. 
Colored lines represent the fitted function (Eq.~{\ref{eq:inverse}}).
}
\label{fig:protonsfinal}
\end{figure}

Transverse momentum spectra of shown in Figs.~\ref{fig:nptkpls}-\ref{fig:nptaprot}
were parametrized by an exponential function~\mbox{\cite{Hagedorn:1968jf,Broniowski:2004yh}}:
\begin{equation}
 \frac{d^{2}n}{dydp_T} = \frac{S~c^{2} p_{T}}{T^{2} + m~T} \exp(-(m_T - m)/T)      ,
\label{eq:inverse}
\end{equation}
where $m$ is the particle mass and  $S$ and $T$ are the yield integral and  the inverse slope parameter, 
respectively.
The functions (Eq.~\ref{eq:inverse}) fitted at mid-rapidity are shown together with the data points for $K^{+}$ 
and $K^{-}$ in Fig.~\ref{fig:kaonsptfinal} and for protons and antiprotons in Fig.~\ref{fig:protonsfinal}.

\begin{figure*}
\begin{center}
\hspace{-1cm}
\includegraphics[width=1\textwidth]{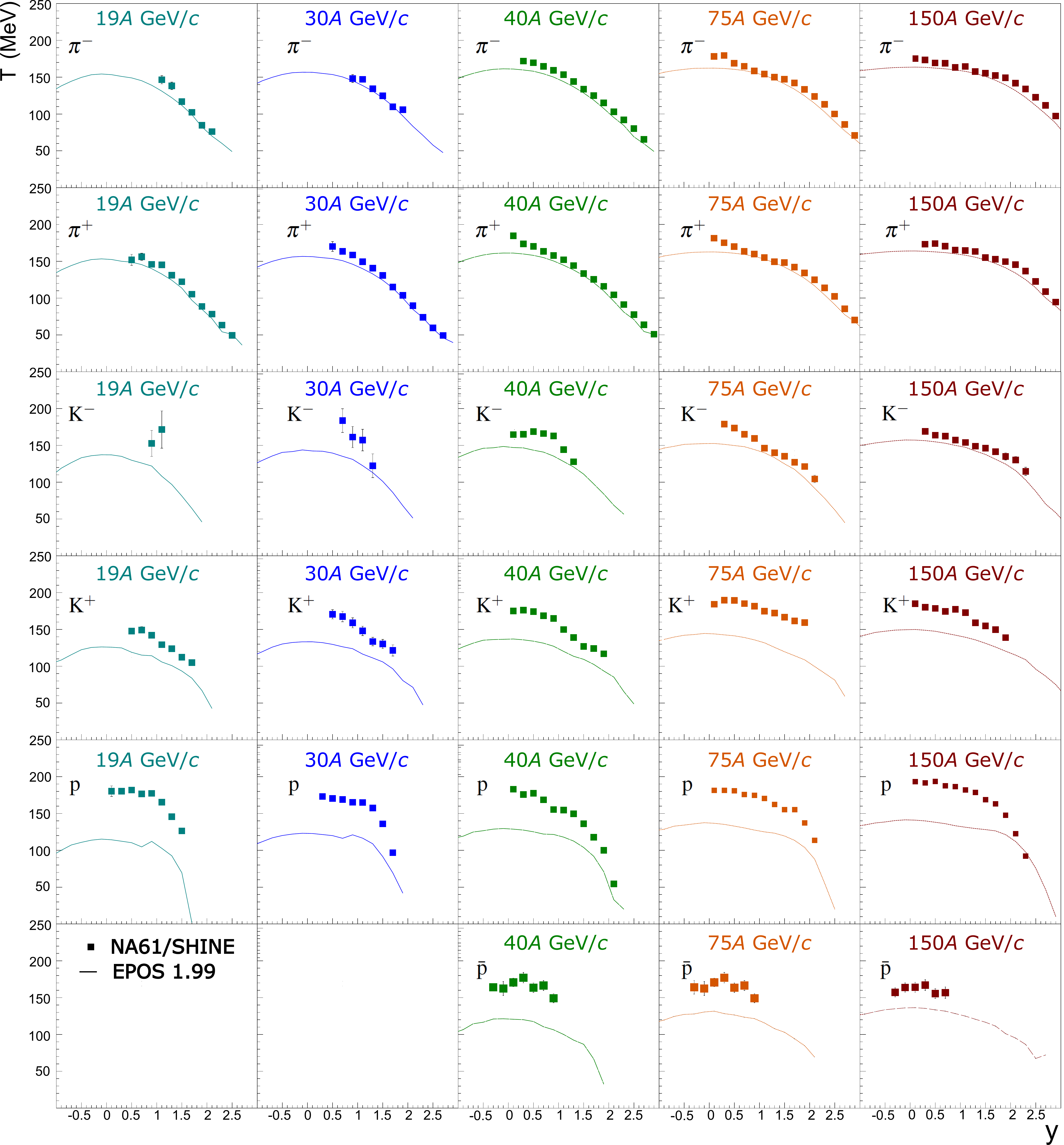}

\end{center}
\caption{Inverse slope parameter $T$ of the $p_T$ distributions as function of rapidity $y$ 
fitted with Eq.~\ref{eq:inverse} in the full available $p_T$ range.
Results for the 20\% most \textit{central} Be+Be collisions at beam momenta of 19$A$, 30$A$ 40$A$, 75$A$ and 150\AGeVc. 
Only statistical uncertainties are shown, systematic uncertainties are estimated at about 10\%. 
Dashed curves show predictions of the \Epos~\cite{Werner:2005jf,Pierog:2018} model for comparison. 
}
\label{fig:Tparam}
\end{figure*}

The fitted inverse slope parameter is plotted in Fig.~\ref{fig:Tparam} as a function of 
the rapidity. The value of $T$ results from an interplay of the kinetic freeze-out temperature 
when final state rescattering stops as well as of the radial expansion flow. Thus the $p_T$ spectra 
need not show a strictly exponential decrease. Note that results are only plotted for 
those rapidity intervals for which there were more than 6 data points in the $p_{T}$-distribution.

Rapidity spectra $dn/dy$ of $K^{+}$ and $K^{-}$ were obtained by integration of the transverse momentum 
distributions shown in Fig.~\ref{fig:kaonsptfinal}. Both $dn/dy$ and the corresponding inverse slope parameter $T$
at mid-rapidity are tabulated in Table~\ref{tab:mid-rapidity_KP_KM_T}.

\begin{table}[ht]
\caption{Mid-rapidity $K^{+}$ and $K^{-}$ multiplicities $dn/dy$ and inverse slope parameters $T$ at in 0 <\y< 0.2 (-0.2 <\y<0.0 for 30\AGeVc)
produced in the 20\% most \textit{central} Be+Be collisions. For each value the statistical and systematic
uncertainties are given as the first and second uncertainty contribution, respectively.}
\vspace{0.5cm}
\small
\centering
\begin{tabular}{|c||c|c|c|c|}
\hline
\multirow{2}{1cm}{ }    & \multicolumn{2}{|c|}{$K^+$} & \multicolumn{2}{|c|}{$K^-$}\\ \cline{2-5}
 & $(\frac{dn}{dy})_{y \approx 0}$ & $T_{K^+}$ & $(\frac{dn}{dy})_{y \approx 0}$ & $T_{K^-}$ \\
 \hline
30\AGeVc & 0.209$\pm$0.025$\pm$0.021 & 0.152$\pm$0.016 $\pm$0.012 & 0.111$\pm$0.019$\pm$0.015 & 0.167$\pm$0.023$\pm$0.014\\
\hline
40\AGeVc & 0.220$\pm$0.009$\pm$0.022 & 0.164$\pm$0.007$\pm$0.006 & 0.136$\pm$0.007$\pm$0.020 & 0.170$\pm$0.011$\pm$0.012 \\
\hline
75\AGeVc & 0.288$\pm$0.009$\pm$0.021 & 0.168$\pm$0.006$\pm$0.005 & 0.182$\pm$0.007$\pm$0.019 & 0.168$\pm$0.009$\pm$0.005 \\
\hline
150\AGeVc & 0.338$\pm$0.008$\pm$0.028 & 0.165$\pm$0.005$\pm$0.007 & 0.254$\pm$0.008$\pm$0.036 & 0.171$\pm$0.007$\pm$0.012\\
\hline
\end{tabular}
\label{tab:mid-rapidity_KP_KM_T}			
\end{table}

\begin{figure}[!ht]
\begin{center}
\includegraphics[width=0.45\textwidth]{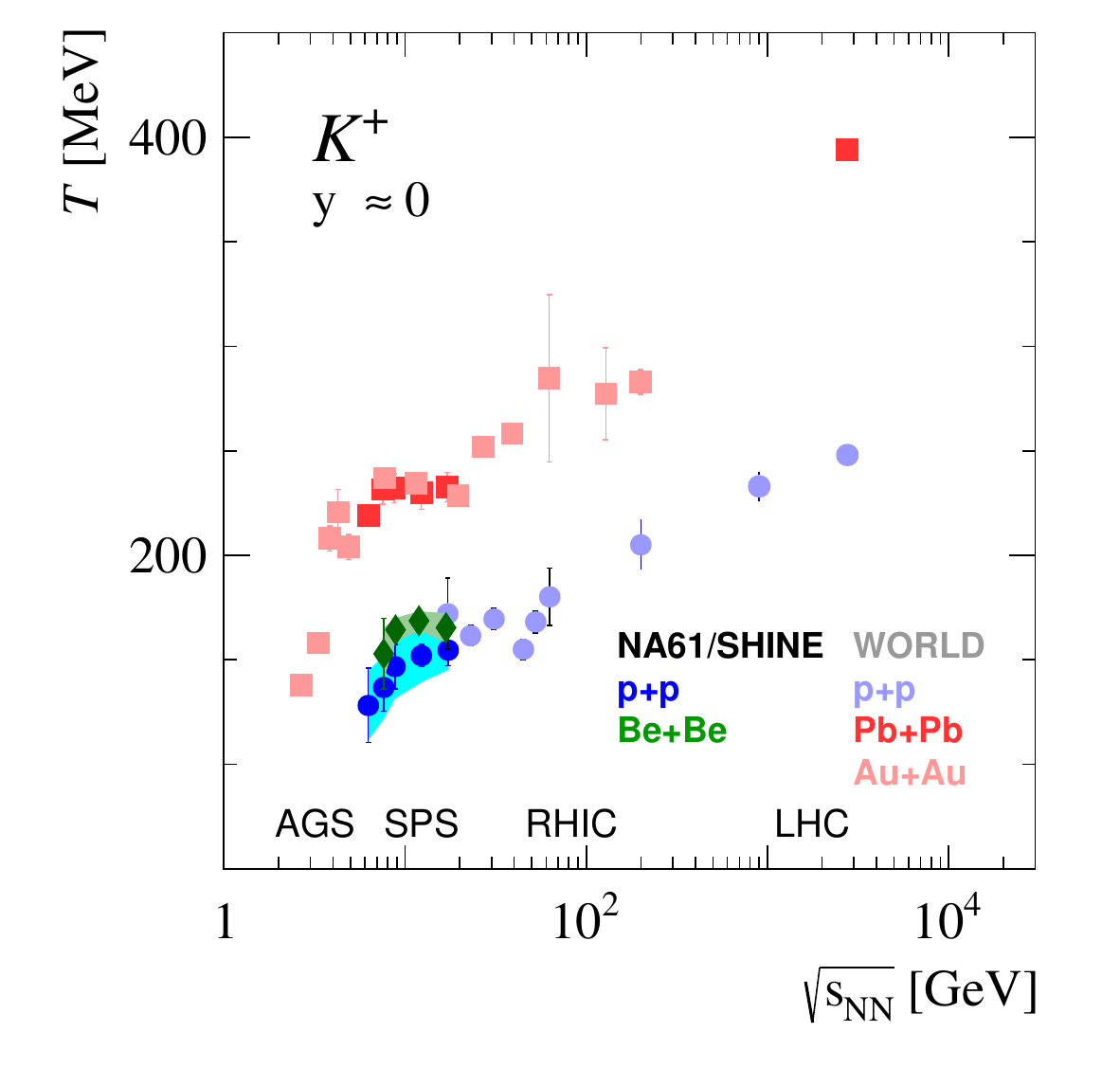}
\includegraphics[width=0.45\textwidth]{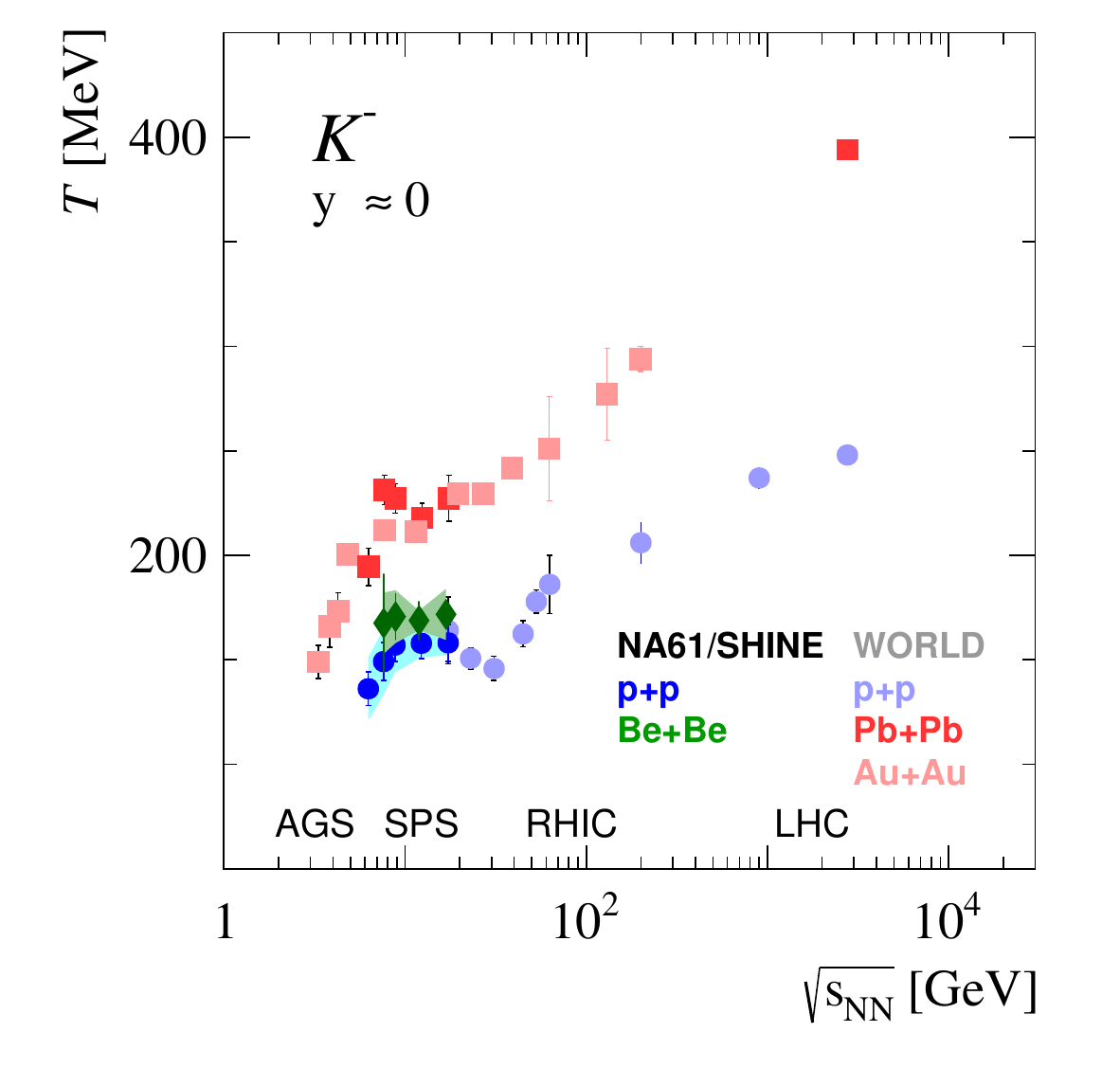}
\end{center}
\caption{The energy dependence of the inverse slope parameter of $p_T$ spectra at mid-rapidity of positively
(\textit{left}) and negatively (\textit{right}) charged $K$ mesons for \textit{central} Be+Be, 
Pb+Pb and Au+Au collisions as well as inelastic \textit{p+p} interactions. 
Both statistical (vertical bars) and systematic uncertainties (shaded bands) are shown.
}
    \label{fig:step_midrapidity}
\end{figure}

A step-like structure in the energy dependence of the inverse slope parameter $T$ of kaons at mid-rapidity 
was predicted~\cite{Gazdzicki:1998vd} at the onset of deconfinement (\textit{step}). 
In this scenario it is caused by the softness of the equation of state of the mixed phase of hadrons and partons 
stalling the expansion of the initial state with rising energy density. Kaon \pt spectra are well described by
a simple exponential because in contrast to pion \pt spectra they are not affected strongly by 
resonance decay products. As seen in Fig.~\ref{fig:step_midrapidity} from a compilation of published results 
such a step is observed in central collisions of heavy nuclei (Au and Pb) in the SPS energy range 
which is consistent with the onset of the phase transition~\cite{Afanasiev:2002mx,Alt:2007aa}. 
The new \NASixtyOne measurements from \textit{central} Be+Be collisions are similar to published results from inelastic \textit{p+p} interactions. This indicates that not much expansion flow is created in the small Be+Be
collision system at SPS energies. Although there appears to be a similar step feature, the values of $T$ are 
much smaller than the ones from collisions of heavy nuclei. The intriguing similarity between the energy dependence
of $T$ between \pp and Pb+Pb collisions is discussed in detail in~Ref.\cite{Aduszkiewicz:2019zsv}.

\subsection{Rapidity spectra and mean multiplicities}
\label{sec:yspectra}

\begin{figure}[!ht]
\begin{center}
\includegraphics[width=0.45\textwidth, trim={2cm 1.5cm 1cm 2cm}, clip]{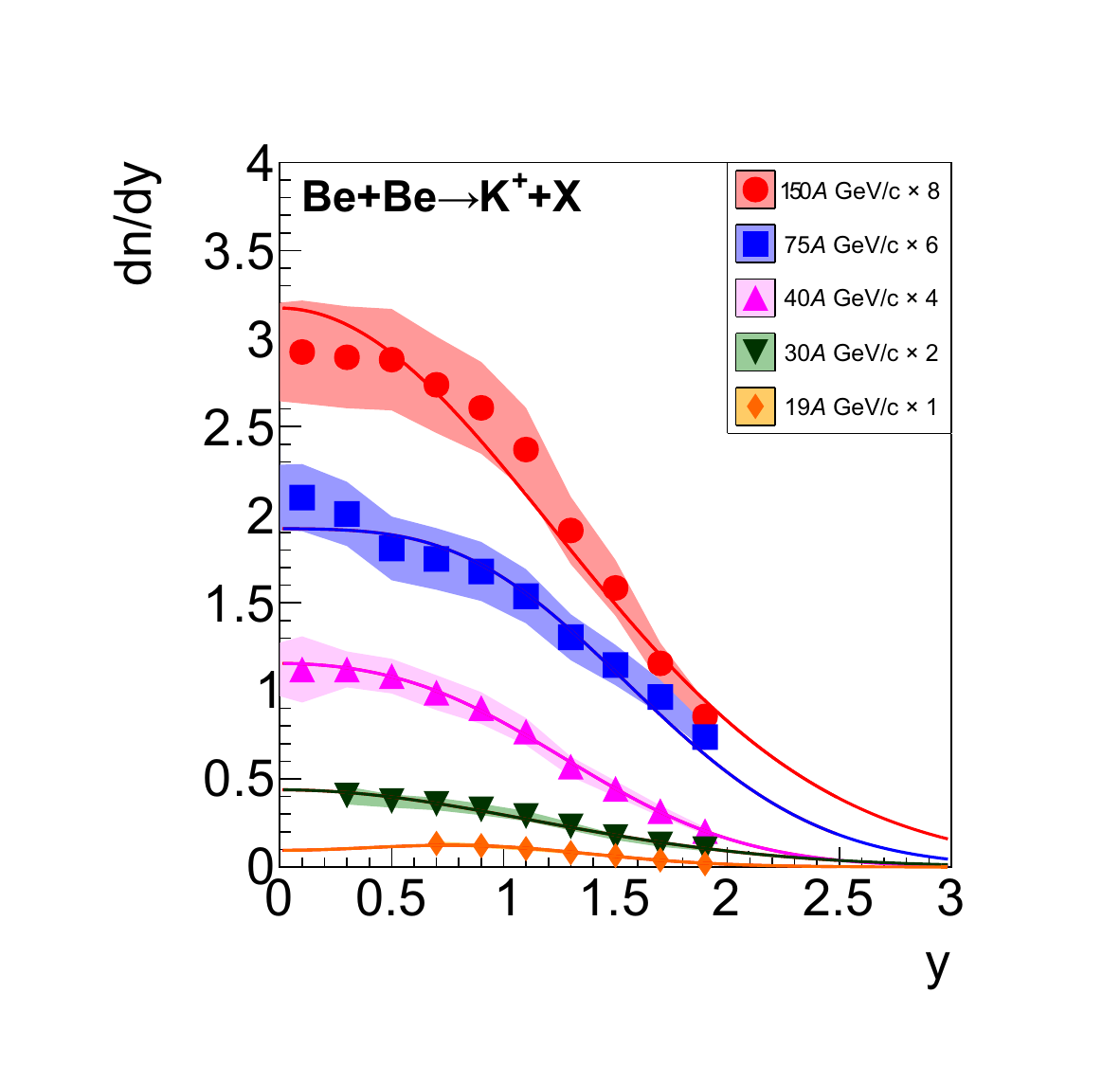}
\includegraphics[width=0.45\textwidth, trim={2cm 1.5cm 1cm 2cm}, clip]{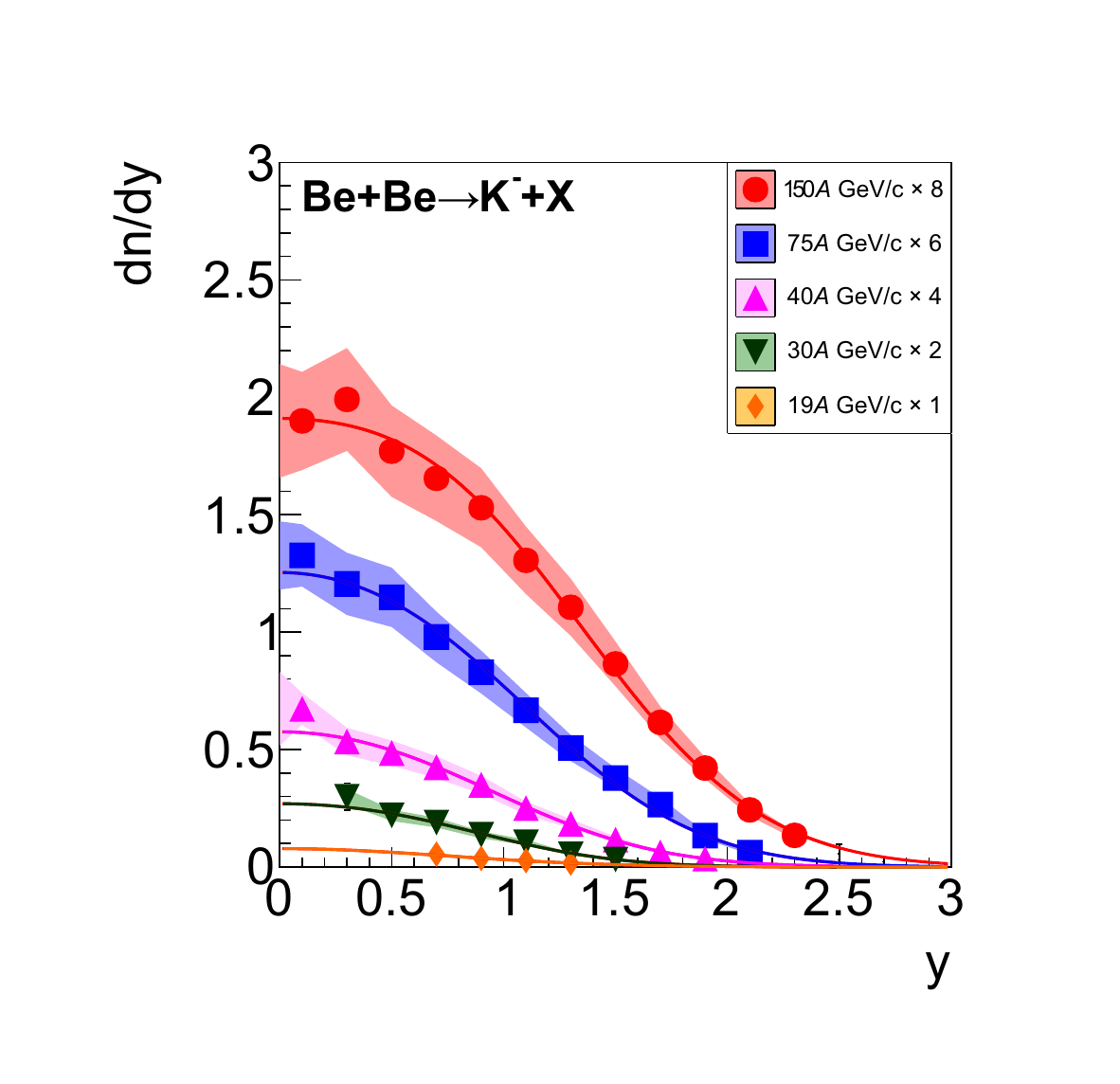}\\
\includegraphics[width=0.45\textwidth, trim={2cm 1.5cm 1cm 2cm}, clip]{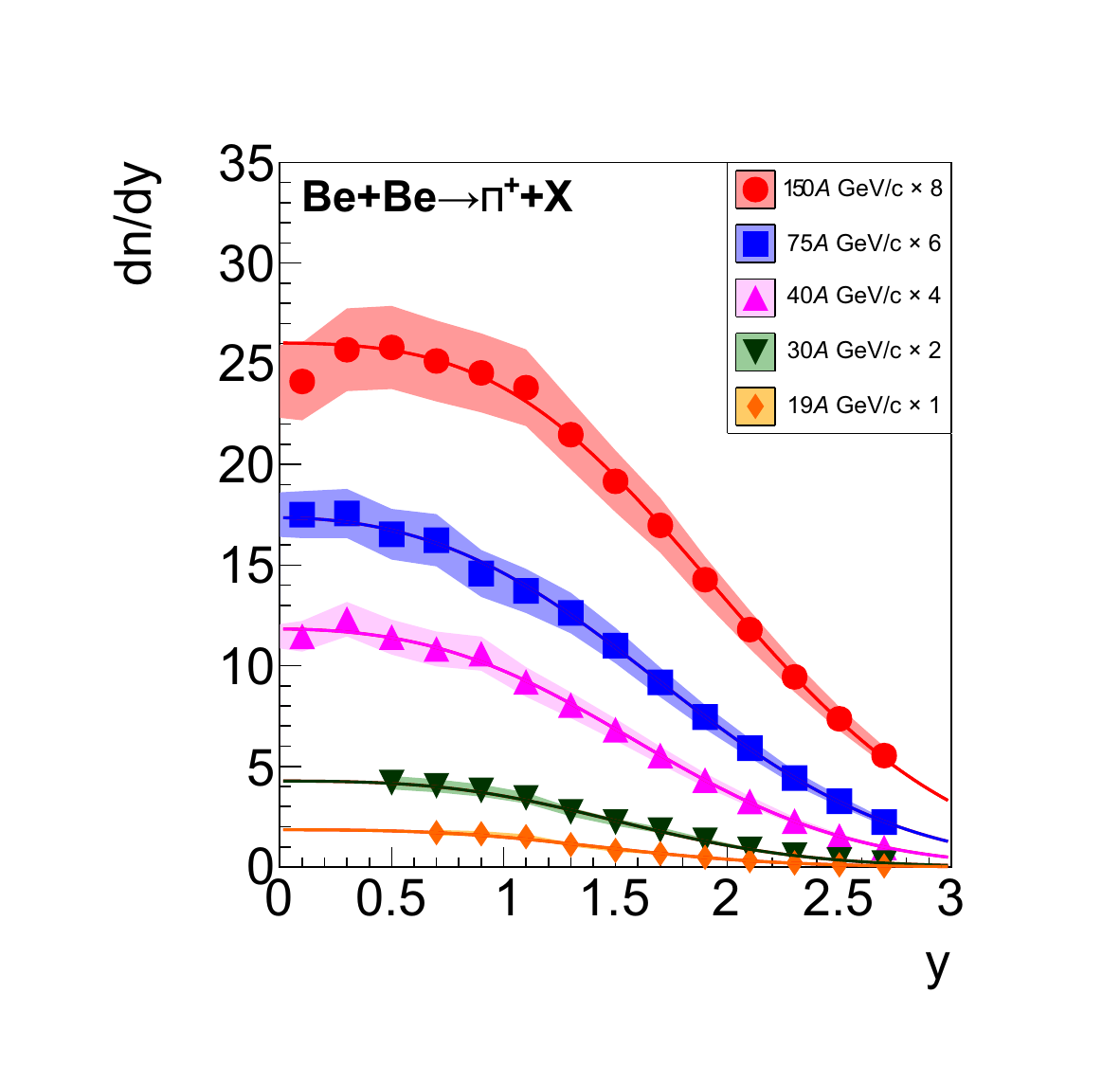}
\includegraphics[width=0.45\textwidth, trim={2cm 1.5cm 1cm 2cm}, clip]{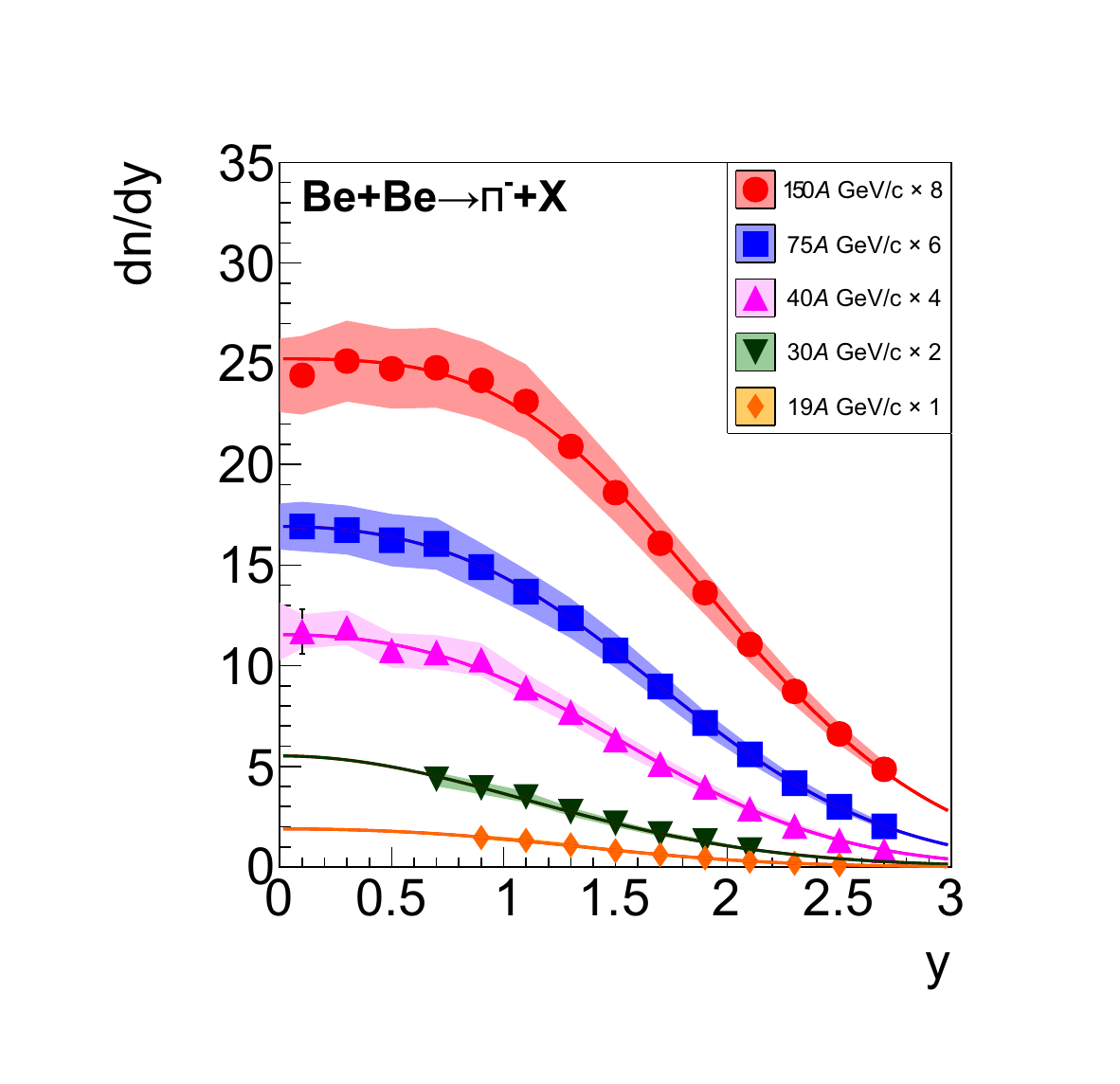}\\
\includegraphics[width=0.45\textwidth, trim={2cm 1.5cm 1cm 2cm}, clip]{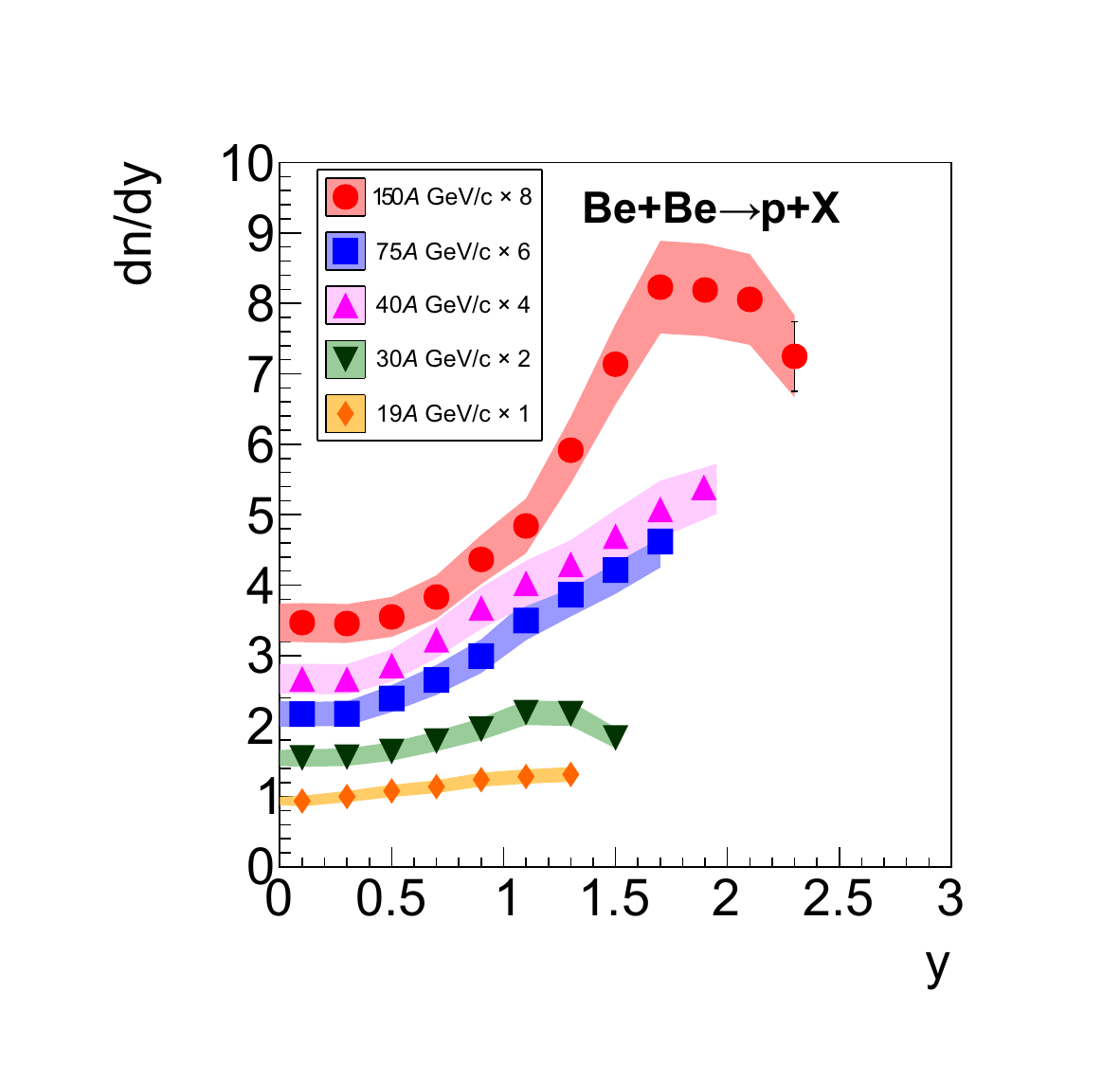}
\includegraphics[width=0.45\textwidth, trim={2cm 1.5cm 1cm 2cm}, clip]{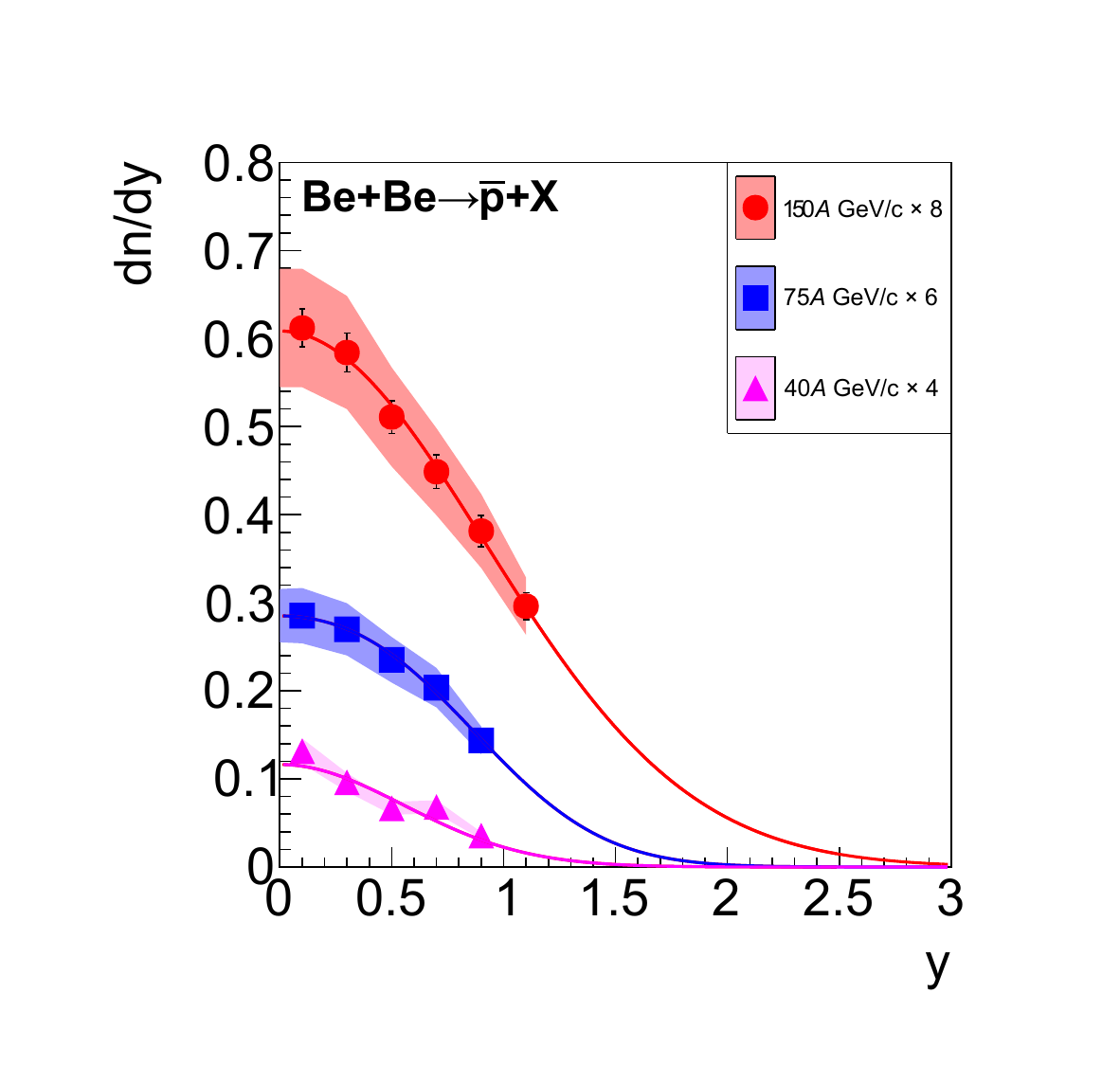}
\end{center}
\caption{Rapidity spectra of $K^{+}$, $K^{-}$, $\pi^{+}$, $\pi^{-}$, protons and antiprotons
produced in the 20\% most \textit{central} Be+Be collisions. Curves depict Gaussian fits used to
determine mean multiplicities. Spectra for different beam momenta were scaled by the following factors for better visibility: 150\AGeVc by factor 8, 75\AGeVc by factor 6, 40\AGeVc by factor 4, 30\AGeVc by factor 2 and 19\AGeVc by factor 1.}
\label{fig:dndy_final}
\end{figure}

Figure~\ref{fig:dndy_final} presents rapidity spectra for all studied particle species at all available 
collision energies. Measurements of the $\pi^-$ rapidity distributions employing the $h^-$ method 
indicated slight forward-backward asymmetry \cite{Acharya:2020cyb}. The acceptance for particle identification
does not extend into the backward hemisphere making such a test impossible. The forward-backward asymmetry of $\pi^-$ rapidity spectra reported in Ref.~\cite{Acharya:2020cyb} was at the level of 5\%. Since this is smaller than the systematic uncertainty no correction was applied.
In fact, Gaussian functions provide a
good description for all particle types except for protons. The latter exhibit a strong leading particle effect.
Mean forward multiplicities (protons excepted), were obtained by summing the measured points and adding an
extrapolation estimated from the fitted Gaussian functions which are shown by the curves 
in Fig.~\ref{fig:dndy_final}. Assuming forward-backward symmetry total mean multiplicities were
calculated and are summarized in Table~\ref{tab:mean_multiplicities}.

The simplest model of nucleus-nucleus collisions, the Wounded Nucleon Model~\cite{Bialas:1976ed}, suggests
that total produced particle multiplicities scale approximately with the respective ratios of wounded
nucleons. This ratio was derived using the \Epos model and the \NASixtyOne \textit{centrality} selection
procedure for Be+Be collisions with the 20\% smallest number of wounded nucleons 
(see Sec.~\ref{sec:centrality}) and inelastic \pp collisions Ref.~\cite{Aduszkiewicz:2019zsv}.
The resulting ratio of about 4 is close to the experimental ratios calculated from the total
multiplicities listed in Table~\ref{tab:mean_multiplicities} for Be+Be and
Ref.~\cite{Aduszkiewicz:2019zsv} for \pp interactions. 

\begin{figure}[!ht]
\begin{center}
\includegraphics[width=0.45\textwidth]{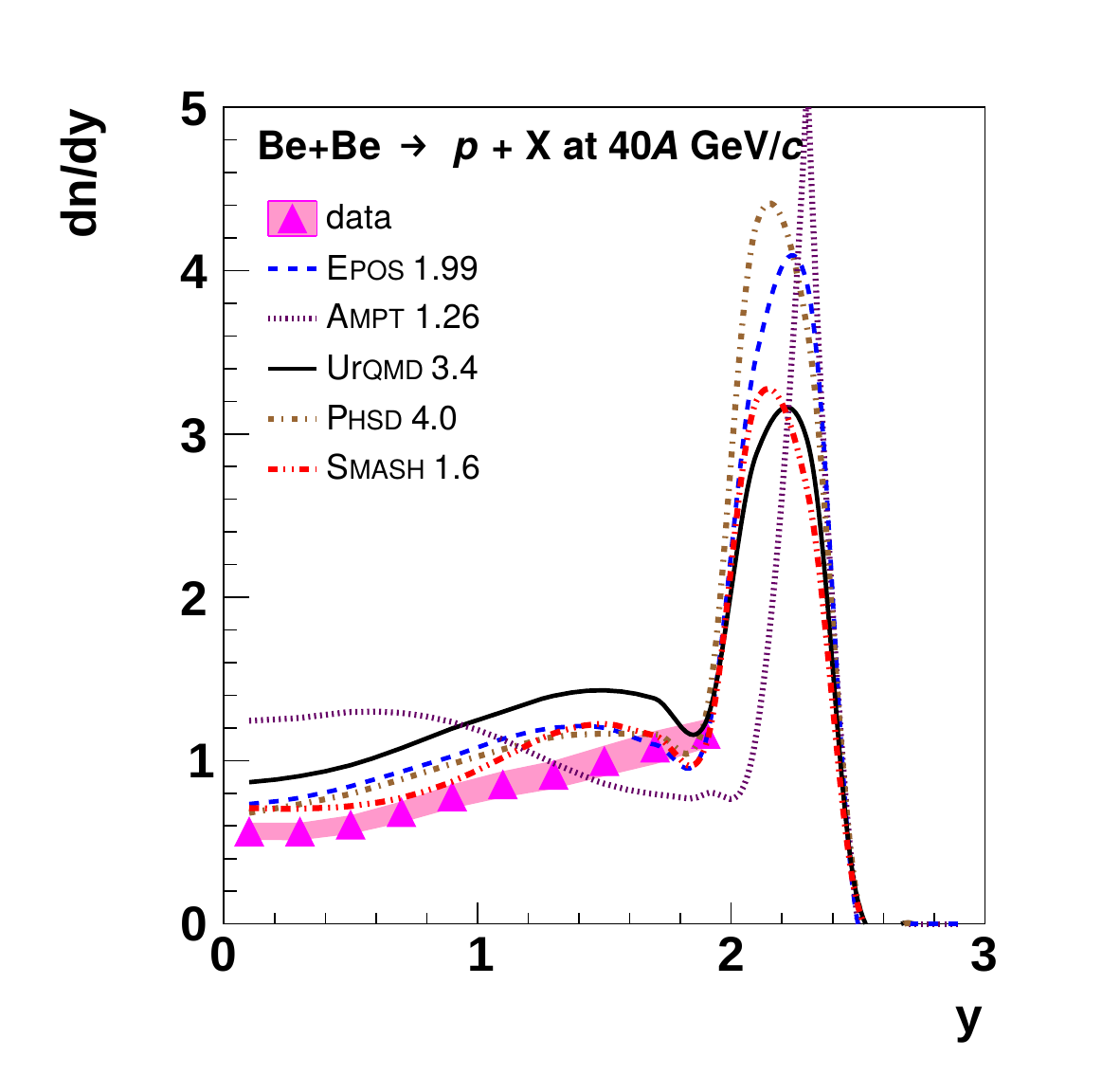}
\includegraphics[width=0.45\textwidth]{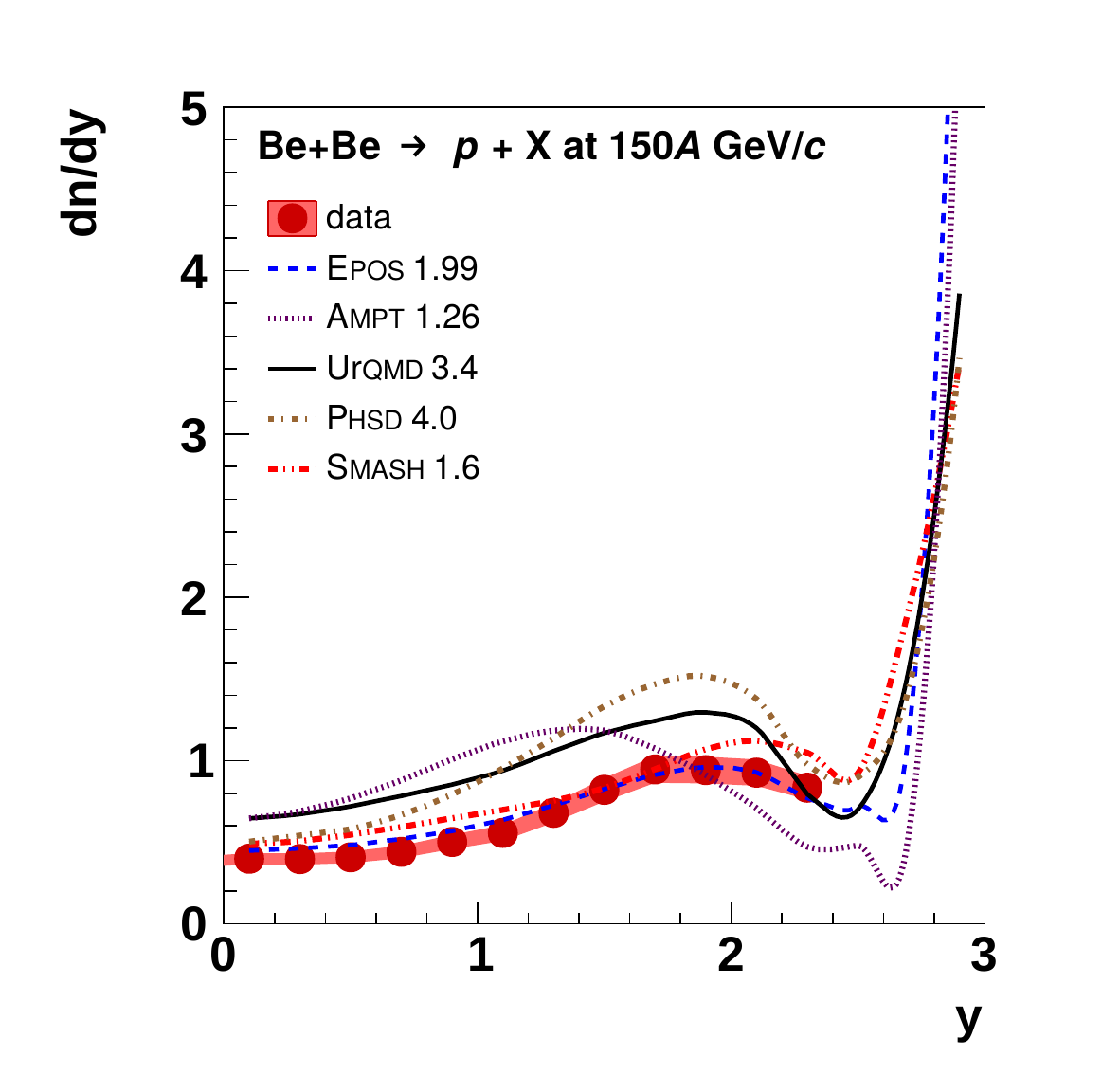}

\end{center}
\caption{Proton rapidity distribution in the 20\% most \textit{central} Be+Be collisions at 40$A$ and 150\AGeVc compared with predictions of the \Epos~1.99~\cite{Werner:2005jf,Pierog:2018} (blue dashed line), \Urqmd~3.4~\cite{Bass:1998ca,Bleicher:1999xi} (black solid line), \Ampt~1.26~\cite{PhysRevC.72.064901, PhysRevC.90.014904, PhysRevC.61.067901} (violet dotted line), \Phsd~4.0~\cite{PhysRevC.78.034919,CASSING2009215} (brown dashed-dotted line) and \Smash~1.6~(red dashed-double dotted line)~\cite{Mohs:2019iee,PhysRevC.94.054905}~models.
}
\label{fig:protonmodel}
\end{figure}

The determination of the mean multiplicity is more complicated for protons due to the rapid rise of 
the yield towards beam rapidity and the lack of measurements in this region. A comparison of the rapidity
distributions obtained in this analysis with predictions
of the \Urqmd~\cite{Bass:1998ca,Bleicher:1999xi}, \Epos~\cite{Werner:2005jf,Pierog:2018},
\Ampt~\cite{PhysRevC.72.064901, PhysRevC.90.014904, PhysRevC.61.067901}, \Phsd~4.0~\cite{PhysRevC.78.034919,CASSING2009215} and \Smash~1.6~\cite{Mohs:2019iee,PhysRevC.94.054905} models at 40$A$ and 150\AGeVc beam momentum is shown in
Fig.~\ref{fig:protonmodel}. Note that protons with $\pt\approx0$ and beam rapidity are assumed to be spectator protons and rejected in the model calculations. To determine the mean proton multiplicity one may calculate extrapolation factors from the models as the ratio of mean multiplicity to that in the region covered by the measurements. The multiplicity of protons in the region of measurements and the extrapolation factor obtained from the \Epos model are given in Table~\ref{tab:proton_multiplicities}. Mean proton multiplicities are not provided since the values will be strongly model dependent.

\begin{figure}[!ht]
\begin{center}
\includegraphics[width=0.45\textwidth, trim={0 1cm 0 2.2cm}, clip]{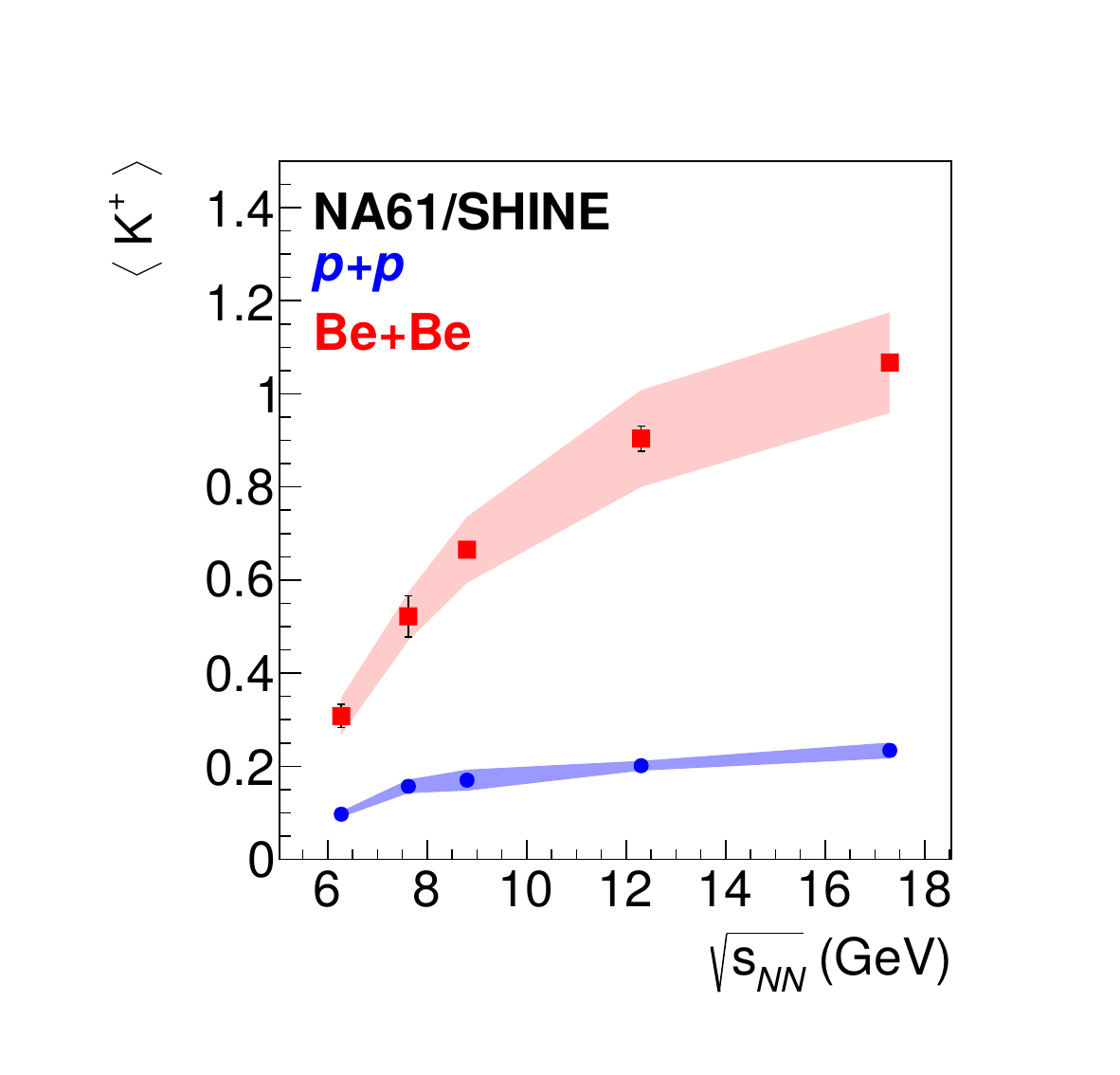}
\includegraphics[width=0.45\textwidth, trim={0 1cm 0 2.2cm}, clip]{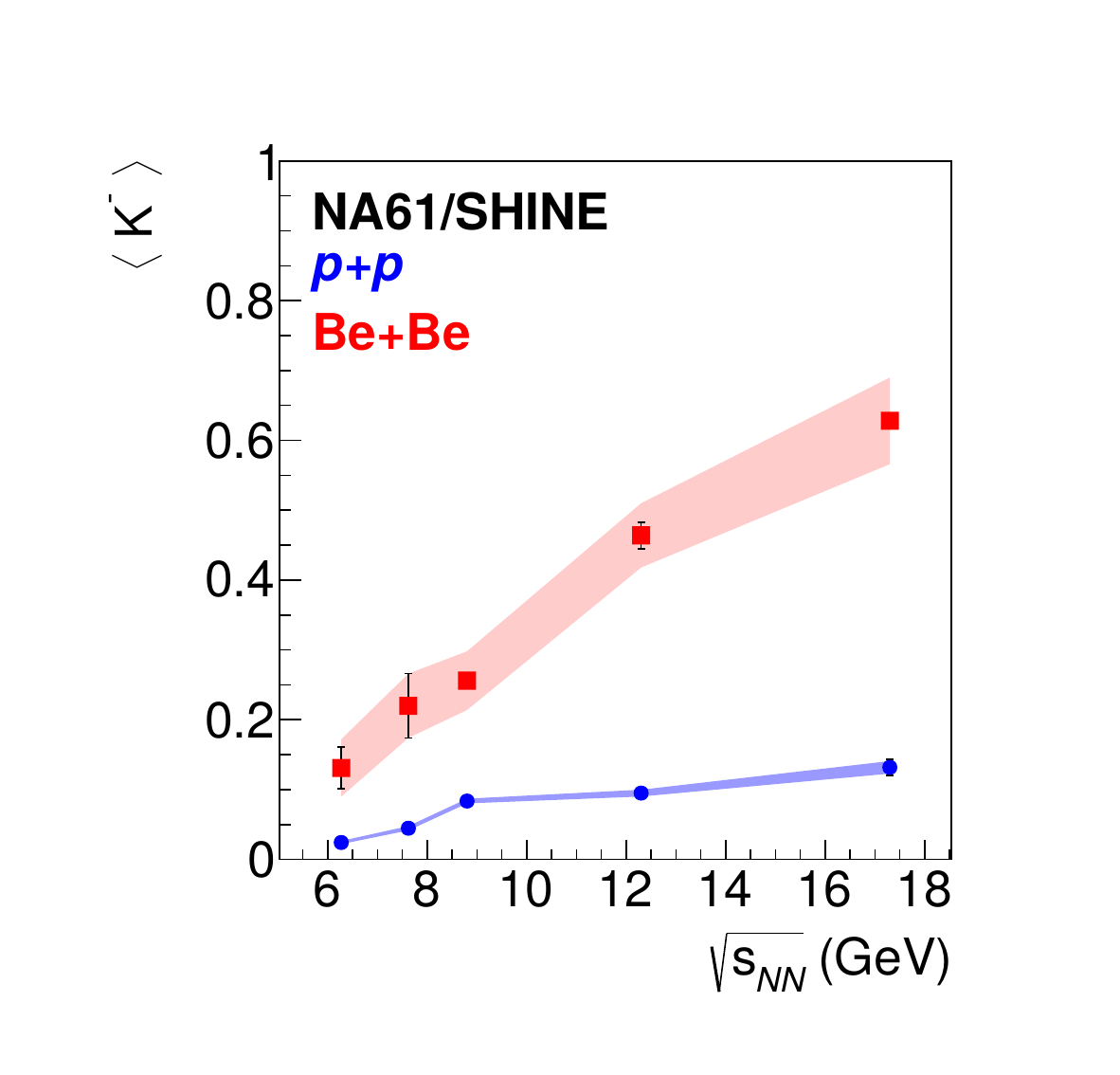}\\
\includegraphics[width=0.45\textwidth, trim={0 1cm 0 2.2cm}, clip]{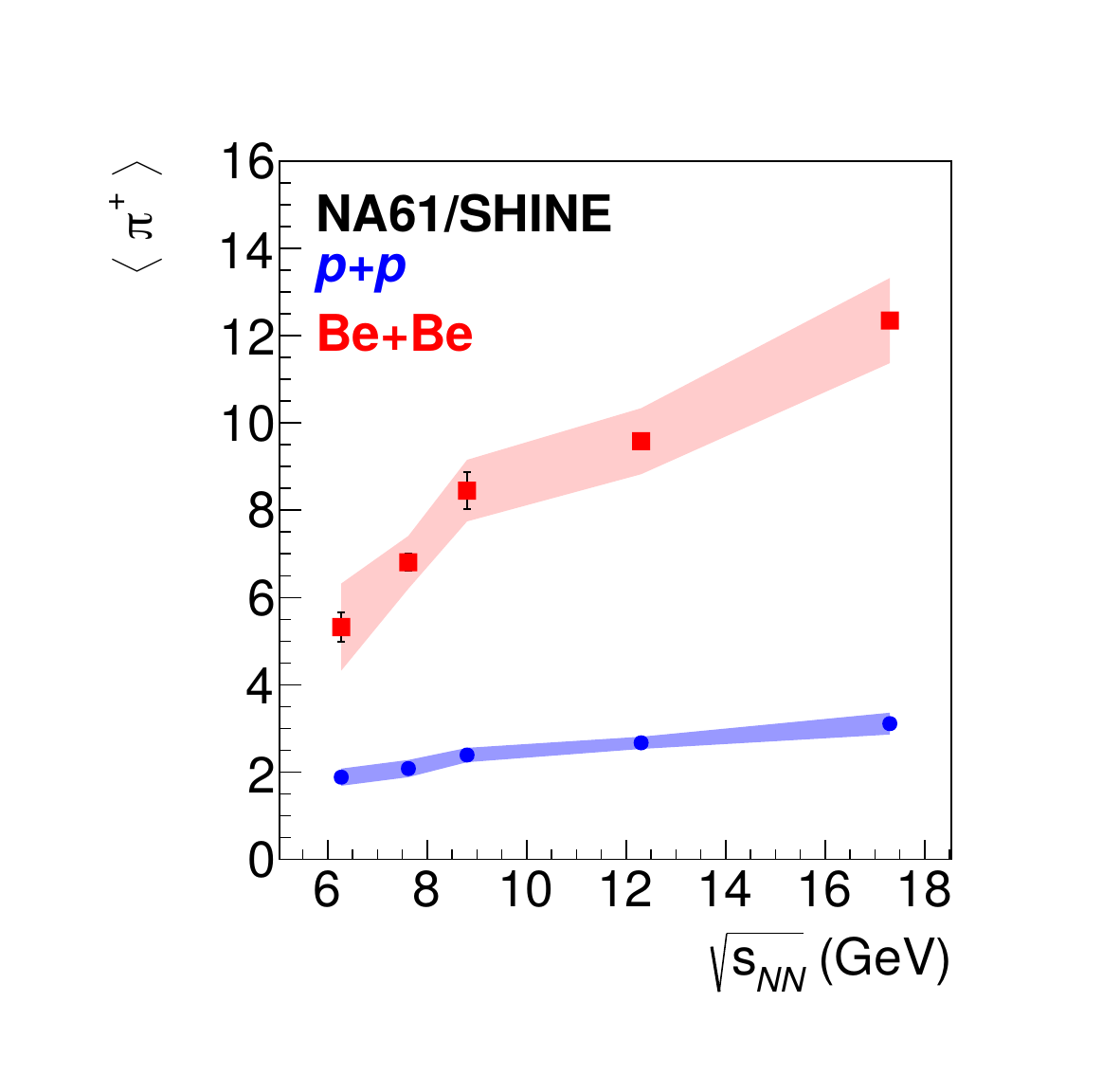}
\includegraphics[width=0.45\textwidth, trim={0 1cm 0 2.2cm}, clip]{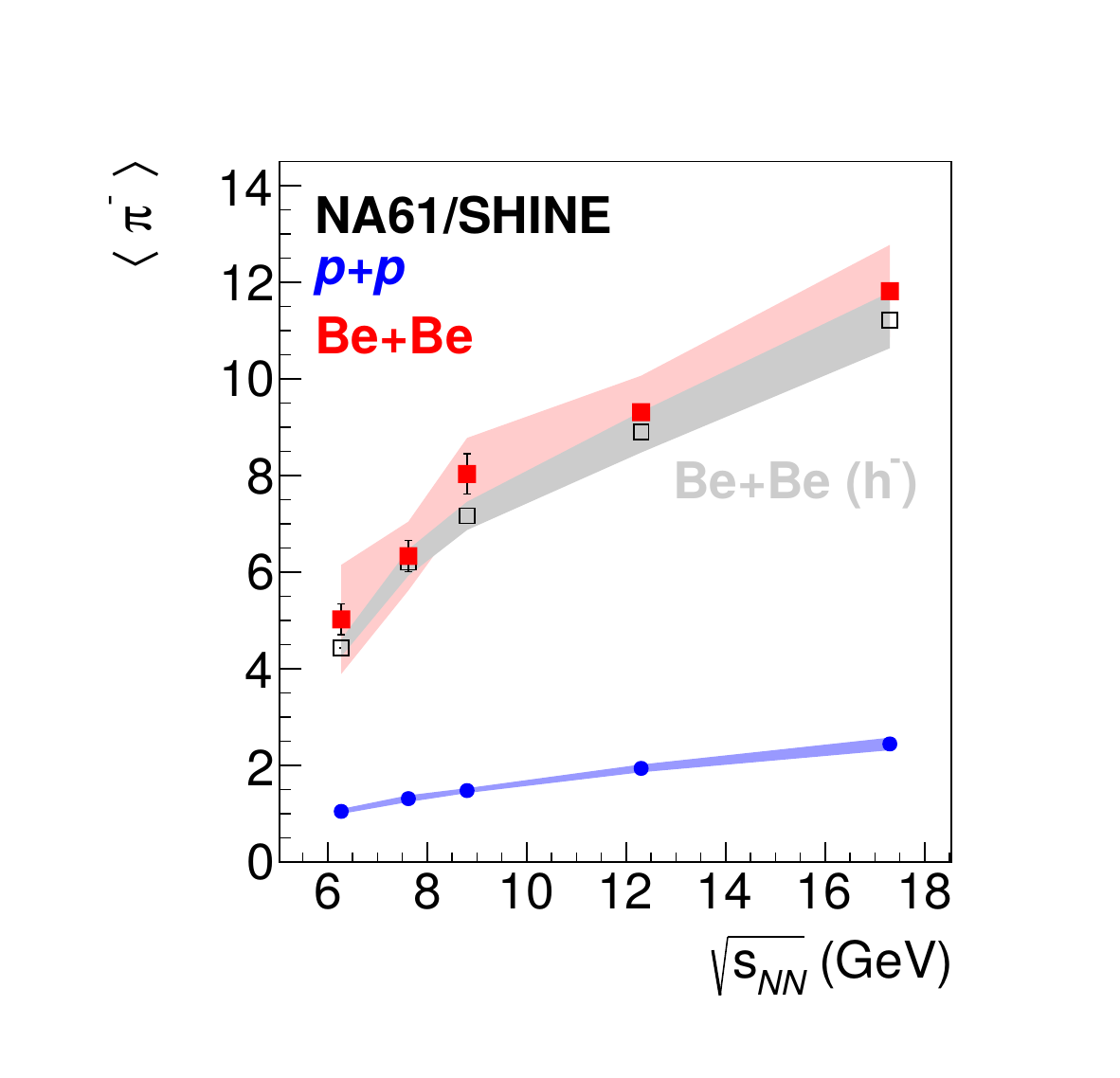}\\
\includegraphics[width=0.45\textwidth, trim={0 1cm 0 2.2cm}, clip]{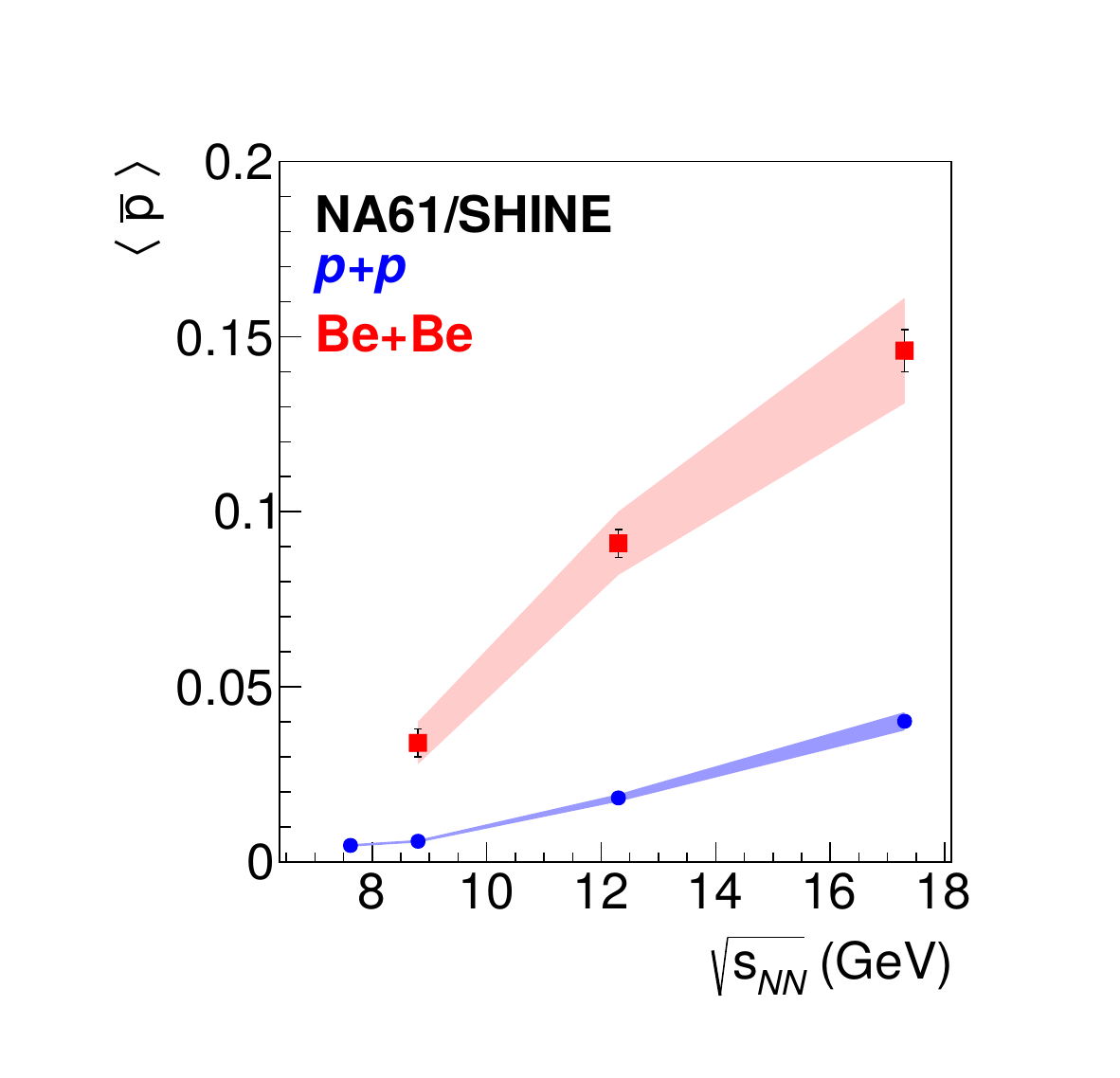}
\end{center}
\caption{Collision energy dependence of mean multiplicities of $K^{+}$, $K^{-}$, $\pi^{+}$, $\pi^{-}$, protons and antiprotons
produced in the 20\% most \textit{central} Be+Be collisions. Results from inelastic \textit{p+p} 
interactions~\cite{Abgrall:2013pp_pim,Aduszkiewicz:2017sei} are plotted for comparison. 
Both statistical (vertical bars) and systematic uncertainties (shaded bands) are shown. Gray points and band correspond to $\pi^{-}$ results obtained via $h^{-}$ method. Results for $\langle \pi^{-} \rangle$ obtained using the $h^{-}$ method ~\cite{Acharya:2020cyb} taking into account backward-forward asymmetry of the rapidity distribution. They were scaled from 5\% \textit{centrality} to 20\% based on the ratio of wounded nucleons estimated using the \Epos model.
}
\label{fig:tot_mult}
\end{figure}

The total multiplicities for all studied particle species are plotted as a function of collision energy 
in Fig.~\ref{fig:tot_mult} and compared to the corresponding results from inelastic \textit{p+p} interactions.
Particle yields in Be+Be collisions are higher by approximately a factor of four consistent with expectations 
from the wounded nucleon model.

\begin{table}[ht]
\caption{Yield of protons in the rapidity acceptance of measurements, the extrapolation factor to mean multiplicity based on the \Epos model.}
\vspace{0.5cm}
\small
\centering
			\begin{tabular}{|c||c|c|c|c|}
			\hline
			&   $\langle p_{y_{acc}} \rangle$ & $y_{acc}$ & $( \langle p_{y_{acc}} \rangle / \langle p_{y>0} \rangle )_{\text{\Epos}}$\\
				\hline
				19\AGeVc  & 1.657 $\pm$ 0.031 $\pm$ 0.191 & 0 $<$ y $<$ 1.5 & 69\%\\
				30\AGeVc  & 1.312 $\pm$ 0.014 $\pm$ 0.063 & 0 $<$ y $<$ 1.5 & 54\%\\
				40\AGeVc  & 1.878 $\pm$ 0.011 $\pm$ 0.057 & 0 $<$ y $<$ 2.1 & 78\%\\
				75\AGeVc  & 1.232 $\pm$ 0.008 $\pm$ 0.026 & 0 $<$ y $<$ 2.1 & 60\%\\
				150\AGeVc & 0.928 $\pm$ 0.029 $\pm$ 0.060 & 0 $<$ y $<$ 2.3 & 48\% \\
				\hline
 			\end{tabular}
\label{tab:proton_multiplicities}			
\end{table}

\begin{table}[ht]
\caption{Mean multiplicities of $K^{+}$, $K^{-}$, $\pi^{+}$, $\pi^{-}$, protons and antiprotons
produced in the 20\% most \textit{central} Be+Be collisions.  For each value the statistical and systematic uncertainties are given as a first and second uncertainty contribution respectively.}
\vspace{0.5cm}
\small
\centering
			\begin{tabular}{|c||c|c|}
				\hline
				          &   $\langle K^{+} \rangle$                       & $\langle K^{-} \rangle$ \\
			    \hline
				19\AGeVc  & 0.308 $\pm$ 0.025 $\pm$ 0.040   & 0.131 $\pm$ 0.030 $\pm$ 0.041 \\
				30\AGeVc  & 0.522 $\pm$ 0.044 $\pm$ 0.052   & 0.220 $\pm$ 0.046 $\pm$ 0.046 \\
				40\AGeVc  & 0.665 $\pm$ 0.012 $\pm$ 0.071   & 0.256 $\pm$ 0.011 $\pm$ 0.042 \\
				75\AGeVc  & 0.904 $\pm$ 0.027 $\pm$ 0.104   & 0.464 $\pm$ 0.019 $\pm$ 0.046 \\
				150\AGeVc & 1.067 $\pm$ 0.012 $\pm$ 0.108   & 0.628 $\pm$ 0.011 $\pm$ 0.062 \\
        
         \hline \hline
                          & $\langle \pi^{+} \rangle$                     & $\langle \pi^{-} \rangle$ \\
			    \hline
				19\AGeVc  & 5.323 $\pm$ 0.340 $\pm$ 0.998 & 5.021 $\pm$ 0.321 $\pm$ 1.129 \\
				30\AGeVc  & 6.807 $\pm$ 0.200 $\pm$ 0.605 & 6.333 $\pm$ 0.321 $\pm$ 0.713 \\
				40\AGeVc  & 8.449 $\pm$ 0.208 $\pm$ 0.705 & 8.033 $\pm$ 0.419 $\pm$ 0.745 \\
				75\AGeVc  & 9.581 $\pm$ 0.078 $\pm$ 0.756 & 9.312 $\pm$ 0.099 $\pm$ 0.754  \\
				150\AGeVc & 12.344 $\pm$ 0.085 $\pm$ 0.975 & 11.817 $\pm$ 0.086 $\pm$ 0.957 \\
				\hline
				
		      \hline \hline
		                  & 
		                  $\langle \overline{p} \rangle$  &                  \\
			    \hline
				19\AGeVc  
				& -&\\
				30\AGeVc  
				& -&\\
				40\AGeVc  
				& 0.034 $\pm$ 0.004 $\pm$ 0.006 &\\
				75\AGeVc  
				&  0.084 $\pm$ 0.004 $\pm$ 0.009 &\\
				150\AGeVc 
				& 0.146 $\pm$ 0.006 $\pm$ 0.015 &\\
				\hline
			\end{tabular}
\label{tab:mean_multiplicities}			
\end{table}

\subsection{$K$/$\pi$ ratio}

\begin{figure}
\begin{center}
\includegraphics[width=0.45\textwidth]{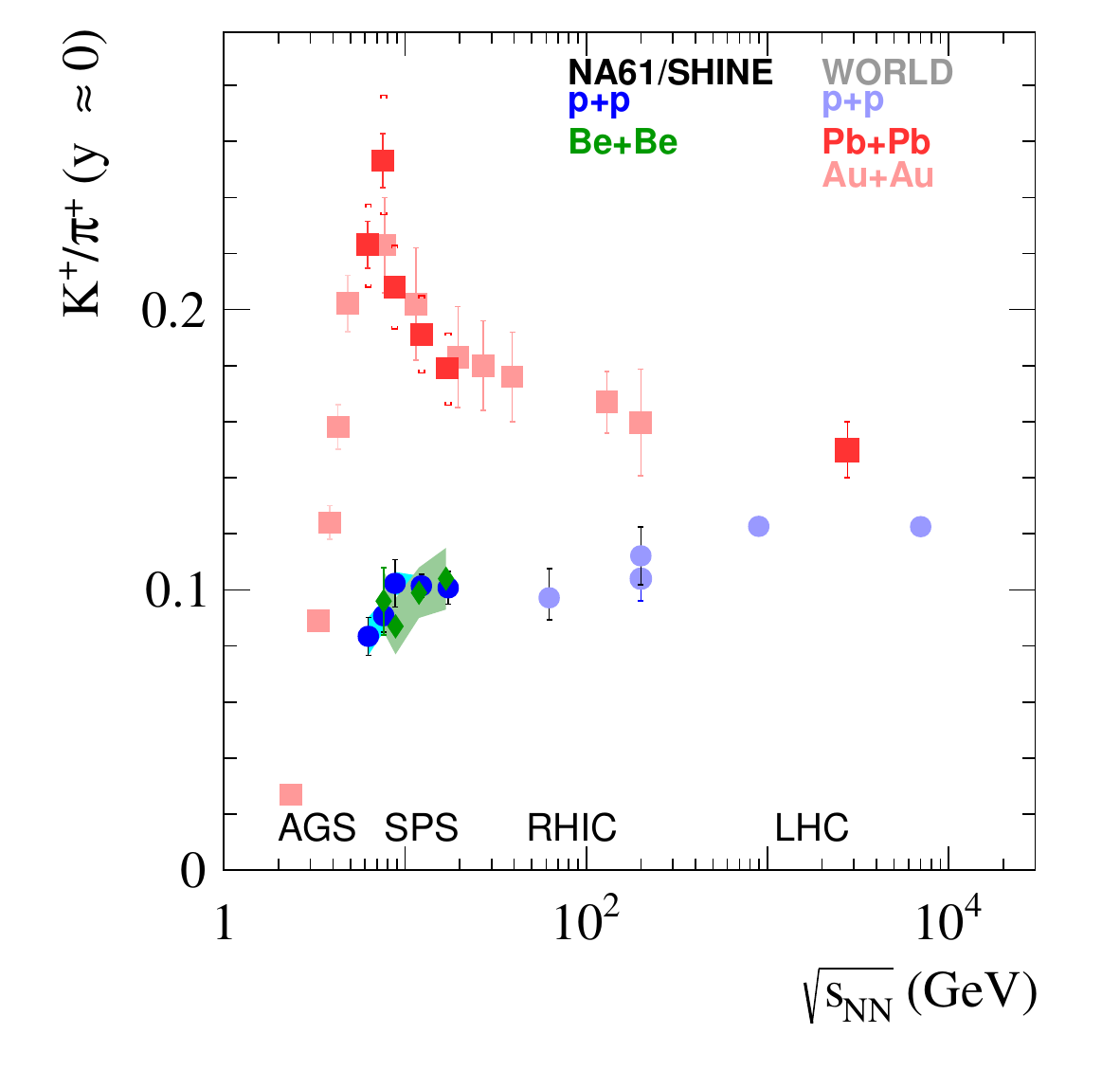} 
\includegraphics[width=0.45\textwidth]{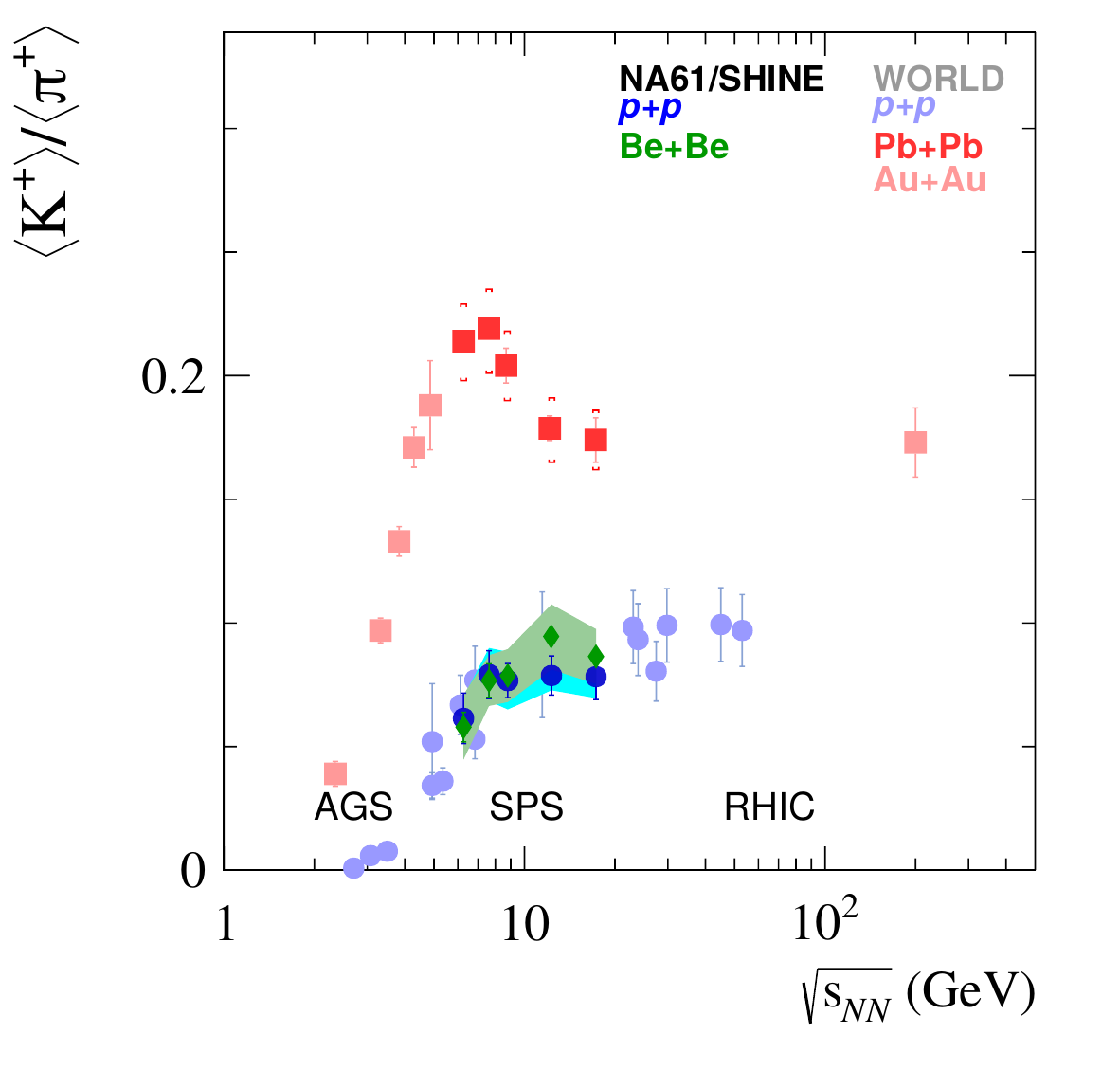}
\end{center}
\caption{The energy dependence of the $K^+$/$\pi^+$ particle yields ratio at mid-rapidity (\textit{left}) 
and full acceptance (\textit{right}) for the 20\% most \textit{central} Be+Be, \textit{central} Pb+Pb 
and Au+Au collisions, as well as inelastic \textit{p+p} interactions. 
Both statistical (vertical bars) and systematic uncertainties (shaded bands) are shown.
}
\label{fig:horn}
\end{figure}

The $K$/$\pi$ ratio at SPS energies was shown to be a good measure of the strangeness to entropy 
ratio~\cite{Gazdzicki:2010iv} which is different in the confined phase (hadrons) and the 
QGP (quarks, anti-quarks and gluons). A maximum (\textit{horn}) and a subsequent plateau in the 
energy dependence of the $K^+$/$\pi^+$ ratio was observed by the NA49 experiment in the SPS energy 
range (see Fig.~\ref{fig:horn}) in central Pb+Pb collisions \cite{Afanasiev:2002mx,Alt:2007aa} and
interpreted as one of the indications of the onset of deconfinement~\cite{Gazdzicki:1998vd}. 

The \NASixtyOne experiment studies the evolution of this signal with respect to the size of 
the collision system in order to find when deconfinement starts occurring. In \textit{p+p} interactions 
and Be+Be collisions the $K^{\pm}$ and $\pi^-$ yield can be measured by \NASixtyOne over all of 
phase space. This is not possible for the $\pi^+$ yield which has to be derived from the $\pi^-$ yield 
using an isospin correction. Be+Be collisions, however, are isospin symmetric and therefore 
the mean multiplicities of $\pi^+$ and $\pi^-$ have to be the same. The $\pi^-$ yields for the 20\% most
\textit{central} Be+Be collisions were calculated by scaling the $\pi^-$ multiplicity published 
in Ref.~\cite{Acharya:2020cyb} by the ratio of $\langle W \rangle$ for the 20\% (see Table~\ref{tab:w}) and 
the 5\% most \textit{central} Be+Be collisions Ref.~ \cite{Acharya:2020cyb}.
The results for the energy dependence of the $K^+$/$\pi^+$ ratio from the 20\% most \textit{central} Be+Be
collisions are shown in Fig.~\ref{fig:horn} together with measurements in inelastic \textit{p+p}, 
central Pb+Pb collisions and other reactions. The results from Be+Be collisions do not yet deviate 
from those in \textit{p+p} reactions and are about a factor two lower for the total yield ratio than 
the values found in central Pb+Pb and Au+Au collisions. In particular, no \textit{horn} structure 
is observed for the small collision systems.

\section{Comparison with models}

This subsection compares observables, sensitive to the onset of deconfinement, between measurements and
model predictions. The \Epos 1.99~\cite{Werner:2008zza}, \Urqmd 3.4~\cite{Bass:1998ca,Bleicher:1999xi}, 
\Ampt~1.26~\cite{PhysRevC.72.064901, PhysRevC.90.014904, PhysRevC.61.067901},
\Phsd~4.0~\cite{PhysRevC.78.034919,CASSING2009215} and \Smash~1.6~\cite{Mohs:2019iee,PhysRevC.94.054905} models were chosen for this study. In \Epos the reaction proceeds from 
the excitation of strings according to Gribov-Regge theory to string fragmentation into hadrons. 
\Urqmd starts with a hadron cascade based on elementary cross sections for resonance production 
which either decay (mostly at low energies) or are converted into strings which fragment into hadrons
(mostly at high energies). \Ampt uses the heavy ion jet interaction generator (\Hijing) for generating 
the initial conditions, Zhang's parton cascade for modeling partonic scatterings and the Lund 
string fragmentation model or a quark coalescence model for hadronization.  \Phsd is a microscopic 
offshell transport approach that describes the evolution of a relativistic heavy-ion collision from 
the initial hard scatterings and string formation through the dynamical deconfinement phase transition 
to the quark-gluon plasma as well as hadronization and the subsequent interactions in the hadronic phase.
\Smash uses the hadronic transport approach where the free parameters of the string excitation 
and decay are tuned to match the experimental measurements in inelastic \textit{p+p} collisions. 
Selection of events in all model calculations follows the procedure for central collisions to which
experimental results correspond to, see Sec.~\ref{sec:centrality}. This is particularly 
important when comparisons of yields with measurements are to be performed.

\begin{figure}[!ht]
    \centering
    \includegraphics[width=0.45\textwidth]{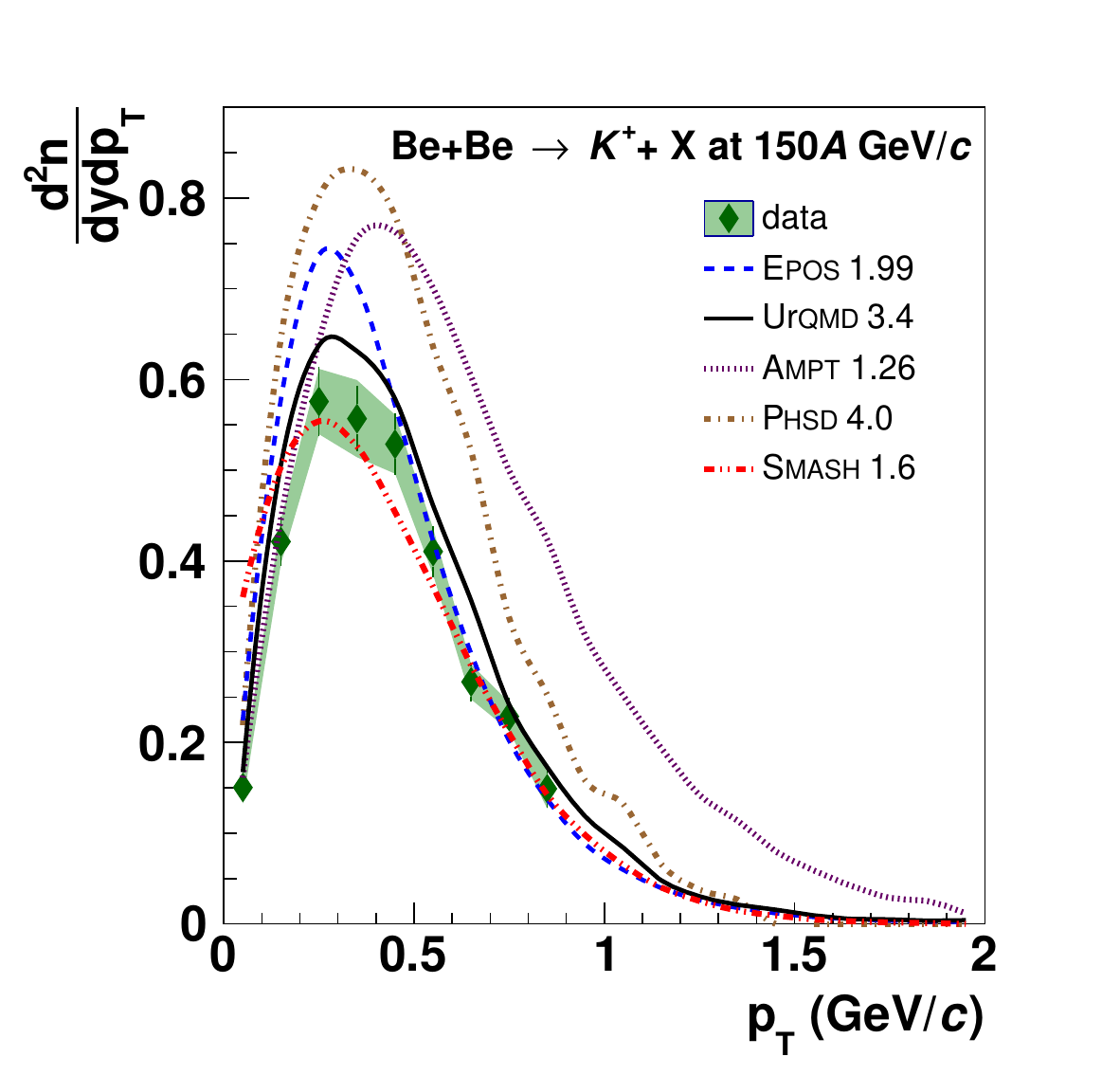}
    \includegraphics[width=0.45\textwidth]{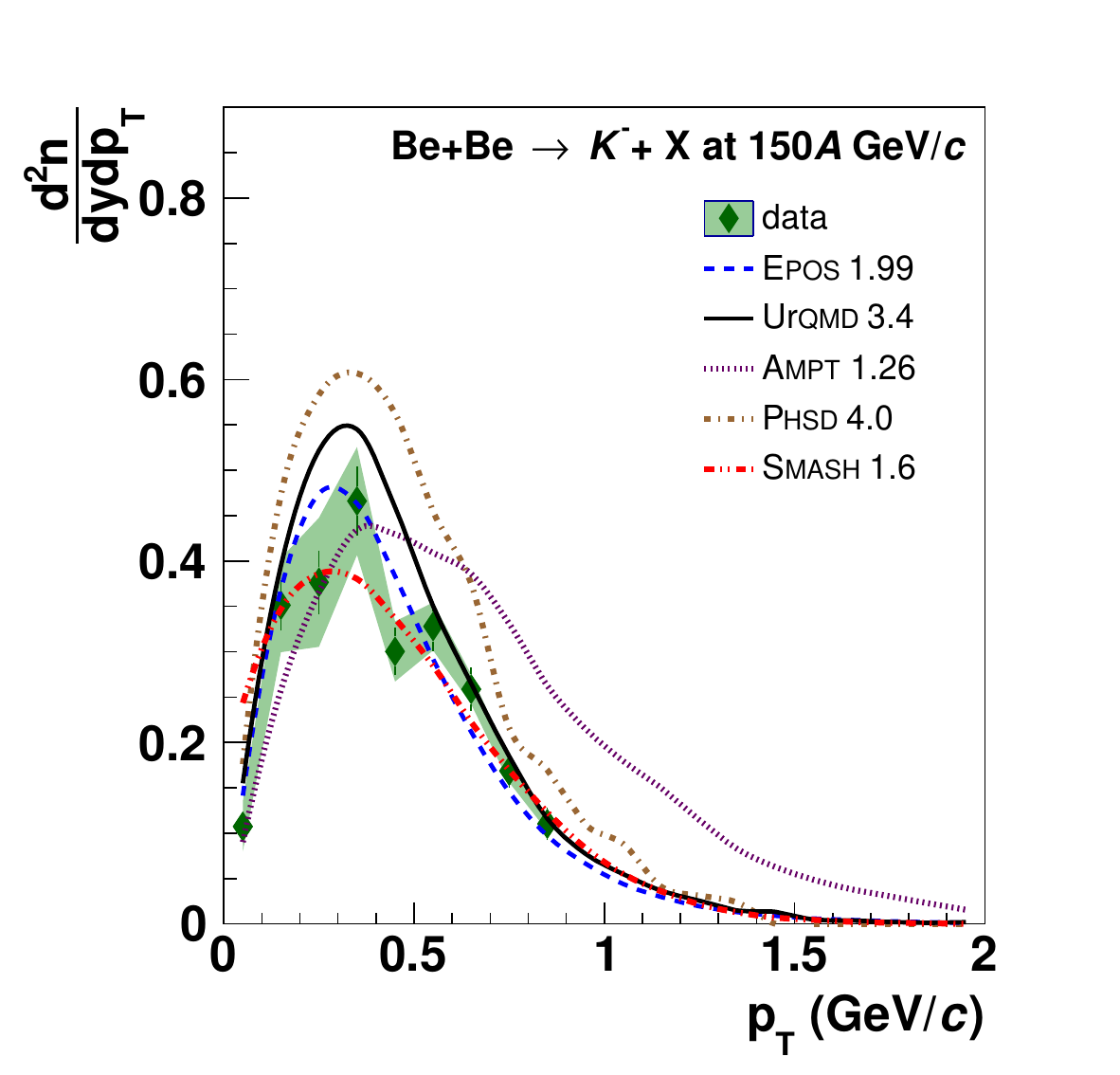}
    \includegraphics[width=0.45\textwidth]{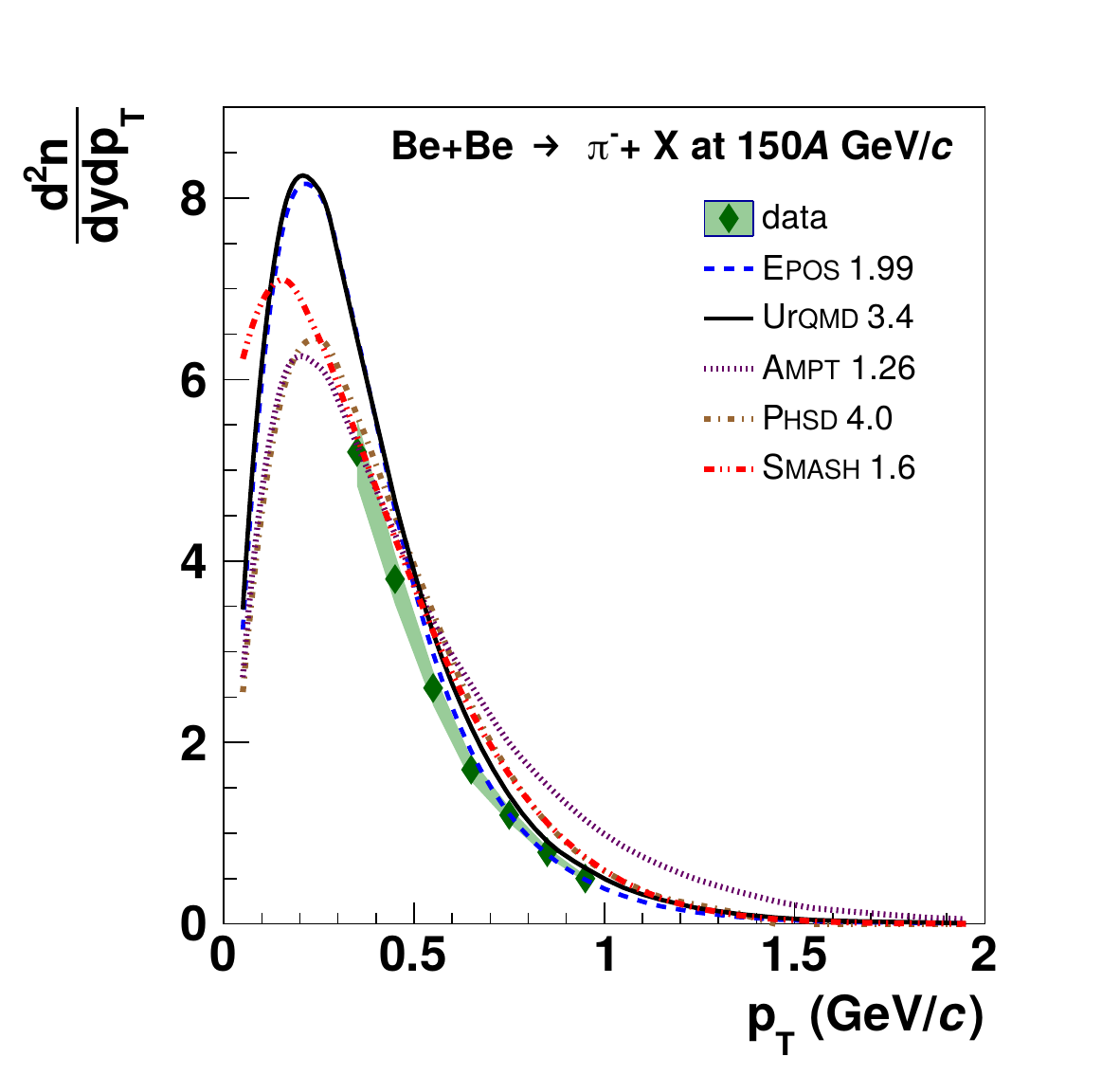}
    \includegraphics[width=0.45\textwidth]{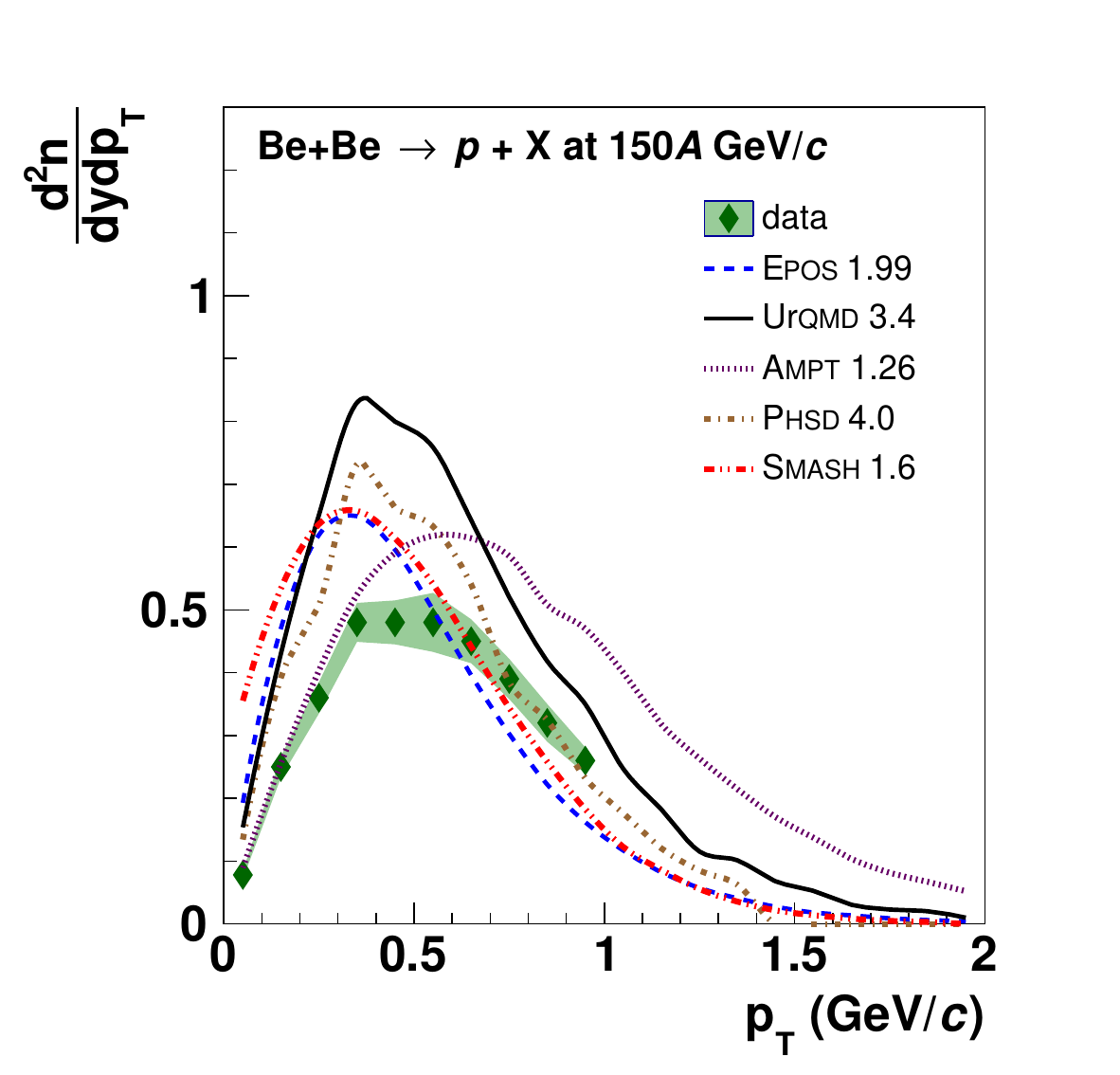}
    \caption{Comparison of the $p_T$ spectra of $K^+$ (\textit{top-left}),  $K^-$ (\textit{top-right}), $\pi^-$ (\textit{bottom-left}) and $p$ (\textit{bottom-right}) at mid-rapidity for the 20\% most \textit{central} Be+Be collisions 
    at 150\AGeVc with models: \Epos 1.99 (blue dashed line), \Urqmd 3.4 (black solid line), 
    \Ampt 1.26 (violet dotted line), \Phsd 4.0 (brown dashed-dotted line) and \Smash 1.6 (red dashed-double dotted line)
    .}
    \label{fig:pT_spectra_vs_models}
\end{figure}

\begin{figure}[!ht]
    \centering
    \includegraphics[width=0.45\textwidth]{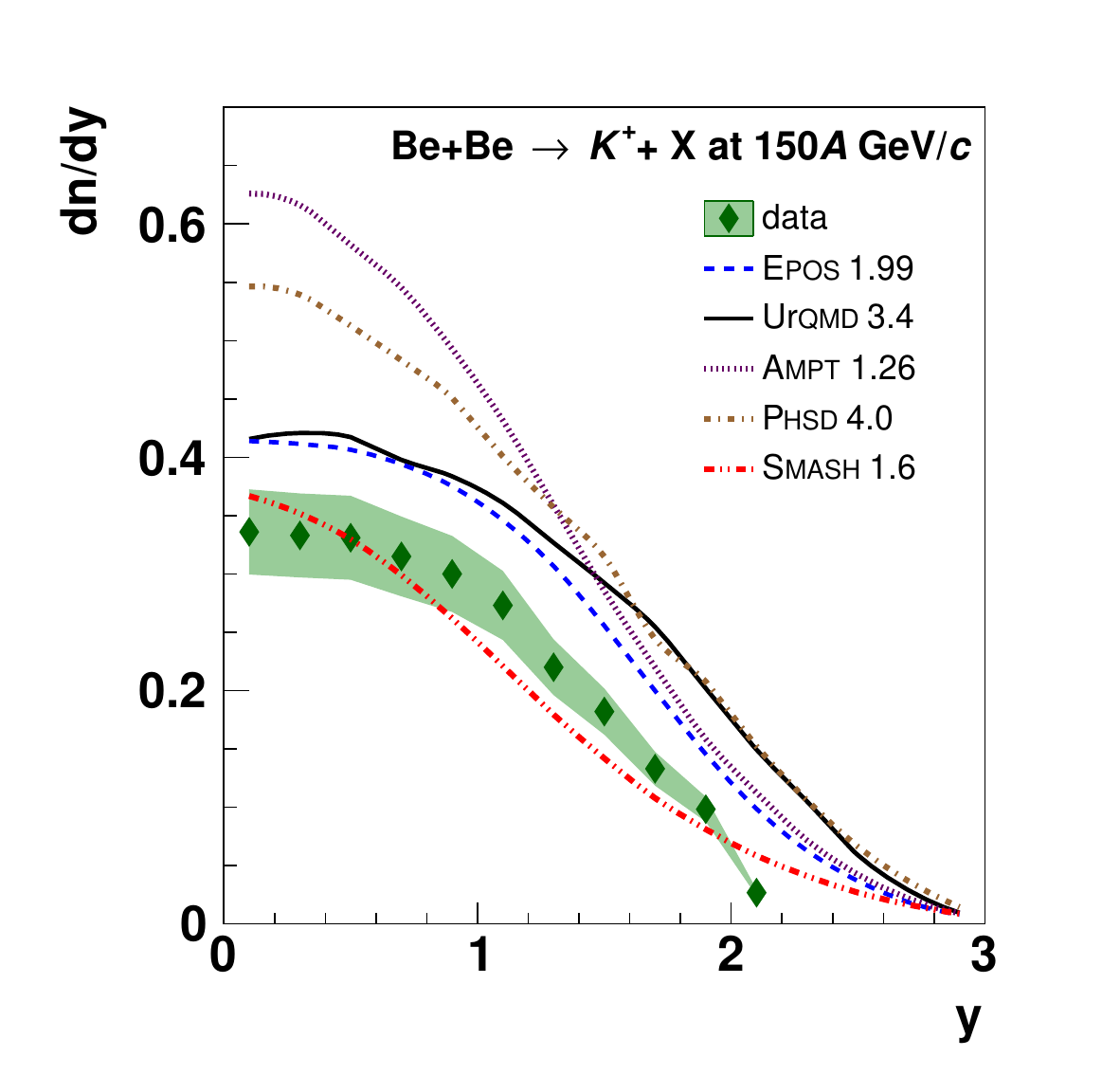}
    \includegraphics[width=0.45\textwidth]{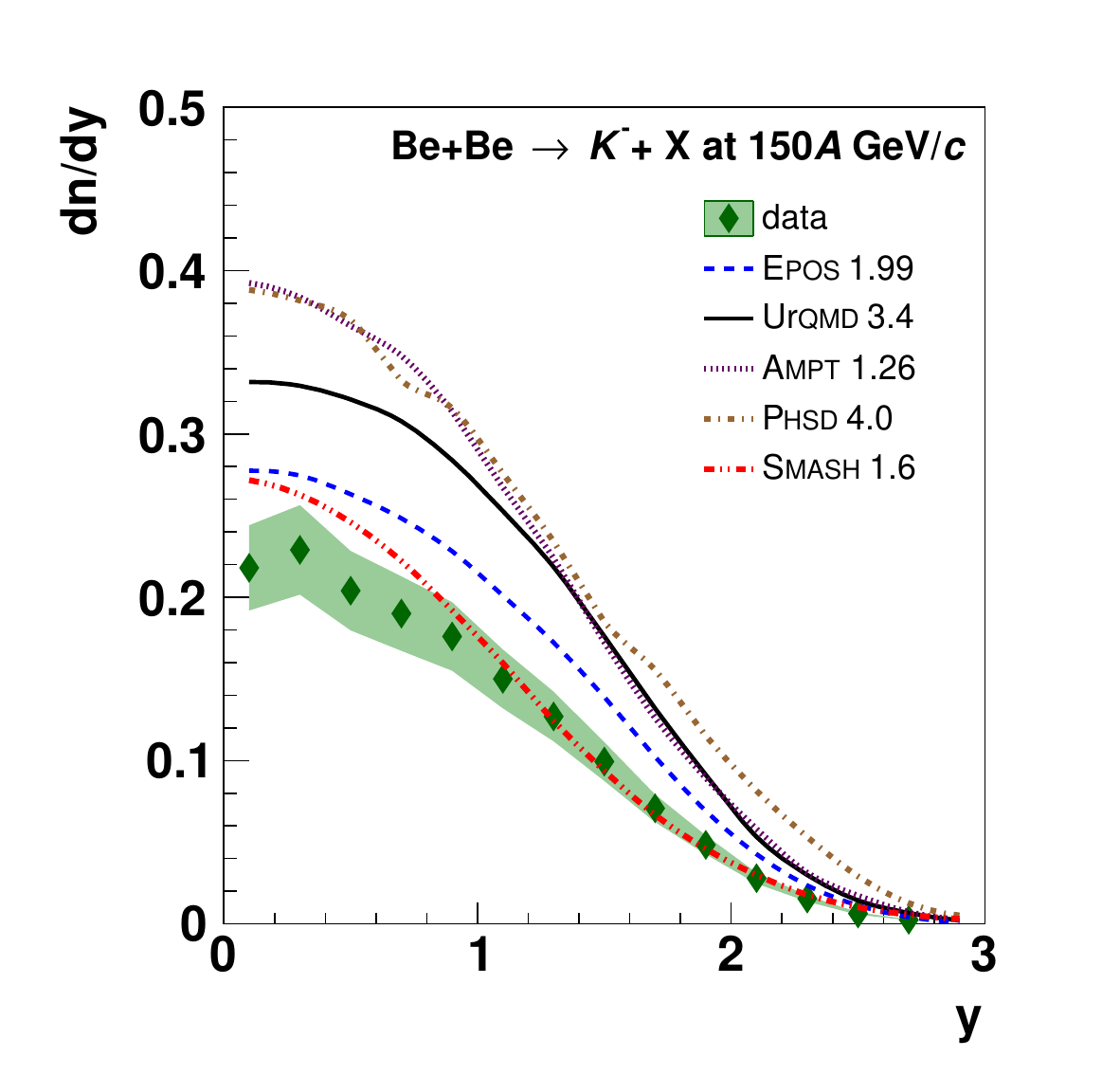}
    \includegraphics[width=0.45\textwidth]{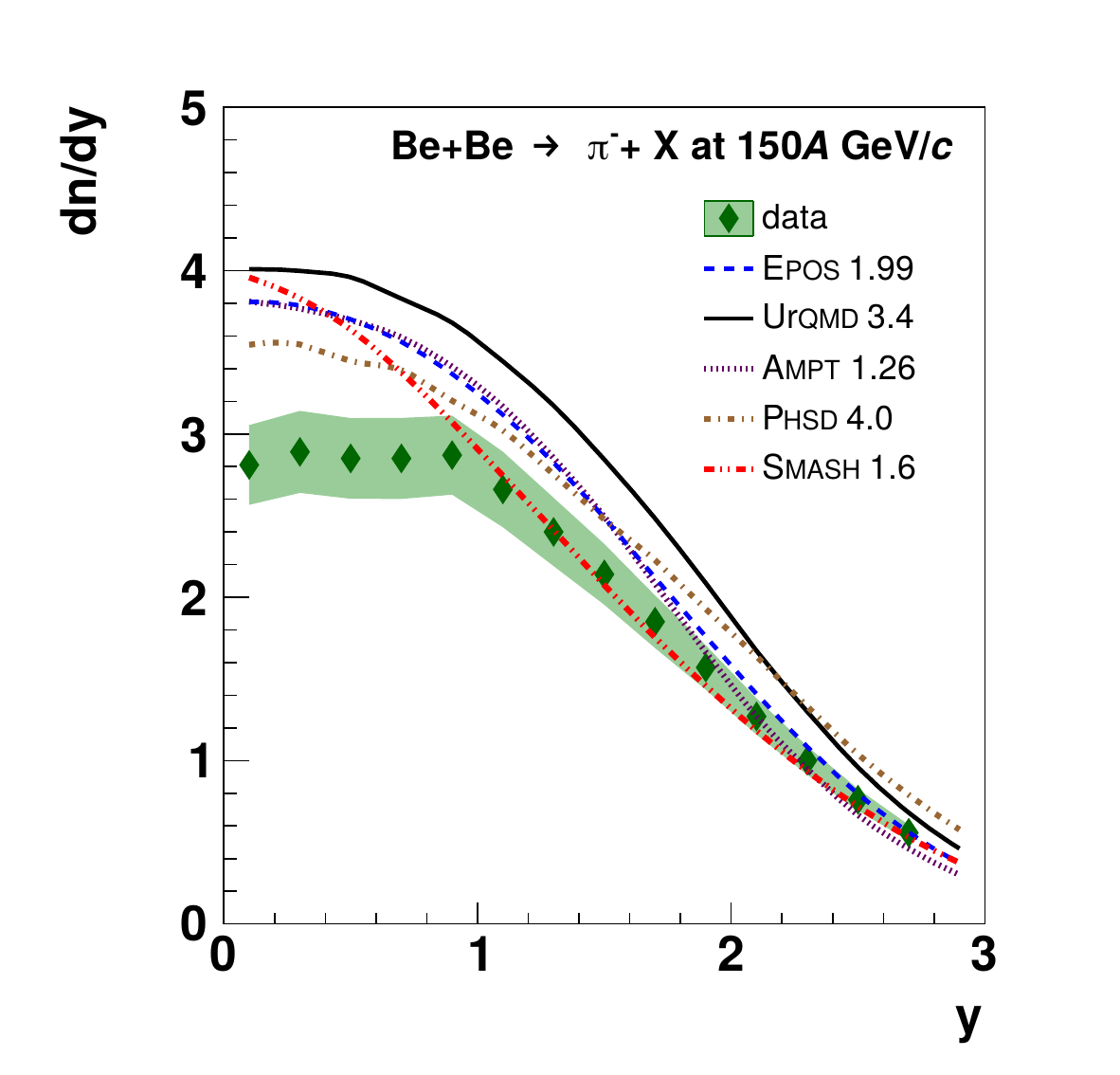}
    \caption{Comparison of the $K^+$ (\textit{left}), $K^-$ (\textit{right}) 
    and $\pi^-$ (\textit{bottom}) rapidity spectra for the 20\% most \textit{central} Be+Be collisions 
    at 150\AGeVc with models: \Epos 1.99 (blue dashed line), \Urqmd 3.4 (black solid line), 
    \Ampt 1.26 (violet dotted line), \Phsd 4.0 (brown dashed-dotted line) and \Smash 1.6 (red dashed-double dotted line)
    .}
    \label{fig:y_spectra_vs_models}
\end{figure}

Comparisons of the $p_T$ spectra at midrapidity and rapidity spectra of $K^+$, $K^-$ and $\pi^-$ in 
the 20\% most \textit{central} Be+Be collisions at 150\AGeVc are shown in 
Figs.~\ref{fig:pT_spectra_vs_models} and~\ref{fig:y_spectra_vs_models} respectively.
Positively charged kaon \pt distributions are overpredicted by \Ampt, \Phsd and \Epos, whereas \Urqmd and \Smash predictions
are closer to the measurements (Fig.~\ref{fig:pT_spectra_vs_models} (\textit{top-left})). In the case of negatively charged $K$ mesons (Fig.~\ref{fig:pT_spectra_vs_models} (\textit{top-right})), \Epos next to \Smash describes \pt the distribution well, while \Urqmd, \Phsd and \Ampt overestimate the measured spectrum.
For $\pi^-$ one finds a similar behaviour (Fig.~\ref{fig:pT_spectra_vs_models} (\textit{bottom-left})). In the case of protons (Fig.~\ref{fig:pT_spectra_vs_models} (\textit{bottom-right})), again all models overestimate the \pt spectrum. Moreover, \Epos, \Smash and \Phsd predict narrower \pt spectra than measured, which result in smaller value of inverse slope parameter than the one obtained by the data fit (see Figs.~\ref{fig:Tparam} and \ref{fig:pT_spectra_vs_models}).

Predictions for the rapidity distributions are too high for all models (Figs.~\ref{fig:y_spectra_vs_models} and \ref{fig:protonmodel}).

\begin{figure}[!ht]
    \centering
    \includegraphics[width=0.45\textwidth]{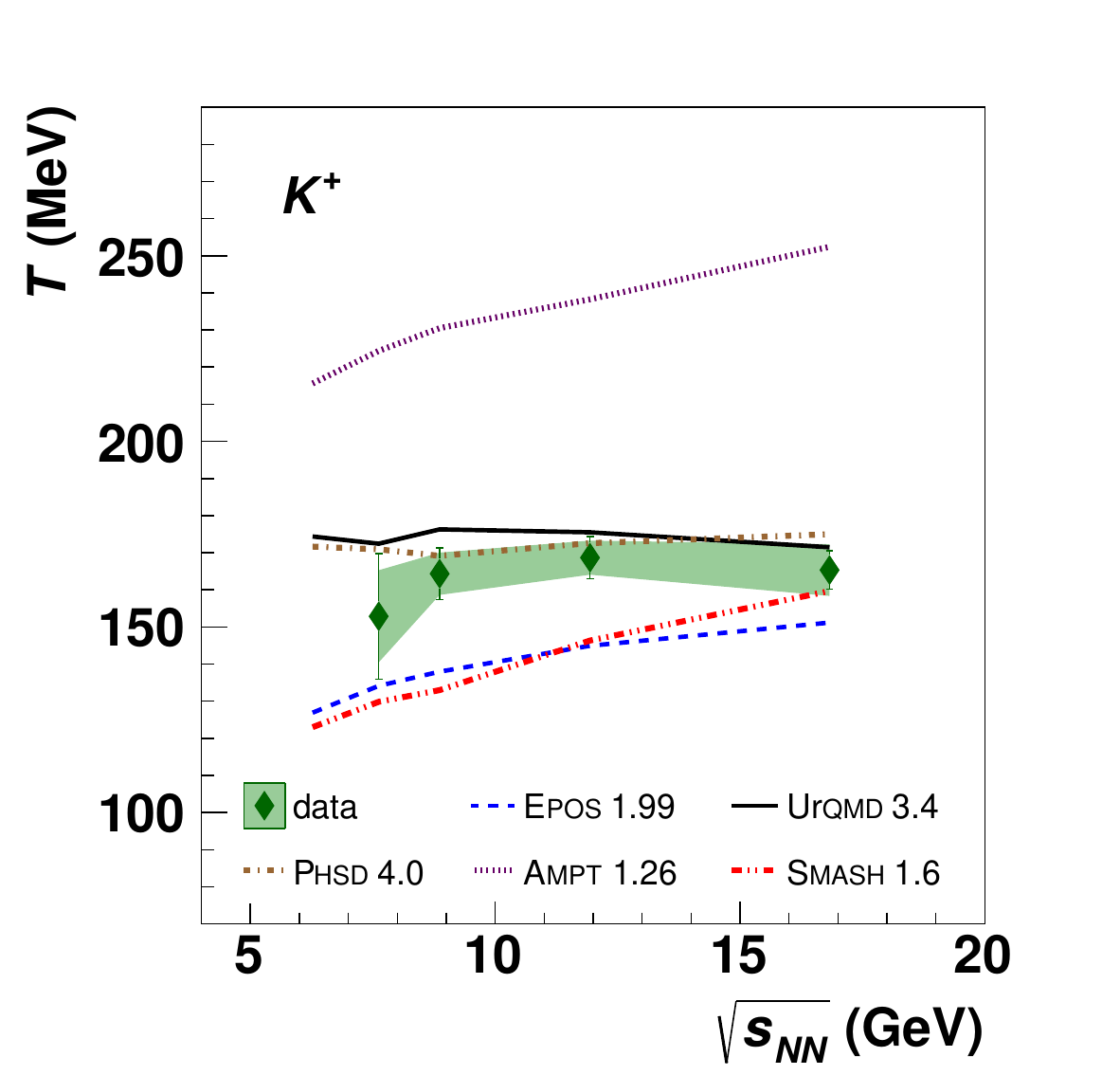}
    \includegraphics[width=0.45\textwidth]{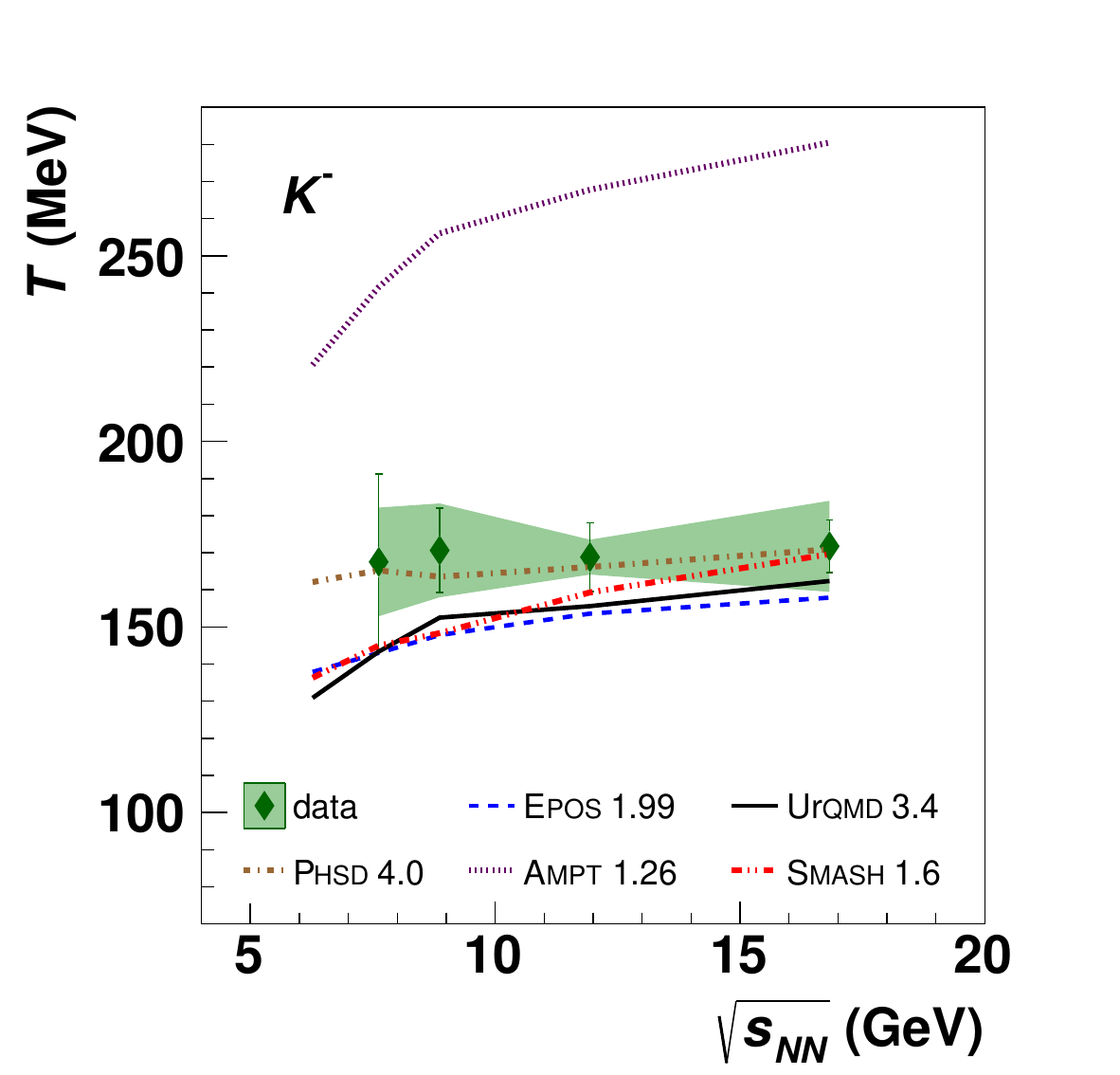}
    \includegraphics[width=0.45\textwidth]{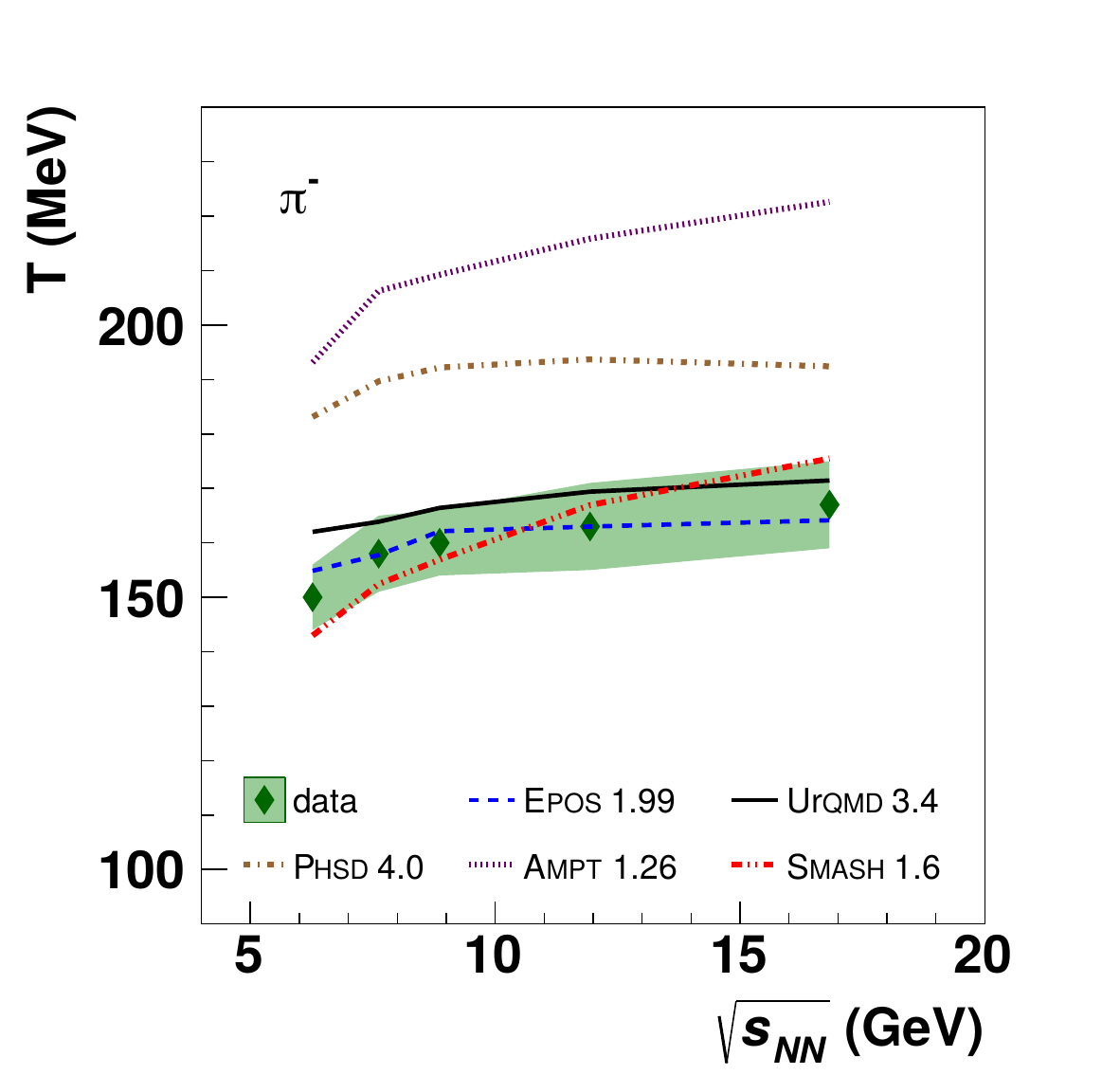}
    \includegraphics[width=0.45\textwidth]{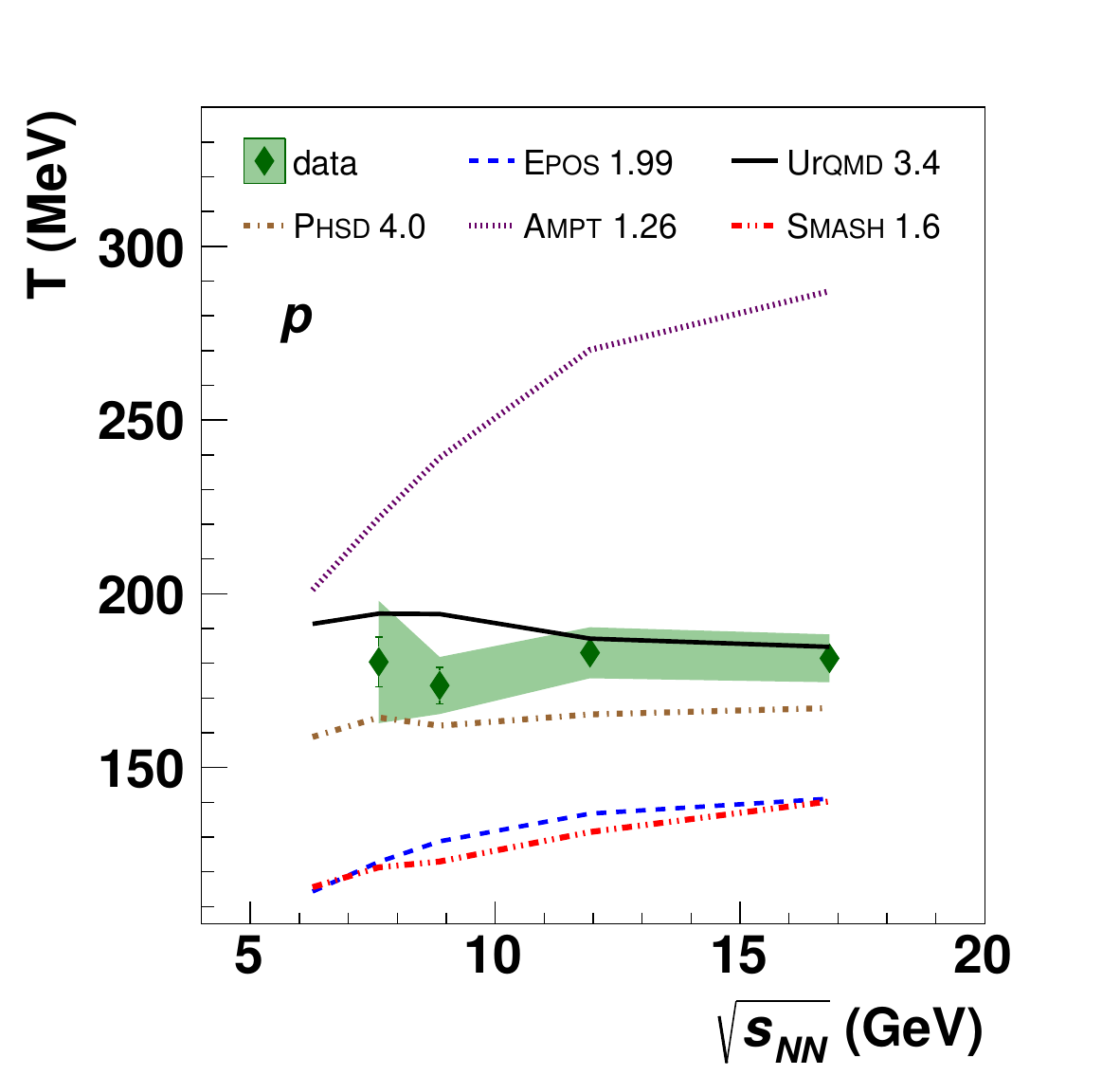}
    \caption{Comparison of the energy dependence of the inverse slope parameter $T$ of $K^+$ (\textit{top-left}), 
    $K^-$ (\textit{top-right}), $\pi^-$ (\textit{bottom-left}) and $p$ (\textit{bottom-right}) spectra at mid-rapidity for the 20\% 
    most \textit{central} Be+Be collisions with models: \Epos 1.99 (blue dashed line), \Urqmd 3.4 (black solid line), 
    \Ampt 1.26 (violet dotted line), \Phsd 4.0 (brown dashed-dotted line) and \Smash 1.6 (red dashed-double dotted line)
    .}
    \label{fig:KP_mid_T_vs_models}
\end{figure}

The measurements of the inverse slope parameter $T$ in midrapidity in the 20\% most \textit{central} Be+Be collisions are shown versus collision energy for $K^+$ and $K^-$ in Fig.~\ref{fig:KP_mid_T_vs_models} (\textit{top})
together with model predictions. Except for \Ampt the predictions cluster around the measurements. 
\Urqmd and \Phsd feature a hadron rescattering phase as the last step of the system's evolution. 
The values of $T$ for $\pi^-$ shown in Fig.~\ref{fig:KP_mid_T_vs_models} (\textit{bottom-left}) are well 
reproduced by \Epos, \Urqmd and \Smash whereas the predictions of \Ampt and \Phsd are too high by 
\mbox{30 - 50 \MeV}. The limited range of studied beam momentum does not allow a definite conclusion on 
a possible \textit{step} structure. \Urqmd predictions for inverse slope parameter of protons reproduce the experimental data quite well. \Ampt model significantly overestimates $T$ value for $p$, while \Phsd, \Epos and \Smash present opposite tendency. Comparison of $T$ as a function of $y$ with \Epos predictions is presented in Fig.~\ref{fig:Tparam} for $K^+$, $K^-$, $\pi^+$, $\pi^-$, $p$ and $\overline{p}$. In general, \Epos underestimates $T$ for all types of identified hadrons except pions, for which simulation follows the experimental results.

\begin{figure}[!ht]
    \centering
    \includegraphics[width=0.45\textwidth]{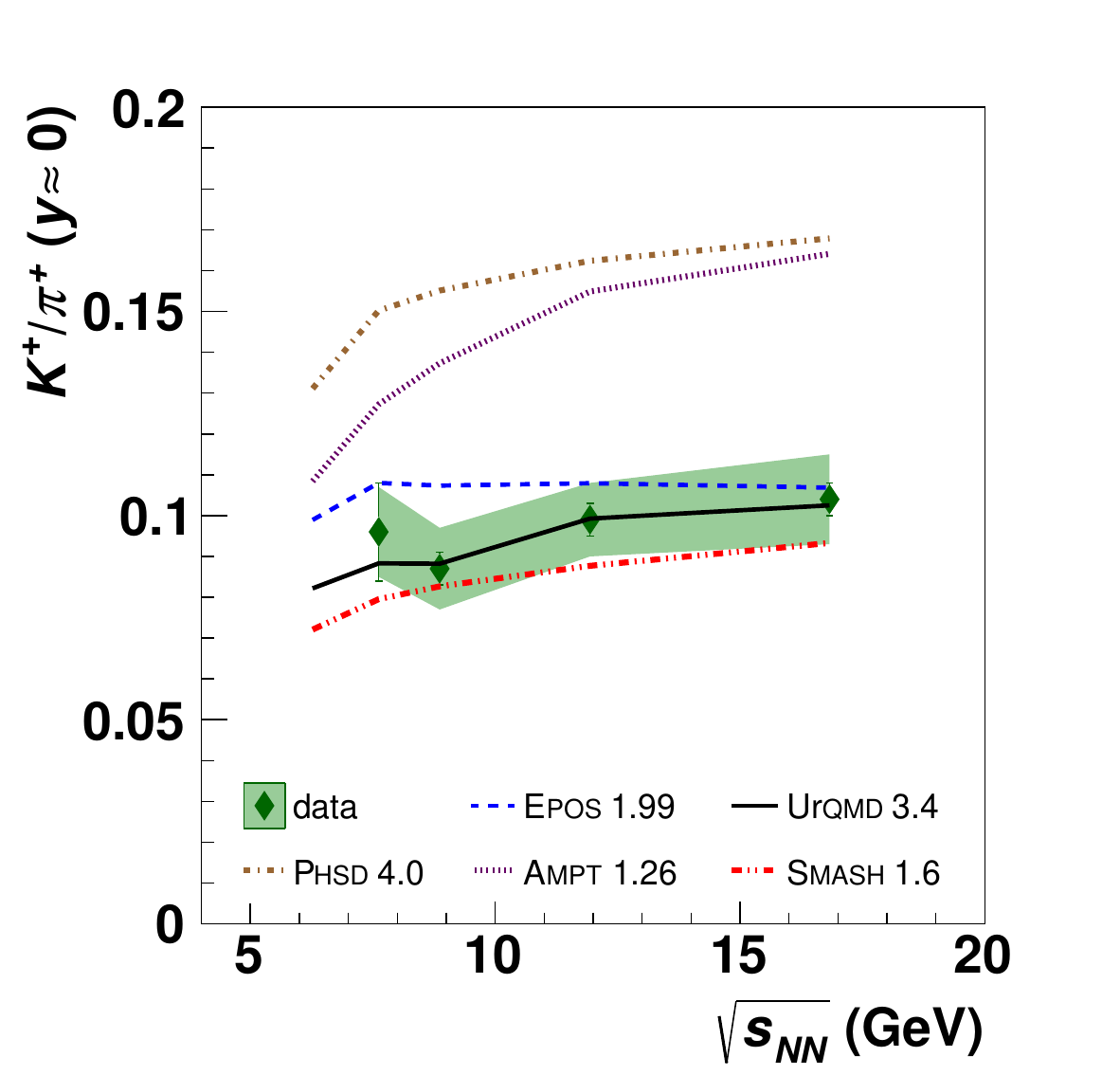}
    \includegraphics[width=0.45\textwidth]{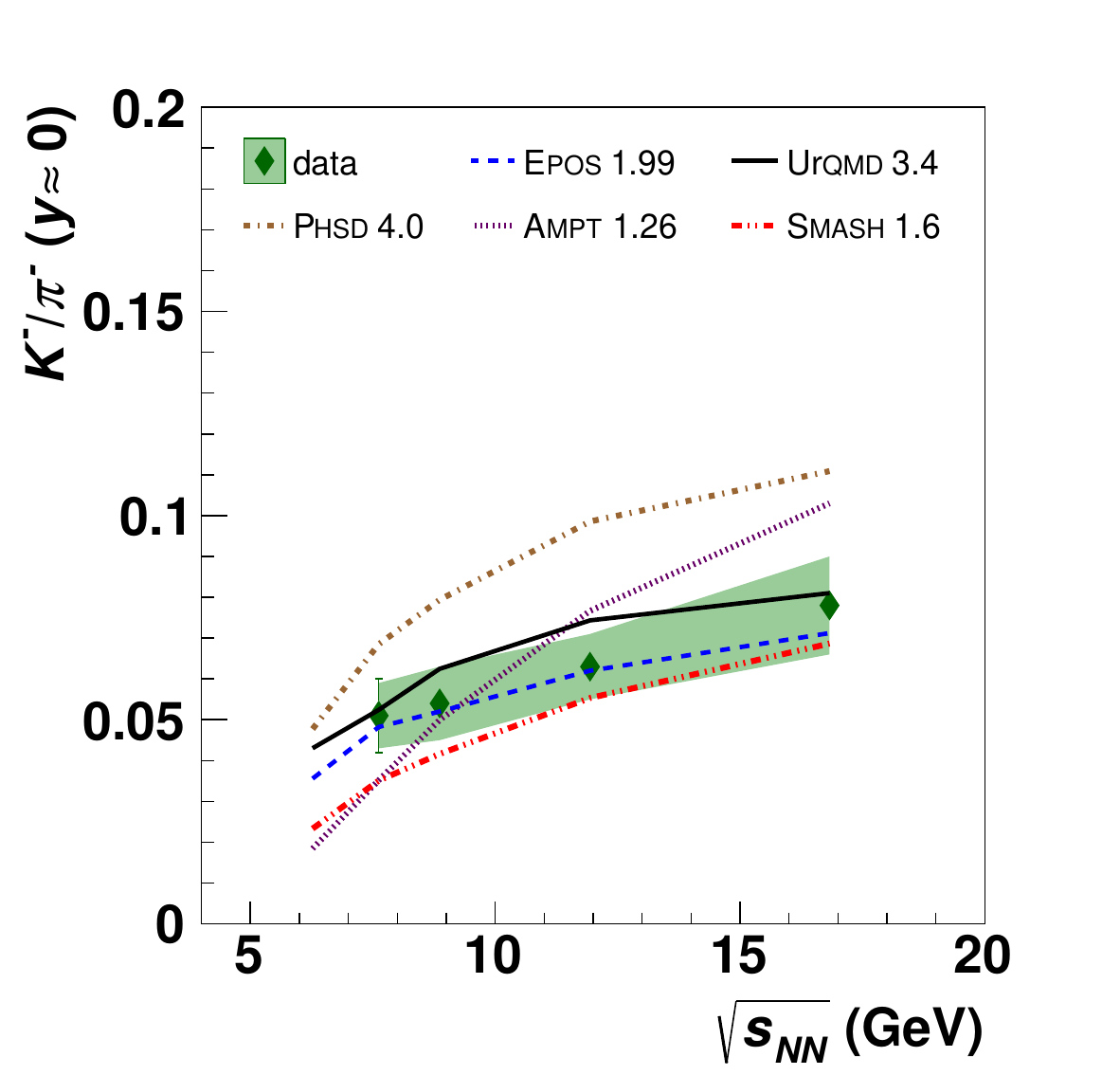}
        \caption{Comparison of the energy dependence of $K^+$/$\pi^+$ (\textit{left}) and $K^-$/$\pi^-$ (\textit{right}) yields ratio at mid-rapidity for the 20\% most \textit{central} Be+Be collisions with models: \Epos 1.99 (blue dashed line), \Urqmd 3.4 (black solid line), \Ampt 1.26 (violet dotted line), \Phsd 4.0 (brown dashed-dotted line) and \Smash 1.6 (red dashed-double dotted line).}
    \label{fig:KP_mid_Kpi_vs_models}
\end{figure}

\begin{figure}[!ht]
    \centering
    \includegraphics[width=0.45\textwidth]{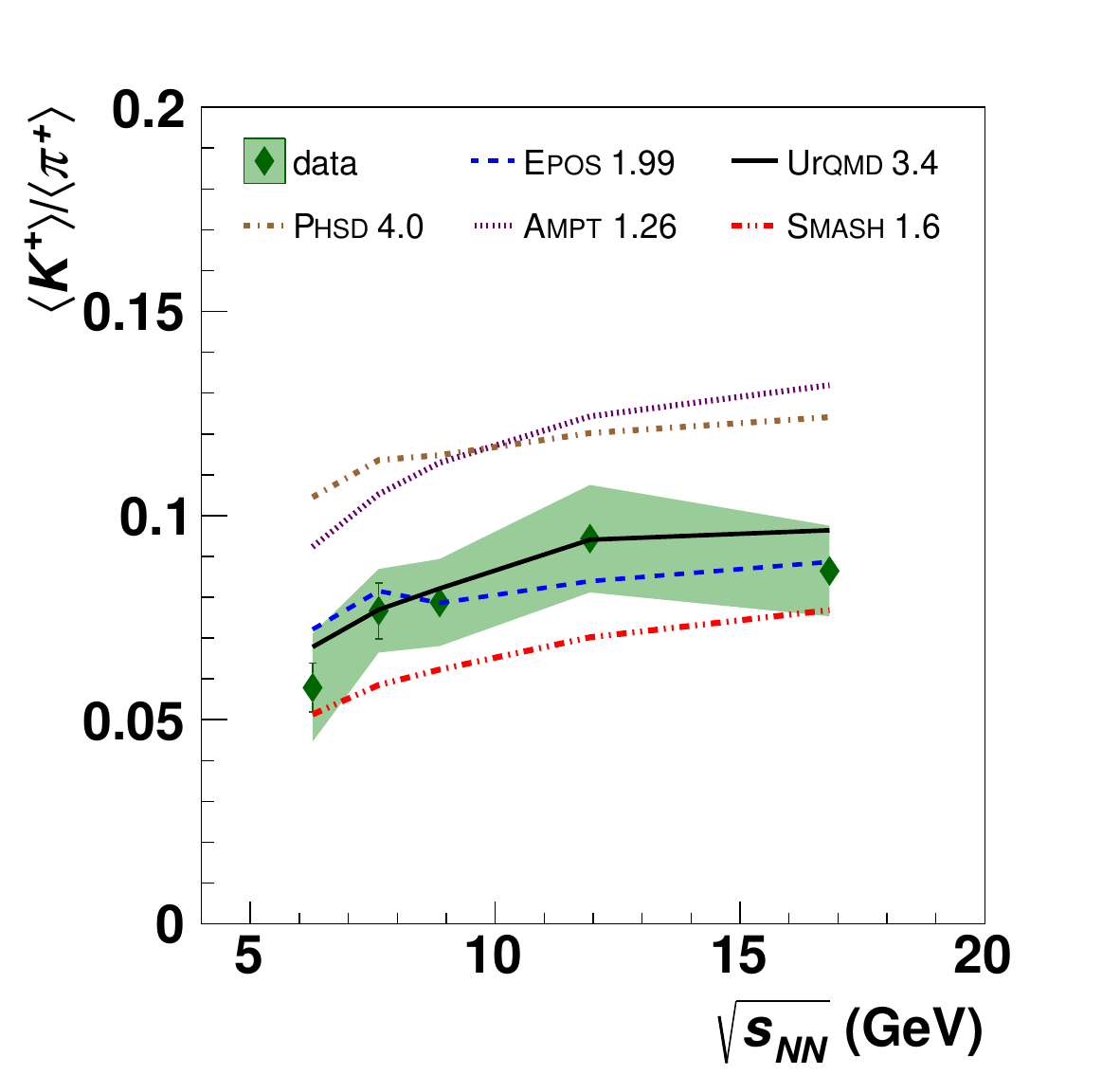}
    \includegraphics[width=0.45\textwidth]{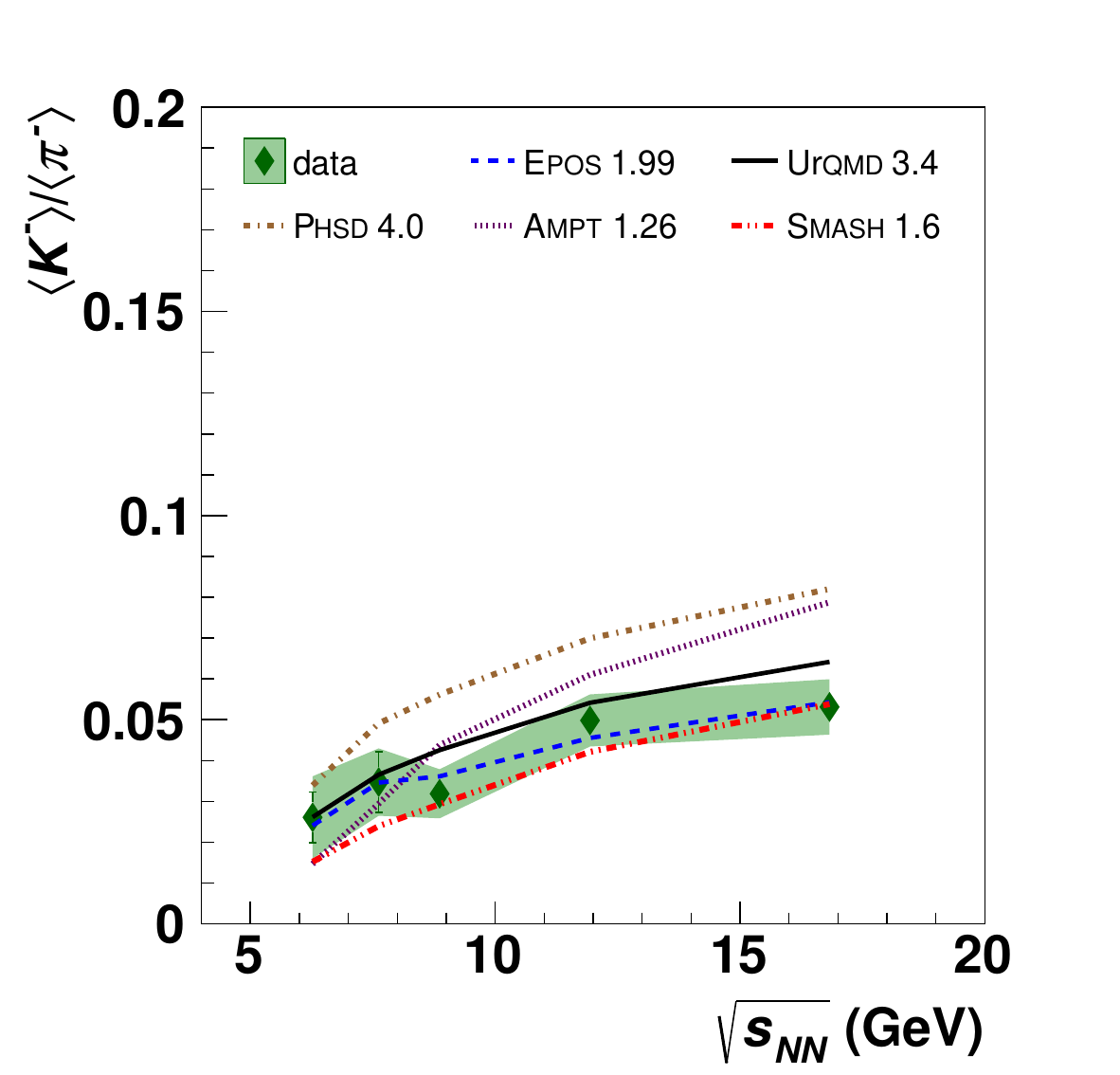}
        \caption{Comparison of the energy dependence of $K^+$/$\pi^+$ (\textit{left}) and $K^-$/$\pi^-$ (\textit{right}) mean multiplicities ratio for the 20\% most \textit{central} Be+Be collisions with models: \Epos 1.99 (blue dashed line), \Urqmd 3.4 (black solid line), \Ampt 1.26 (violet dotted line), \Phsd 4.0 (brown dashed-dotted line) and \Smash 1.6 (red dashed-double dotted line).}
    \label{fig:KP_yield_Kpi_vs_models}
\end{figure}

Finally, the energy dependence of the ratios of kaon to pion yields are compared to model predictions. 
Figure~\ref{fig:KP_mid_Kpi_vs_models} shows the mid-rapidity results for $K^+$/$\pi^+$ (left)
and $K^-$/$\pi^-$ (right), respectively. Figure~\ref{fig:KP_yield_Kpi_vs_models} displays the ratios of
total mean multiplicities. Unlike particle yields, particle ratios are not sensitive
to the details of the event selection assuming that the shapes of the spectra do not change significantly 
in the studied centrality range. For $K^+$/$\pi^+$ \Epos, \Urqmd and \Smash provide a good description
whereas \Ampt and \Phsd overpredict the data. For $K^-$/$\pi^-$ all models are close to the 
experimental results.

A detailed investigation of the effects included in the models and their impact on the predictions of the 
experimental results is beyond the scope of this paper. However, the new \NASixtyOne measurements
provide useful input for future refinements of the models.
\FloatBarrier
\section{Summary and conclusions}

This paper reports measurements by the \NASixtyOne experiment at the CERN SPS of spectra 
and mean multiplicities of $\pi^\pm, K^\pm, p$ and $\bar{p}$  produced in 
the 20~\% most \textit{central} $^7$Be+$^9$Be collisions at beam momenta of 19$A$, 30$A$, 40$A$, 75$A$ 
and 150\AGeVc.  This is the lightest nucleus-nucleus system investigated
in the system size scan of \NASixtyOne. In this program data were also recorded 
from Ar+Sc, Xe+La and Pb+Pb collisions for which the analysis is ongoing.
Results on \textit{central} $^7$Be+$^9$Be collisions were found to be similar to those from 
inelastic \textit{p+p} interactions for shapes of transverse momentum and rapidity spectra 
as well as the $K^+$/$\pi^+$ ratio. However, particle yields are higher by approximately 
a factor of four consistent with expectations from the wounded nucleon model. Summarising, neither measurements nor models show indications of a horn structure at low SPS energy
for small collision systems in contrast to the results from \textit{central} Pb+Pb interactions, for
detailed discussion on possible indications of onset of deconfinement in small systems see Ref.~\cite{Aduszkiewicz:2019zsv}.

The results were compared with predictions of the models: \Epos~1.99, \Urqmd~3.4, \Ampt~1.26, \Phsd~4.0 and \Smash~1.6.
None of the models reproduces all features of the presented results.

\clearpage
\section*{Acknowledgments}
We would like to thank the CERN EP, BE, HSE and EN Departments for the
strong support of NA61/SHINE.

This work was supported by
the Hungarian Scientific Research Fund (grant NKFIH 123842\slash123959),
the Polish Ministry of Science
and Higher Education (grants 667\slash N-CERN\slash2010\slash0,
NN\,202\,48\,4339, NN\,202\,23\,1837 and DIR\slash WK\slash 2016\slash 2017\slash 10-1), the National Science Centre Poland (grants~2014\slash14\slash E\slash ST2\slash00018, 2014\slash15\slash B\slash ST2 \slash\- 02537 and
2015\slash18\slash M\slash ST2\slash00125, 2015\slash 19\slash N\slash ST2 \slash01689, 2016\slash23\slash B\slash ST2\slash00692,
2017\slash\- 25\slash N\slash\- ST2\slash\- 02575,
2018\slash 30\slash A\slash ST2\slash 00226,
2018\slash 31\slash G\slash ST2\slash 03910),
the Russian Science Foundation, grant 16-12-10176 and 17-72-20045,
the Russian Academy of Science and the
Russian Foundation for Basic Research (grants 08-02-00018, 09-02-00664
and 12-02-91503-CERN),
the Russian Foundation for Basic Research (RFBR) funding within the research project no. 18-02-40086,
the National Research Nuclear University MEPhI in the framework of the Russian Academic Excellence Project (contract No.\ 02.a03.21.0005, 27.08.2013),
the Ministry of Science and Higher Education of the Russian Federation, Project "Fundamental properties of elementary particles and cosmology" No 0723-2020-0041,
the European Union's Horizon 2020 research and innovation programme under grant agreement No. 871072,
the Ministry of Education, Culture, Sports,
Science and Tech\-no\-lo\-gy, Japan, Grant-in-Aid for Sci\-en\-ti\-fic
Research (grants 18071005, 19034011, 19740162, 20740160 and 20039012),
the German Research Foundation (grant GA\,1480/8-1), the
Bulgarian Nuclear Regulatory Agency and the Joint Institute for
Nuclear Research, Dubna (bilateral contract No. 4799-1-18\slash 20),
Bulgarian National Science Fund (grant DN08/11), Ministry of Education
and Science of the Republic of Serbia (grant OI171002), Swiss
Nationalfonds Foundation (grant 200020\-117913/1), ETH Research Grant
TH-01\,07-3 and the Fermi National Accelerator Laboratory (Fermilab), a U.S. Department of Energy, Office of Science, HEP User Facility managed by Fermi Research Alliance, LLC (FRA), acting under Contract No. DE-AC02-07CH11359 and the IN2P3-CNRS (France).

\clearpage

\bibliographystyle{include/na61Utphys}
{\footnotesize\raggedright
\bibliography{include/na61References}
}

\end{document}